\documentclass[fleqn,usenatbib]{mnras}

\usepackage{newtxtext,newtxmath}
\usepackage[T1]{fontenc}
\usepackage{ae,aecompl,times}

\usepackage{graphicx}	% Including figure files
\usepackage{amsmath}	% Advanced maths commands
\usepackage{amssymb}	% Extra maths symbols
\usepackage{multirow}
\usepackage[normalem]{ulem}
\usepackage{xcolor}
\usepackage{bm}
\newcommand{\Mpch}{\,h^{-1}{\rm Mpc}}
\newcommand{\Mpchc}{\,h^{-3}{\rm Mpc}^3}
\newcommand{\hMpc}{\,h\,{\rm Mpc}^{-1}}
\newcommand{\hMpcc}{\,h^3{\rm Mpc}^{-3}}
\newcommand{\Msh}{\,h^{-1}M_\odot}
\newcommand{\Pmill}{{\sc Pmill}}
\newcommand{\glam}{{\sc glam}}
\newcommand{\Galform}{{\sc Galform}}
\newcommand{\SU}{{Skies \& Universes}}
\newcommand{\lla}{\left\langle}
\newcommand{\rra}{\right\rangle}

\definecolor{mred}{rgb}{0.058, 0.588, 0.778}

\title[Properties and clustering of DESI-like LRGs]{Building a digital twin of a luminous red galaxy spectroscopic survey: galaxy properties and clustering covariance}
\author[C.~Hern\'andez-Aguayo et al.]
{
{C\'esar Hern\'andez-Aguayo$^{1}$\thanks{E-mail: 
cesar.hernandez-aguayo@durham.ac.uk (CH-A)}, 
Francisco Prada$^{2}$\thanks{E-mail: f.prada@csic.es (FP)}, 
Carlton M. Baugh$^{1}$, 
Anatoly Klypin$^{3}$}
\\
\\
$^{1}$Institute for Computational Cosmology, Department of Physics, Durham University, South Road, Durham, DH1 3LE, UK.\\
$^{2}$Instituto de Astrof\'isica de Andaluc\'ia (CSIC), Glorieta de la Astronom\'ia, E-18080 Granada, Spain.\\
$^{3}$Astronomy Department, New Mexico State University, Las Cruces, 88001, New Mexico, USA.
}

\date{Accepted XXX. Received YYY; in original form ZZZ}

\pubyear{2020}

\begin{document}
\label{firstpage}
\pagerange{\pageref{firstpage}--\pageref{lastpage}}
\maketitle

\begin{abstract}
Luminous red galaxies (LRGs) are one of the key tracers of the large-scale structure of the Universe used by galaxy surveys. Hence, it is important to make accurate predictions for their properties and clustering, including the errors on these statistics. Here, we describe a novel technique which uses the semi-analytical model of galaxy formation {\sc Galform}, embedded in the high-resolution $N$-body Planck-Millennium simulation, to populate a thousand halo catalogues generated using the Parallel-PM $N$-body {\sc glam} code. Our hybrid scheme allows us to make clustering predictions on scales that cannot be modelled in the original $N$-body simulation. LRGs are selected in the redshift range $z=0.6-1$ from the {\sc Galform} output using similar  colour-magnitude cuts in the $r$, $z$ and $W1$ bands to those that will be applied in the Dark Energy Spectroscopic Instrument (DESI) survey. We find that the LRG-halo connection is non-trivial, leading to the prediction of a non-standard halo occupation distribution; in particular, the occupation of central galaxies does not reach unity for the most massive haloes, and drops with increasing mass. The {\sc glam} catalogues reproduce the abundance and clustering of the LRGs predicted by {\sc Galform}, and show  good agreement with recent measurements of the clustering of DESI-like LRGs using photometric redshifts. We use the \glam{} mocks to compute the covariance matrices for the two-point correlation function and power spectrum of the LRGs and their background dark matter density field, revealing important differences. We also make predictions for the linear-growth rate and the baryon acoustic oscillations distances at $z=0.6$, $0.74$ and $0.93$. All DESI-like LRG catalogues are made publicly available.
\end{abstract}


%%%%%%%%%%%%%%%%%%%%%%%%%%%%%%%%%%%%%%%%%%%%%%%%%%

%%%%%%%%%%%%%%%%% BODY OF PAPER %%%%%%%%%%%%%%%%%%
%Suggested colour scheme for comments and corrections: \Paco{Paco}, \Carlton{Carlton}, \Anatoly{Anatoly}

%---------------------------------------------------------------
\section{Introduction}
\label{sec:intro}
%---------------------------------------------------------------
Luminous red galaxies (LRGs) have played an important role in the study of the large-scale structure of the Universe. As expected from their bright intrinsic luminosity and large stellar masses, LRGs display a strong clustering signal that make them an ideal tracer of the large-scale structure of the Universe \citep{Zehavi:2005}. LRGs were used to extract the scale of the baryon acoustic oscillations (BAO) in the local large-scale structure from the Sloan Digital Sky Survey (SDSS) redshift-space correlation function \citep{Eisenstein:2005su}. 
LRGs have also been used to study the impact of redshift-space distortions (RSDs) on their small and large scale clustering \citep[see e.g.][]{Zehavi:2005,Cabre:2008sz,Cabre:2008ta,Wake:2008,Crocce:2011lrgs,Samushia:2011cs}. Additionally, the large-scale clustering of LRGs has also been used to constrain the cosmological parameters \citep{Eisenstein:2005su,Tegmark:2006az,Sanchez:2009jq,Troster:2019ean}, and to test modified gravity models \citep[see e.g.,][]{Barreira:2016ovx,Hernandez-Aguayo:2018oxg}.

LRGs are the main targets of the SDSS-III Baryon Oscillation Spectroscopic Survey (BOSS) \citep{Dawson:2013}, in the redshift range $0.2 < z < 0.75$. This survey has provided the most precise measurements to date of cosmological distances using BAO and the growth rate using RSDs at effective redshifts $z=0.38$, $0.51$, and $0.61$ \citep[see][and references therein]{BOSS-DR12:2017sqa}. Recently, the SDSS-IV extended-BOSS survey (eBOSS) \citep{Dawson:2016,Prakash:2015trd} has presented the first clustering measurements of LRGs at $z\sim 0.7$ \citep{Zhai:2016gyu,Bautista:2017wwp,Miguel:2020}.

The Dark Energy Spectroscopic Instrument\footnote{https://www.desi.lbl.gov} (DESI) survey aims to measure BAO scales and the growth of structure through RSDs at an unprecedented level of precision \citep{DESI:2016zmz}. This imminent survey will target luminous red galaxies in the redshift range from $z=0.4$ to $z=1$, [OII] emission-line galaxies (ELGs) in the range $0.6 < z < 1.6$, QSOs (tracers) up to $z=2.1$, and QSOs (Ly-$\alpha$) at higher redshifts $(2.1 < z < 3.5)$. In addition to a bright galaxy sample at low redshifts $z < 0.4$, DESI will provide a total of $\sim 35$ million biased tracers of the large-scale structure of the Universe over $14\,000\,{\rm deg}^2$ \citep[see][for details]{Kitanidis:2019rzi}. The LRG target selection at $z<0.6$ will be complementary to that performed in the SDSS-IV/eBOSS \citep{Prakash:2015trd}; hence we will focus here on the DESI LRGs at $z \geq 0.6$.

Our aim here is to provide a qualitative study of the properties and clustering of LRGs which meet the selection requirements of a real survey such as DESI. We select DESI-like LRGs from the output of the semi-analytic model (SAM) of galaxy formation \Galform{} \citep{Cole:2000ex} run on the Planck-Millennium $N$-body simulation \citep{Baugh:2018kkh}, and provide estimates of the large-scale galaxy clustering using the \glam{} code, which allows us to generate a substantial number of large galaxy mock catalogues \citep{Klypin:2017jwl}.
This hybrid approach takes the SAM calculations made using a high-resolution, moderate volume $N$-body simulation and uses the results to populate a large number ($\mathcal{O}(10^3)$) of larger volume low-resolution simulations run with \glam{}. This allows us to make predictions for the large-scale clustering of LRGs on scales, such as the BAO scale, that were inaccessible in the simulation used to run the SAM. Furthermore, by being able to generate a large number of independent realisations of the density field at relatively low computational cost, we can estimate the covariance on two-point statistics of the large-scale structure.

The use of SAMs to study the properties and clustering of LRGs is not new. \citet{Almeida:2006mv,Almeida:2007ef} presented predictions for the abundance, structural and photometric properties of LRGs using two earlier versions of \Galform{} \citep{Baugh:2005,Bower:2005vb}. The authors found that their predictions were in good agreement with different observations from the SDSS \citep{Bernardi:2003rb,Bernardi:2004zx,Wake:2006qf}. More recently,  \citet{Stoppacher:2019ssr} used the {\sc Galacticus} SAM \citep{Benson:2010kx} run on the MultiDark Planck 2 simulation \citep{Klypin:2014kpa} to study the galaxy-halo connection and clustering of the BOSS-CMASS DR12 sample \citep{Alam:2015mbd}, finding good agreement between predictions and observations.

Recently, \citet{Rongpu:2020} presented small-scale $(r \lesssim 20\Mpch)$ clustering measurements of DESI-like LRGs selected from the DESI Legacy Imaging Surveys\footnote{http://www.legacysurvey.org} \citep{Dey:2019} and fitted their results using the halo occupation distribution (HOD) framework. Since spectroscopic redshifts are not yet available for these targets, these authors estimated photometric redshifts (photo-$z$) using the Dark Energy Camera Legacy Survey (DECaLS) imaging. There are a number of differences between the work carried out by \citet{Rongpu:2020} and our paper: first, we are interested in providing a study of the impact of the DESI-LRG target selection on galaxy properties and the galaxy-halo connection using a physical model of galaxy formation, \Galform{}; and second, we focus on the large-scale galaxy clustering and in the generation of a large number of mock catalogues to provide an accurate estimate of the covariance of the clustering measurements. Both of these objectives are beyond the reach of the original simulation used to run the SAM and mark a key advantage of our hybrid approach.

In order to extract the cosmological information from our \glam{} mock catalogues for the DESI LRG tracers, it is necessary to meet the requirements of the expected error budget for DESI. Hence, it is imperative to construct covariance matrices for our clustering measurements \citep[see e.g.][and references therein]{Baumgarten:2018jcn,Blot:2018oxk,Colavincenzo:2018cgf,Lippich:2018wrx}.
Here, we make predictions of the linear-growth rate through a linear theory description of RSDs \citep{Kaiser:1987,Hamilton:1992zz}, and an isotropic analysis of the BAO scale \citep[see e.g.,][]{Anderson:2013zyy} in configuration and Fourier space using the covariance matrices constructed from our \glam{} catalogues.

The outline of the paper is as follows. In Section~\ref{sec:sims} we present the simulations used in our analysis. Section~\ref{sec:selection} describes the selection of DESI-like LRGs from \Galform{}. In Section~\ref{sec:hod} we provide a detailed study of the galaxy-halo connection of DESI-like LRGs. Our results for the galaxy clustering and covariance errors are presented in Section~\ref{sec:clustering}. Finally, in Section~\ref{sec:conc} we give our summary and conclusions.
%---------------------------------------------------------------
\section{Simulations and Galaxy formation in semi-analytical models}\label{sec:sims}
%---------------------------------------------------------------

%--------- Figure --------------
\begin{figure*}
 \centering
\includegraphics[width=0.33\textwidth]{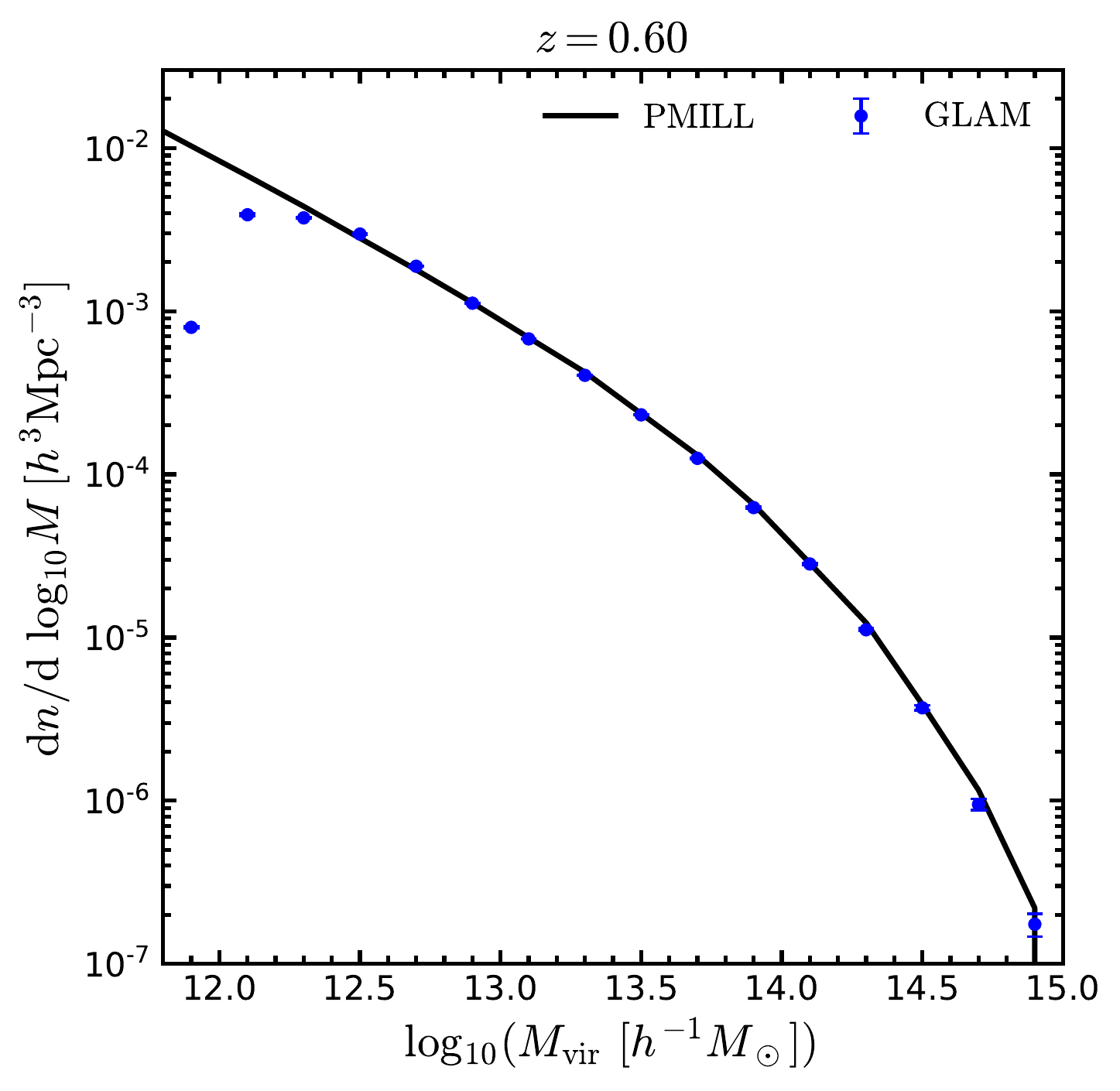}
\includegraphics[width=0.33\textwidth]{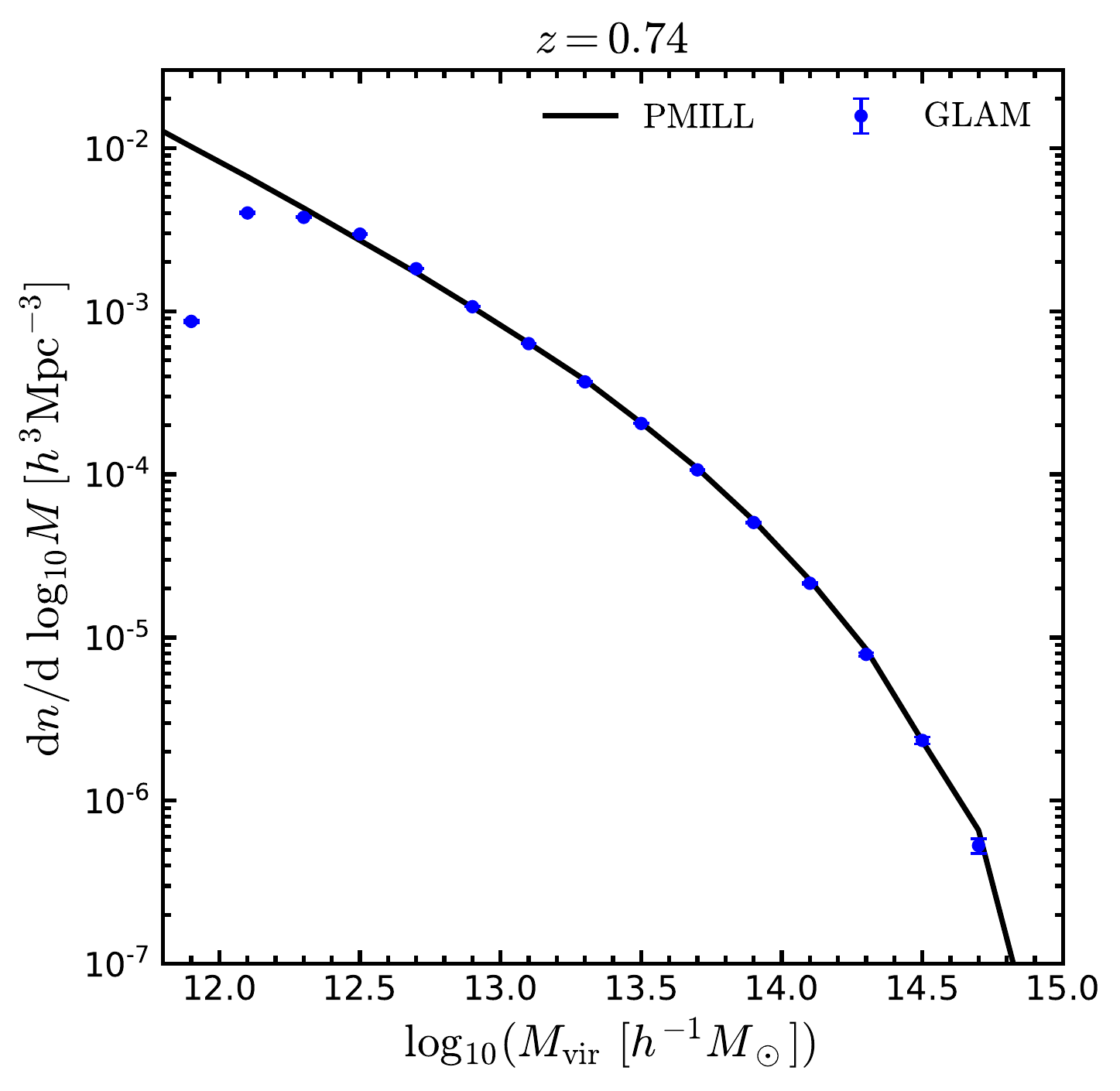}
\includegraphics[width=0.33\textwidth]{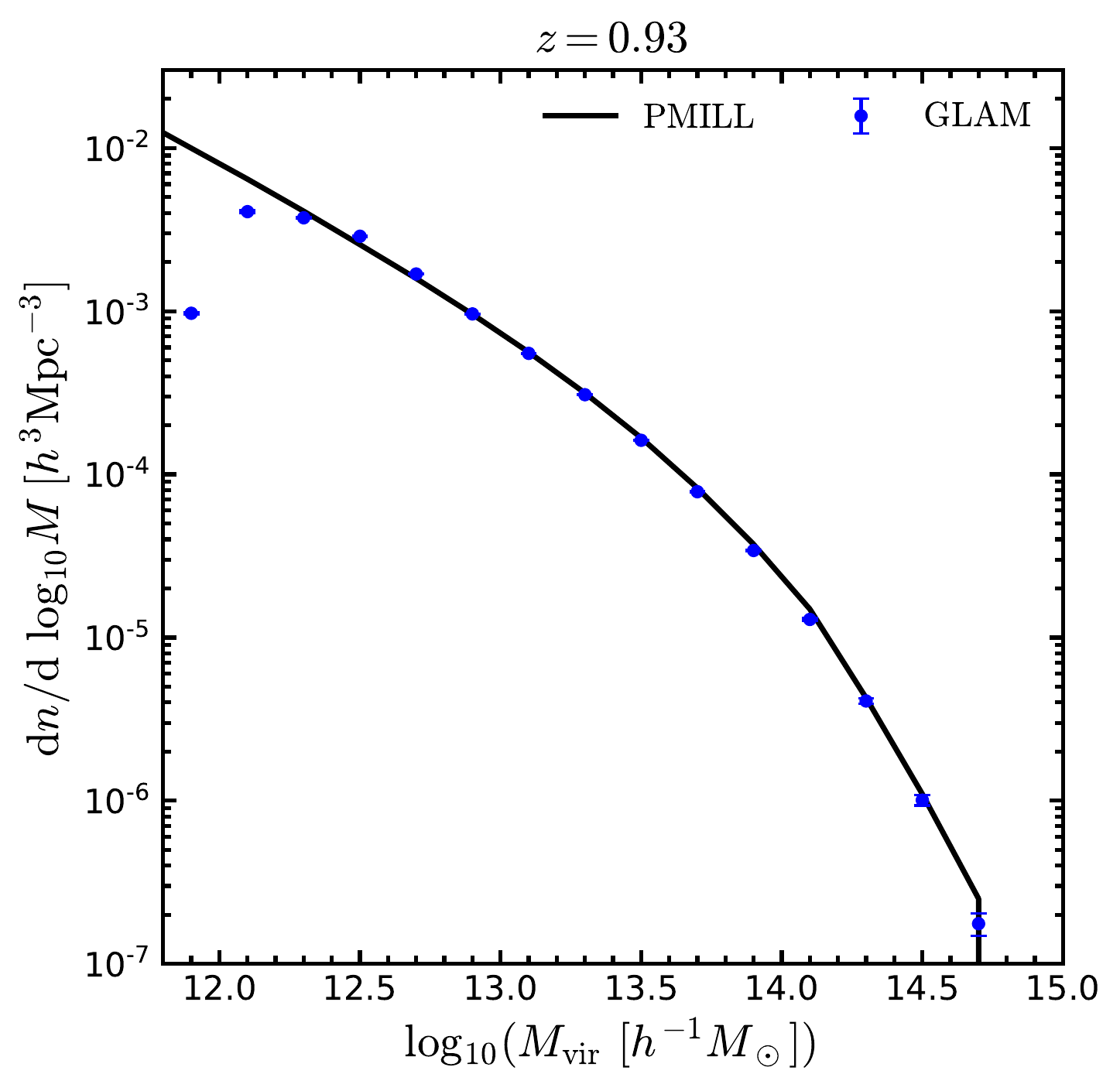}
\includegraphics[width=0.33\textwidth]{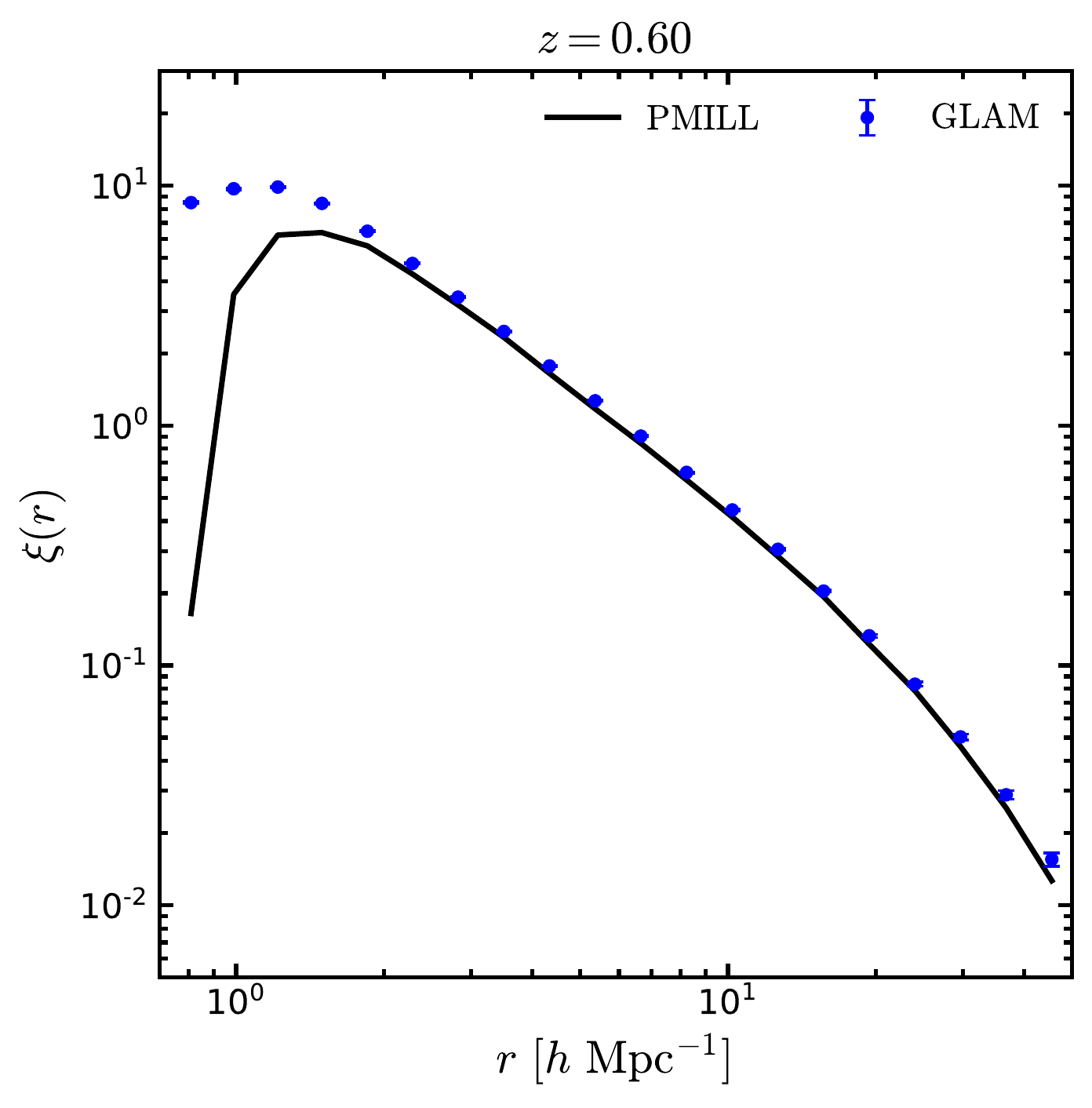}
\includegraphics[width=0.33\textwidth]{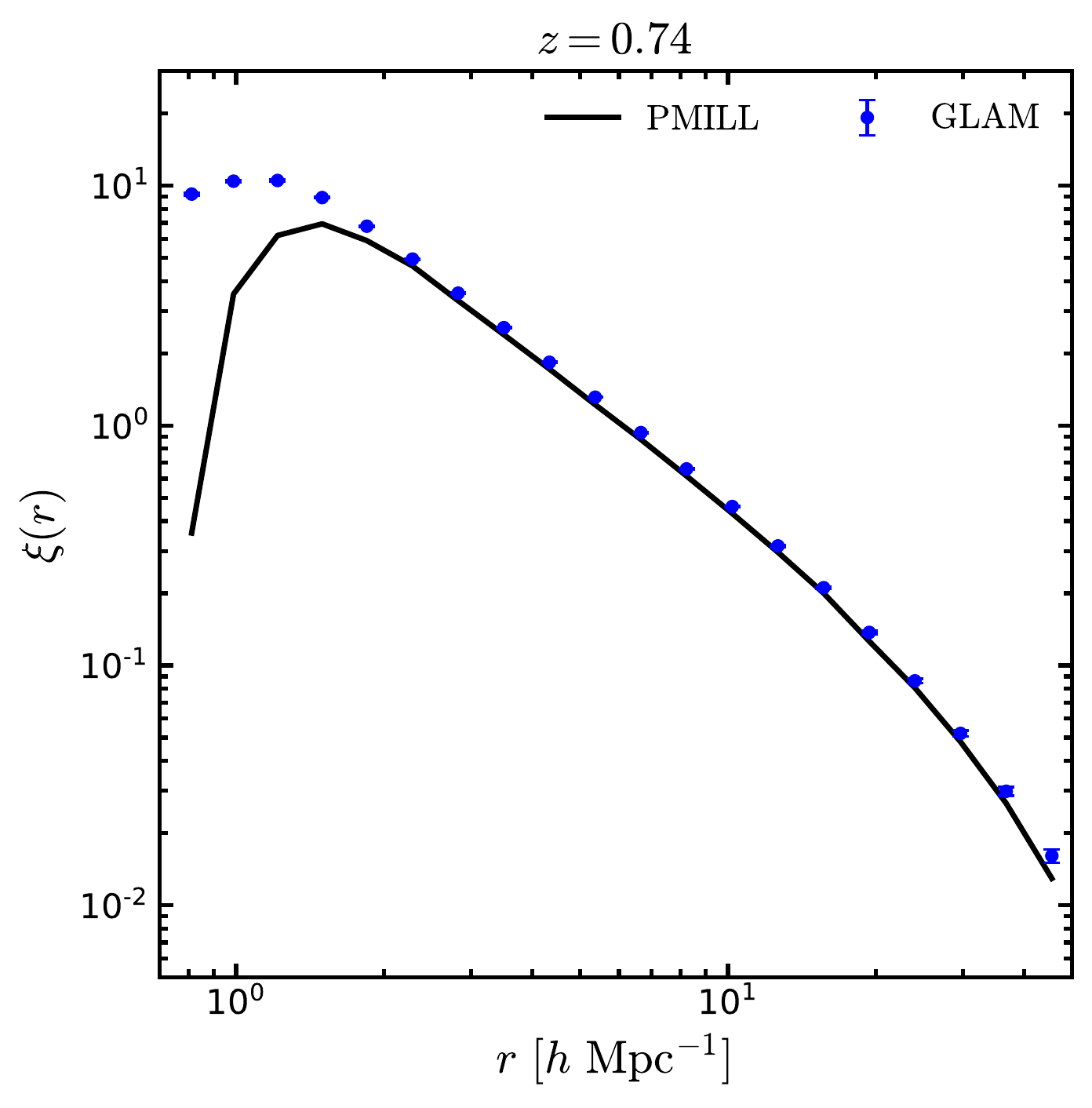}
\includegraphics[width=0.33\textwidth]{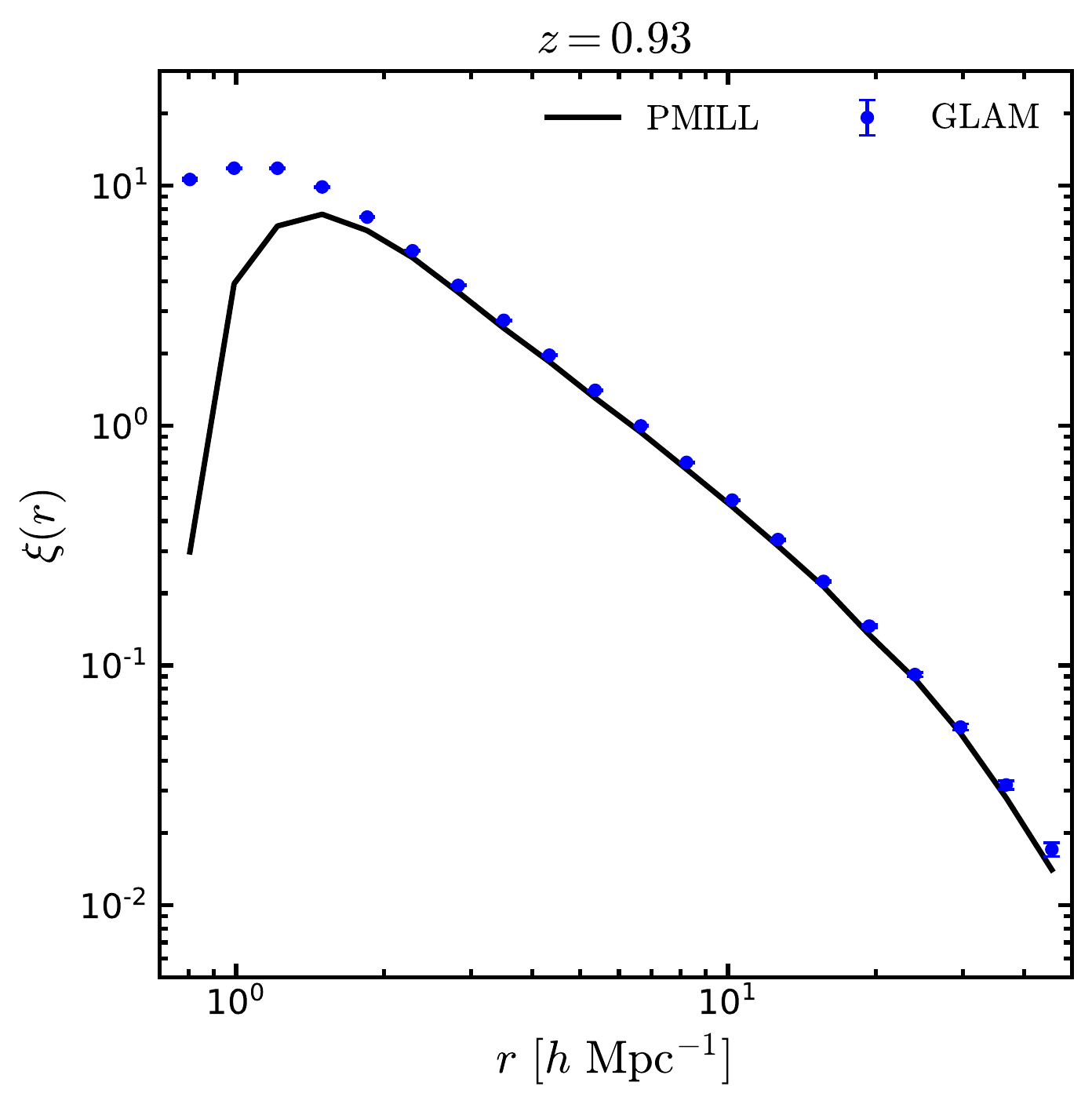}
\caption{{\it Top row}: differential halo mass function in the \Pmill{} (black solid lines) and the mean of 1000 \glam{} simulations (blue dots) as a function of $M_{\rm vir}$. {\it Bottom row}: Real-space halo two-point correlation function measured from the \Pmill{} (black solid lines) and the mean of 1000 \glam{} simulations (blue dots) for haloes with mass $M_{\rm vir} > 10^{12.5} \Msh$. We show measurements at $z=0.60$ ({\it left column}), $z=0.74$ ({\it middle column}) and $z=0.93$ ({\it right column}). Errobars correspond to the $1\sigma$ standard deviation over 1000 \glam{} realisations.}
\label{fig:hmf}
\end{figure*}

Here we introduce the Planck Millennium $N$-body simulation and the galaxy formation model (Sec.~\ref{sec:Pmill}). The {\sc GLAM} simulations are described in Section~\ref{sec:GLAM}. In Section~\ref{sec:halo} we show the halo mass function and halo clustering of our simulations.
%---------------------------------------------------------------
\subsection{Galaxy formation in the Planck Millennium simulation}\label{sec:Pmill}
%---------------------------------------------------------------
The Planck Millennium $N$-body simulation \citep[hereafter the \Pmill{} simulation;][]{Baugh:2018kkh} follows the evolution of $5040^3$ dark matter particles in a cosmological volume of $542.16^3\Mpchc$ $(800^3\rm{Mpc}^3)$. The simulation was run using a reduced memory version of the {\sc Gadget-2} $N$-body code \citep{Springel:2005mi}, employing the cosmological parameters corresponding to the 2014 results from the Planck collaboration \citep{Ade:2013zuv}:
$$\{\Omega_{\rm b}, \Omega_{\rm m}, h, n_s, \sigma_8\} = \{0.04825,0.307,0.6777,0.9611,0.8288\}.$$

The large number of dark matter particles used in the \Pmill{} simulation gives a mass resolution of $1.06\times 10^{8}\Msh$ and a halo mass limit, corresponding to 20 particles, of $2.12\times 10^9 \Msh$. The simulation starts at $z=127$, with initial conditions  generated using second-order Lagrangian perturbation theory \citep{Jenkins:2009um} and the {\sc panphasia} code \citep{Jenkins:2013raa}. The halo properties and selected particle information are saved in $271$ snapshots. Haloes and sub-haloes were identified with {\sc subfind} \citep{Springel:2000qu}. {\sc subfind} first identifies haloes  using a friend-of-friends ({\sc FoF}) algorithm with a linking length of $b=0.2$ times the mean interparticle separation. Then, these {\sc FoF} groups (main or distinct haloes) are split into subhaloes of bound particles. {\sc subfind} uses several definitions of halo mass; we use $M_{200m}$ which is the mass enclosed within a radius where the average overdensity is 200 times the mean density of the Universe. The subhalo mass is just the sum of the mass of the particles that are gravitationally bound to that subhalo. The haloes and subhaloes are used to build halo merger trees using the {\sc dhalo} code \citep{Jiang:2013dda}.

Here, we use the \Galform{} semi-analytical model of galaxy formation  \citep{Cole:2000ex,Baugh:2006pf,V.:2013dht,Lacey16} to populate the dark matter haloes in the \Pmill{} simulation with galaxies. We use the recalibration of the \cite{V.:2013dht} model presented by \citet{Baugh:2018kkh} to identify LRGs and study their clustering. In order to match local observations of galaxies, just two of the parameters describing the physical processes modelled in \Galform{} were changed slightly by \citeauthor{Baugh:2018kkh}, from the values adopted by \citeauthor{V.:2013dht}, to take into account the change in cosmology and mass resolution in the \Pmill{} compared with the original $N$-body simulation used by \citeauthor{V.:2013dht}, and an improvement to the treatment of galaxy mergers (see \citealt{Baugh:2018kkh} for further details of these changes; we note that \citealt{Gonzalez-Perez:2017mvf} used an updated version of their model, which also included the new galaxy merger scheme first implemented by \citealt{Campbell:2015} and explained in full by \citealt{Simha:2017}).

%---------------------------------------------------------------
\subsection{{\sc GLAM} simulations}\label{sec:GLAM}
%---------------------------------------------------------------
\glam{} is a new $N$-body Parallel Particle-Mesh (PM) code developed for the massive production of large volume mock galaxy catalogues \citep{Klypin:2017jwl}. \glam{} first generates the density field at an early epoch, including peculiar velocities, for a particular cosmological model and initial conditions.
The code uses a regularly spaced three-dimensional mesh of size $N^3_{\rm g}$ that covers the cubic domain $L^3$ of a simulation box using $N^3_{\rm p}$ particles. The size of a cell, $\Delta x = L/N_{\rm g}$, and the mass of each particle, $m_{\rm p}$, define the force and mass resolutions, respectively \citep[see Appendix A of][for details]{Klypin:2017jwl}.

We generate 1000 \glam{} simulations using the same cosmology and linear perturbation theory power spectrum as used in the \Pmill{} simulation. Because our goal is to study the clustering of LRGs, the \glam{} simulations follow the evolution of $2000^3$ particles of mass $1.06\times 10^{10}\Msh$ in a cubic box of size $1\,h^{-1}\rm{Gpc}$ with $N_{\rm s} = 136$ time-steps, and mesh of $N_{\rm g} = 4000$. This numerical set-up yields a spatial resolution of $\Delta x = 0.25\Mpch$. The initial conditions were generated using the Zeldovich approximation starting at $z_{\rm ini}= 100$. %The particle data is saved in 21 snapshots between $0 < z < 1.2$ for each realisation. 

Haloes in \glam{} are identified with the bound density maximum ({\sc BDM}) halo finder \citep{Klypin:1997sk}. Only distinct haloes are saved in our catalogues. In {\sc BDM} the virial mass, $M_{\rm vir}$, is adopted as the definition of halo mass. The virial mass of a halo corresponds to the mass enclosed within a spherical overdensity of radius $R_{\rm vir}$, such that the mean overdensity within this radius is $\Delta_{\rm vir} \approx 330$ times the mean matter density of the Universe at the present time. The virial overdensity, $\Delta_{\rm vir}(z)$, is computed using the approximation of \citet{Bryan:1998A}. Only halo catalogues are saved in 21 snapshots between $0 < z < 1.2$ for each realisation. 

%--------- Figure --------------
\begin{figure*}
 \centering
\includegraphics[width=0.46\textwidth]{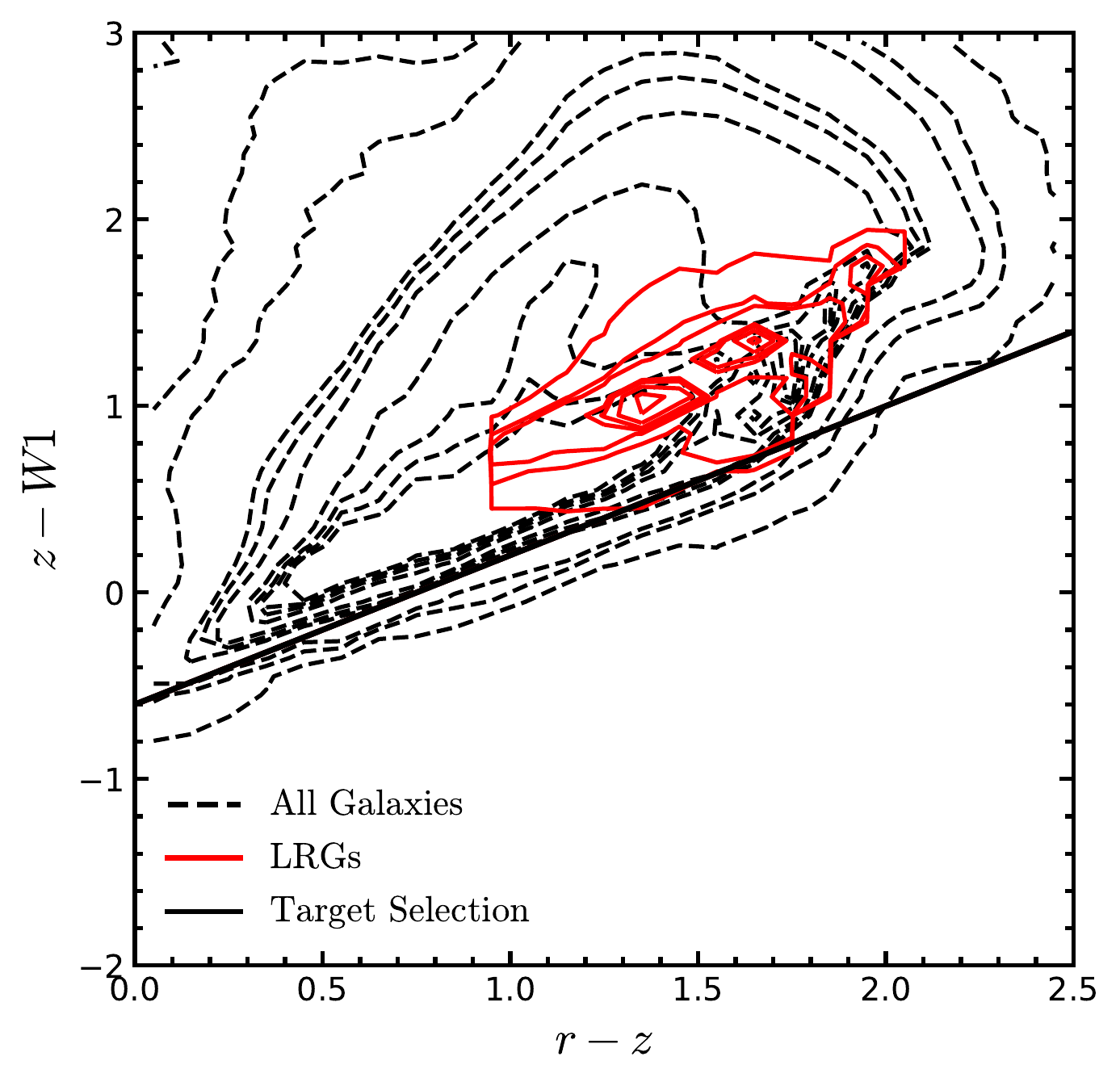}
\includegraphics[width=0.45\textwidth]{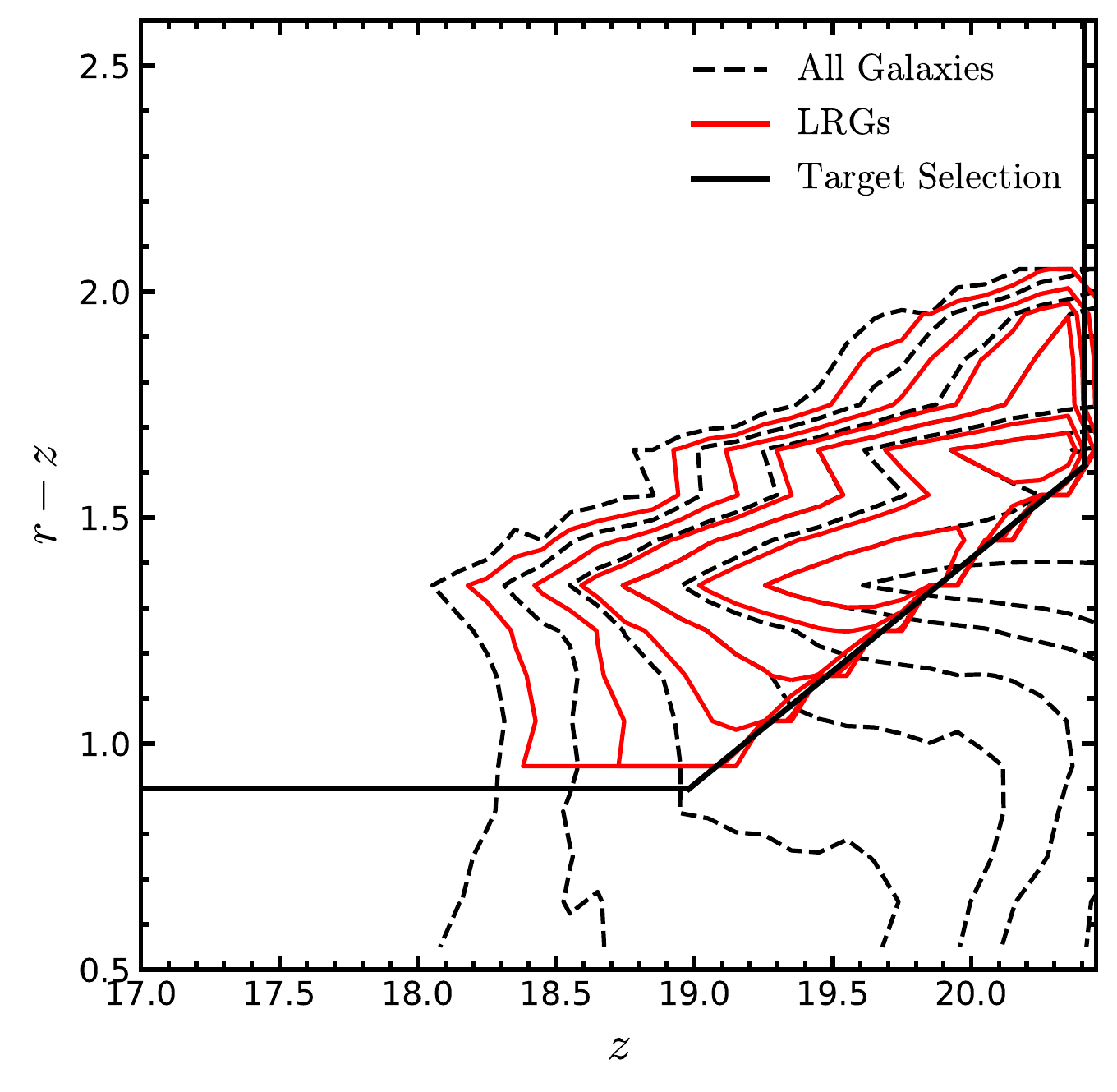}
\caption{Colour-colour ({\it left}) and colour-magnitude ({\it right}) diagrams predicted using the \Galform{} snapshots at $z=0.6$ to $z=1$ and using the $r$, $z$ and $W1$ bands. Dashed black lines represent the distribution of all galaxies with stellar mass $M_* > 10^9\Msh$ from the \Galform{} output. Red solid lines show the locus of \Galform{} galaxies which remain after applying the DESI LRG selection cuts. The solid black polygons indicate the DESI LRG photometric selection given by Eqs.~\eqref{eq:c1}-\eqref{eq:c4}, the same used by \citet{Rongpu:2020}.}
\label{fig:selection}
\end{figure*}
%---------------------------------------------------------------
\subsection{Halo mass function and halo clustering}\label{sec:halo}
%---------------------------------------------------------------
To check the performance of our \glam{} simulations we compare the halo mass function and the halo two-point correlation function measured from them with those obtained from the \Pmill{} simulation. Since we are interested in LRGs at $z\geq 0.6$ we use halo catalogues corresponding to snapshots at $z=0.6$, $0.74$ and $0.93$, where $z=0.74$ corresponds to the median redshift of the expected $n(z)$ distribution of LRGs in DESI \citep{DESI:2016zmz,Rongpu:2020}. In a future work we plan to build proper light-cones using all the \glam{} halo catalogues available in the relevant redshift range.

The upper panels in Fig.~\ref{fig:hmf} show the differential halo mass function measured at $z=0.6$, $0.74$ and $0.93$ from the \Pmill{} run (black solid lines) and the \glam{} simulations (blue dots with errobars) using $M_{\rm vir}$ as the halo mass definition. We use the mass conversion algorithm of \cite{Hu:2002we} to convert $M_{200m}$ into $M_{\rm vir}$ for the \Pmill{} measurements.
We find good agreement between the \glam{} and \Pmill{} results, with a difference of less than 10\% for haloes with mass $\log_{10}(M_{\rm vir}/\Msh) > 12.5$ at all redshifts. This mass value is well below the typical LRG host halo mass (see below). The differences seen between the results from \glam{} and \Pmill{} for lower mass haloes are due to the lower resolution in the \glam{} simulations. The differences seen at the high-mass end are due to the much smaller volume of the \Pmill{} simulation compared with that used in the \glam{} simulations.

The real-space clustering of haloes of mass $\log_{10}(M_{\rm vir}/\Msh) > 12.5$ is shown in the lower panels of Fig.~\ref{fig:hmf} at different redshifts. We find good agreement in the clustering measured on scales $r>2\Mpch$ between the two types of simulations. There is a 10 per cent difference over the separation range $2< r/\Mpch <40$.
Nevertheless, \glam{} predicts a higher clustering amplitude for $r \sim 1 \Mpch$ with respect to that measured in the \Pmill{} simulation. This effect is due to the different algorithms used to find dark matter haloes, i.e. {\sc BDM} predicts more halo pairs at small separations, hence resulting in a higher clustering amplitude on small scales. As we will see in Section~\ref{sec:clustering}, the difference in the halo clustering does not affect the clustering of LRGs when an appropriate HOD is applied to the \glam{} catalogues.

%---------------------------------------------------------------
\section{Selection of luminous red galaxies}\label{sec:selection}
%---------------------------------------------------------------
The DESI team plan to use the $3.4\,\mu{\rm m}$ band $(W1)$ from the space-based Wide-Field Infrared Survey Explorer (WISE), in combination with the $r$ and $z$ bands from the DESI Legacy Imaging Surveys \citep{Dey:2019}, to select LRGs efficiently in the redshift range $0.6 < z < 1.0$ \citep{DESI:2016zmz}. \cite{Rongpu:2020} described an updated version of the DESI LRG target selection, which we adopt here:
\begin{eqnarray}
&&  z < 20.41 \label{eq:c1}\\
&& -0.6 < (z-W_1) - 0.8(r-z)\,,\label{eq:c2}\\
&& r-z > 0.9\,,\label{eq:c3}\\
&& r-z > (z - 17.18)/2\,.\label{eq:c4}    
\end{eqnarray}

\Galform{} outputs observer frame absolute magnitudes with dust attenuation, $M_{\rm AB}$, so we need to convert these into apparent magnitudes, $m_{\rm AB}$, in order to apply the above cuts: 
\begin{equation}\label{eq:m_ab}
m_{\rm AB} = M_{\rm AB} + 5\log_{10}(d_L(z)/\Mpch) + 25 - 2.5\log_{10}(1 + z)\,,
\end{equation}
where the magnitudes are on the AB-magnitude system, $d_L(z)$ is the cosmological luminosity distance in units of $\Mpch$, and the factor $-2.5\log_{10}(1 + z)$ is from the band shifting of the filter width.

%--------- Figure --------------
\begin{figure}
 \centering
\includegraphics[width=0.45\textwidth]{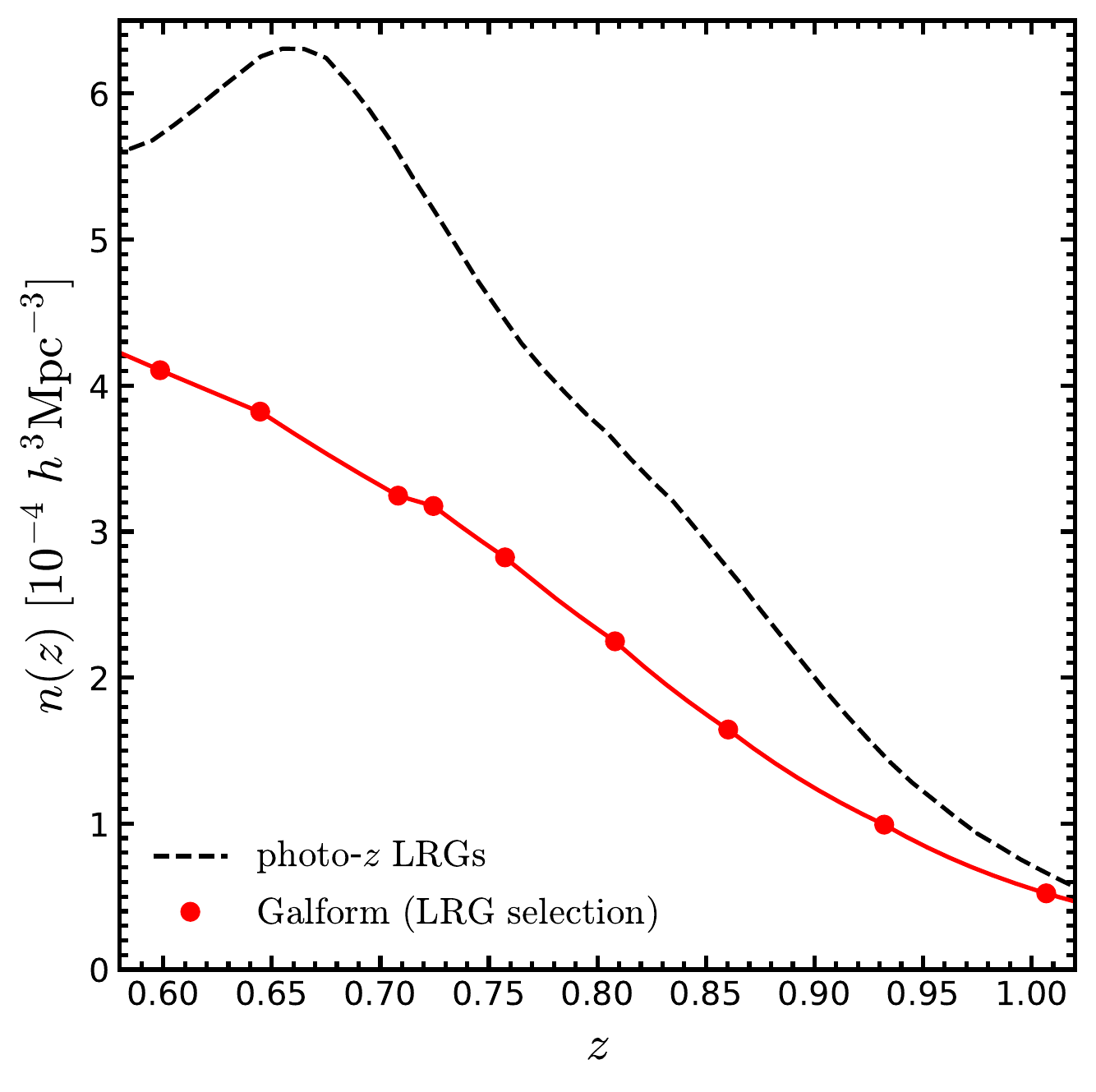}
\caption{The space density of LRGs meeting the DESI selection criteria, as predicted using \Galform, as a function of redshift. We show the nine \Pmill{} snapshots between $0.6 < z <1$ (red dots). The red solid line simply connects the points. The dashed black line shows the space density of DESI-LRGs estimated observationally using photometric redshifts by  \citet{Rongpu:2020}.}
\label{fig:nz}
\end{figure}

%--------- Figure --------------
\begin{figure*}
 \centering
\includegraphics[width=0.33\textwidth]{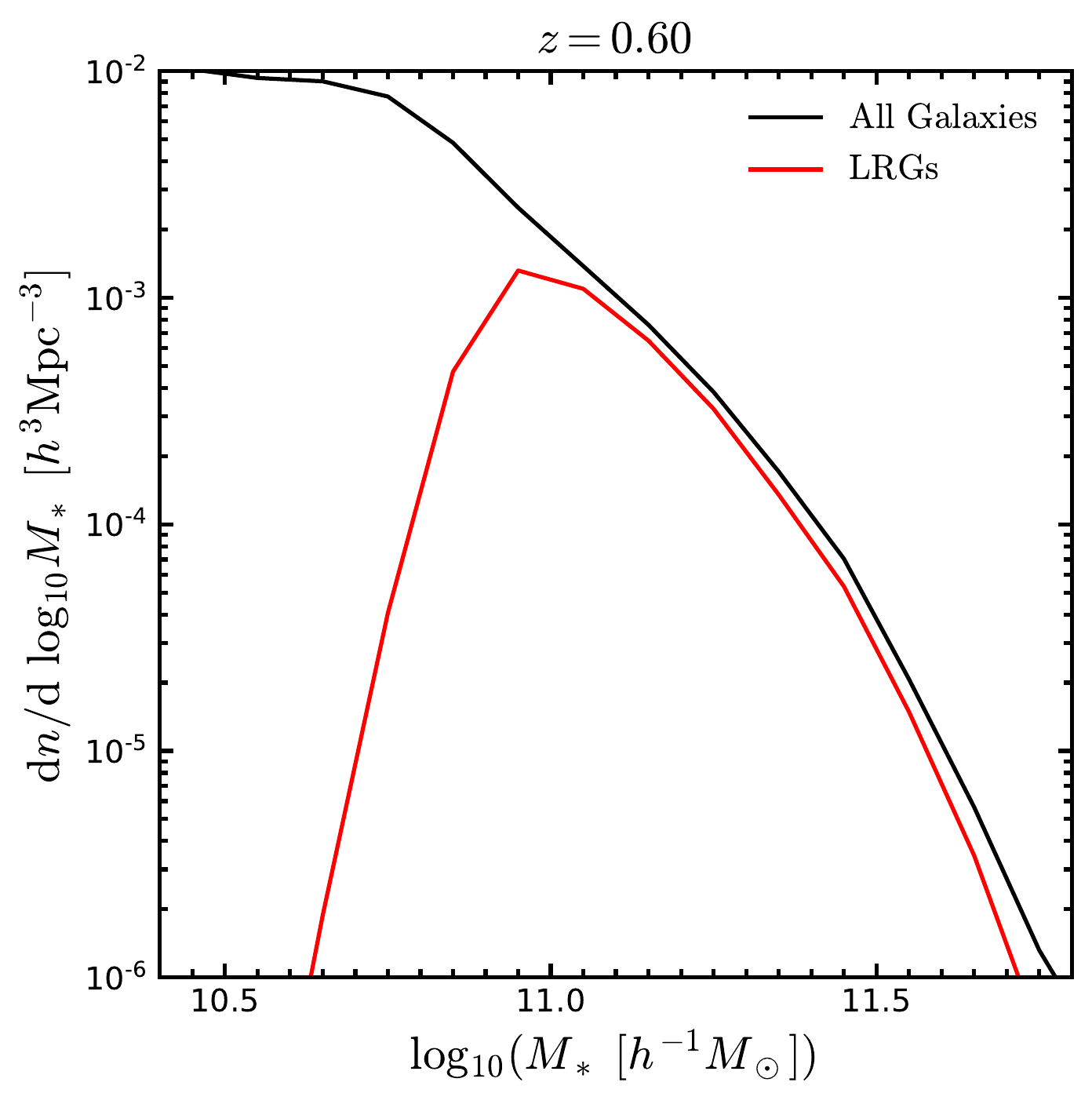}
\includegraphics[width=0.33\textwidth]{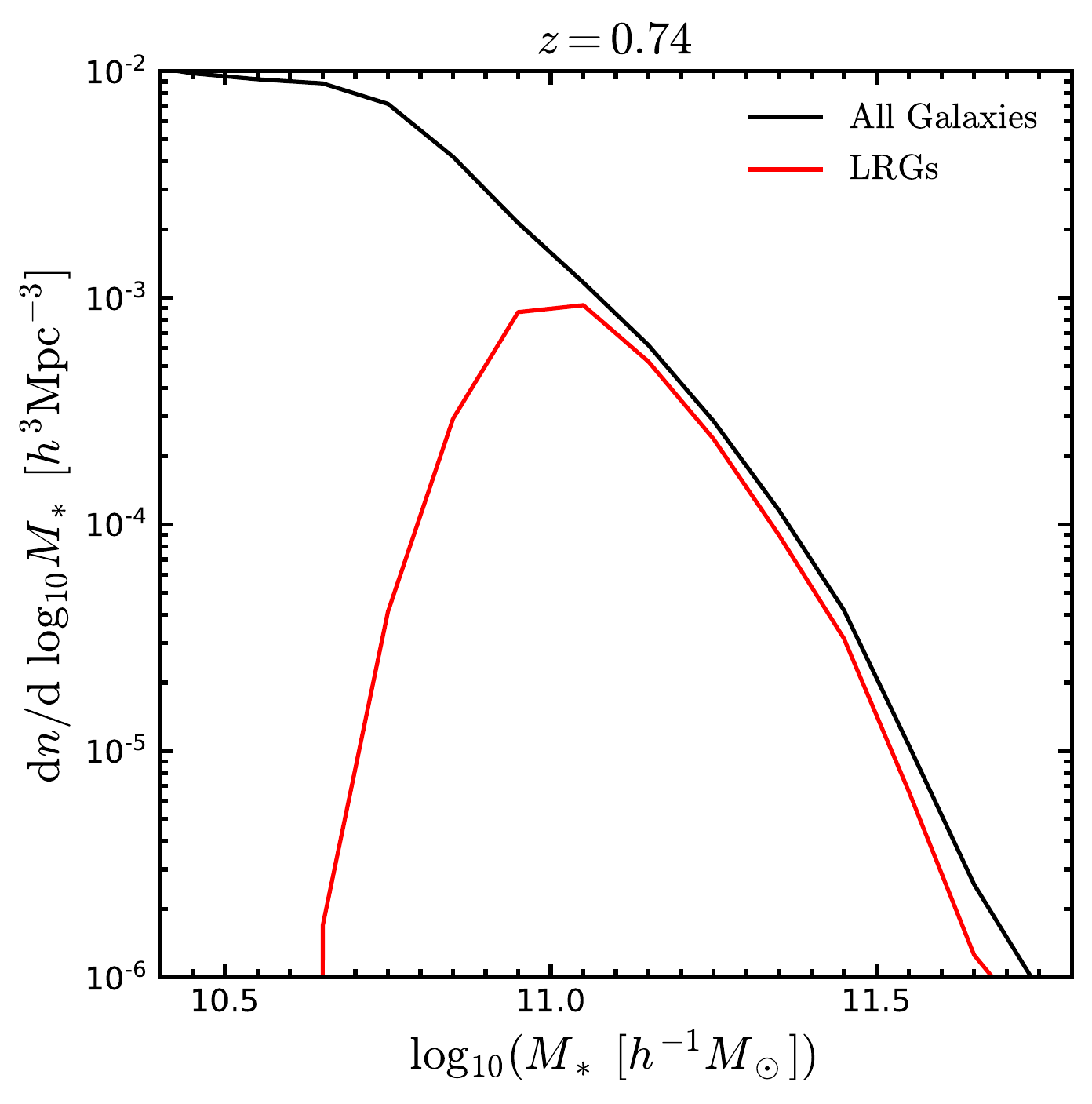}
\includegraphics[width=0.33\textwidth]{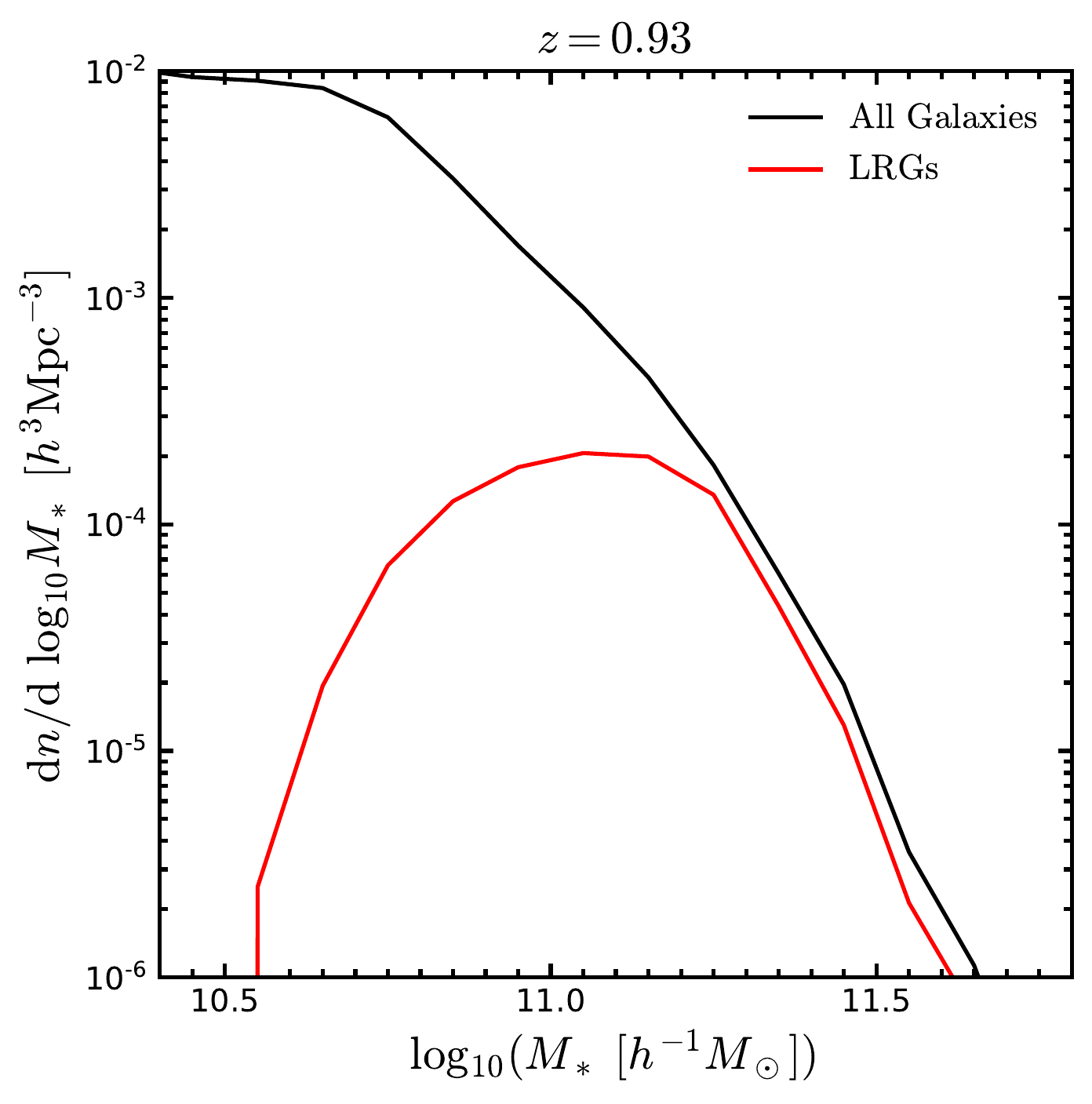}
\includegraphics[width=0.33\textwidth]{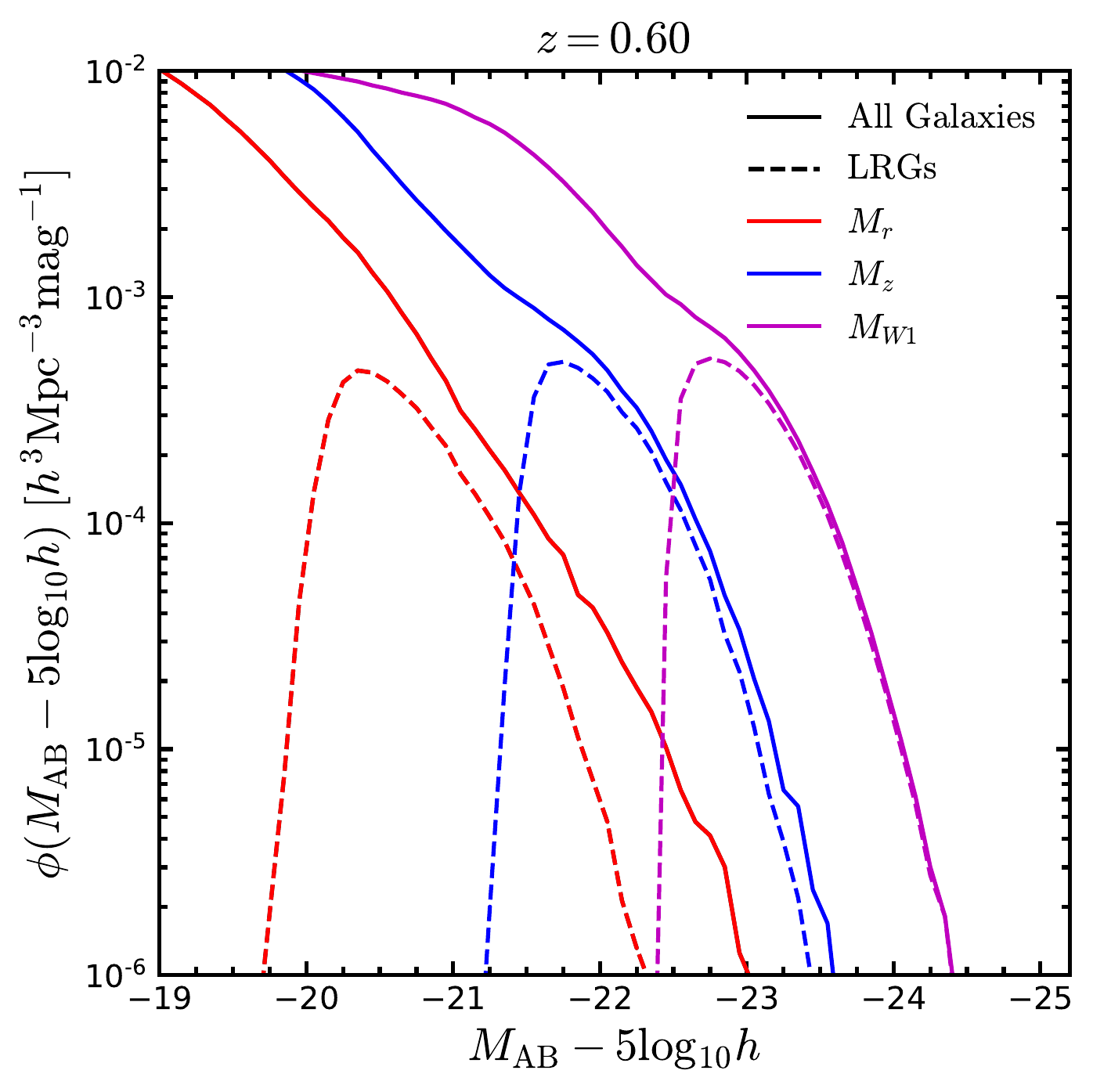}
\includegraphics[width=0.33\textwidth]{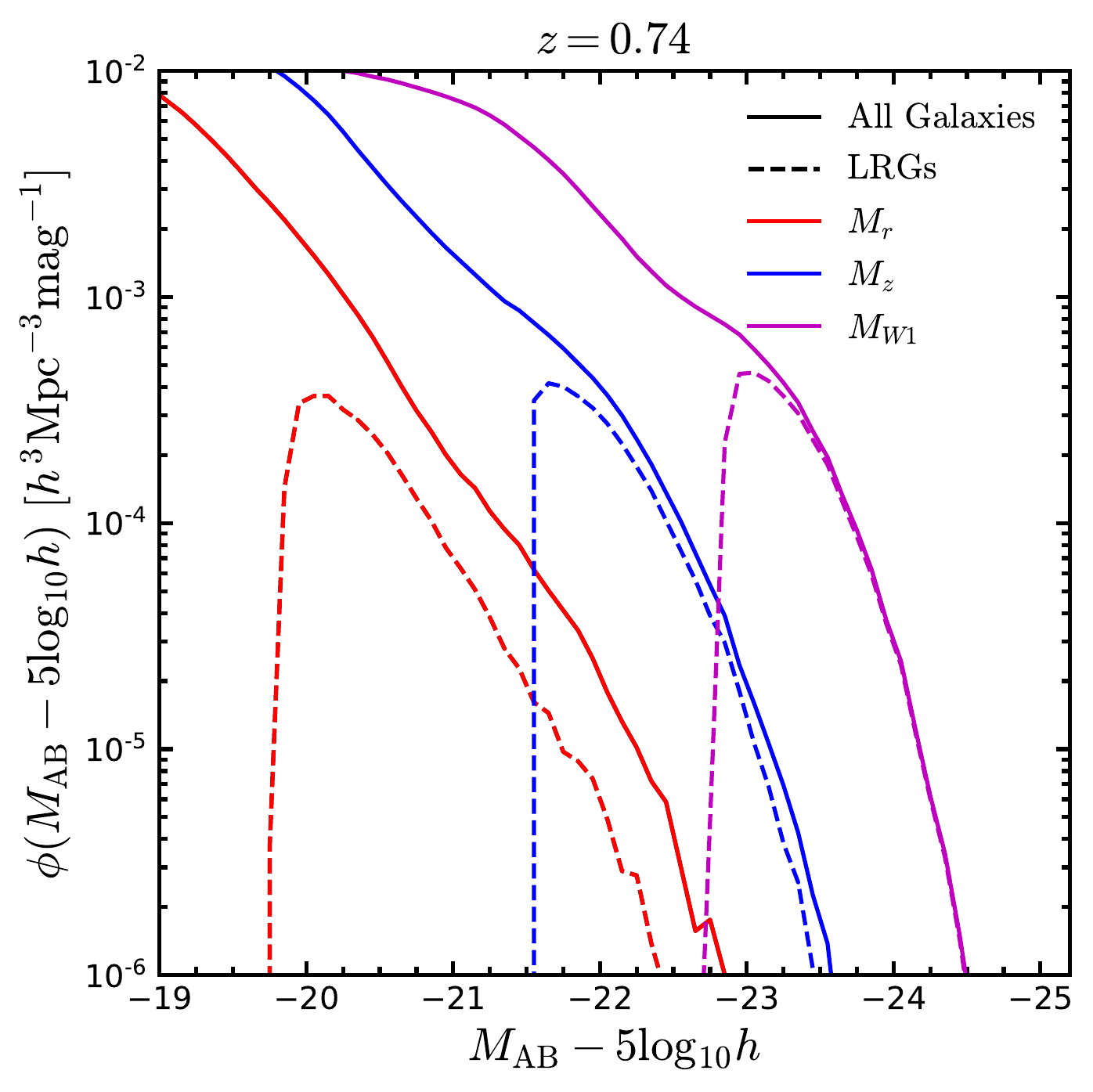}
\includegraphics[width=0.33\textwidth]{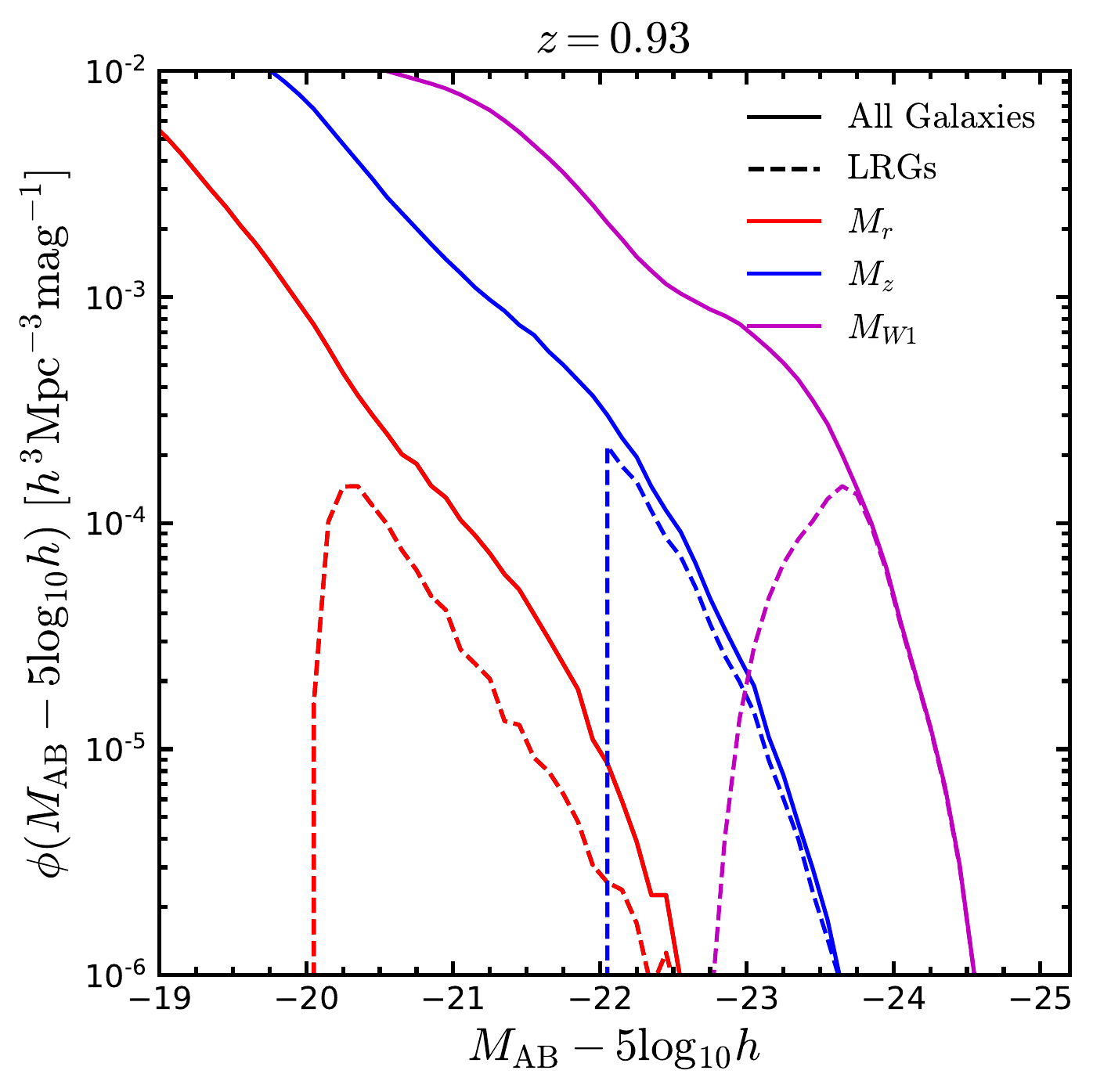}
\caption{Stellar mass ({\it upper panels}) and luminosity ({\it lower panels}) functions predicted by \Galform{} at $z=0.6$ ({\it left }), $z=0.74$ ({\it middle}) and $z=0.93$ ({\it right}) for all galaxies from the \Galform{} output and LRGs. Different colours and line styles indicate different properties and selections as indicated in the legend.}
\label{fig:sMF_LF}
\end{figure*}

The left panel of Fig.~\ref{fig:selection} shows \Galform{} galaxies in the redshift range $0.6 < z < 1$ in the $(r-z)-(z-W_1)$ colour-colour plane. The black contours show the locus of galaxies with stellar mass in excess of $10^{9}\Msh$ and the red contours show the galaxies that meet the DESI LRG selection criteria set out in Eqs.~\eqref{eq:c2} and \eqref{eq:c3}. The right panel of Fig.~\ref{fig:selection} shows the distribution of galaxies in the $z-(r-z)$ colour-magnitude plane, again showing all galaxies with stellar mass above $10^{9}\Msh$ (black contours) along with those which satisfy the LRG selection (red contours). The stellar mass cut of $10^{9}\Msh$ is much lower than we expect for the stellar mass of LRGs (see below), but is applied for illustrative purposes, to allow us to see the locus of the \Galform{} galaxies in the colour-magnitude planes, before the photometric LRG selection is applied. Note that in these panels we simply show all of the galaxies that pass the stellar mass cut or LRG selection from {\it each} of the nine \Pmill{} snapshots that fall within the redshift interval. As such, we are mainly interested in the locus of the \Galform{} galaxies in these colour-magnitude planes, rather than the detailed changes in the density of points. 

Reassuringly, the red contours in the $(r-z)-(z-W_1)$ colour-colour plane are well within the black polygons denoting the selection boundaries; the blue colour boundary of the $r-z$  vs. $z$ selection box is a key component in setting the space density of LRGs, as the red contours touch this cut. At $z=0.6$ \Galform{} predicts that around 6.2 million galaxies in the \Pmill{} volume have stellar mass $M_* > 10^9\Msh$ but only a small fraction ($0.84$ per cent) of these galaxies are selected as LRGs.

%--------- Figure --------------
\begin{figure*}
 \centering
\includegraphics[width=0.33\textwidth]{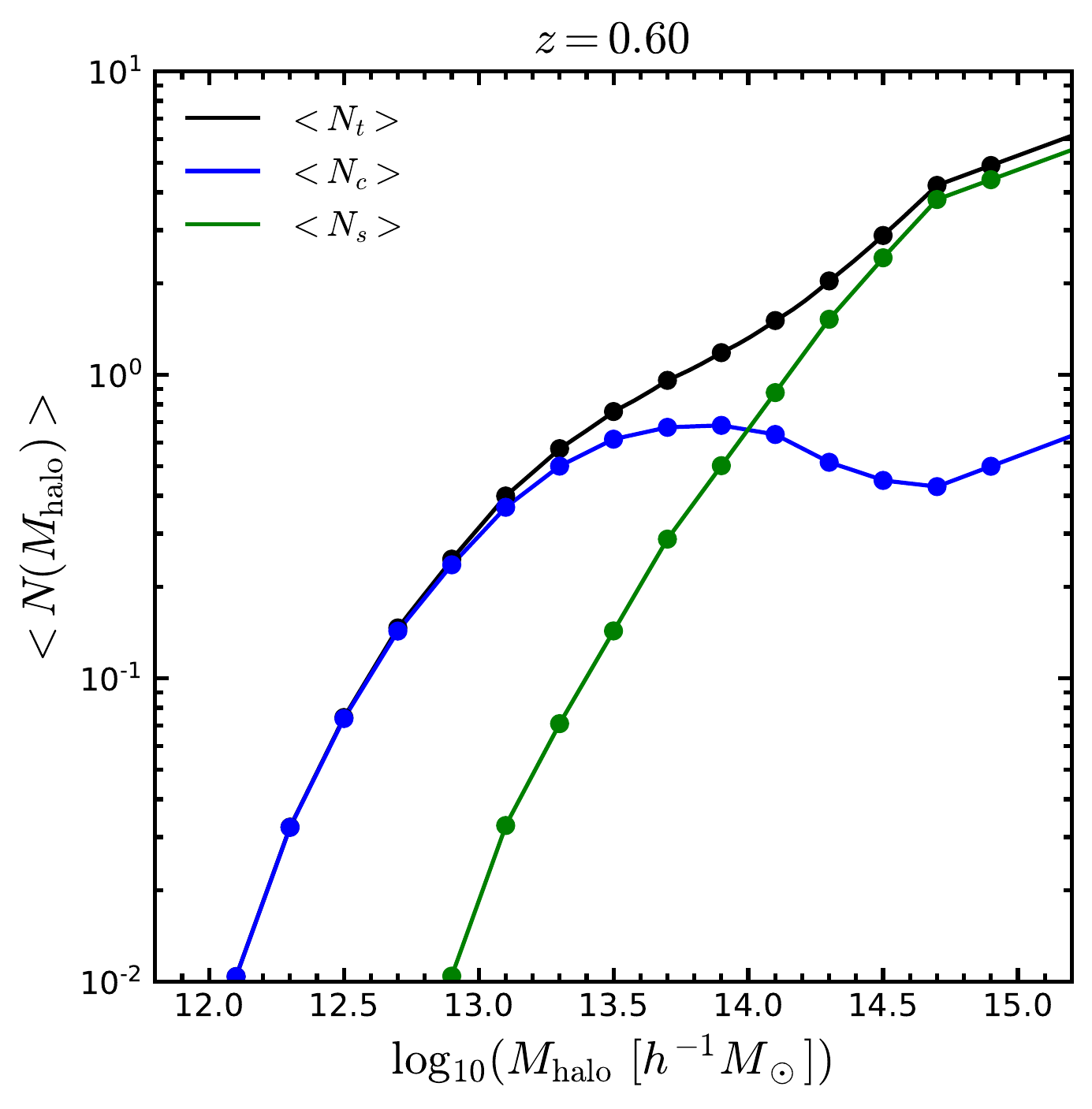}
\includegraphics[width=0.33\textwidth]{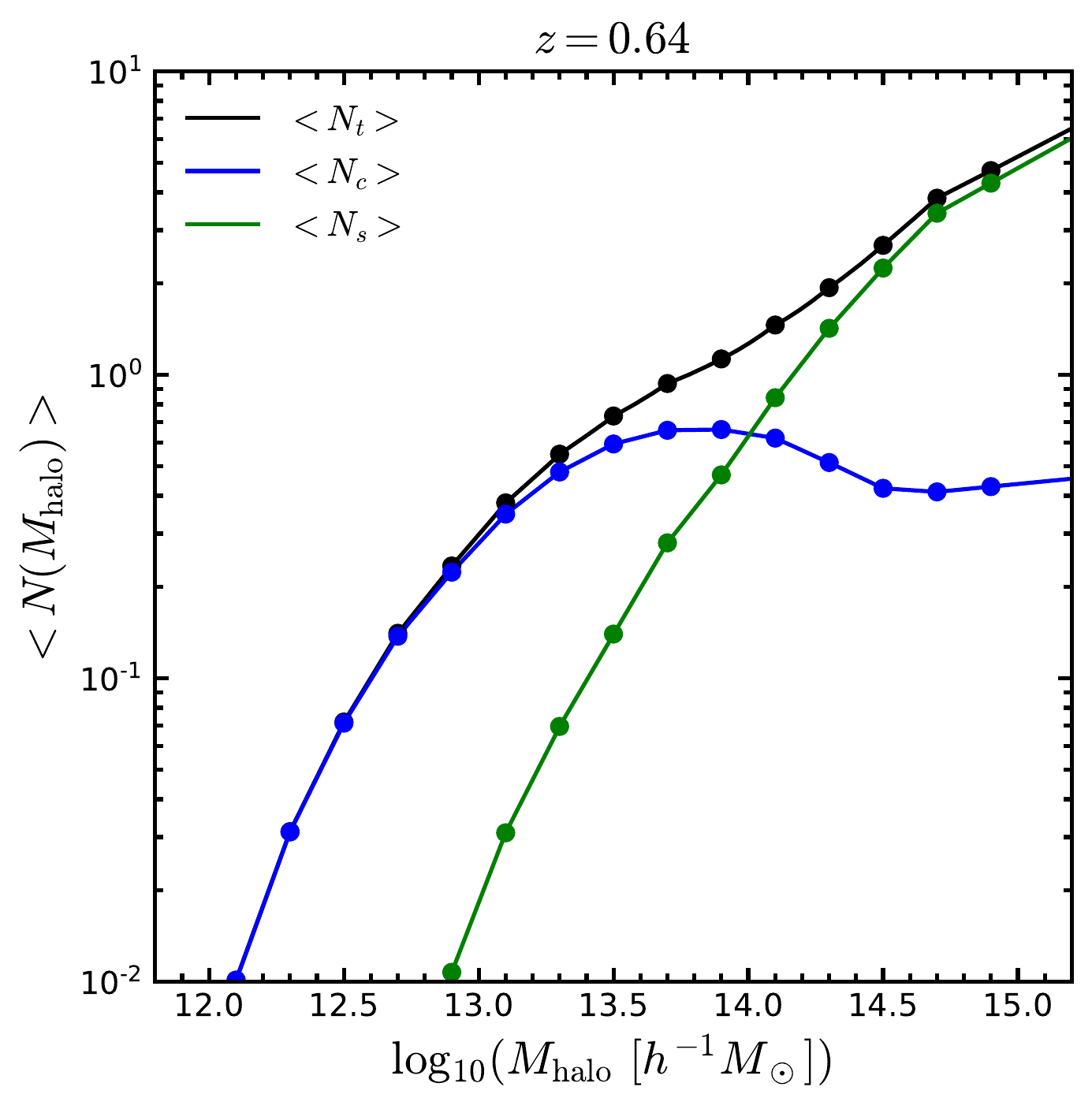}
\includegraphics[width=0.33\textwidth]{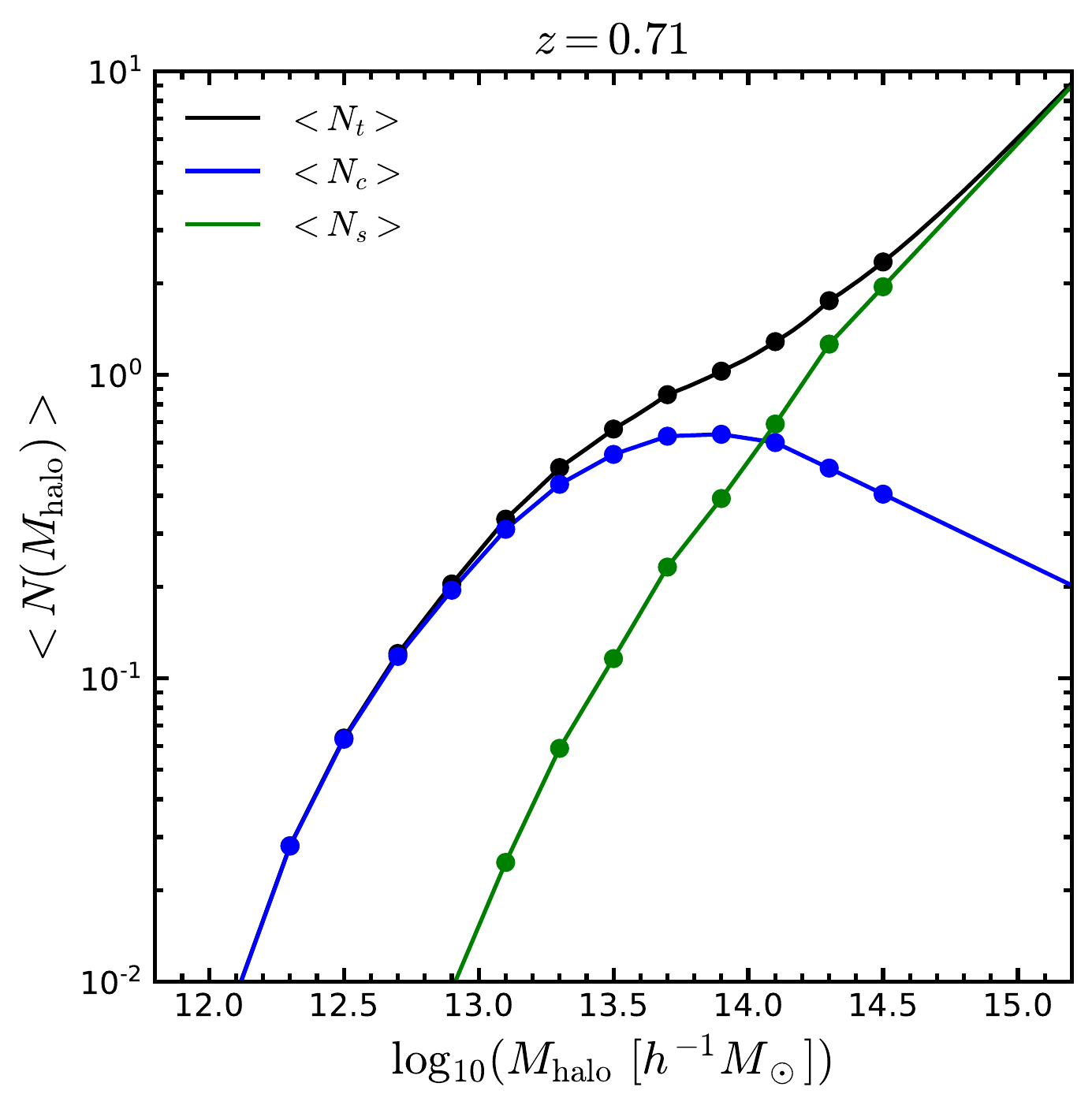}
\includegraphics[width=0.33\textwidth]{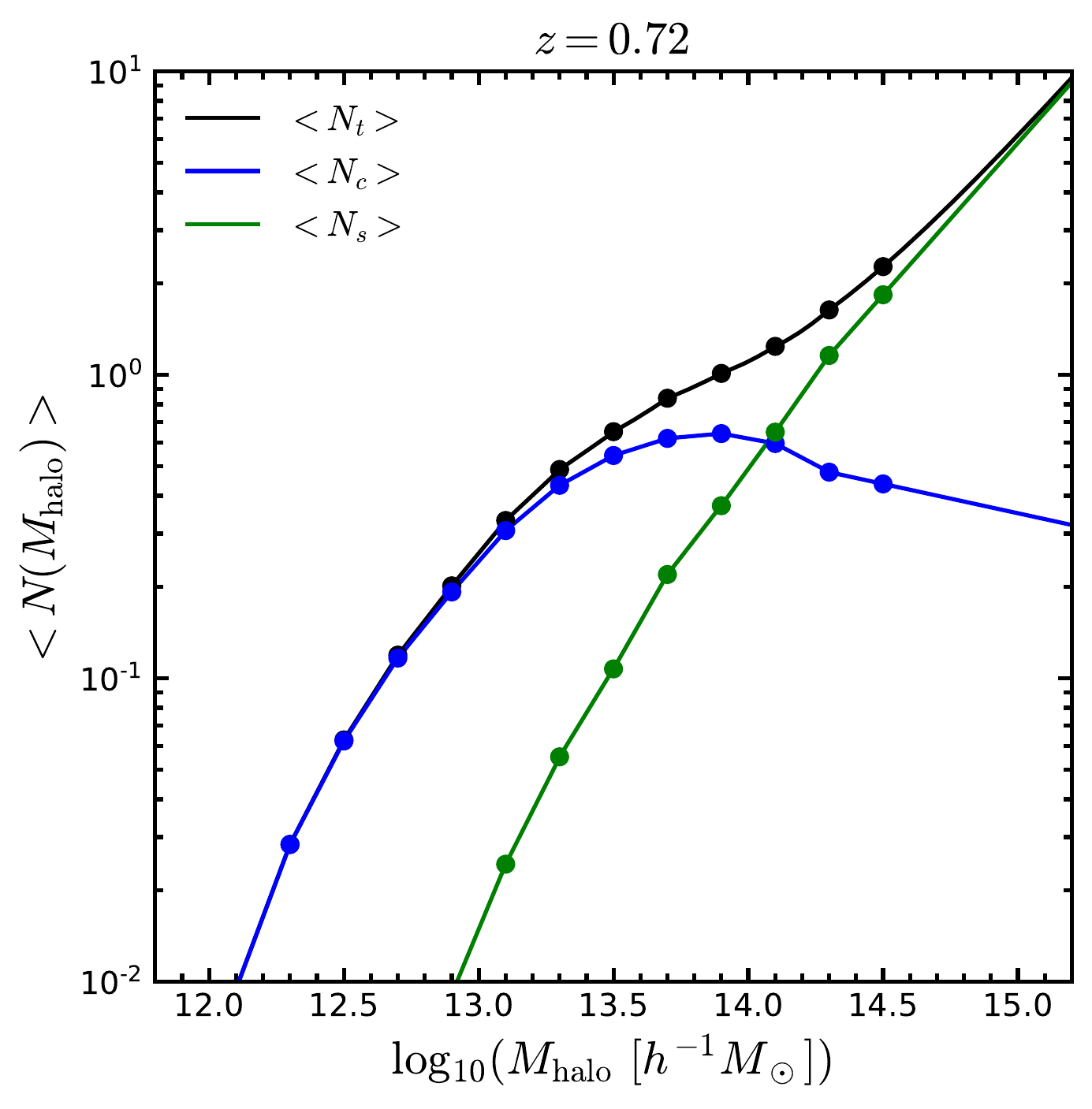}
\includegraphics[width=0.33\textwidth]{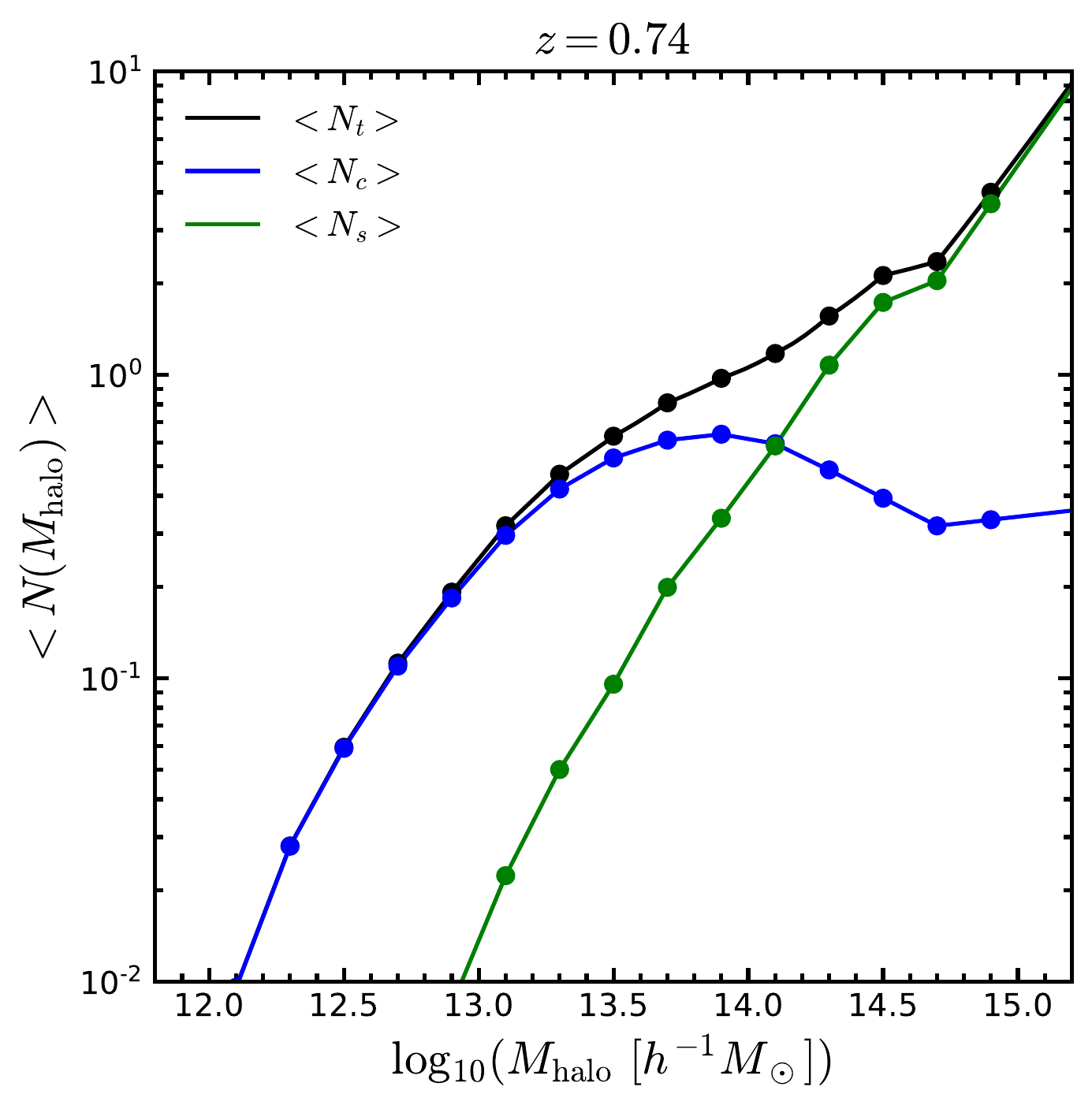}
\includegraphics[width=0.33\textwidth]{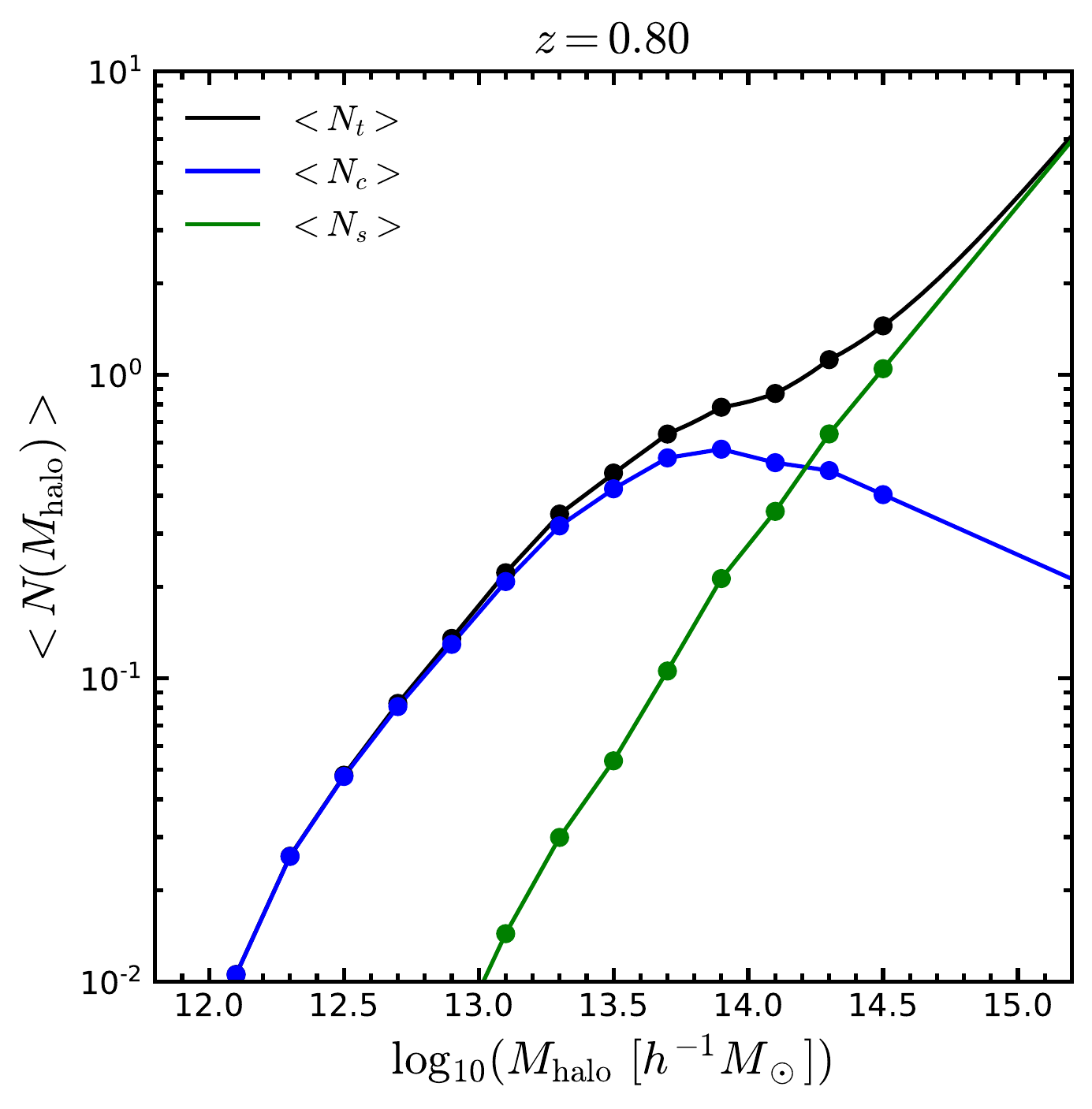}
\includegraphics[width=0.33\textwidth]{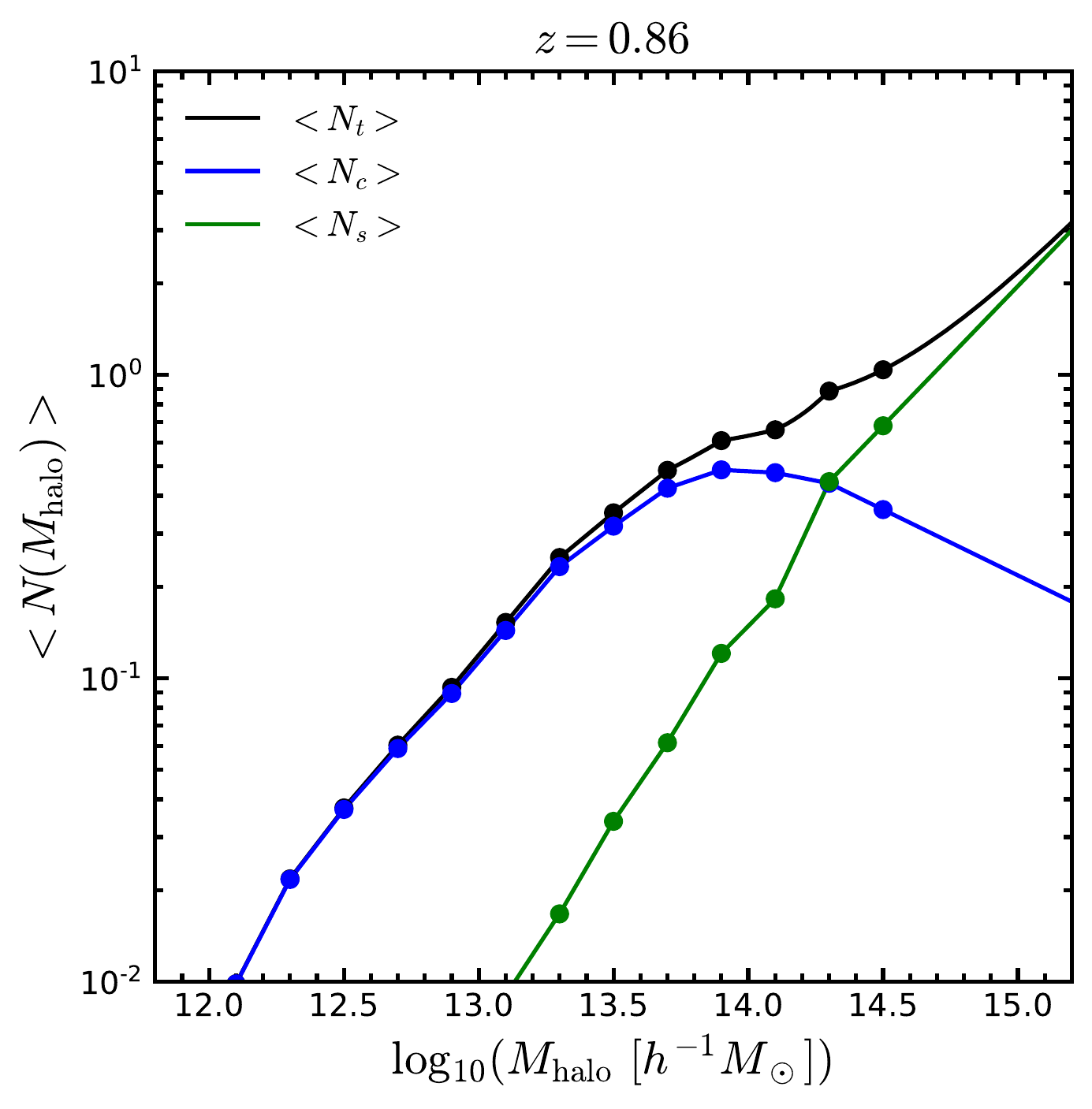}
\includegraphics[width=0.33\textwidth]{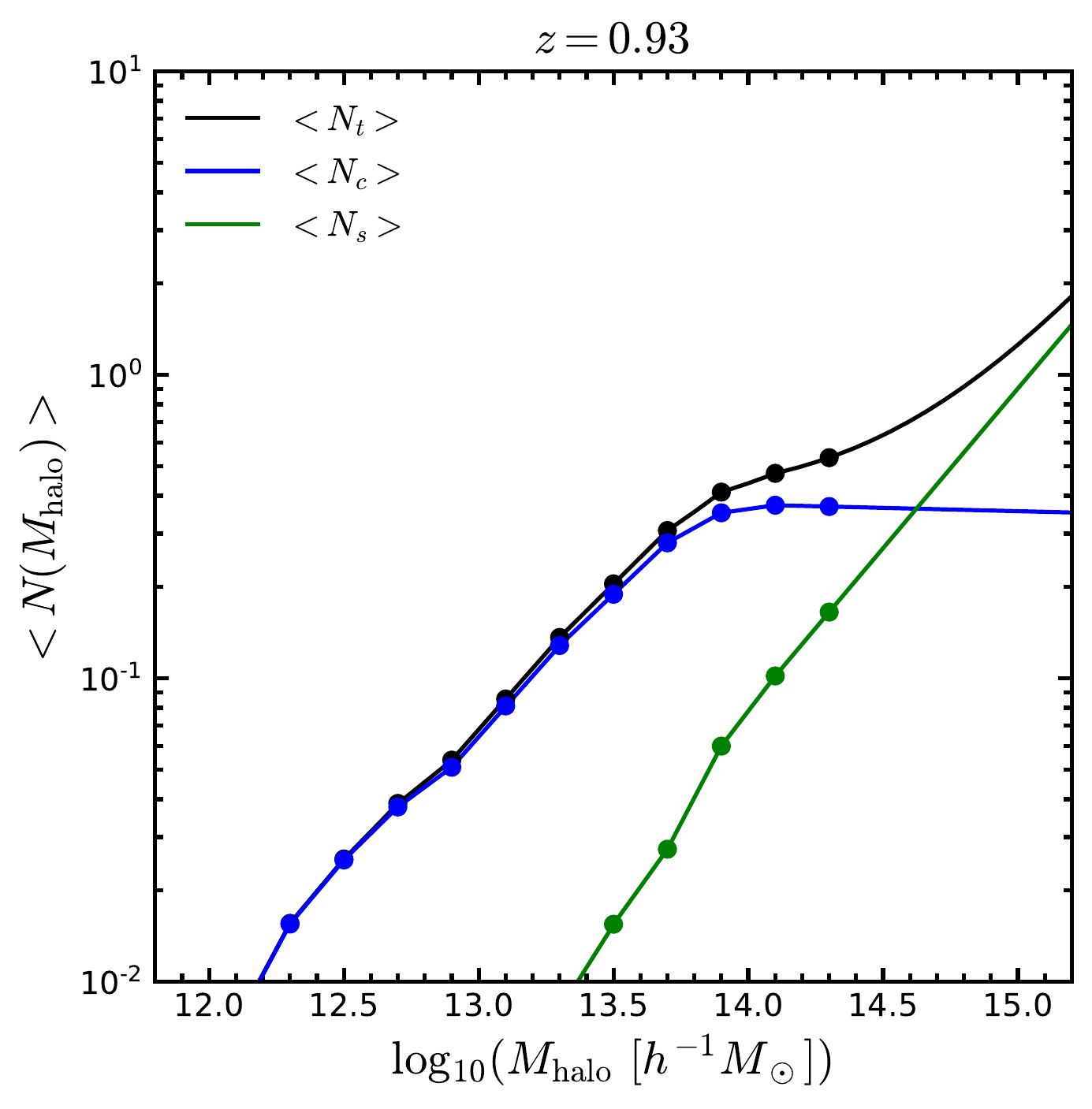}
\includegraphics[width=0.33\textwidth]{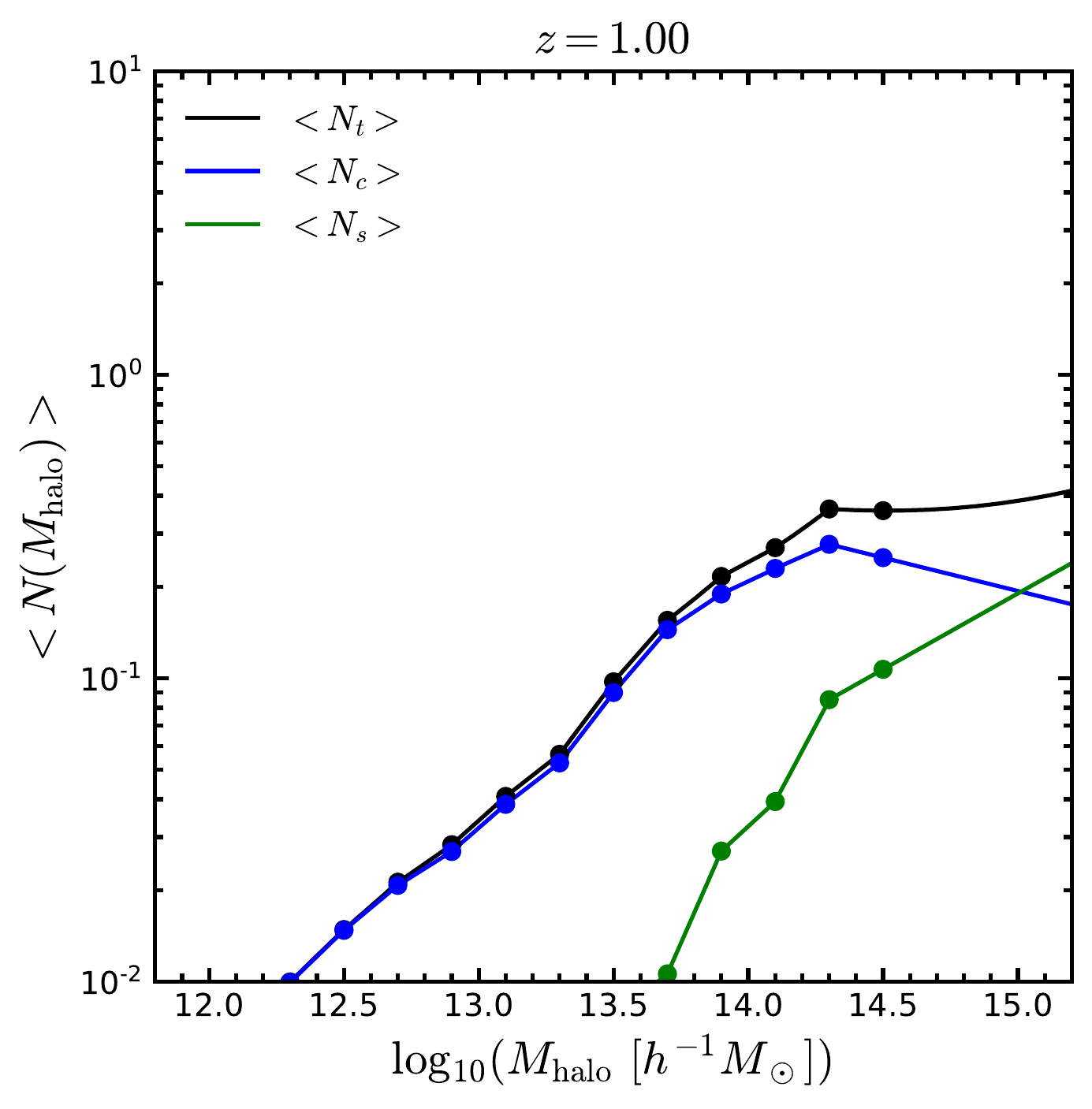}
\caption{Halo occupation distribution of DESI-like LRGs predicted by \Galform{} (symbols) as a function of their host halo masses. Each panel shows a different redshift between $z=0.6$ and $z=1$ as labelled. The solid lines connect the symbols. Outside the mass range for which model predictions are available, the solid lines show a power-law extrapolation of the HOD for centrals and satellites, based on the last measured points. The total, central and satellite galaxy occupancy is shown in black, blue and green, as labelled.}
\label{fig:hod}
\end{figure*}

%--------- Figure --------------
\begin{figure*}
 \centering
\includegraphics[width=0.33\textwidth]{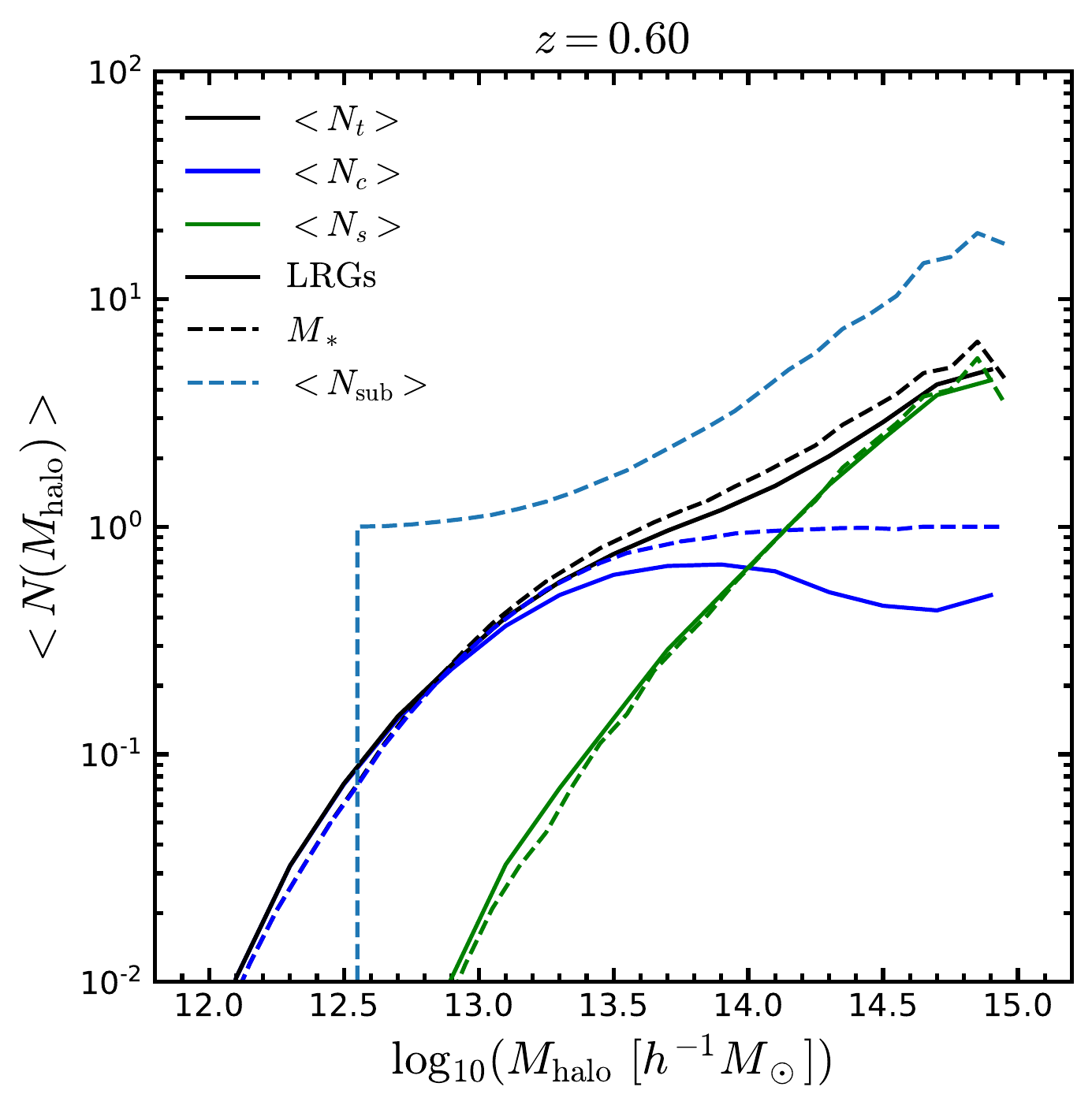}
\includegraphics[width=0.33\textwidth]{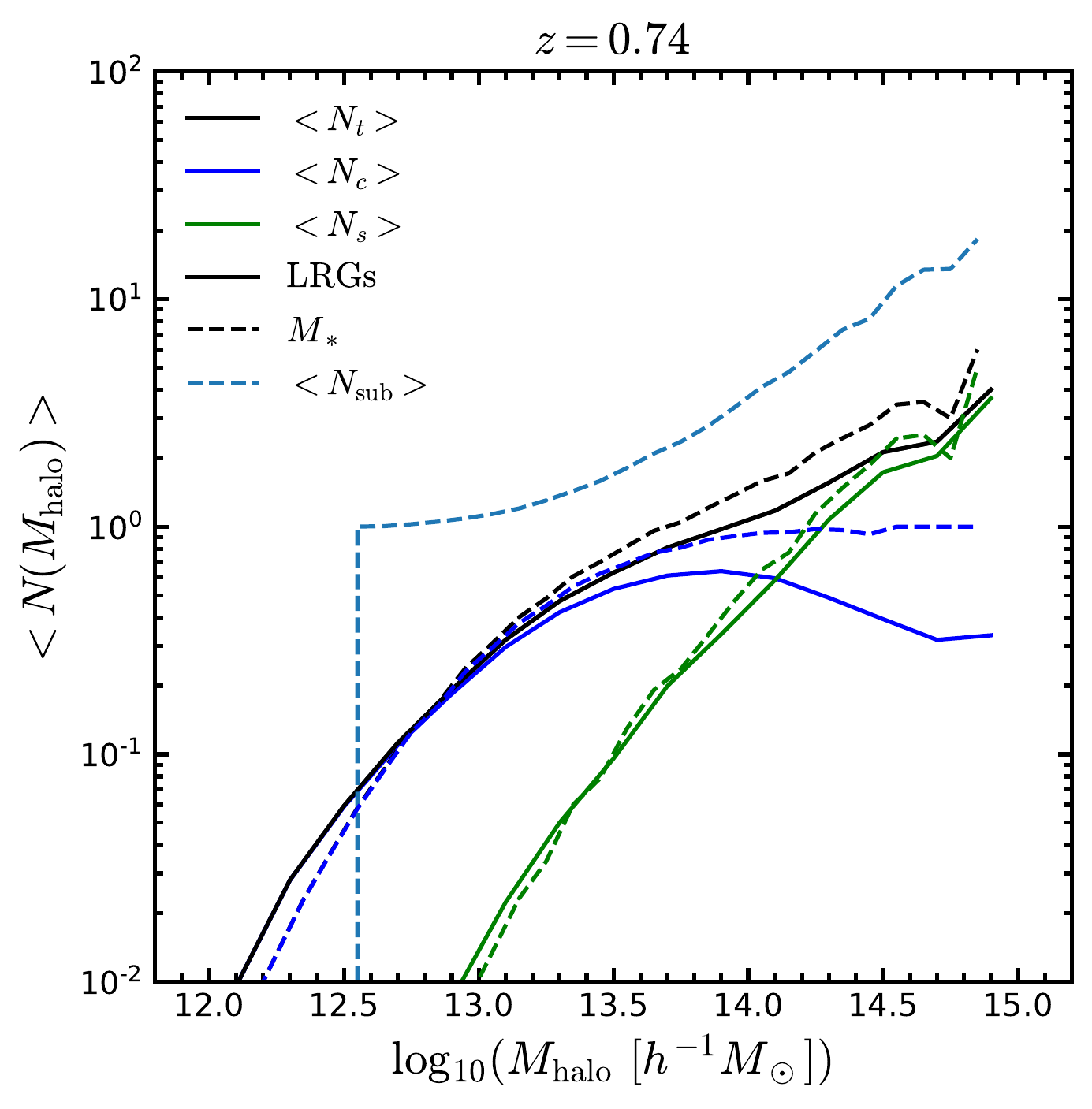}
\includegraphics[width=0.33\textwidth]{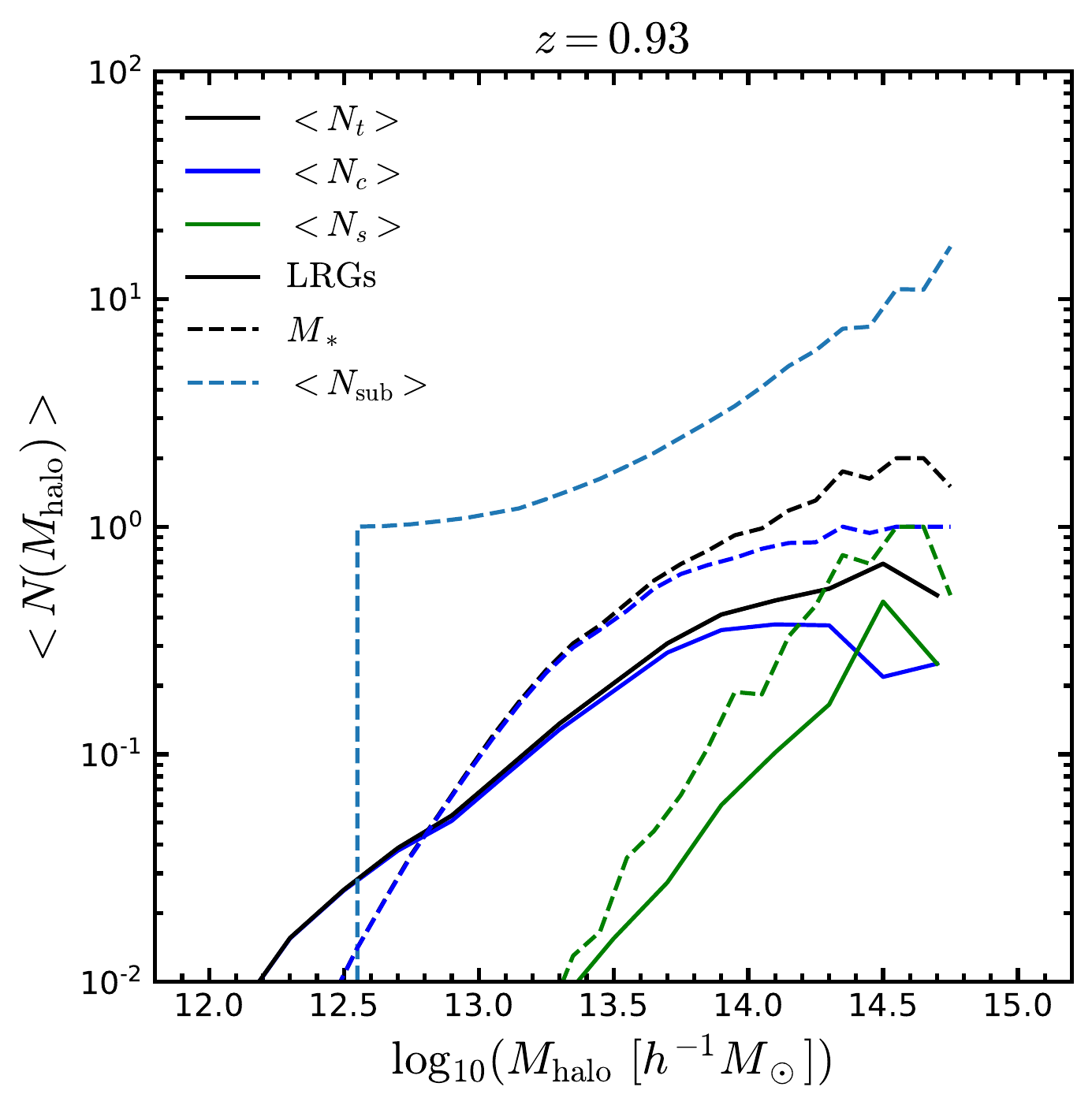}
\includegraphics[width=0.33\textwidth]{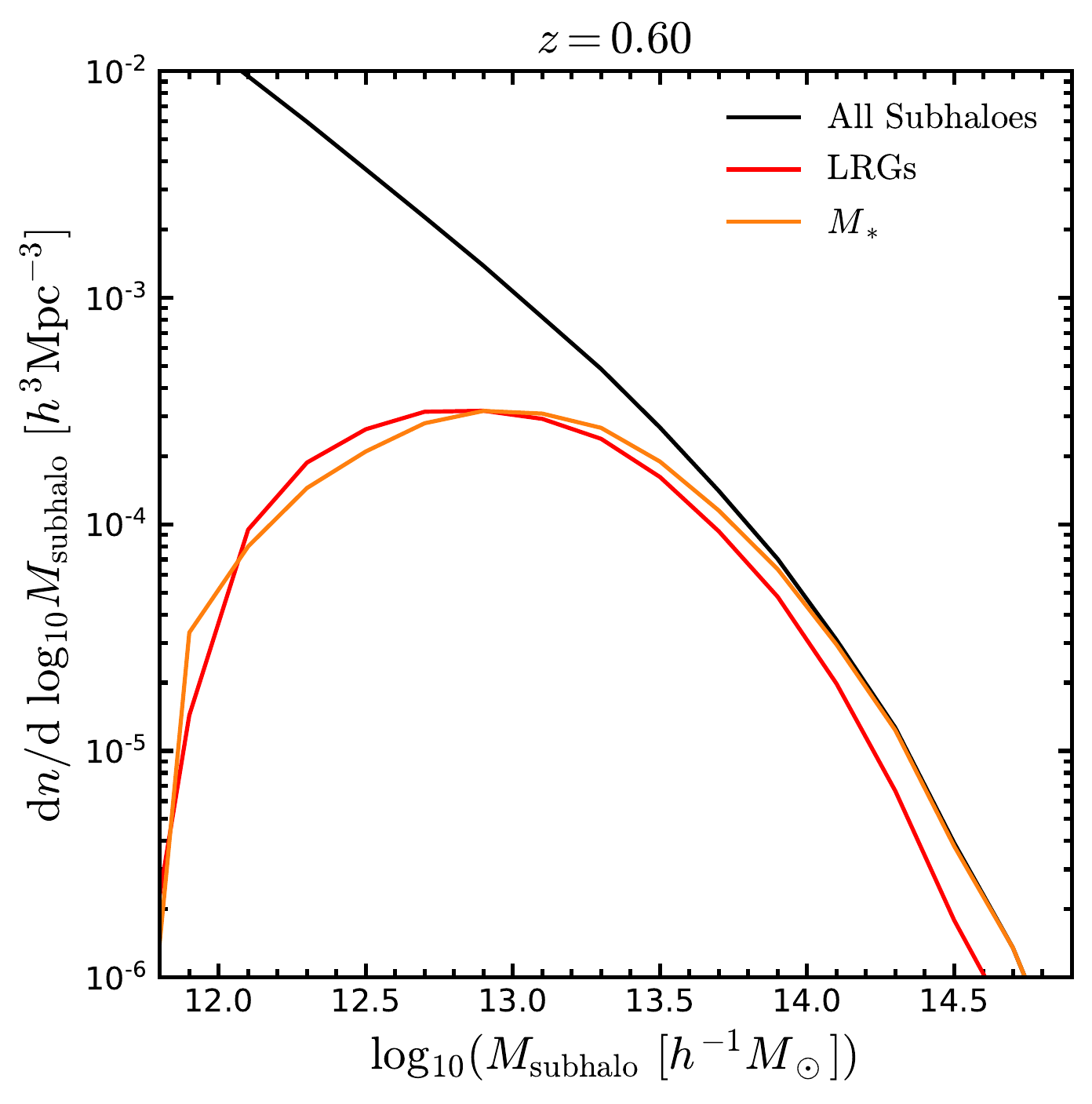}
\includegraphics[width=0.33\textwidth]{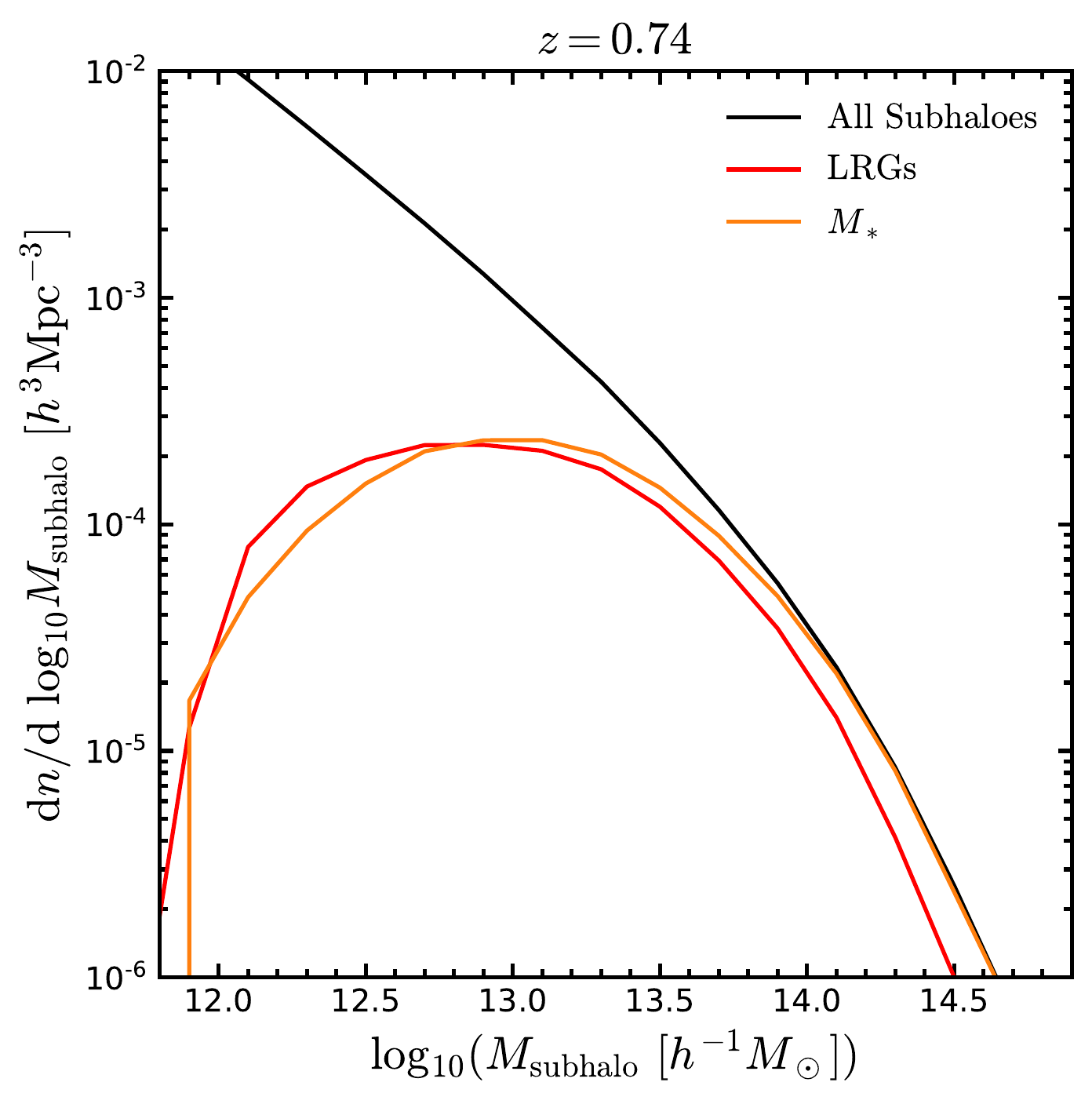}
\includegraphics[width=0.33\textwidth]{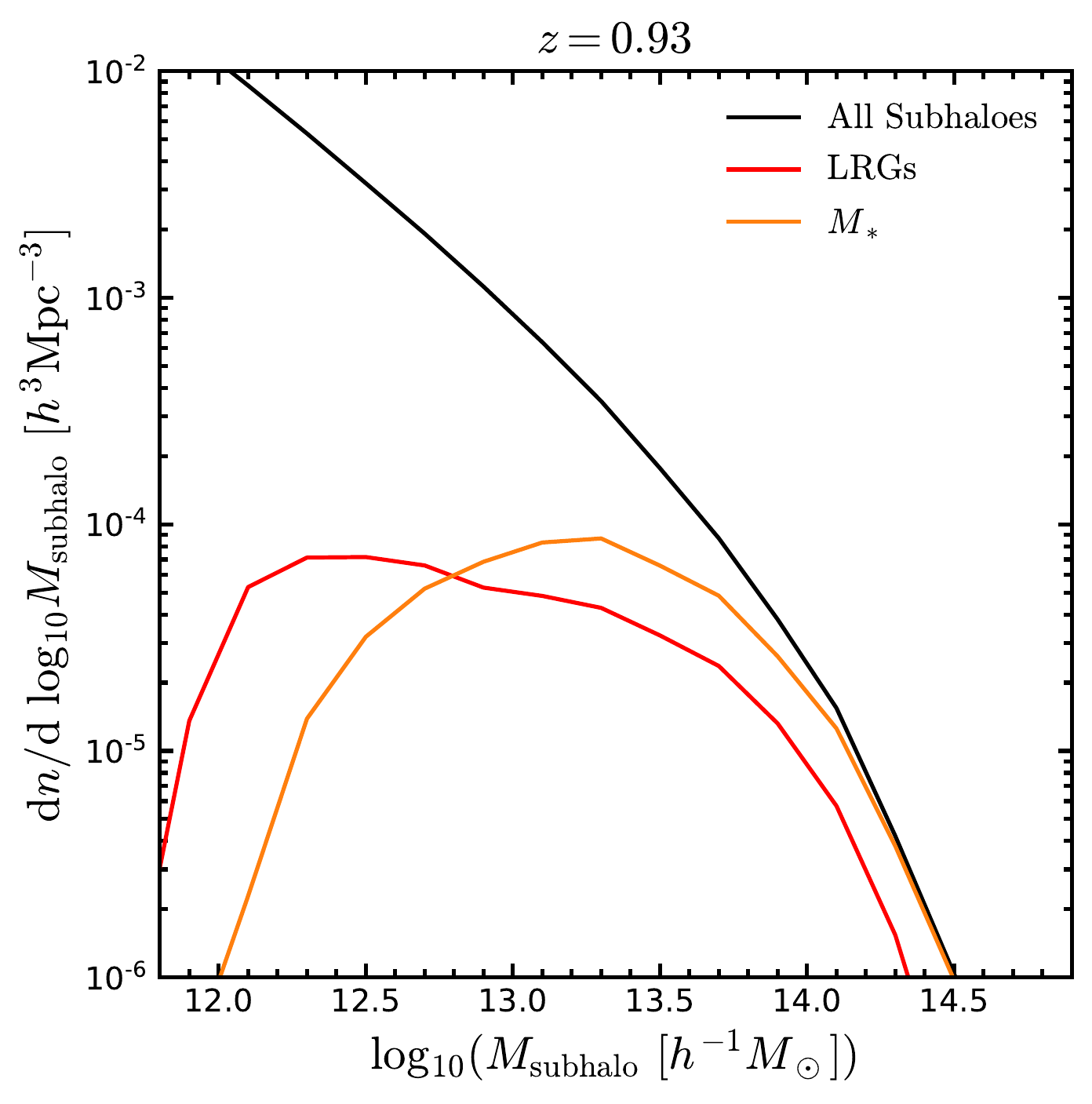}
\caption{{\it Upper panels:} Halo occupation distribution of subhaloes (dashed cyan line), LRGs (solid lines) and galaxies ranked by stellar mass (dashed lines) at $z=0.6$ ({\it left panel}), $z=0.74$ ({\it middle panel}) and $z=0.93$ ({\it right panel}). For galaxies, the occupation of total, centrals and satellites are specified as black, blue and green lines respectively. {\it Lower panels:} Subhalo mass functions measured using all galaxies (black solid lines), LRGs (red solid lines) and galaxies ranked by stellar mass (orange solid lines) at at $z=0.6$ ({\it left panel}), $z=0.74$ ({\it middle panel}) and $z=0.93$ ({\it right panel}).}
\label{fig:sub_hod}
\end{figure*}

Fig.~\ref{fig:nz} shows the space density, $n(z)$, of DESI LRGs predicted using \Galform{}. We have applied the colour-magnitude cuts (Eq.~\eqref{eq:c1} to Eq.~\eqref{eq:c4}) to nine \Pmill{} snapshots in the redshift range $0.6 < z < 1$ to obtain the abundance of LRGs -- the redshift of the snapshots is indicated by the points in Fig.~\ref{fig:nz}. 
In the same figure, we show the number density of DESI-like LRGs inferred from observations using photometric redshifts  from \citet{Rongpu:2020} (black dashed line). We note that \Galform{} underpredicts the abundance of LRGs at all redshifts, with the discrepancy reaching a factor of $\approx 1.7$ at $z\sim 0.66$.
The predicted space densities could be reconciled with those inferred observationally using photometric redshifts by perturbing, for example, the $r-z$ selection to a bluer colour in Fig.~\ref{fig:selection}. 
Nevertheless, here we are interested in showing the theoretical predictions from the \Galform{} model and the applications on the large-scale clustering of our \glam{} catalogues. 

To further investigate the impact of the LRG colour-magnitude selection on the galaxy population predicted by \Galform{} we present, in Fig.~\ref{fig:sMF_LF}, the stellar mass and luminosity functions for all galaxies and for those selected as DESI LRGs. The top panels of Fig.~\ref{fig:sMF_LF} show the evolution with redshift of the stellar mass function (sMF) for all galaxies and for LRGs, for $z=0.6$, $z=0.74$ and $z=0.93$. Given the halo mass resolution of the \Pmill{}, robust predictions can be made using \Galform{} for galaxies with stellar masses $M_* > 10^7\Msh$ \citep{Baugh:2018kkh}.
As expected, the LRG sample is dominated by massive galaxies, although not {\it all} massive galaxies are LRGs. These massive galaxies are predicted to be in massive dark matter haloes above the mass at which heating by active galactic nuclei suppresses gas cooling \citep{Contreras:2015, Mitchell:2016}. Some massive galaxies, however, have recent star formation driven by the cold gas accreted in galaxy mergers, making their $r-z$ colour too blue to be selected as LRGs. 
The predicted stellar mass function of LRGs drops sharply below $\log_{10}(M_*/\Msh) = 11.1$, but is similar to the overall SMF for larger stellar masses. The amplitude of the LRG SMF is similar at $z=0.6$ and $z=0.74$, which reflects the lack of evolution seen in the overall SMF.
As we can see from Fig.~\ref{fig:nz}, the number density of LRGs drops from $4.11\times 10^{-4}\hMpcc$ at $z=0.6$ to $3.02\times 10^{-4}\hMpcc$ at $z=0.74$, while at $z=0.93$ the abundance of LRGs is $0.99\times 10^{-4}\hMpcc$. 

Similar to the plots showing the galaxy stellar mass function, in the lower panels of Fig.~\ref{fig:sMF_LF}, we show, at the same redshifts as used in the top row, the luminosity functions for the $r$, $z$ and $W1$ bands for all galaxies and for LRGs. We find a similar trend as that discussed for the stellar mass functions. The fraction of bright galaxies that are selected as LRGs increases with the wavelength of the band: above a threshold luminosity, all galaxies in the $W_{1}$-band are LRGS, whereas only a fraction, around a half, of galaxies that are bright in the $r$-band are LRGs. Below the threshold luminosity, the fraction of galaxies that are LRGs plunges dramatically.

%---------------------------------------------------------------
\section{The galaxy$-$(sub)halo connection of DESI luminous red galaxies}\label{sec:hod}
%---------------------------------------------------------------
To explore the galaxy-(sub)halo connection of the DESI-like LRGs predicted by \Galform{} we first examine their halo occupation distribution (HOD). The HOD is an useful tool to understand the galaxy-halo connection, clustering and evolution of galaxies in general (see the review by \citealt{Wechsler:2018}). The HOD specifies the average number of galaxies (centrals and satellites) hosted by a dark matter halo. Previous observational studies have described the HOD of LRGs using a functional form that distinguishes between central and satellite galaxies \citep[see e.g.,][]{Blake:2007xp,Brown:2008eb,Padmanabhan:2008dda,Zheng:2008np}. In a traditional HOD there is a transition in the mean number of central galaxies from $\lla N_c \rra = 0$ to $\lla N_c \rra = 1$ with increasing halo mass and the occupation by satellites $(\lla N_s \rra)$ follows a power-law in halo mass \citep{Zheng:2004id}.

Fig.~\ref{fig:hod} shows the evolution of the HOD of DESI LRGs as predicted by \Galform{} in the redshift range $z=0.6-1$. We show the predicted HOD for the nine redshifts we used to measure the evolution of the LRG number density distribution in Fig.~\ref{fig:nz}. At first glance we see that the occupancy of central galaxies ($\lla N_c \rra$) does not reach the canonical value of unity at high halo masses, and even begins to decline after a peak at intermediate halo masses. This behaviour is typically seen in the models when galaxies are selected by their star-formation rate instead of a property that correlates more closely with stellar mass \citep{Contreras:2013kr,Cowley:2015nqa,Jimenez:2019moq}. More recently \cite{Gonzalez-Perez:2017mvf} found similar behaviour for the HOD of emission-line galaxies selected by the colour-magnitude cuts that will be used by the DESI emission-line galaxy survey (see also \citealt{Merson:2019vfr} and \citealt{Violeta:2020}). The LRG population is dominated by central galaxies and contains a satellite fraction of $f_{\rm sat} \sim 0.10$ to $f_{\rm sat} \sim 0.04$ in the redshift range $z=0.6-1$, where the mean number of satellites $\lla N_s \rra$ is close to a power-law.

Fig.~\ref{fig:hod} shows that there is a clear turnover in the HOD predicted by \Galform{} for central galaxies at intermediate redshifts ($z=0.74, 0.80)$. At higher redshifts than this the trend is less clear due to the evolution in the halo mass function and the resulting lack of high mass haloes. One might have expected that the mean number of centrals would reach unity in massive haloes, due to the suppression of gas cooling through the heating of the hot gas halo by active galactic nuclei. However, some central galaxies in massive haloes can become too blue to be selected as LRGs due to star formation triggered by mergers, which use the cold gas brought in by the merging galaxy.

To develop a deeper understanding of the galaxy-(sub)halo connection we now explore which {\it subhaloes} are able to host an LRG. To do so, we consider the number of subhaloes in haloes of different mass and the subhalo mass function, including a version that shows only those subhaloes that host an LRG. We also define a new galaxy sample for comparison purposes by ranking galaxies in order of decreasing stellar mass, and choosing a stellar mass cut to match the number density of the LRG sample. This comparison sample allows us to understand the impact of the selection cuts on the haloes and subhaloes that host LRGs; we call this the stellar mass selected sample. 

The upper panels of Fig.~\ref{fig:sub_hod} show the HOD for the LRG and stellar mass selected galaxy samples, which we compare to the total number of subhaloes available to host an LRG (see below for how this is defined). Focusing on the galaxy HODs first, the black, blue and green lines in Fig.~\ref{fig:sub_hod} show the number, respectively, of all galaxies, central galaxies and satellites galaxies as a function of halo mass; solid lines show the model predictions for the LRG sample and the dashed lines for the stellar mass selected sample. The light blue dashed lines show the number of subhaloes more massive than $M_{\rm subhalo} > 10^{12.5}\Msh$ as a function of the mass of their main host halo. This mass cut is arbitrary but was chosen because the HODs for the galaxy samples are significant for halo masses above this value. In an illustrative sense, a subhalo mass of $M_{\rm subhalo} \approx 10^{12.5}\Msh$, based on the mass coverage of the galaxy sample HODs, could be loosely thought of as the minimum subhalo mass needed to host an LRG or a galaxy in the comparator stellar mass selected sample.

Fig.~\ref{fig:sub_hod} shows us that only a small fraction of subhaloes with masses above $M_{\rm subhalo} > 10^{12.5}\Msh$ host an LRG: this fraction reduces from $22\%$ to $8\%$ as the redshift increases from $z=0.6$ to $z=0.93$. The shape of the total (centrals+satellites) and satellite-only HOD is similar for LRGs and the stellar mass selected sample at $z=0.6$ and $0.74$. However, at $z=0.93$ the DESI LRG selection cuts modify the form of the LRG HOD away from that of the stellar mass selected sample. The HODs of central galaxies in the two samples are markedly different at all redshifts shown in Fig.~\ref{fig:sub_hod}. The HOD of stellar mass selected central rises to unity with increasing halo mass, but for the LRGs it turns over after reaching a maximum below unity. This behaviour is swamped by the satellite HOD so that the overall HODs for the LRG and stellar mass samples differ less than the central HODs. At the highest redshift shown in Fig.~\ref{fig:sub_hod}, the transition from zero to peak occupancy fraction for centrals is slower for the LRGs than for the stellar mass sample. As centrals dominate the overall sample at lower halo masses, this produces a significant difference in the HOD for LRGs and the stellar mass selected sample.

To gain further insight into the LRG subhalo population, we show the subhalo mass function in the lower panels of Fig.~\ref{fig:sub_hod}. Two versions of the subhalo mass function are shown: one is the `dark matter view' in which we include all subhaloes and the other is the `galaxy view', in which case a subhalo is only included if it contains a galaxy in the sample. If the `galaxy view' version of the subhalo mass function coincides with the `dark matter view', then all subhaloes at that mass that could host a galaxy do so. In the case of the stellar mass selected samples shown in the bottom row of Fig.~\ref{fig:sub_hod}, we see that the most massive subhaloes all host a galaxy. As we move to lower masses, the galaxy-view subhalo mass function falls below the dark-matter view version; for these masses only a fraction of the available subhaloes host a galaxy. Eventually, as we continue to mover towards even lower subhalo masses, there is a dramatic downturn in the galaxy-view subhalo mass function, with only a tiny fraction, less than one in a thousand subhaloes hosting a galaxy. Qualitatively, the galaxy-view subhalo mass functions for the LRGs are similar to those for the stellar mass selected sample, with one exception: at the massive end, not all subhaloes host an LRG. This difference becomes more pronounced with increasing redshift. The conclusion of this comparison is that it is essential to perform the full colour-magnitude selection to define the LRG sample. Applying a stellar mass cut to attain a target number density of objects is a fair approximation to performing the full photometric selection at low redshifts, but results in a fundamentally different set of subhaloes being chosen with increasing redshift.

%--------- Figure --------------
\begin{figure*}
 \centering
\includegraphics[width=0.33\textwidth]{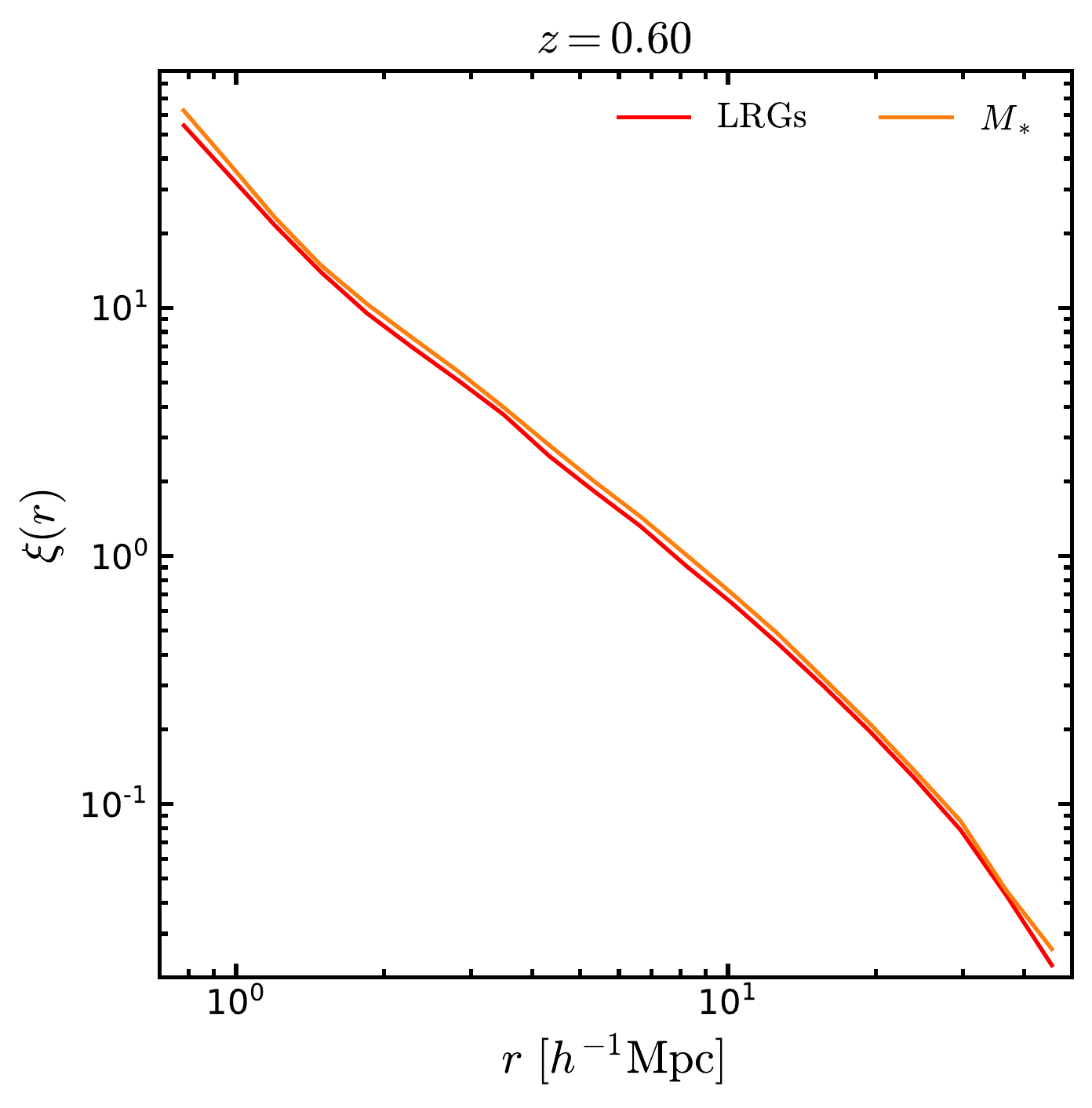}
\includegraphics[width=0.33\textwidth]{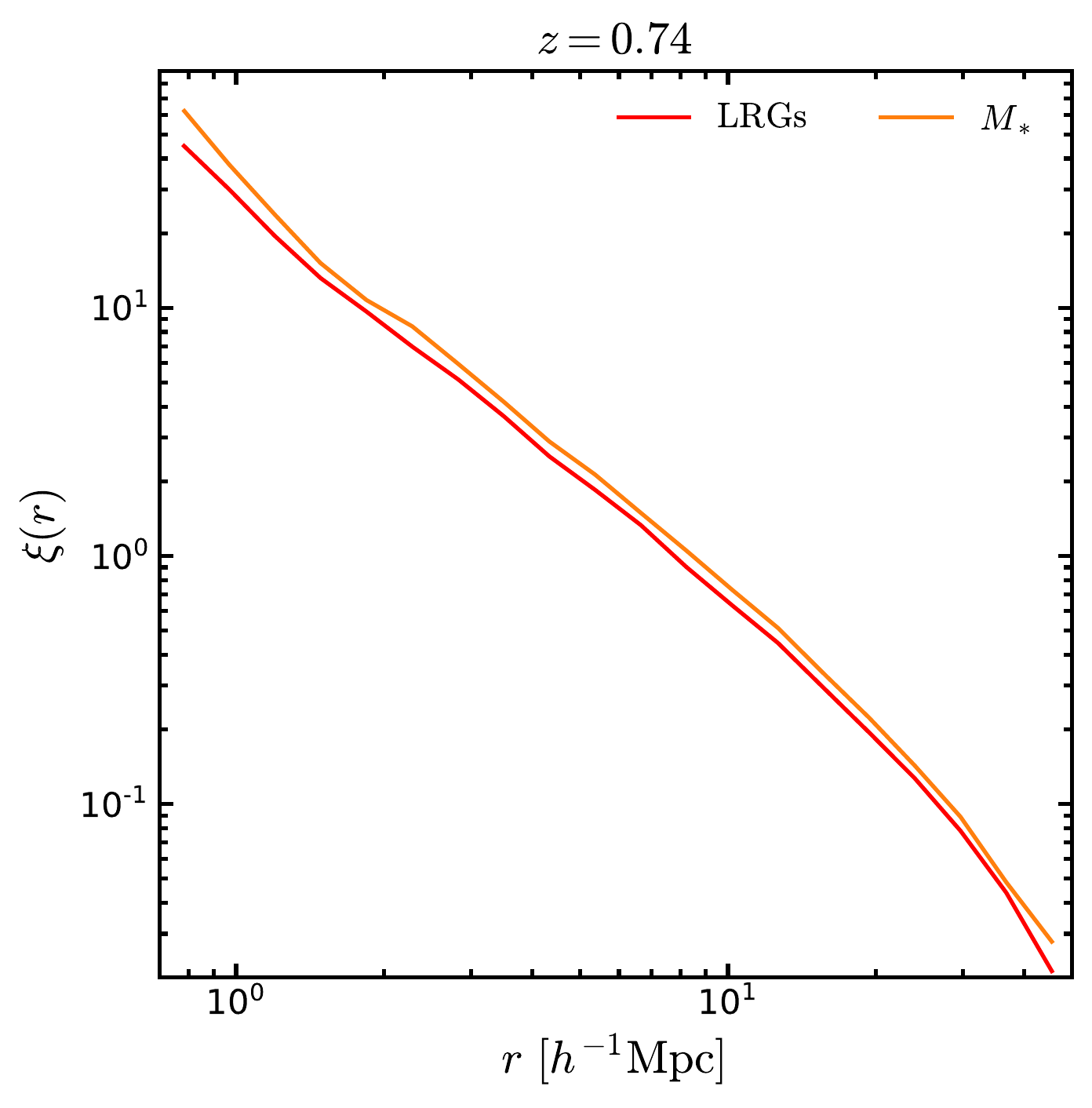}
\includegraphics[width=0.33\textwidth]{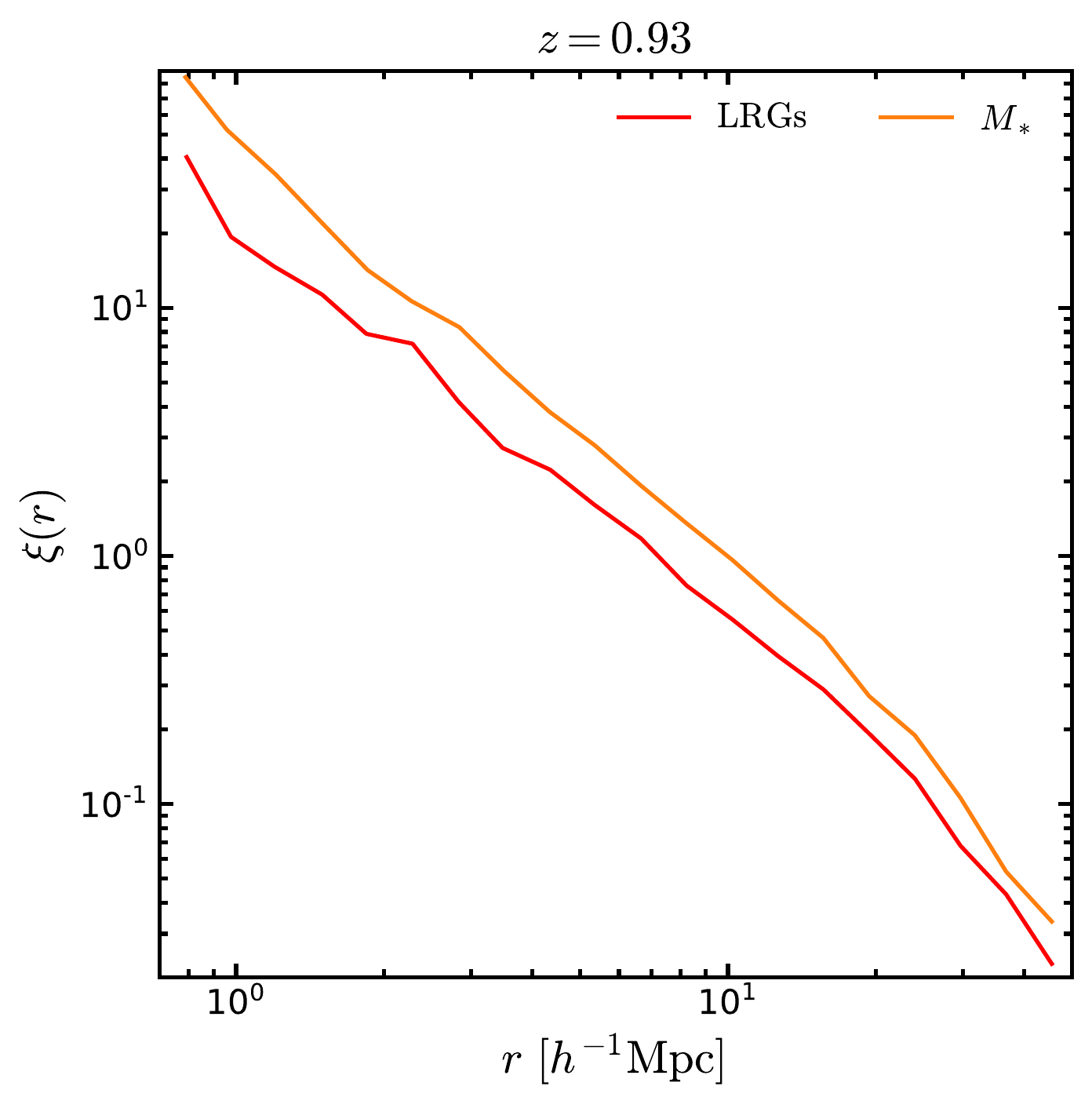}
\caption{The real-space galaxy correlation functions predicted by \Galform{} for LRGs (red lines) and the stellar mass selected sample (orange line) at $z=0.6$ ({\it left panel}), $z=0.74$ ({\it middle panel}) and $z=0.93$ ({\it right panel}).}
\label{fig:xir}
\end{figure*}

%---------------------------------------------------------------
\section{Galaxy Clustering}\label{sec:clustering}
%---------------------------------------------------------------
In previous sections we explored the impact of the DESI LRG colour-magnitude selection on galaxy statistics such as the stellar mass function and the luminosity functions at different wavelengths. We also presented predictions for which haloes and subhaloes contain LRGs. Here we take this a step further by investigating  the evolution of the clustering in configuration and Fourier space, in both real- and redshift-space. We measure the clustering from the simulations with the {\sc Nbodykit} toolkit \citep{Hand:2017pqn}. 

%---------------------------------------------------------------
\subsection{Galaxy clustering in the {\sc Pmill} and {\sc Glam} simulations}\label{sec:comparison}
%---------------------------------------------------------------
First, we present in Fig.~\ref{fig:xir} a comparison of the predicted real-space galaxy two-point correlation function for pair separations in the range $0.7 < r/[\Mpch] < 50$  at redshifts $0.6$, $0.74$ and $0.93$ for LRGs and the stellar mass selected sample. Since LRGs do not populate all of the most massive (sub)haloes, as seen in the lower panels of Fig.~\ref{fig:sub_hod}, the LRG sample is less biased than the stellar mass selected one, leading to a smaller clustering amplitude on all scales. We find a constant offset in the clustering amplitude of around $10\%$ between the samples at $z=0.6$ and $z=0.74$ on all scales. At higher redshifts, where the DESI-LRG colour-magnitude cuts have a bigger impact on which subhaloes host LRGs, we find that the difference in clustering amplitude increases to  $50\%$ on large scales, rising to $\sim 150\%$ on small scales. The larger difference on small-scales at $z=0.93$ is due to the abundance of satellite galaxies in the different galaxy samples; as seen in the upper right panel of Fig.~\ref{fig:sub_hod}, the stellar mass selected sample has a larger satellite fraction than the DESI-LRG sample. This comparison shows that selecting LRGs using stellar mass as a proxy for the full colour-magnitude selection leads to a significant change in the predicted clustering signal.
%--------- Figure --------------
\begin{figure*}
 \centering
\includegraphics[width=0.33\textwidth]{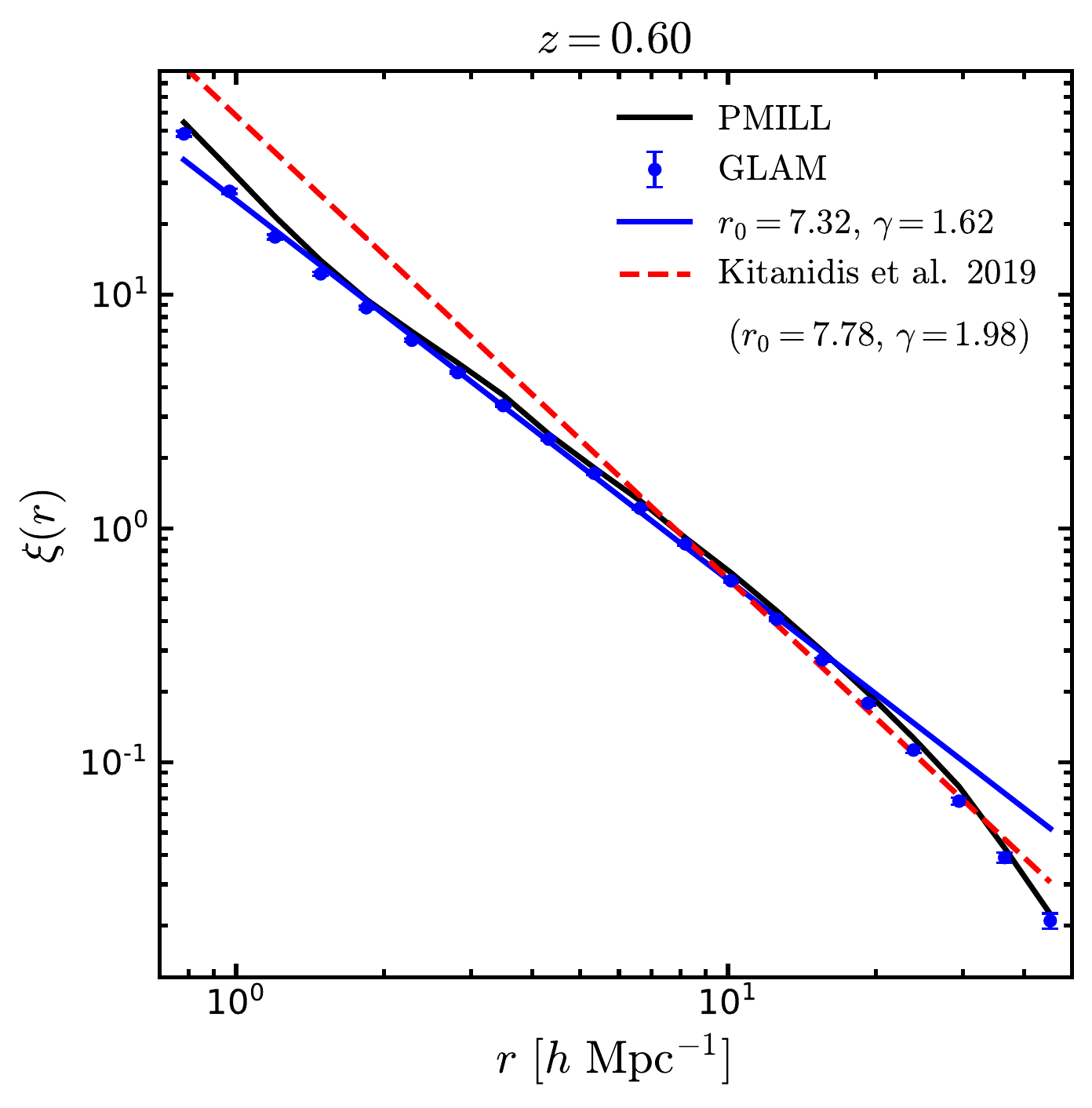}
\includegraphics[width=0.33\textwidth]{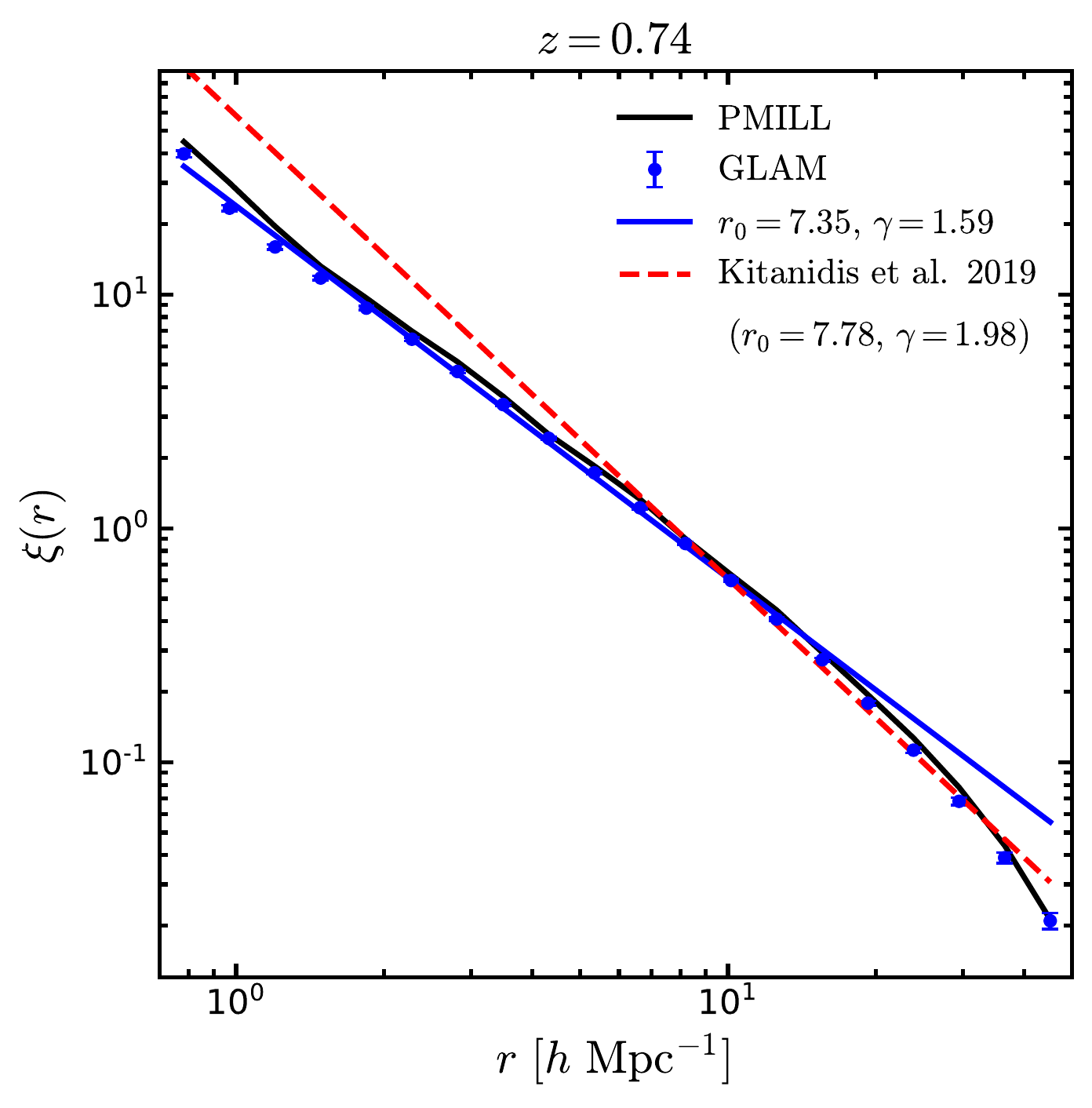}
\includegraphics[width=0.33\textwidth]{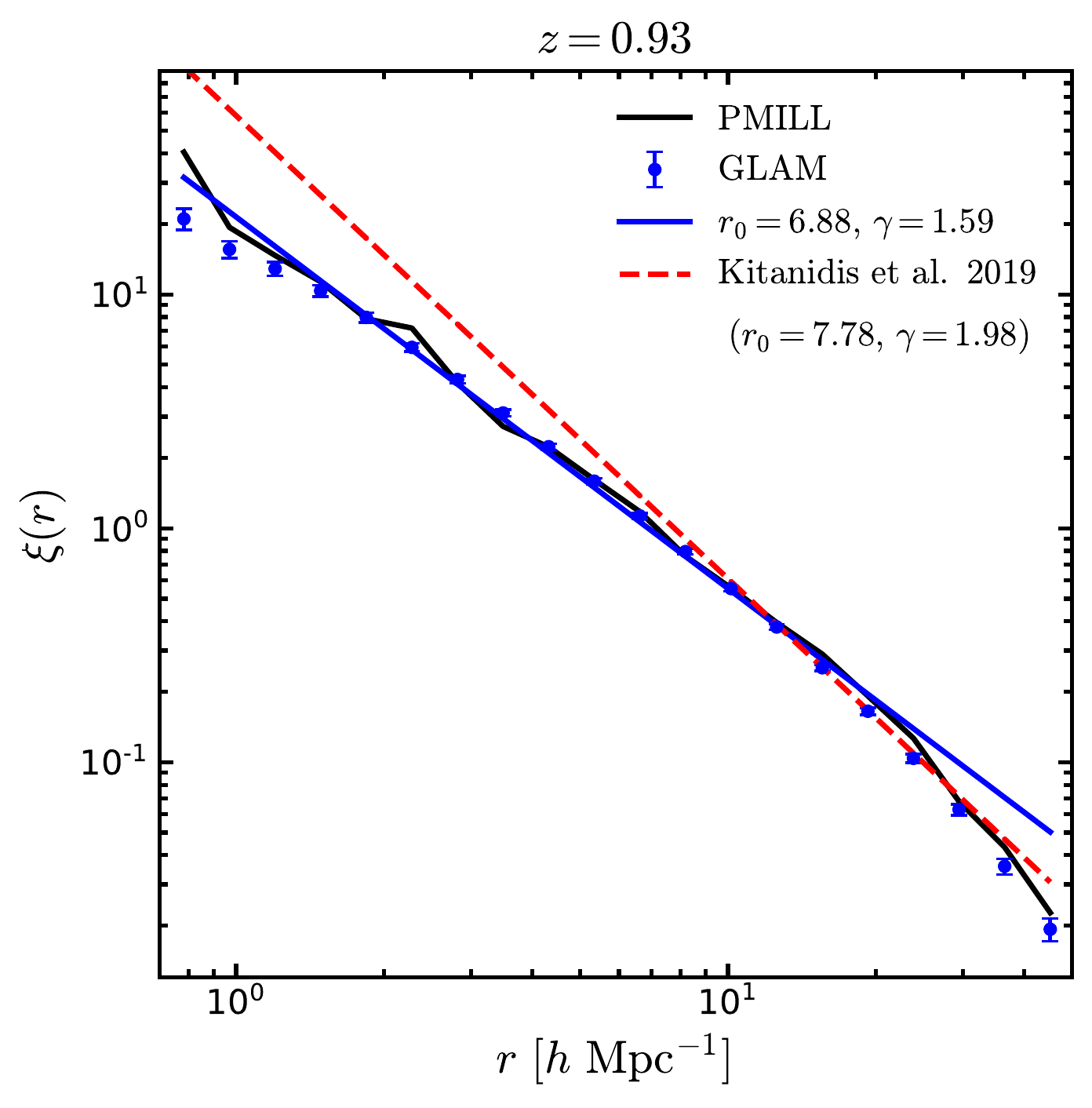}
\includegraphics[width=0.33\textwidth]{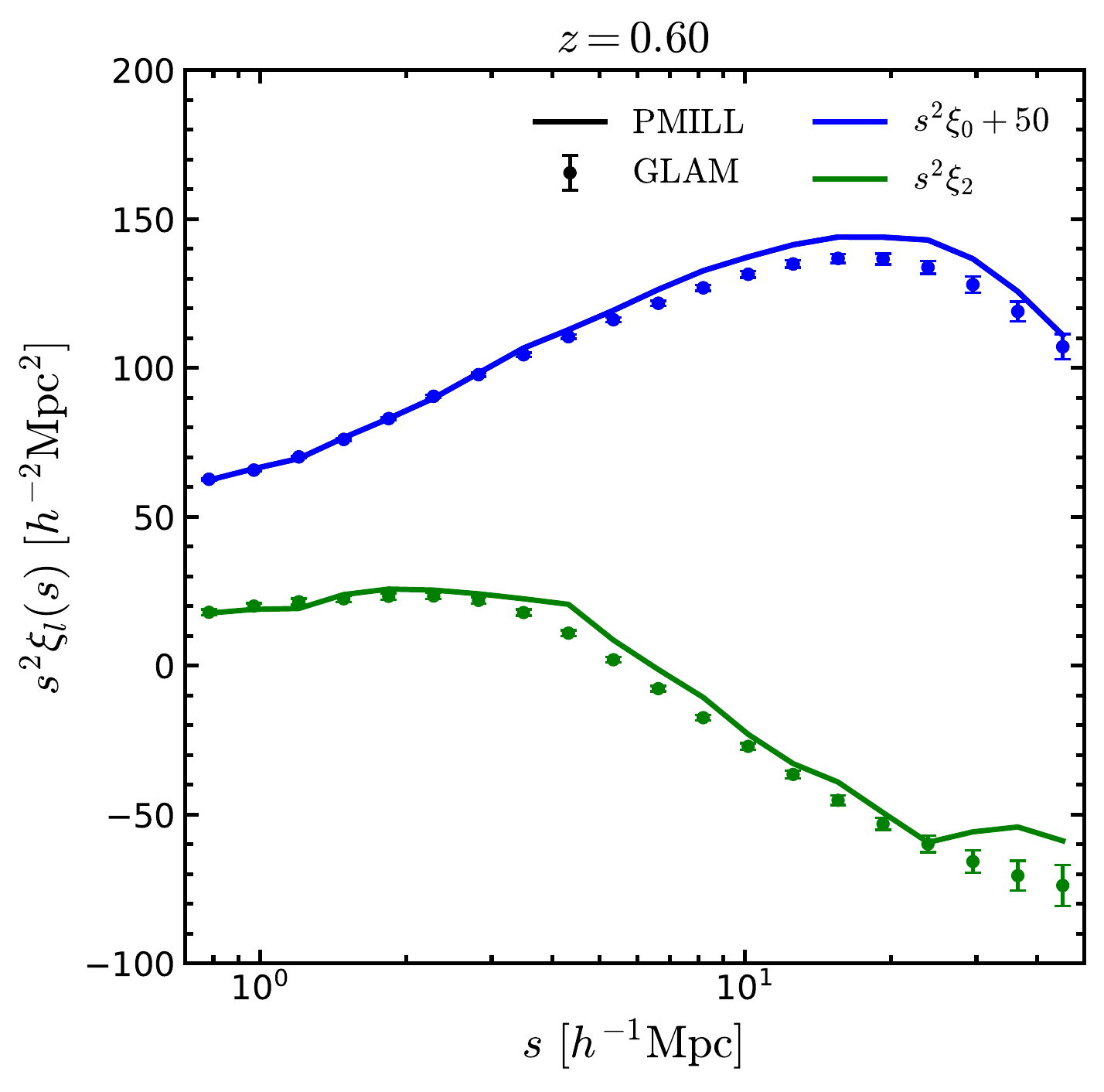}
\includegraphics[width=0.33\textwidth]{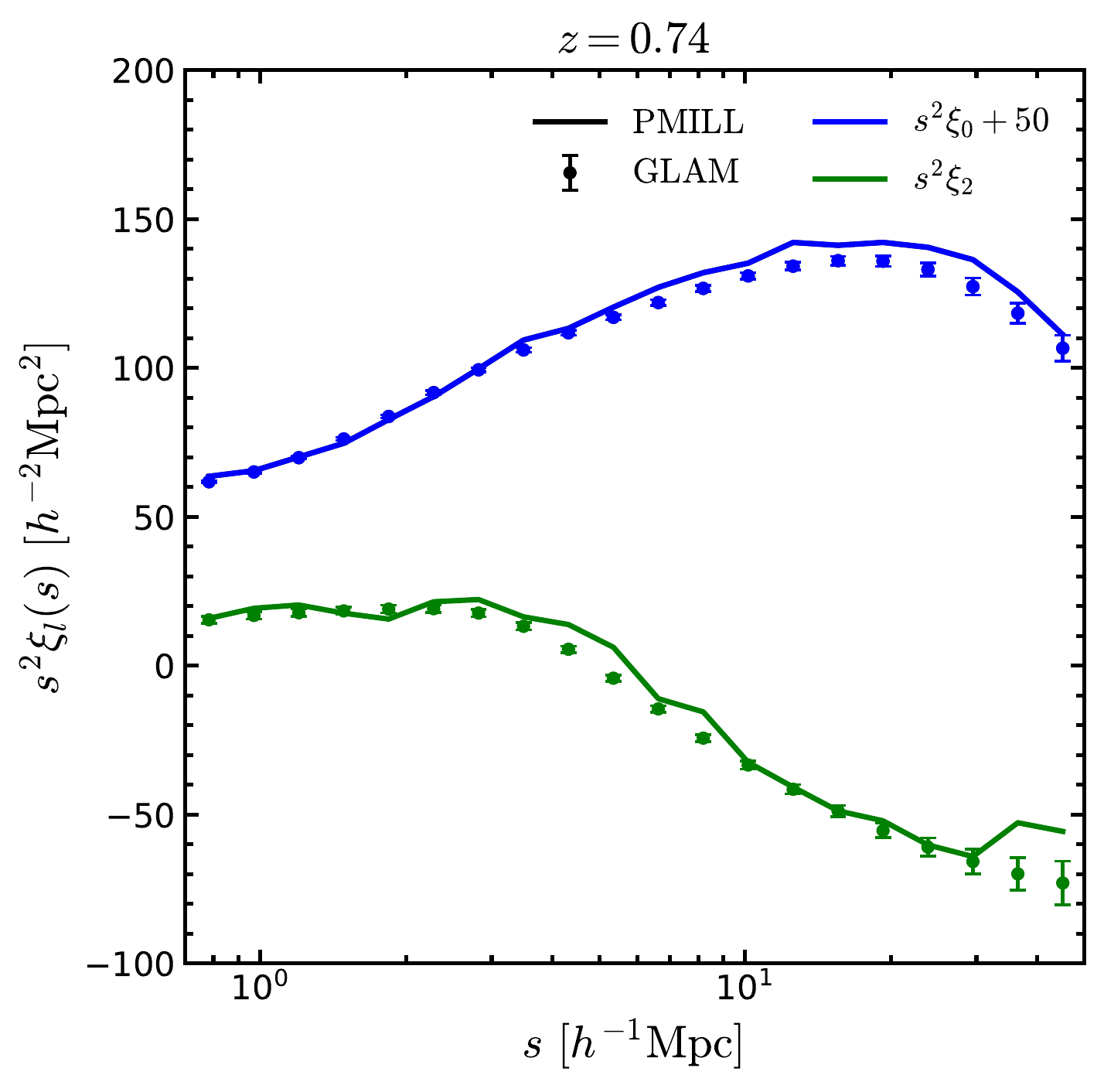}
\includegraphics[width=0.33\textwidth]{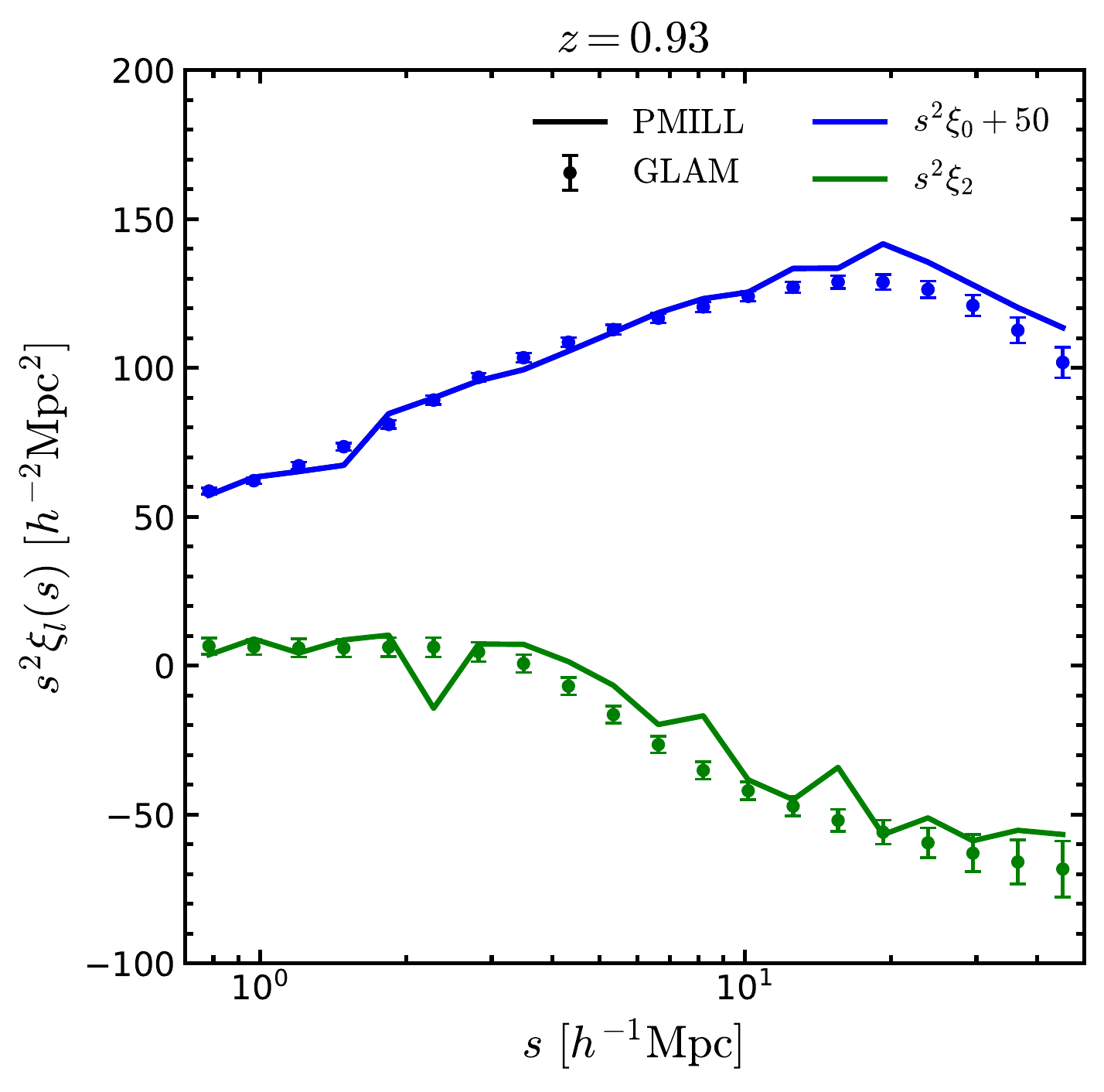}
\caption{{\it Upper panels:} Real-space galaxy correlation function of the \Galform{}-\Pmill{} LRGs (black lines) and the HOD-\glam{} LRGs (blue symbols with error bar). We also show the best fitting power-law form, $\xi(r) = (r/r_0)^{-\gamma}$, to the DESI-LRG measurements reported by \citet[][red dashed lines]{Kitanidis:2019rzi} and to our measurements (blue solid lines). {\it Lower panels:} Redshift-space monopole and quadropole moments of the correlation function for \Galform{}-\Pmill{} LRGs (solid lines) and \glam{}-HOD LRGs (symbols with error bar). Note that the monopole has been shifted upwards  for clarity. In the case of the \glam{}-HOD LRGs measurements, we show the mean and standard deviation over 1000 realisations. The measurements are made at $z=0.6$, $0.74$ and $0.93$, as labelled at the top of each panel.}
\label{fig:xi_glam}
\end{figure*}

%--------- Figure --------------
\begin{figure*}
\centering
\includegraphics[width=0.33\textwidth]{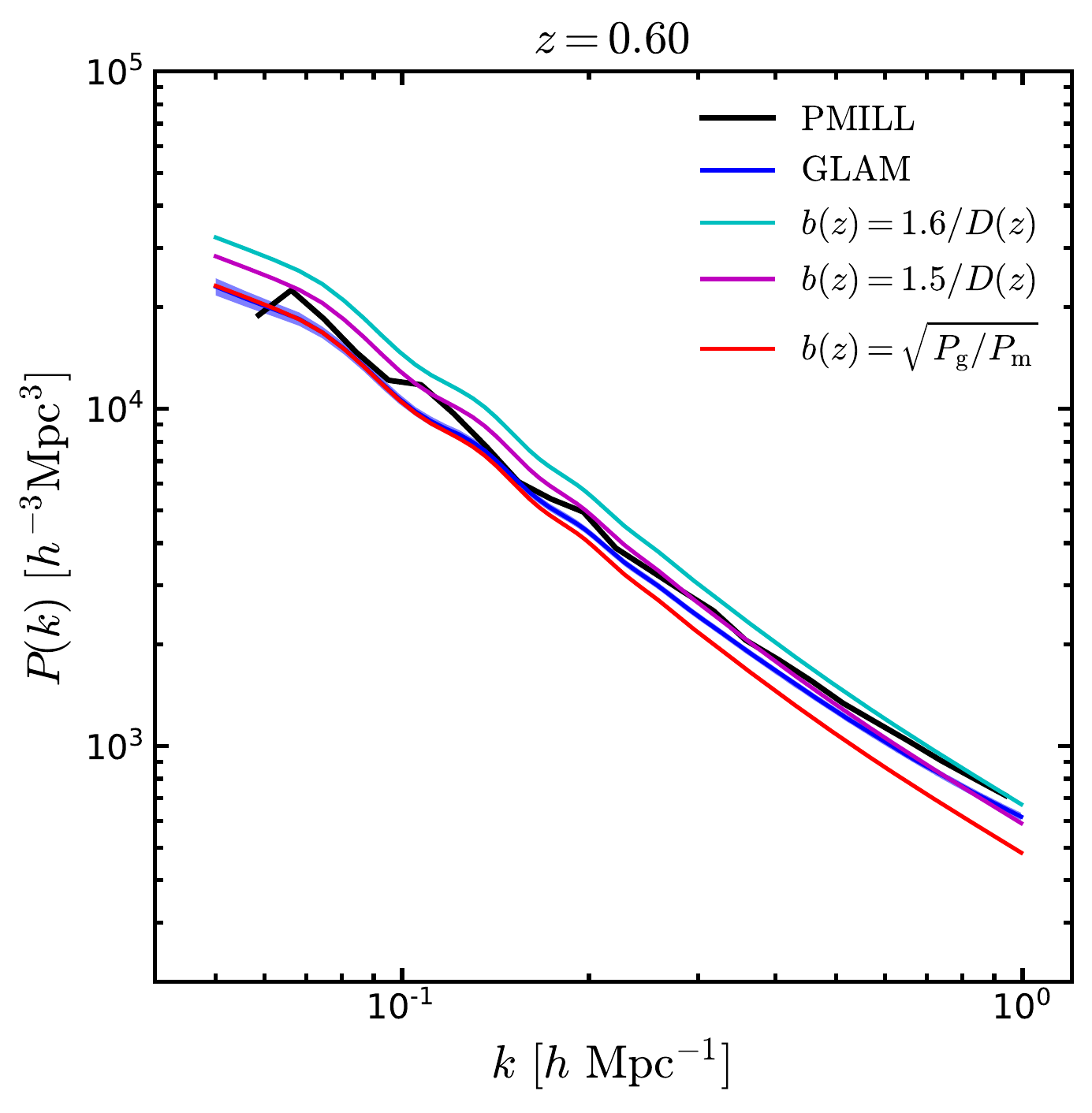}
\includegraphics[width=0.33\textwidth]{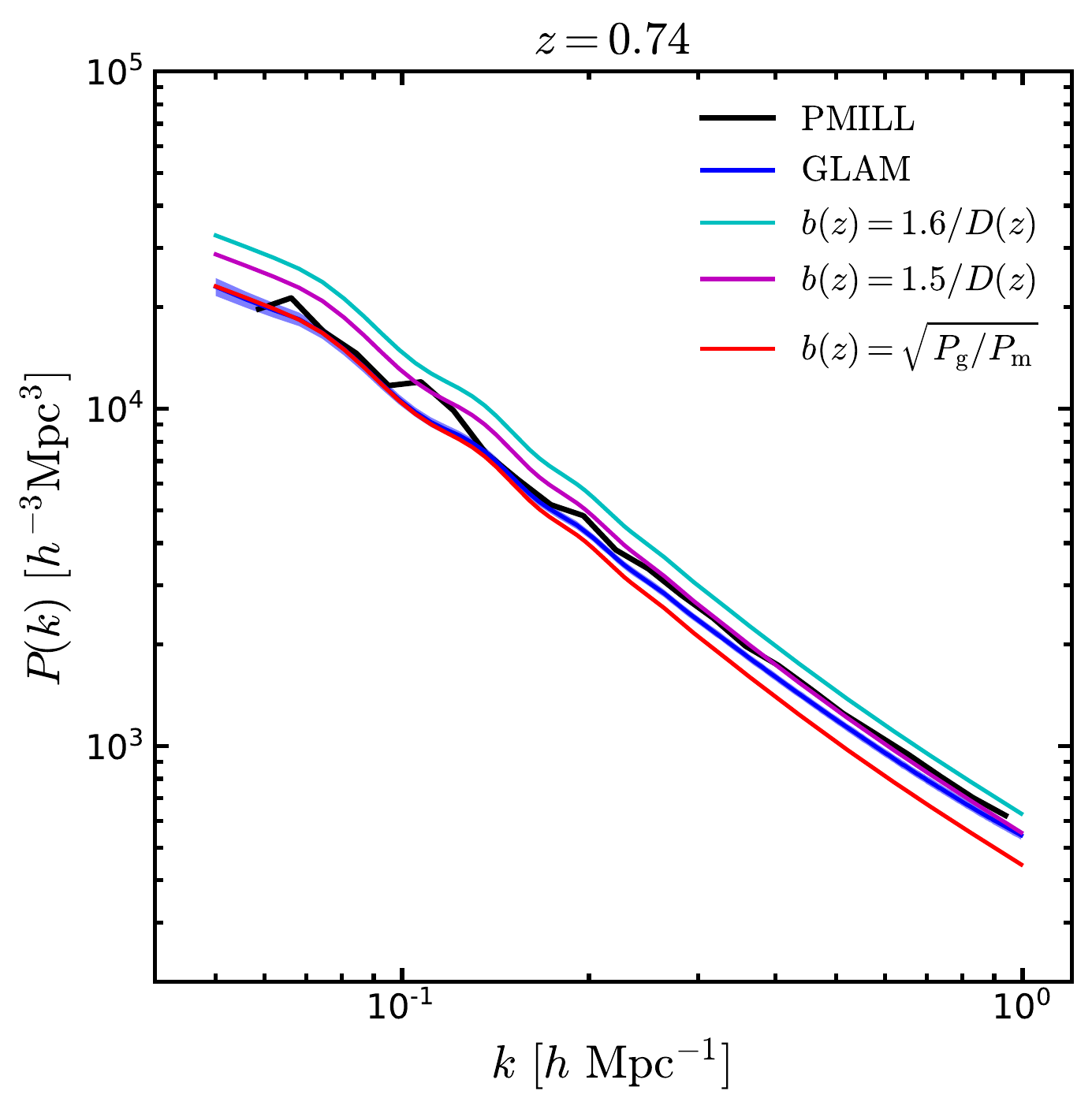}
\includegraphics[width=0.33\textwidth]{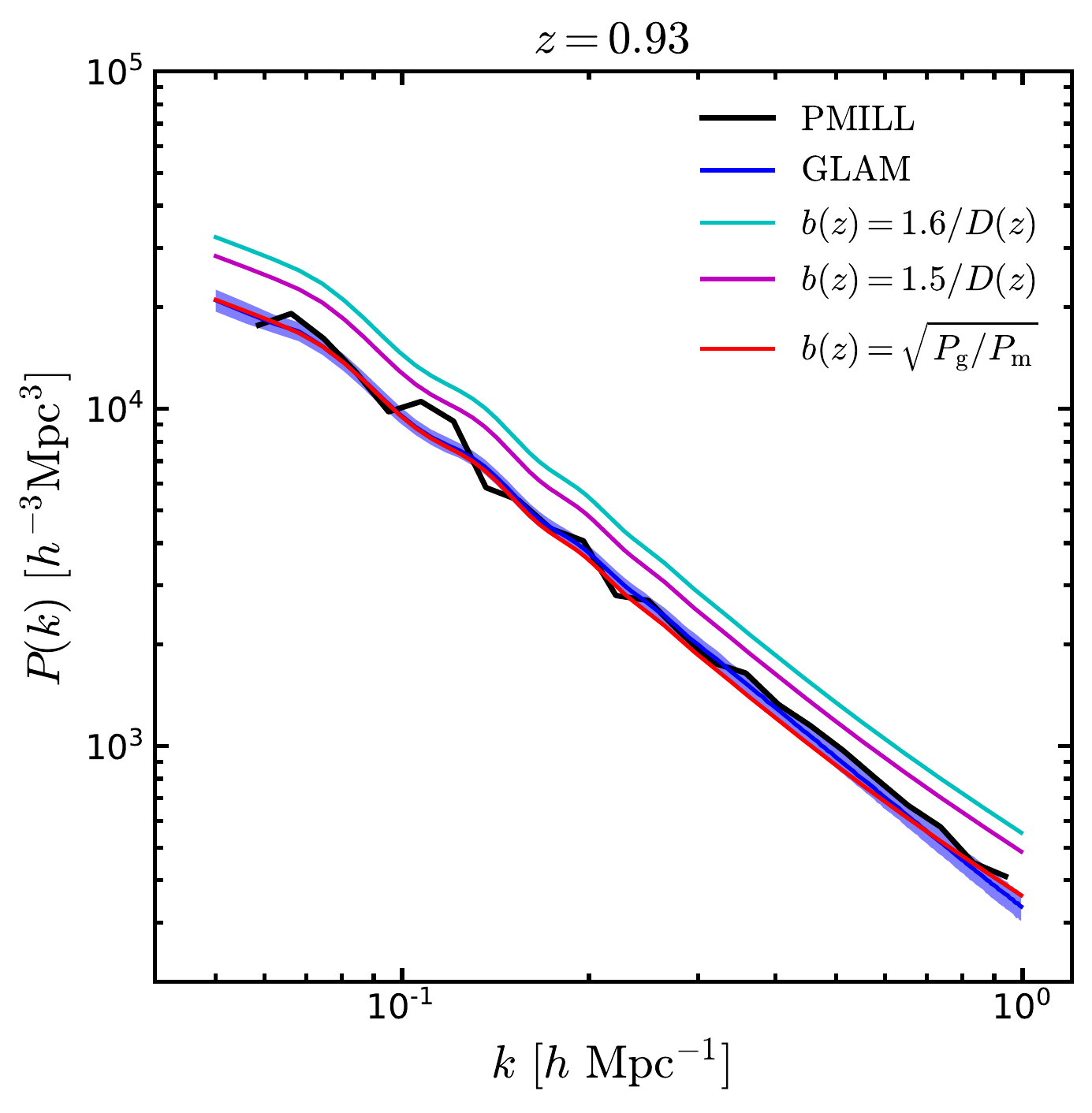}
\includegraphics[width=0.33\textwidth]{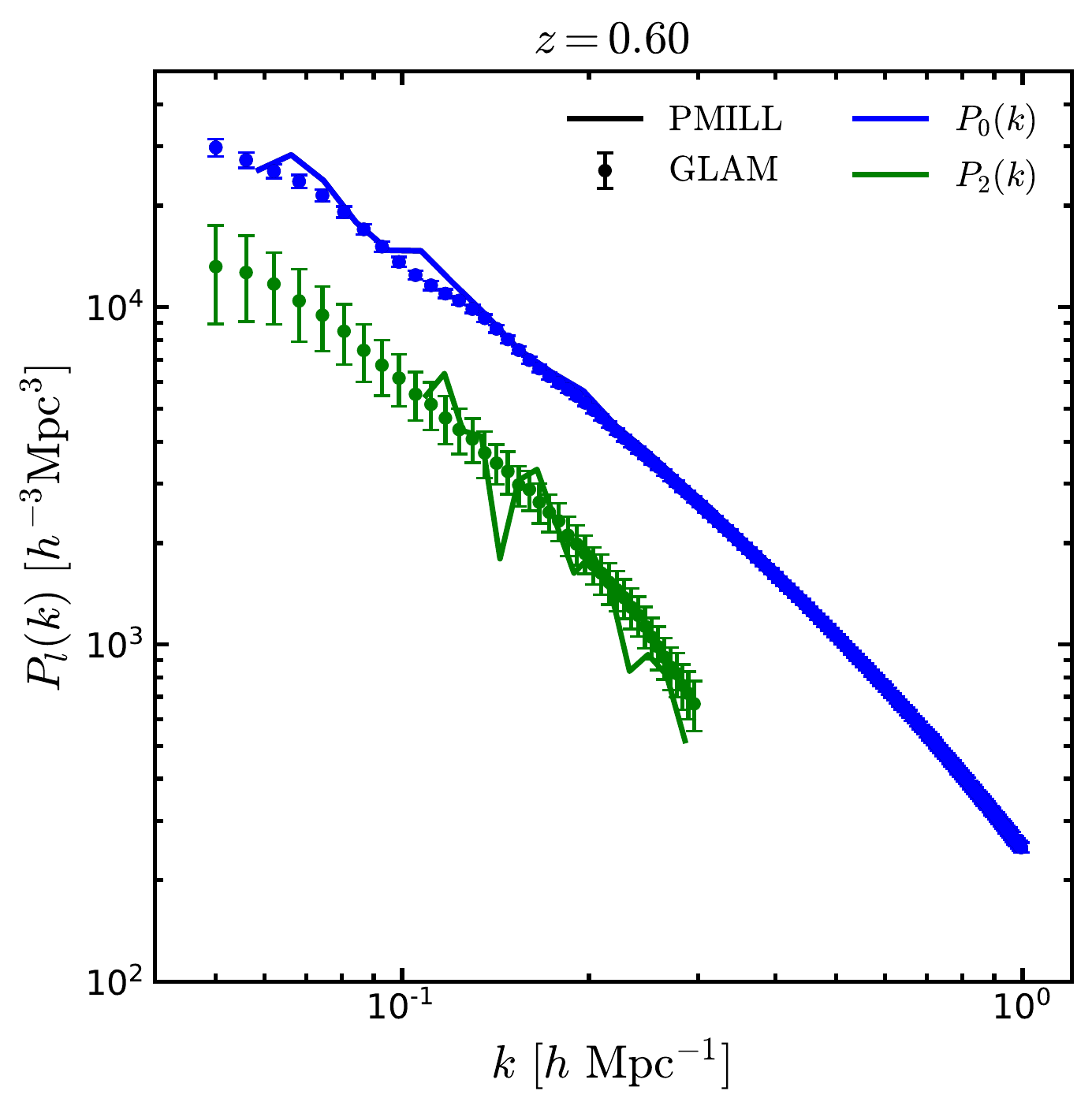}
\includegraphics[width=0.33\textwidth]{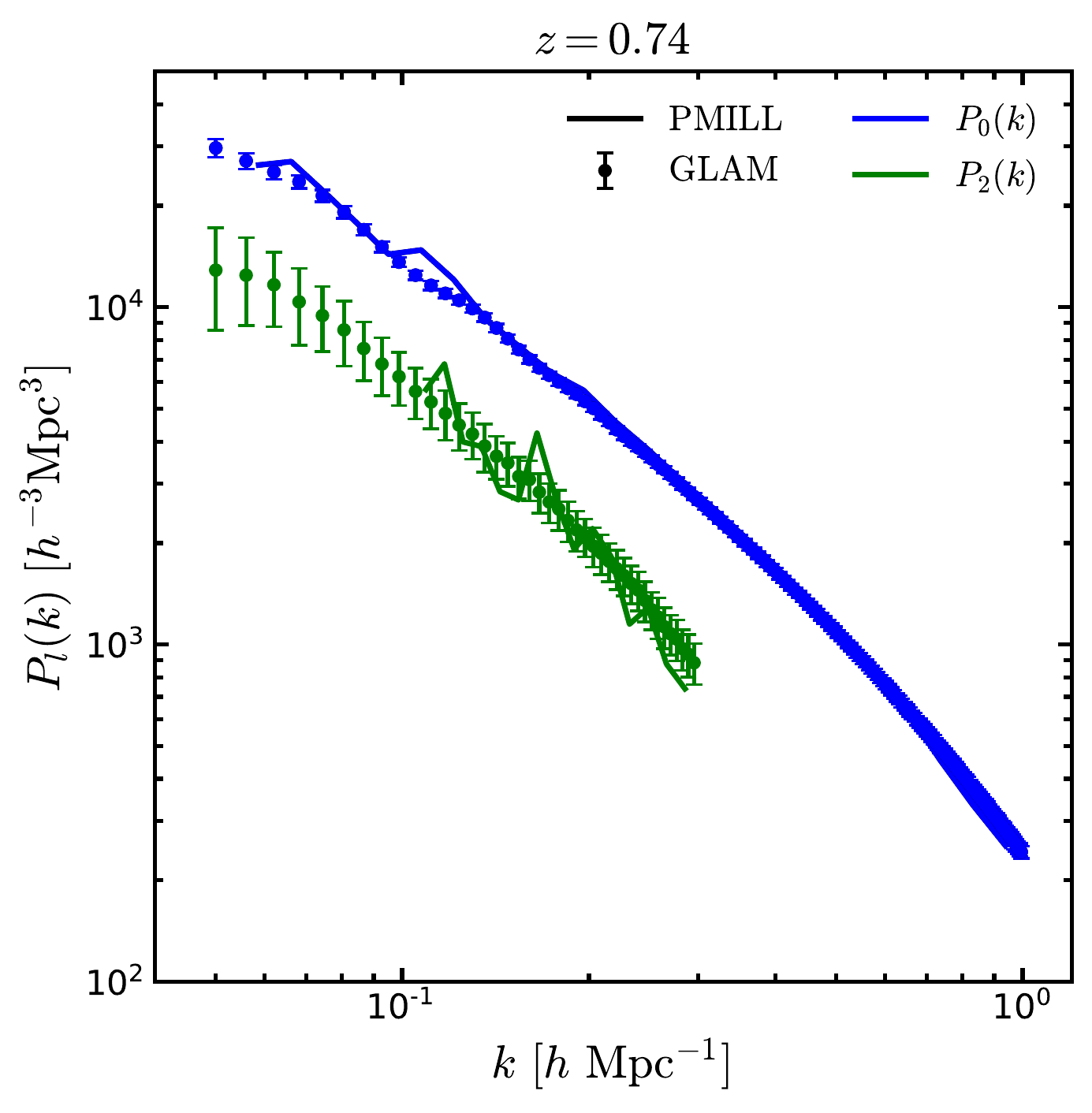}
\includegraphics[width=0.33\textwidth]{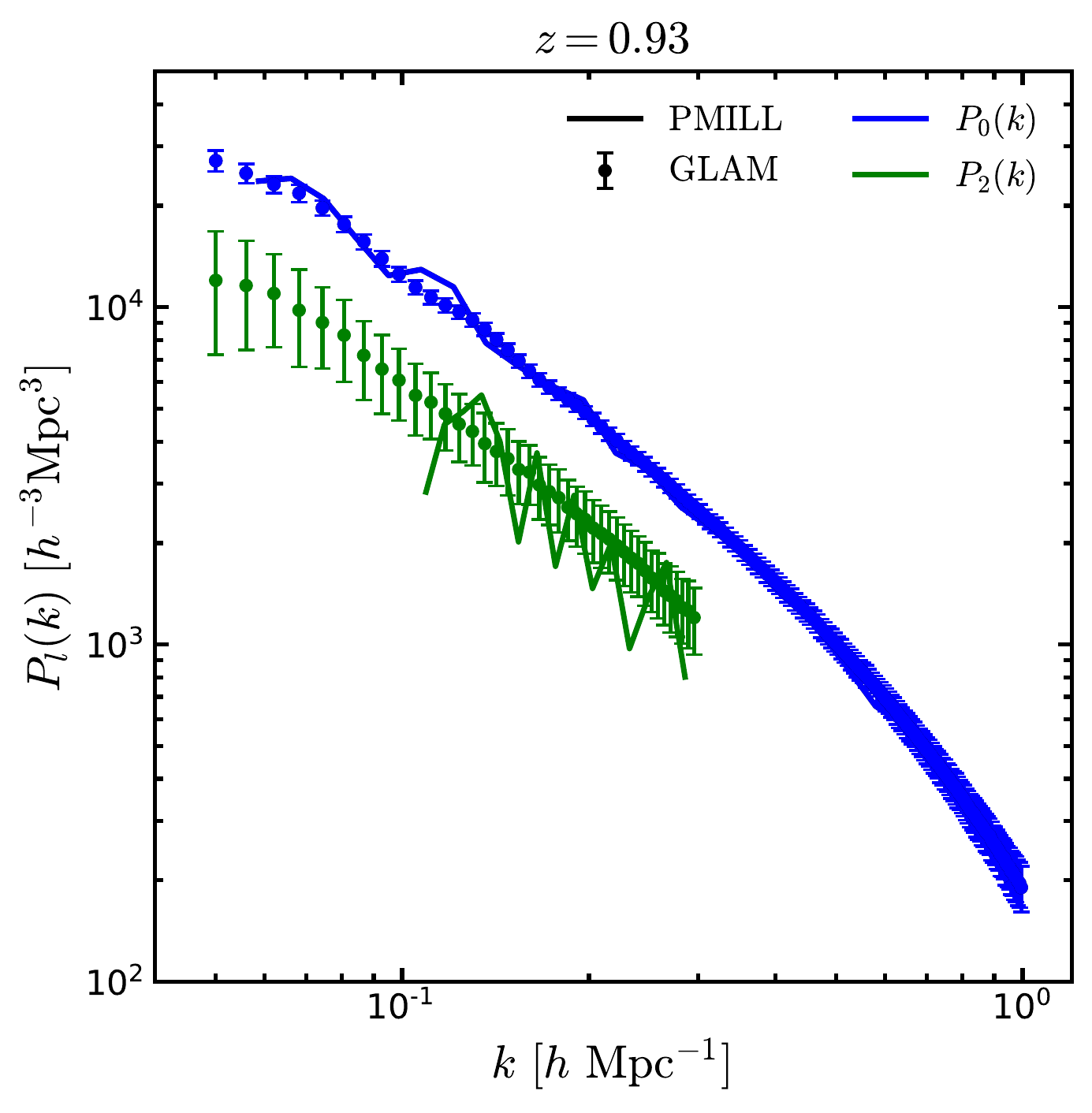}
\caption{{\it Upper panels:} Real-space galaxy power spectrum of the \Galform{}-\Pmill{} LRGs (black lines) and the \glam{}-HOD LRGs (blue lines, the shaded region represents the $1\sigma$ error over 1000 realisations), we also show the dark-matter power spectrum multiplied by the galaxy bias squared estimations of \citet[][cyan solid lines]{Kitanidis:2019rzi}, \citet[][magenta solid lines]{Rongpu:2020} and from Eq.~\eqref{eq:bias} (red solid line). {\it Lower panels:} Redshift-space monopole and quadropole moments of the power spectrum for \Galform{} LRGs (solid lines) and \glam{} LRGs (symbols with error bar). For the \glam{}-HOD LRGs measurements, we show the mean and standard deviation over 1000 realisations. The measurements are made at $z=0.6$, $0.74$ and $0.93$, as labelled at the top of each panel.}
\label{fig:Pk_glam}
\end{figure*}

As we mentioned before, one of our aims is to produce a large number of mock DESI LRG catalogues using the \glam{} code to give an accurate estimate of the galaxy clustering signal and its full covariance matrix of errors. For this reason, we populate our 1000 \glam{} simulations with LRGs using the tabulated HOD predicted by \Galform{} (see Fig.~\ref{fig:hod}), as explained below.

Since \Galform{} predicts an HOD for DESI-LRGs that does not appear to follow any of the popular parametric forms in the literature (see Appendix A of \citealt{Contreras:2013kr}), we bypass carrying out a fit altogether and instead use the tabulated model predictions for the HOD directly to populate \glam{} haloes with LRGs.
Hence, in order to populate a given \glam{} halo we interpolate between the HOD values predicted by \Galform{} to the \glam{} halo mass (see below for further details). In the case of the most massive haloes we extrapolate beyond the halo mass range of the HOD values; we do not have robust predictions for these haloes from the \Pmill{} simulation due to its smaller volume compared to the \glam{} boxes. This method was used recently by \cite{Merson:2019vfr}, where the authors extracted the HOD of ${\rm H}\alpha$ galaxies from the {\sc Galacticus} SAM catalogue \citep{Benson:2010kx,Merson:2017efv}, and used this to populate the Millennium-MXXL halo light-cone from \cite{Smith:2017tzz}.

In detail our HOD method is as follows. We assign a central galaxy to a \glam{} halo if $\lla N_c \rra > \mathcal{U}(0,1)$, where $\lla N_c \rra$ is the mean number of central galaxies that could be found in a \glam{} halo and $\mathcal{U}(0,1)$ is a uniform random number between $0$ and $1$. Recall that the \Galform{} predictions for the HOD of central galaxy LRGs never reach unity. We place the central galaxy at the centre of mass of the host halo, and give it the velocity of the centre of mass. The number of satellite galaxies is drawn from a Poisson distribution with mean equal to $\lla N_s \rra$, as derived from the tabulated HOD predicted using \Galform{}. Satellite galaxies are radially distributed within the virial radius, $(0 < r < R_{\rm vir})$, following a Navarro-Frenk-White (NFW) density profile \citep{Navarro:1995iw,Navarro:1996gj}, with a uniform angular distribution. The satellite is assigned a velocity that is made up of the halo velocity plus a perturbation along the $x$, $y$ and $z$ coordinates drawn from a Gaussian distribution with variance equal to the 1D velocity dispersion of the host halo.

We measure the real- and redshift-space clustering in configuration and Fourier space from the \glam{}-HOD catalogues and compare these with their \Pmill{} counterparts to corroborate the precision of our method. 
In addition, the real-space clustering measurements provide us a relation between the distribution of galaxies and the underlying dark-matter density field via the galaxy bias \citep{Peebles:1980}. The galaxy bias is directly measured from our \glam{} LRG mocks as
\begin{equation}\label{eq:bias}
b(k,z) = \sqrt{\frac{P_{\rm g}(k,z)}{P_{\rm m}(k,z)}} \quad {\rm or} \quad 
b(r,z) = \sqrt{\frac{\xi_{\rm g}(r,z)}{\xi_{\rm m}(r,z)}}\,,
\end{equation}
where $P_{\rm g}(k,z)$ $(\xi_{\rm g}(r,z))$ and $P_{\rm m}(k,z)$ $(\xi_{\rm g}(r,z))$ are the real-space galaxy and dark matter power spectra (correlation functions) at a given redshift, respectively. We tried both approaches to estimating the bias and found consistent answers, $b(z)=1.84,\,1.96,\,2.06$ at $z=0.6$, $0.74$ and $0.93$, respectively. The DESI-like LRG bias has been estimated from the measured angular power spectrum and from the halo model of the photo-$z$ LRGs giving the following relations, $b(z)=1.6/D(z)$  \citep{Kitanidis:2019rzi}  and $b(z)=1.5/D(z)$ \citep{Rongpu:2020}. Note these relations are slightly different to the value of  $b(z)=1.7/D(z)$ reported in \citet{DESI:2016zmz}, where $D(z)$ is the linear growth factor at redshift $z$, with $D(z=0) = 1$.
For the cosmological parameters used in the \Pmill{} simulation, the linear growth factor is $D(z=0.6)=0.73$, $D(z=0.74)=0.69$ and $D(z=0.93) = 0.63$, which means that the values we recover for the bias are slightly lower than those inferred from the observations, more similar to $b(z) = (1.3-1.4)/D(z)$.

We use the distant-observer approximation to shift the positions of galaxies from real- to redshift-space, treating the $z$-axis as the line of sight,
\begin{equation}\label{eq:coord_z}
{\bf s} = {\bf r} + \frac{(1+z)v_z}{H(z)}\hat{e}_z\,,
\end{equation}
where ${\bf r}$ is the coordinate vector in real space, ${\bf s}$ is the equivalent of this in redshift-space, and $z$ is the redshift of the simulation snapshot used to generate the galaxy catalogue. $H(z)$ is the Hubble parameter, $v_z$ and $\hat{e}_z$ are the components of the velocity and the unit vector along the $z$-direction.
%--------- Figure --------------
\begin{figure*}
 \centering
\includegraphics[width=0.33\textwidth]{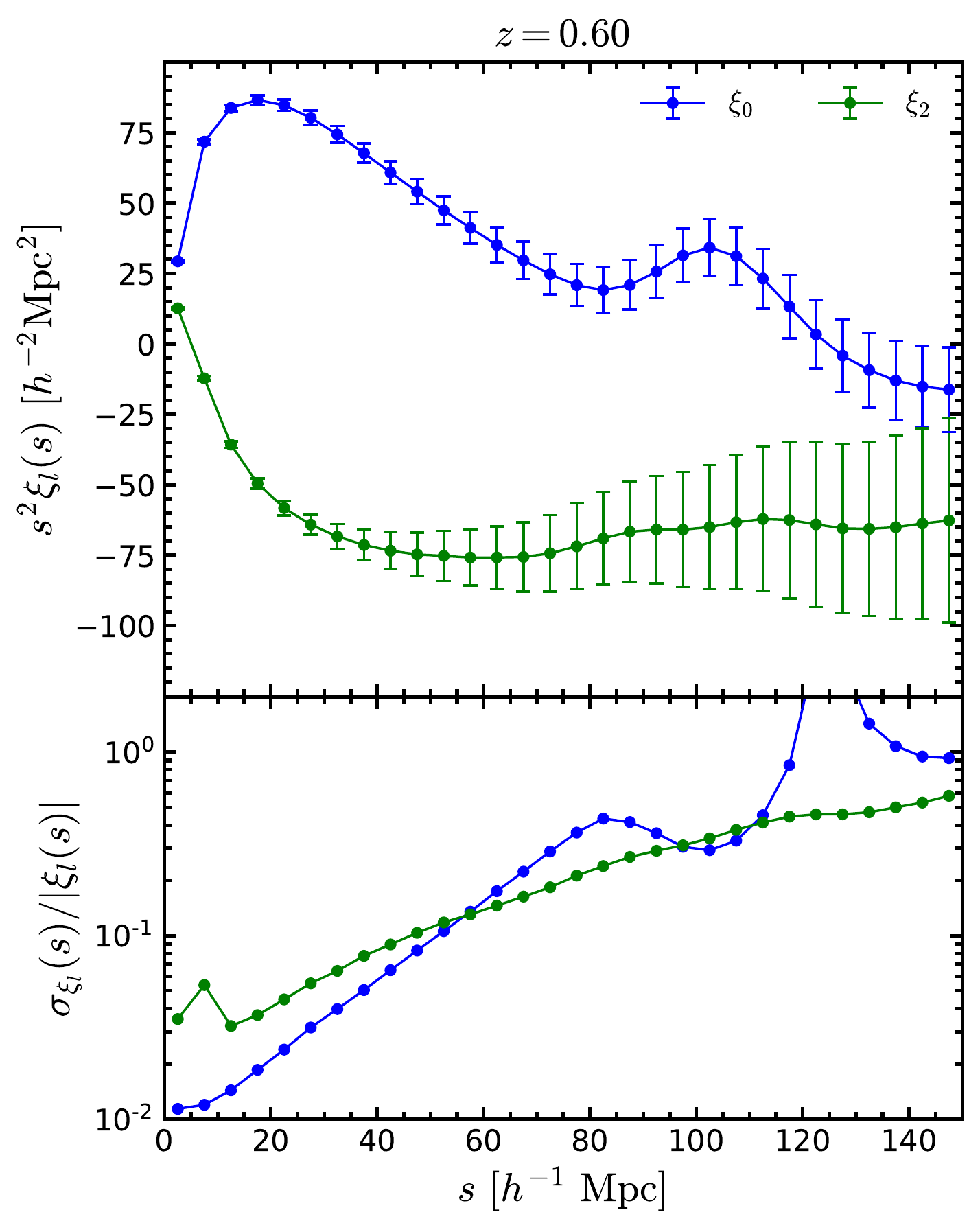}
\includegraphics[width=0.33\textwidth]{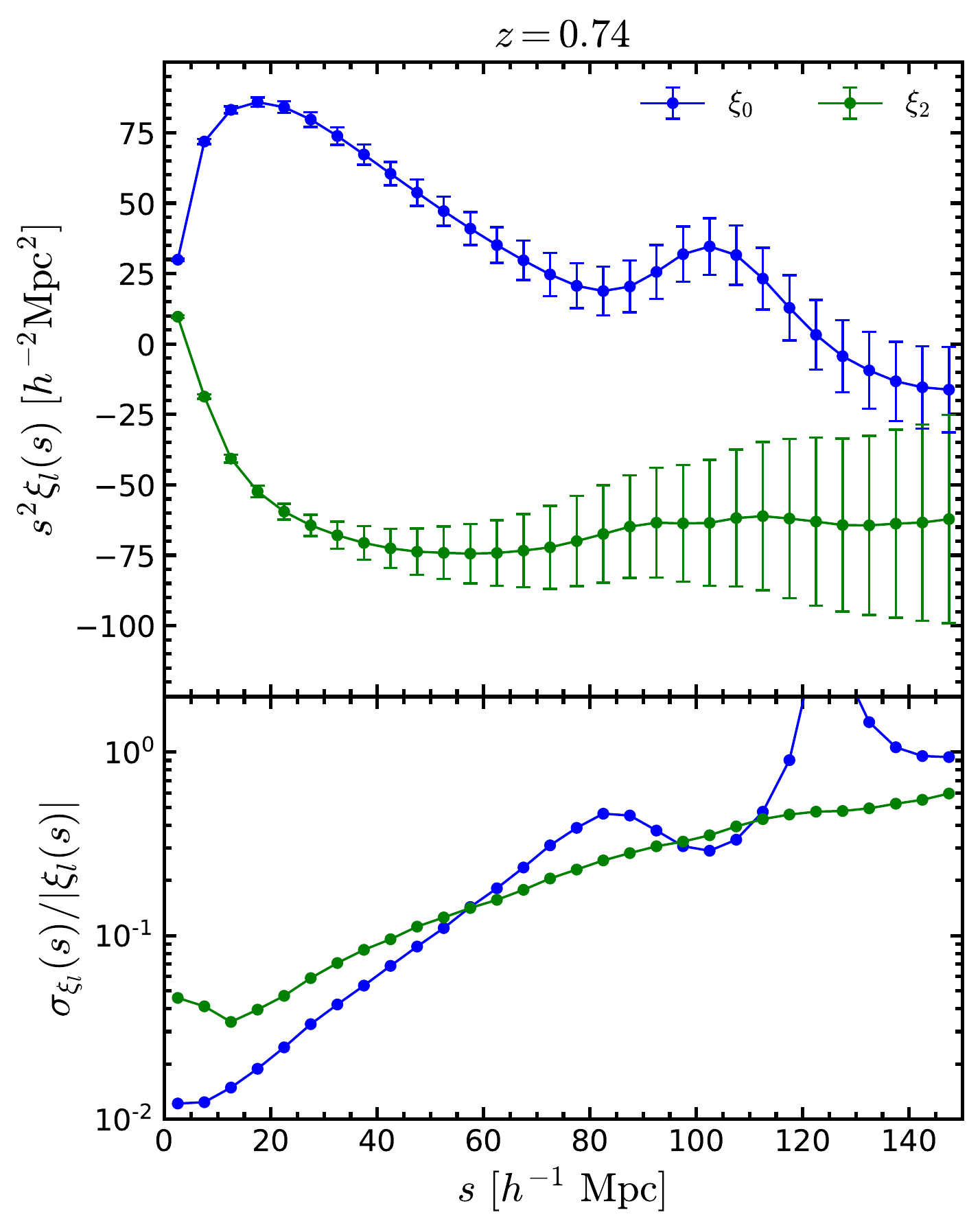}
\includegraphics[width=0.33\textwidth]{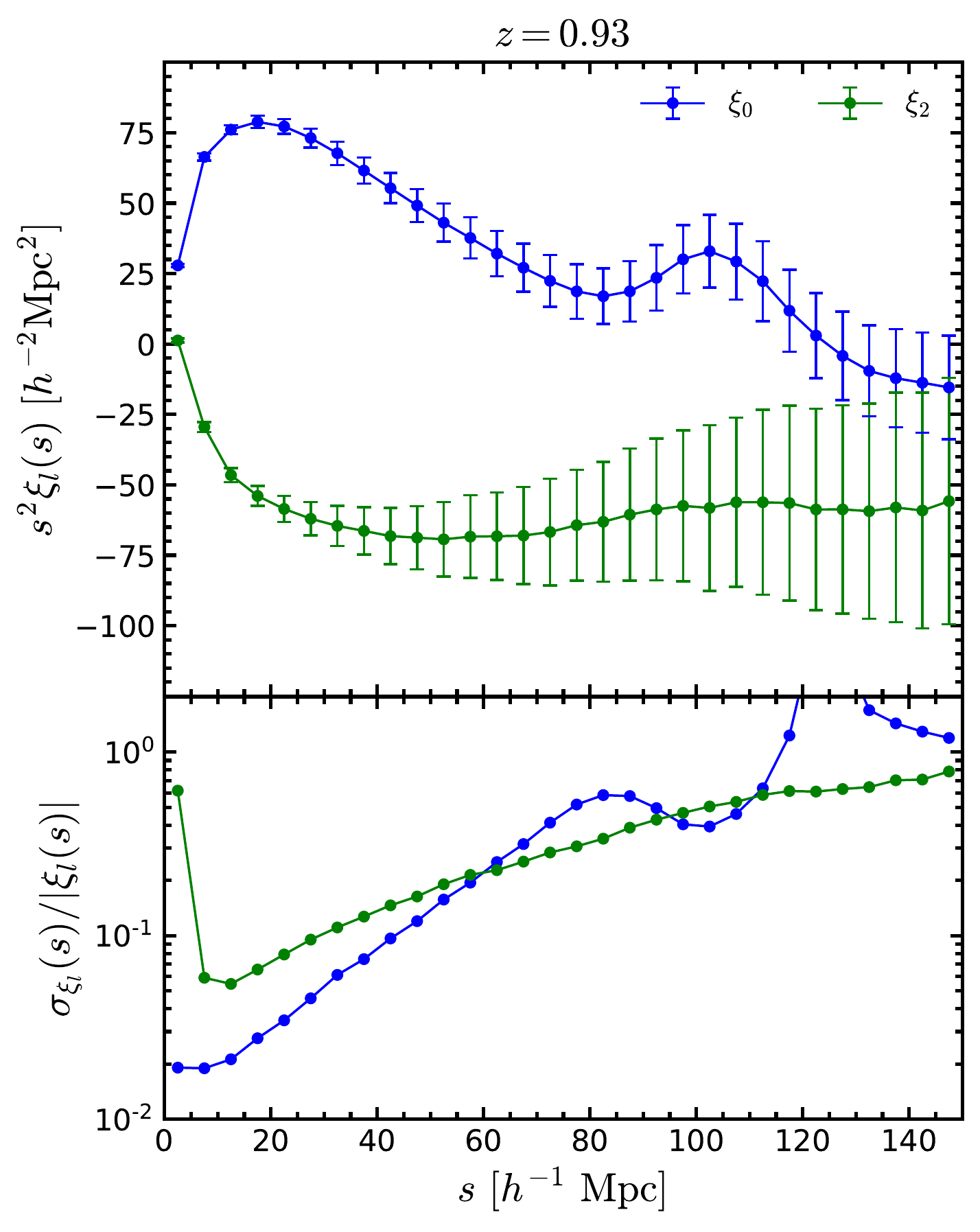}
\includegraphics[width=0.33\textwidth]{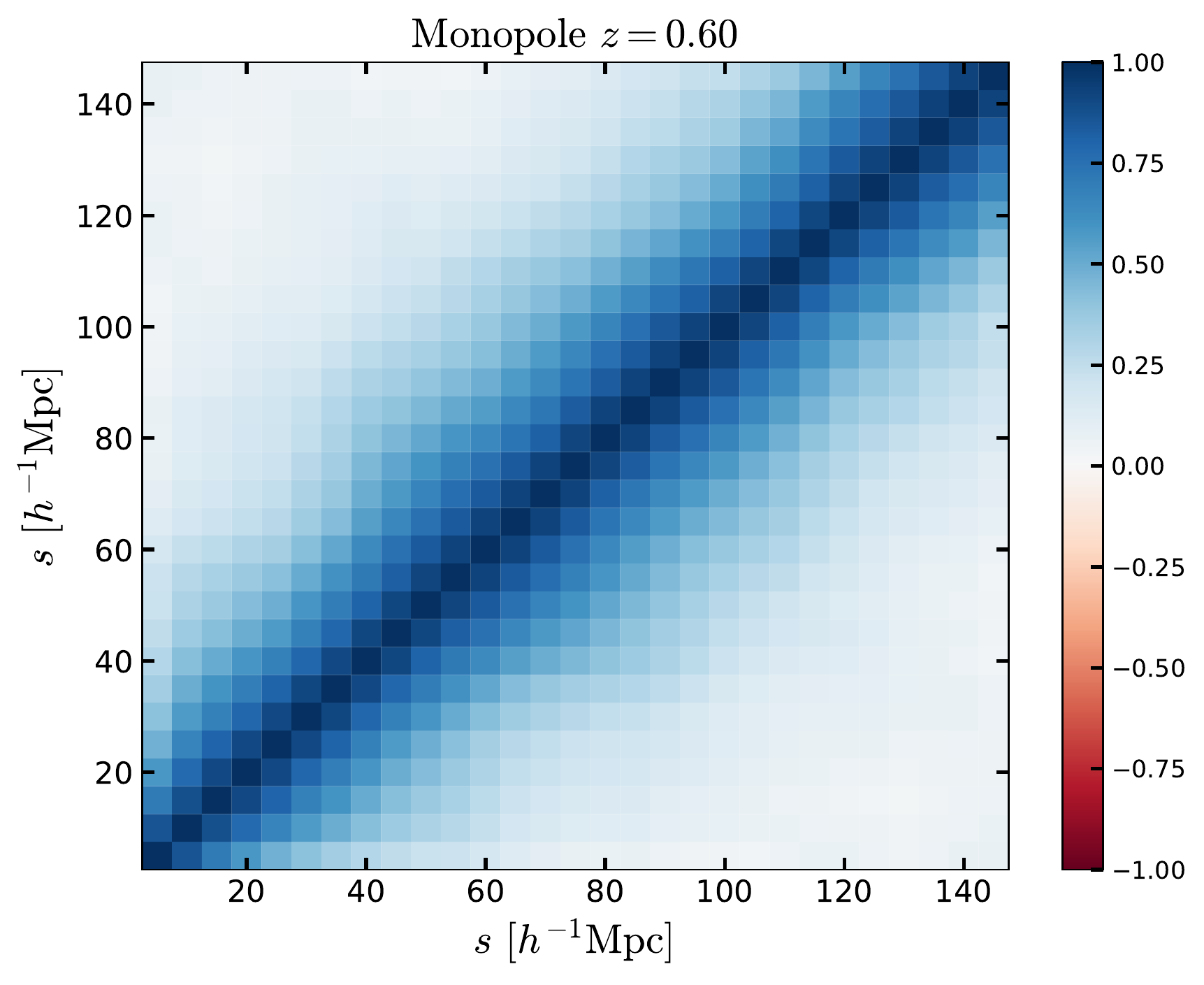}
\includegraphics[width=0.33\textwidth]{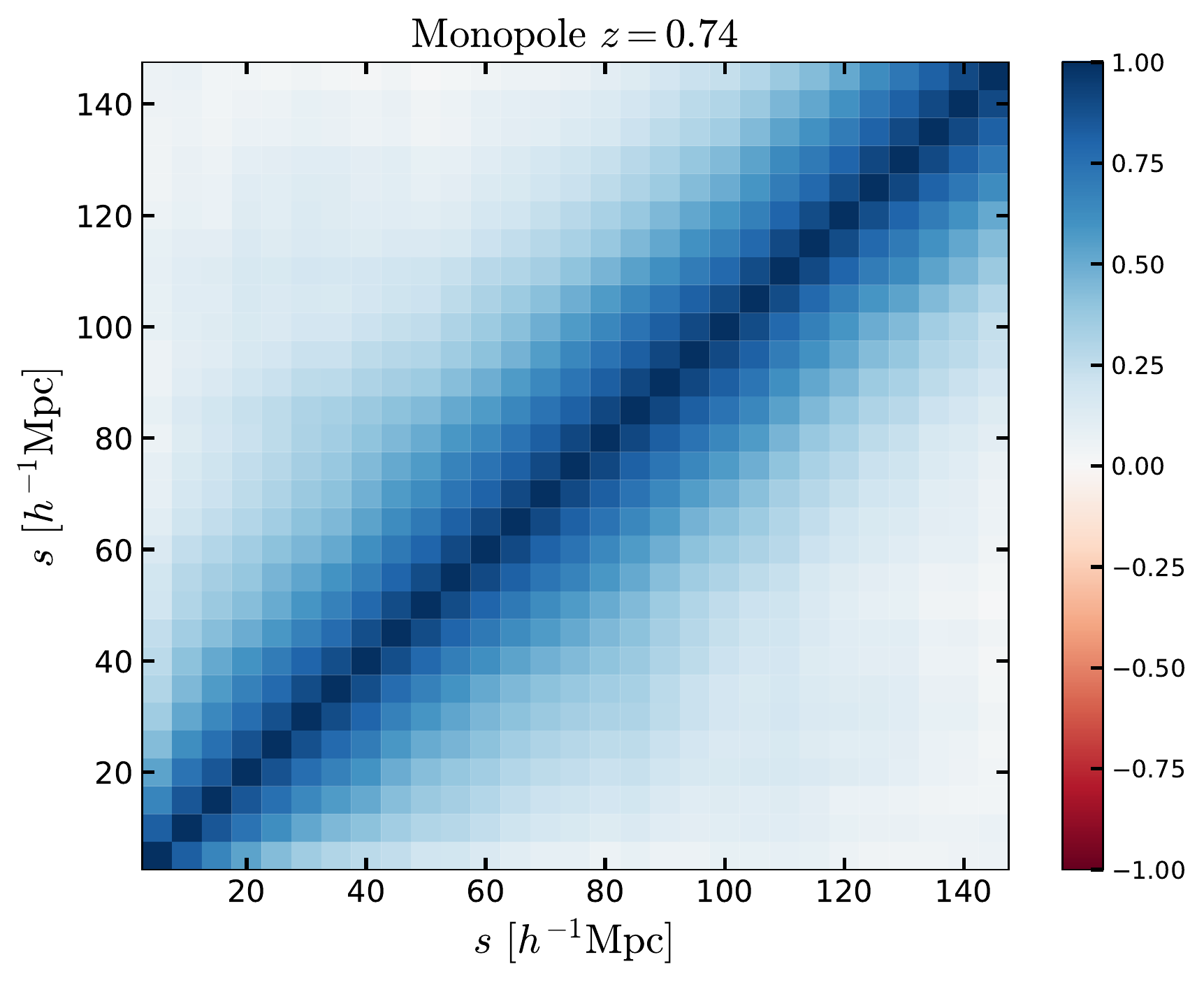}
\includegraphics[width=0.33\textwidth]{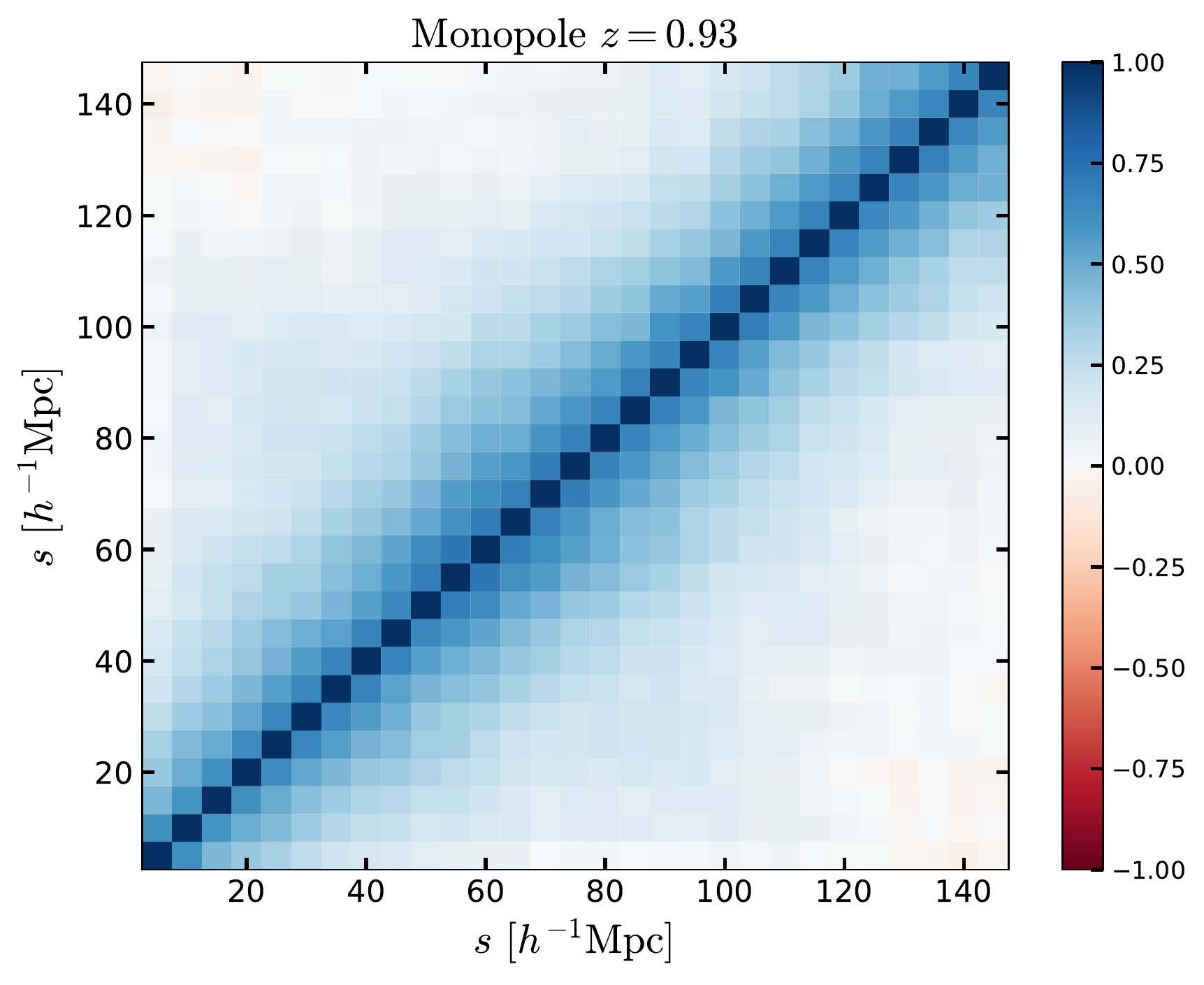}
\includegraphics[width=0.33\textwidth]{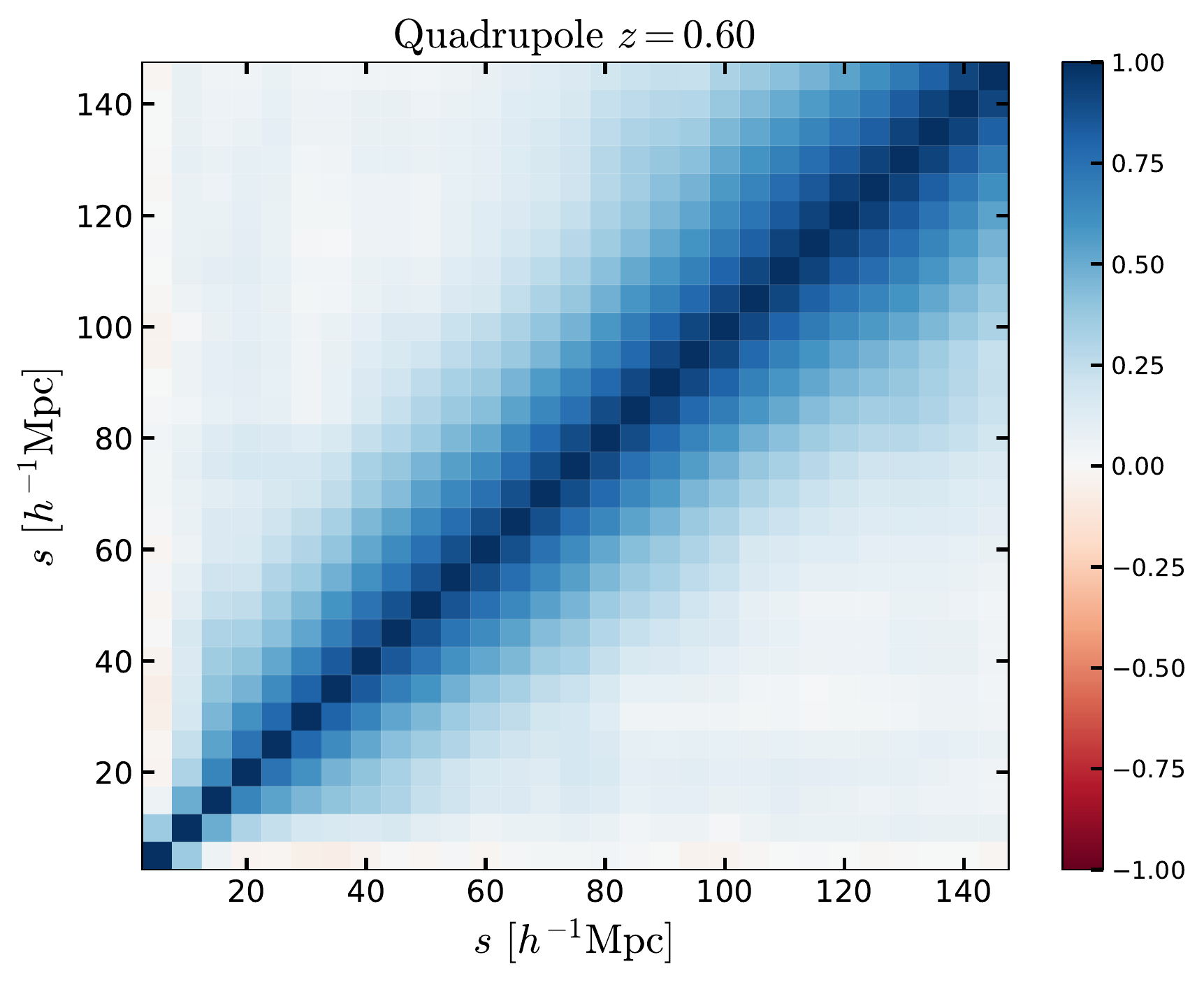}
\includegraphics[width=0.33\textwidth]{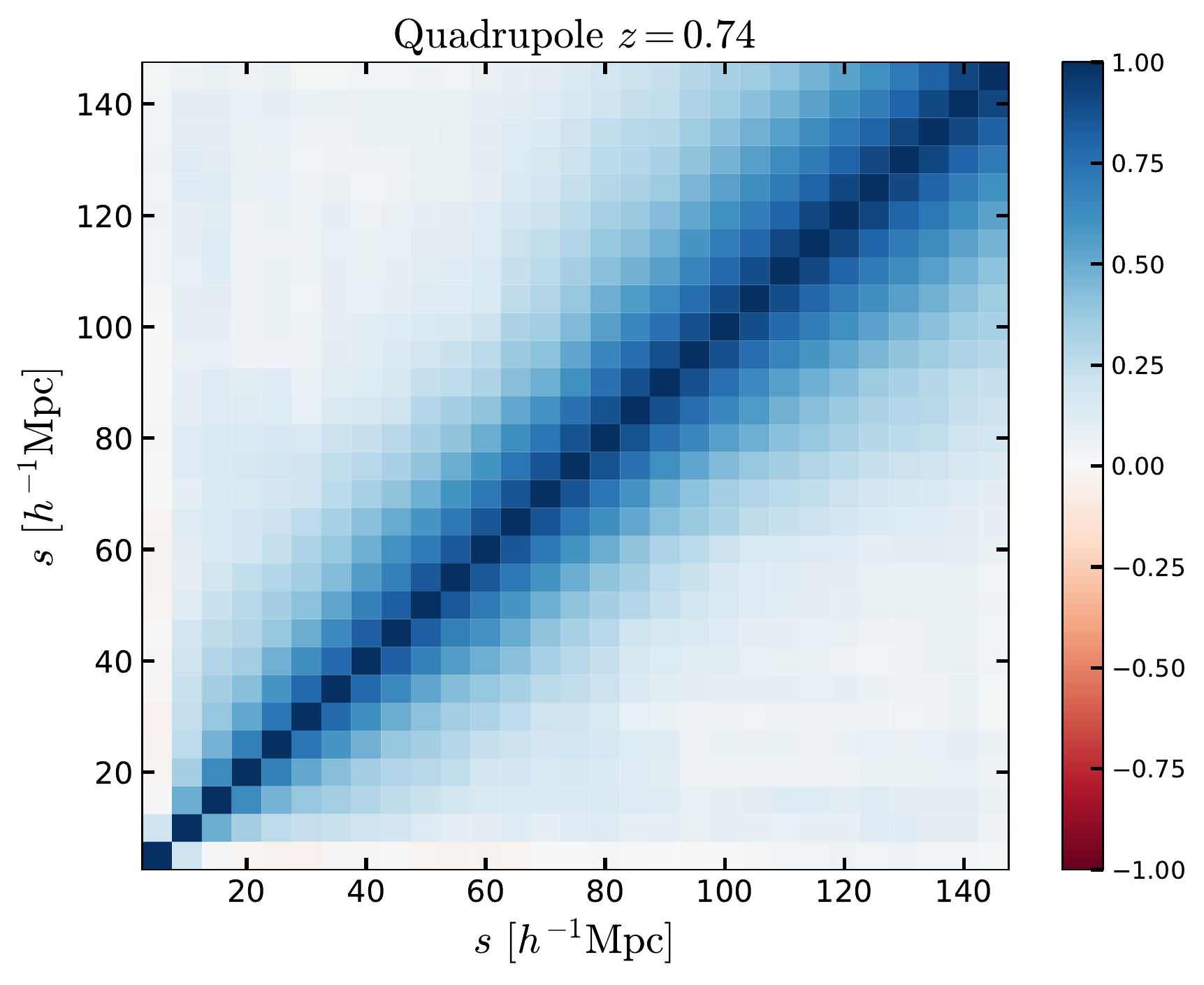}
\includegraphics[width=0.33\textwidth]{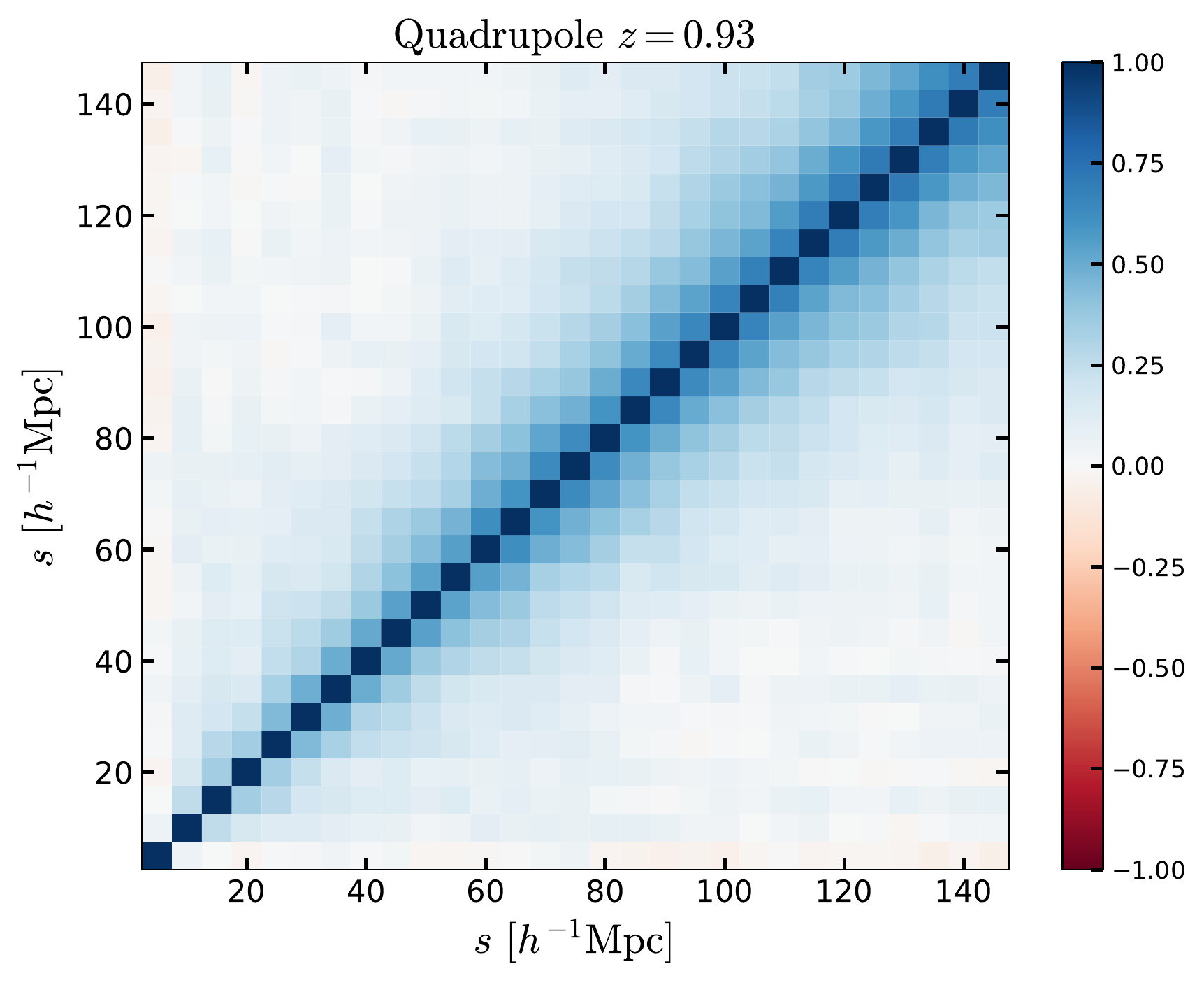}
\caption{{\it Upper panels:} Measured monopole (blue lines) and quadrupole (green lines) of the redshift-space two-point correlation function of DESI-like LRGs from our \glam{}-HOD catalogues, the lower subpanels show the diagonal error contribution to the monopole and the quadrupole. The error is calculated using Eq.~\eqref{eq:sig}. {\it Middle panels:} Correlation matrix, Eq.~\eqref{eq:Rij}, of the monopole. {\it Bottom panels:} Correlation matrix, Eq.~\eqref{eq:Rij}, of the quadrupole. The colour bar in the correlation matrices display values from $-1 \leq R(s_i,s_j) \leq 1$. The measurements are made at $z=0.6$, $0.74$ and $0.93$, as labelled at the top of each panel.}
\label{fig:xil_glam}
\end{figure*}

%--------- Figure --------------
\begin{figure*}
 \centering
\includegraphics[width=0.33\textwidth]{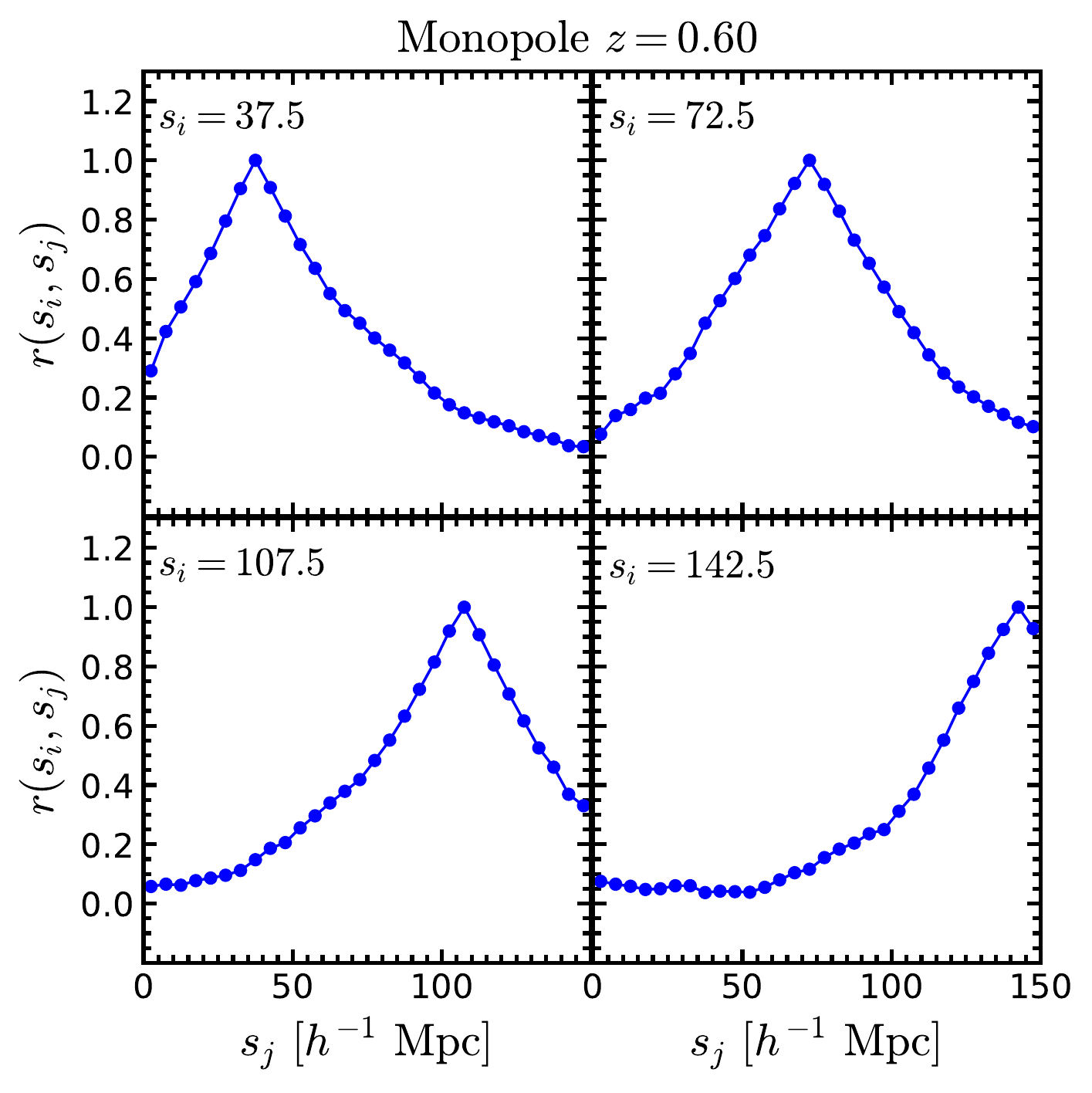}
\includegraphics[width=0.33\textwidth]{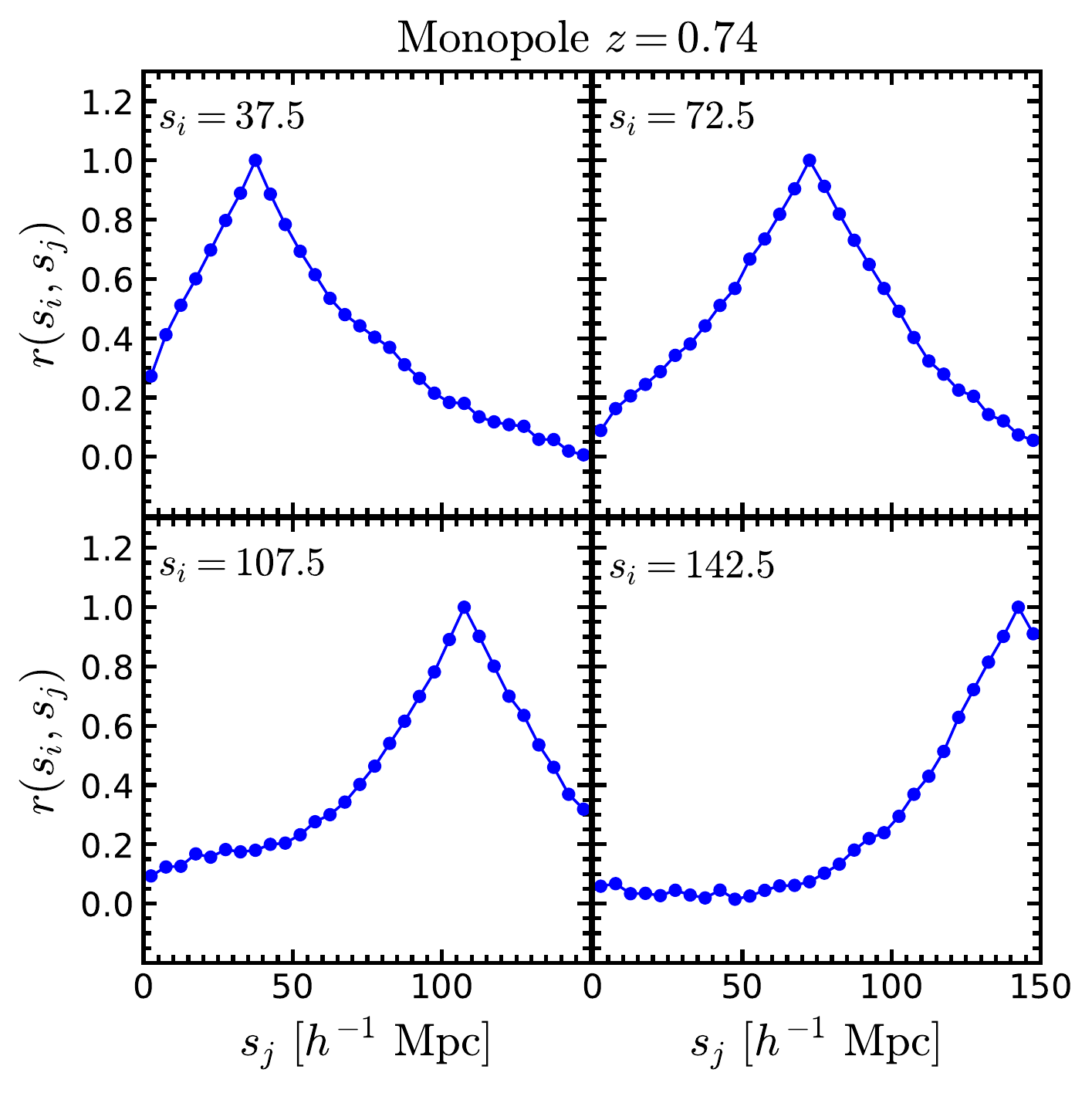}
\includegraphics[width=0.33\textwidth]{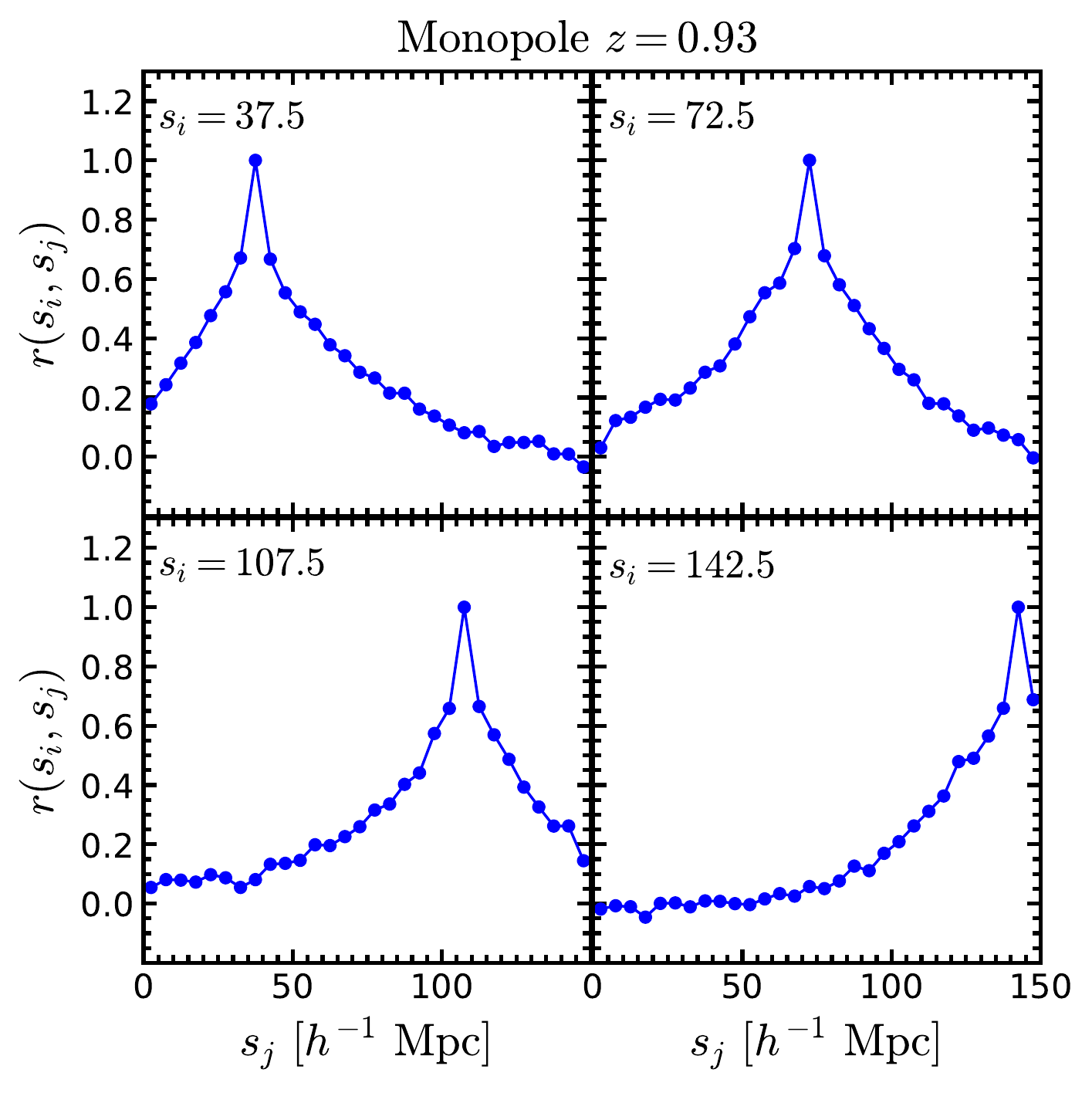}
\includegraphics[width=0.33\textwidth]{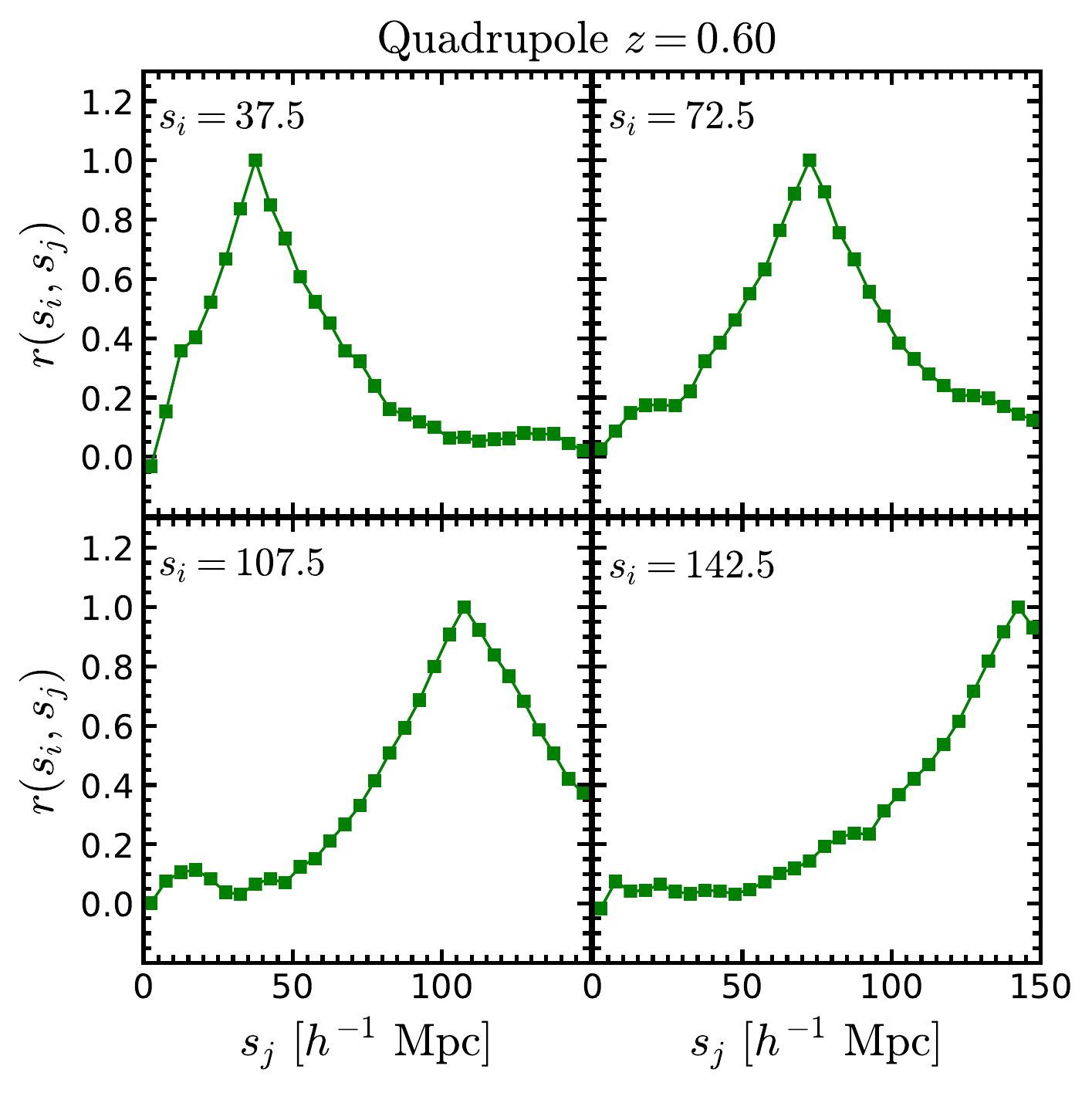}
\includegraphics[width=0.33\textwidth]{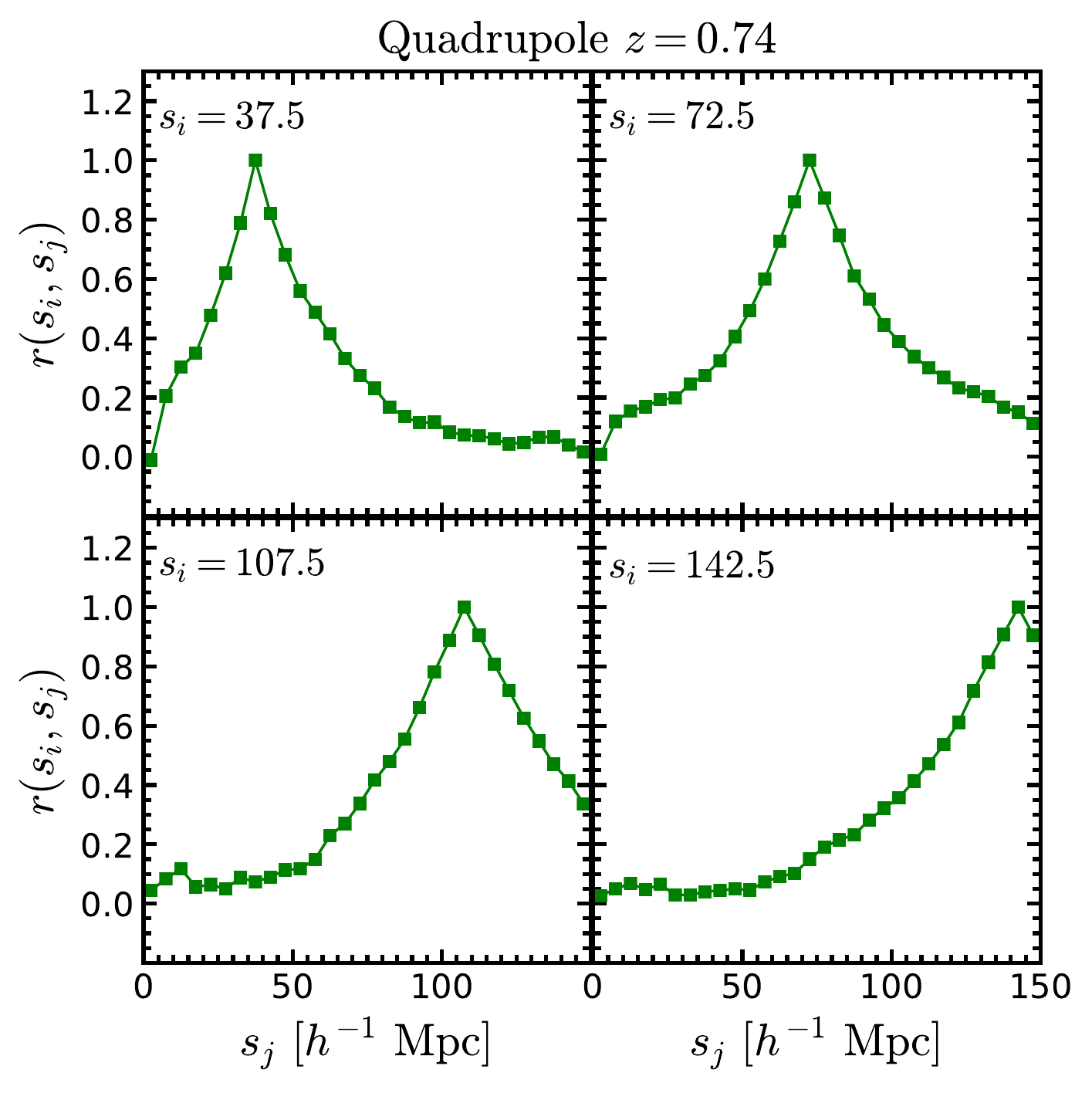}
\includegraphics[width=0.33\textwidth]{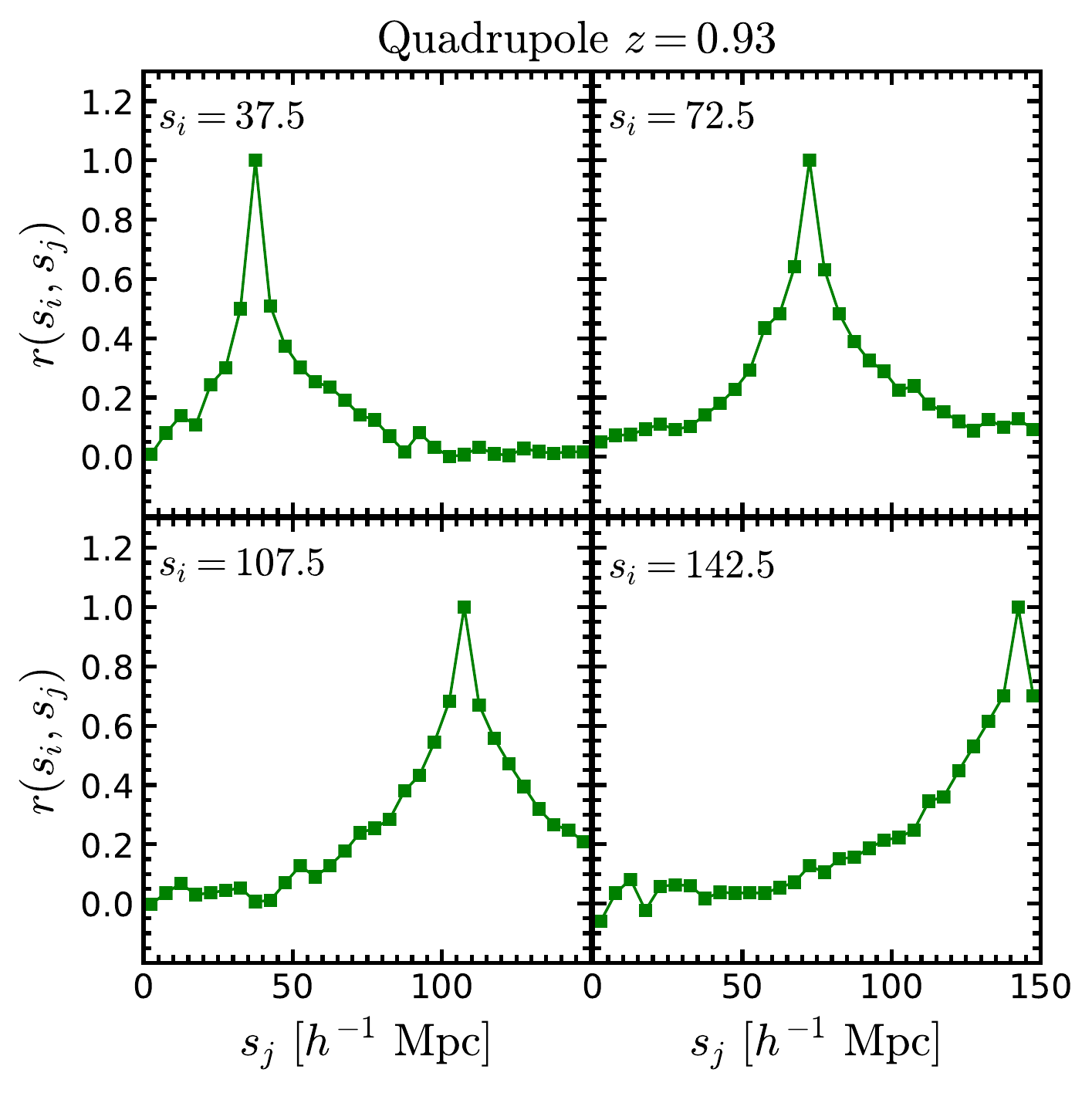}
\caption{Cuts through the correlation matrix of the monopole ({\it upper panels}) and the quadrupole ({\it bottom panels}) of the redshift-space correlation function at four different values of $s_i$ in units of $[\Mpch]$ as indicated in the panels. The measurements are made at $z=0.6$, $0.74$ and $0.93$, as labelled at the top of each panel.}
\label{fig:xil_error}
\end{figure*}

We measure the monopole and quadrupole moments of the redshift-space correlation function, $\xi_l(s)$, and power spectrum, $P_l(k)$, using
\begin{eqnarray}
\xi_l(s) &=& (2l + 1) \int^1_{0}{\xi(s,\mu)} \mathcal{L}_l(\mu)~{\rm d}\mu\,,\label{eq:xi_l}\\
P_l(k) &=& (2l + 1) \int^1_{0}{P(k,\mu)} \mathcal{L}_l(\mu)~{\rm d}\mu\,\label{eq:Pk_l},
\end{eqnarray}
where $\xi(s,\mu)$ and $P(k,\mu)$ are the full two-dimensional correlation function and power spectrum, $\mu$ is the cosine of the angle between the separation vector, $\mathbf{s}$ or $\mathbf{k}$, and the line-of-sight in configuration or Fourier space, respectively. The $\mathcal{L}_l(\mu)$ are the Legendre polynomials where $l=0$ is the monopole and $l=2$ is the quadrupole. We use 20 bins logarithmically spaced over the separation range $0.7 < s/[\Mpch] < 50$ in which to measure the correlation function. The power spectrum is measured in the range $0 < k/[\hMpc] < k_{\rm Nyq}$ using linear bins in $k$ with separation $\Delta k = 0.006\hMpc$, where $k_{\rm Nyq} = \pi N_{\rm mesh}/L_{\rm box}$ is the 1D Nyquist frequency, $N_{\rm mesh}=512$ and $L_{\rm box}$ is the box size of the \Pmill{} or \glam{} simulations. In all cases we adopt 30 linearly spaced bins between $0$ and $1$ for $\mu$. 

In the upper panels of Fig.~\ref{fig:xi_glam} we display the real-space clustering measured from the \Galform{} output (black line) and the \glam{} LRG mock catalogues (blue symbols with errorbars). Additionally, we show the best-fitting power law fit to the correlation function reported by \citet{Kitanidis:2019rzi} $(r_0 = 7.78\Mpch,\, \gamma=1.98)$ which agrees well with our measurements, especially on scales $r \geq r_0$. Note that \citeauthor{Kitanidis:2019rzi} fitted the angular correlation function in the range $0.001^{\circ} < \theta < 1^{\circ}$ which translates to comoving separation $\theta_{\rm min}D_{\rm A}(z) < r/[\Mpch] < \theta_{\rm max}D_{\rm A}(z)$, where $D_{\rm A}$ is the angular-diametre distance. 
We also show results when fitting our \glam{} measurements with a power-law using the range mentioned above, $\xi(r) = (r/r_{0})^{-\gamma}$, finding $r_0/[\Mpch] = (7.316\pm0.022,7.346\pm0.024,6.883\pm0.04)$ and $\gamma = (1.623\pm0.006,1.592\pm0.007,1.589\pm0.012)$ at $z=0.6$, $0.74$ and $0.93$.

The lower panels of Fig.~\ref{fig:xi_glam} shows the predicted multipoles of the redshift-space correlation function of the \glam{}-HOD LRGs (symbols with errorbars), plotted in comparison with their \Galform{} counterparts (black line). We find excellent agreement between the clustering measured in both real- and redshift-space for the \glam{} and \Galform{} LRGs at all scales and all redshifts. 

In Fig.~\ref{fig:Pk_glam} we display the clustering measurements in Fourier space. First, we note the good agreement between the \Galform{} and \glam{} measurements on all scales. In the upper panels of Fig.~\ref{fig:Pk_glam} we also show the measured dark-matter power spectrum scaled by the galaxy bias squared relations of \citet[][cyan lines]{Kitanidis:2019rzi}, \citet[][magenta lines]{Rongpu:2020} and from our simulations, Eq.~\eqref{eq:bias}. We find that our measurements slightly underpredict the bias value compare to the measured relations estimated by \citet{Kitanidis:2019rzi} and \citet{Rongpu:2020}. 
In the lower panels of Fig.~\ref{fig:Pk_glam} we show the multipole moments of the redshift space power spectrum, finding almost perfect agreement between the \Galform{} and \glam{} measurements on scales $k > 0.1\hMpc$. Nevertheless, there is a noisy signal for the \Galform{} quadrupole of the redshift-space power spectrum, due to the smaller box size of the \Pmill{}. Nevertheless, this signal is in good agreement with the predictions from \glam{} over the range $0.1 < k/[\hMpc] < 0.3$. 

We conclude that populating \glam{} haloes using our interpolated-HOD method reproduces accurately the clustering of LRGs predicted directly by \Galform{} on all scales of interest. 
%---------------------------------------------------------------
\subsection{Large-scale galaxy clustering and covariance matrices}\label{sec:cov}
%---------------------------------------------------------------
%--------- Figure --------------
\begin{figure*}
 \centering
\includegraphics[width=0.33\textwidth]{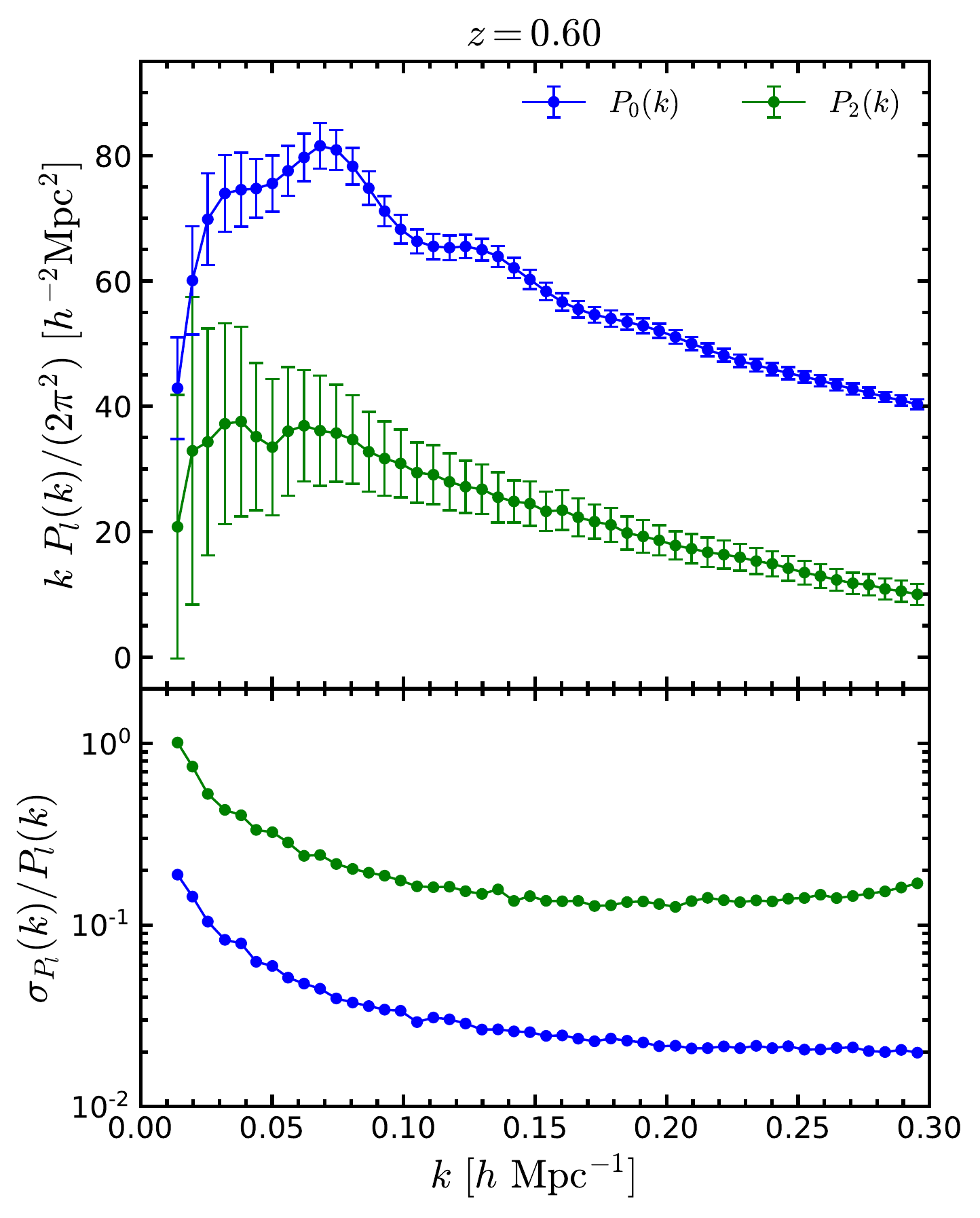}
\includegraphics[width=0.33\textwidth]{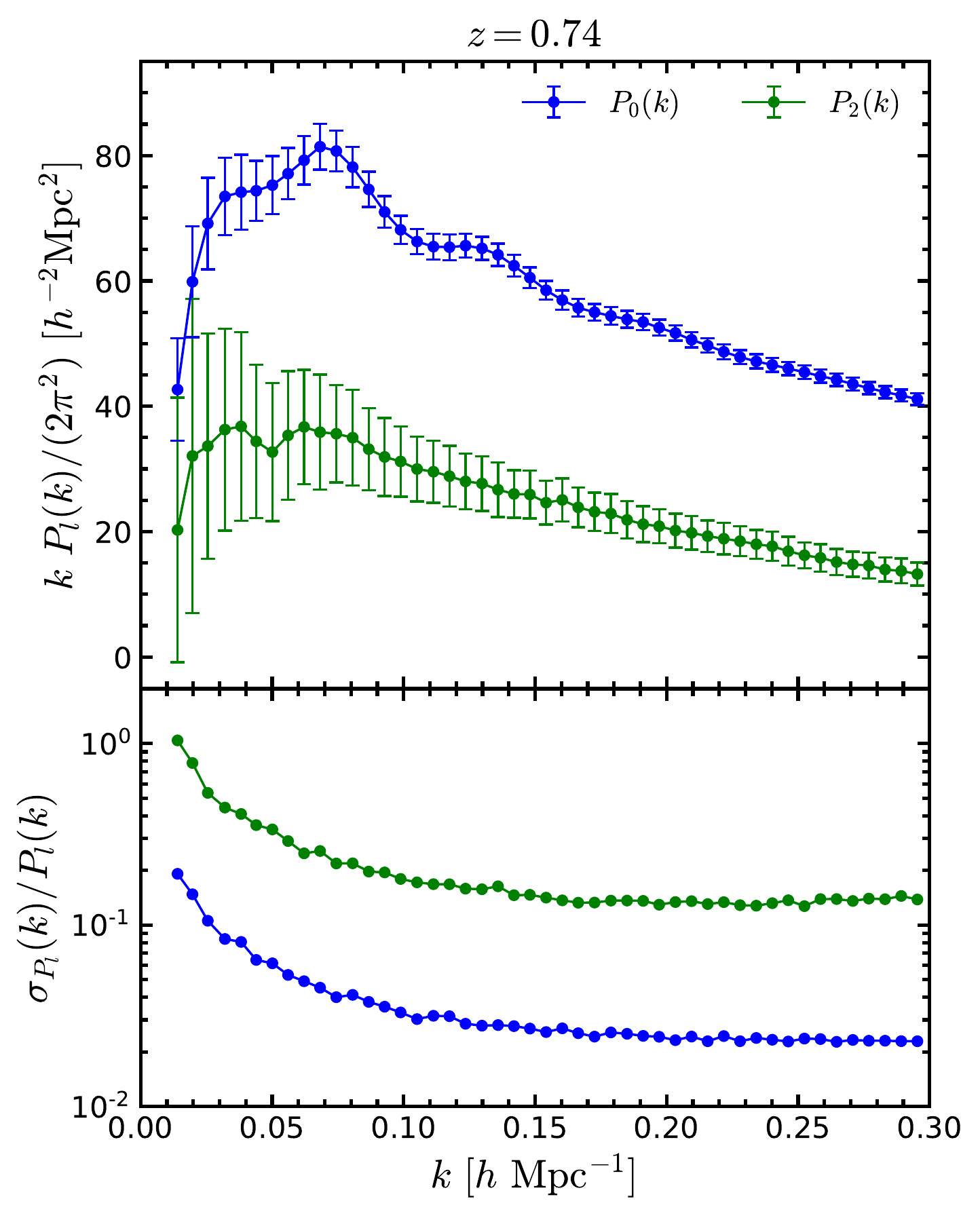}
\includegraphics[width=0.33\textwidth]{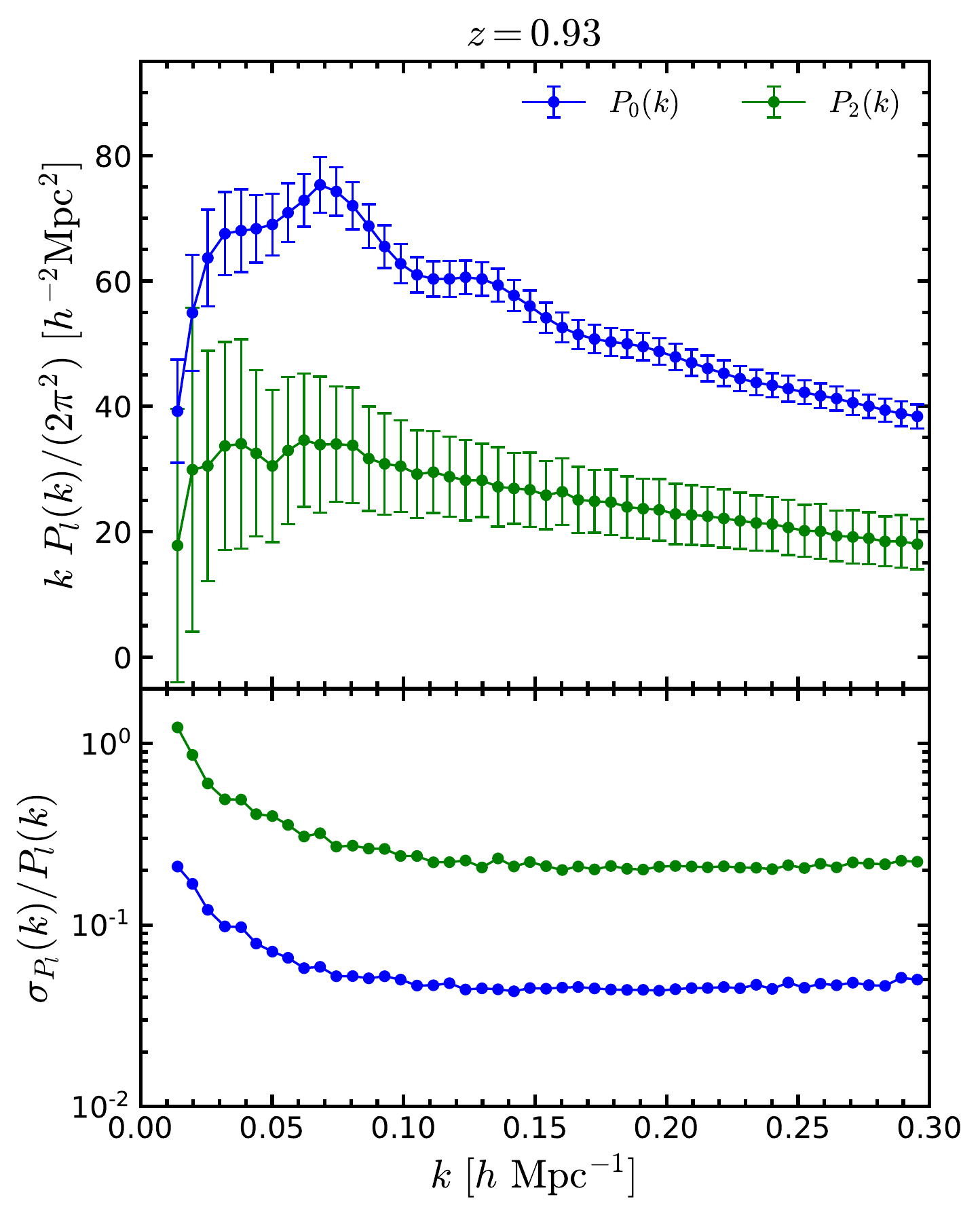}
\includegraphics[width=0.33\textwidth]{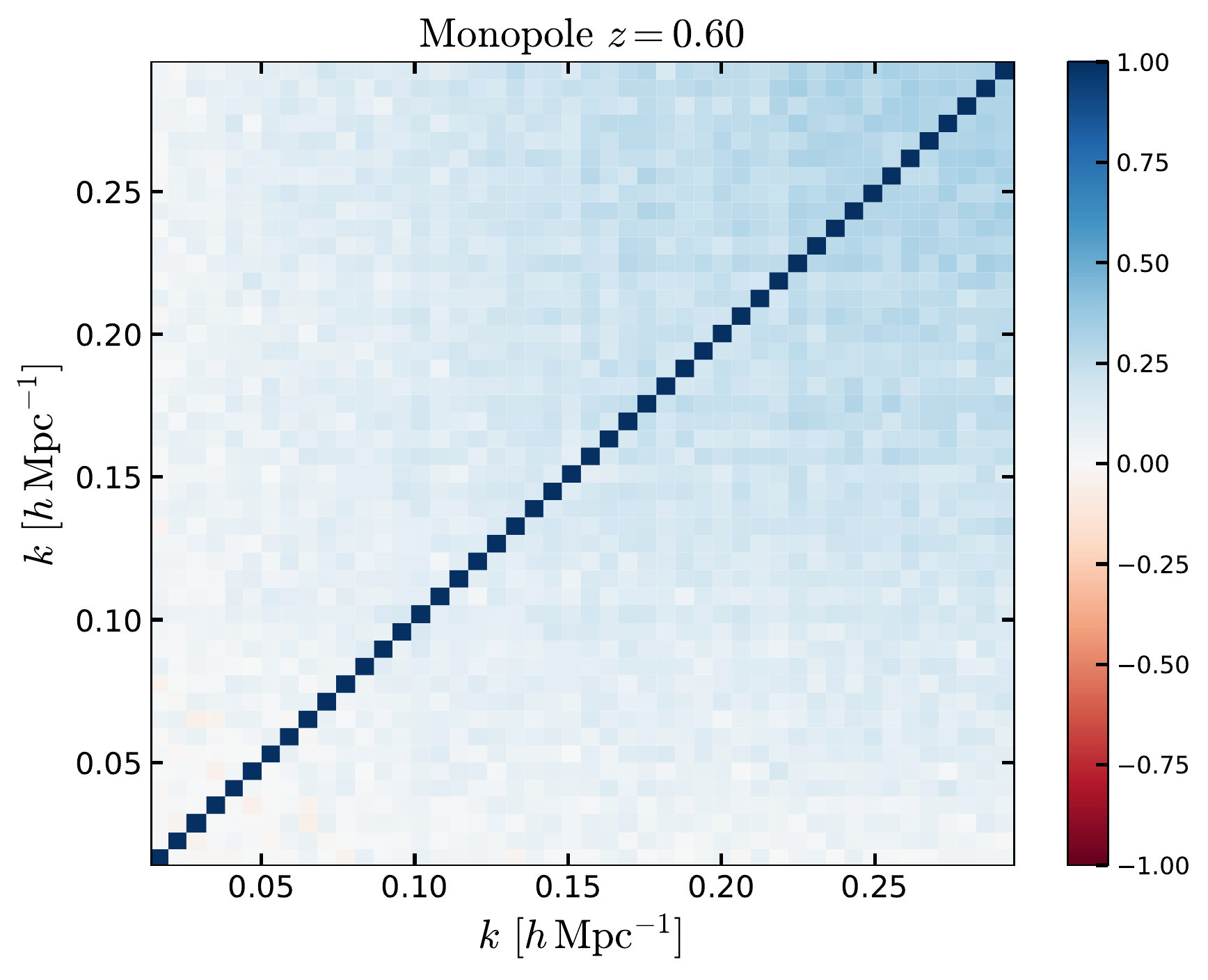}
\includegraphics[width=0.33\textwidth]{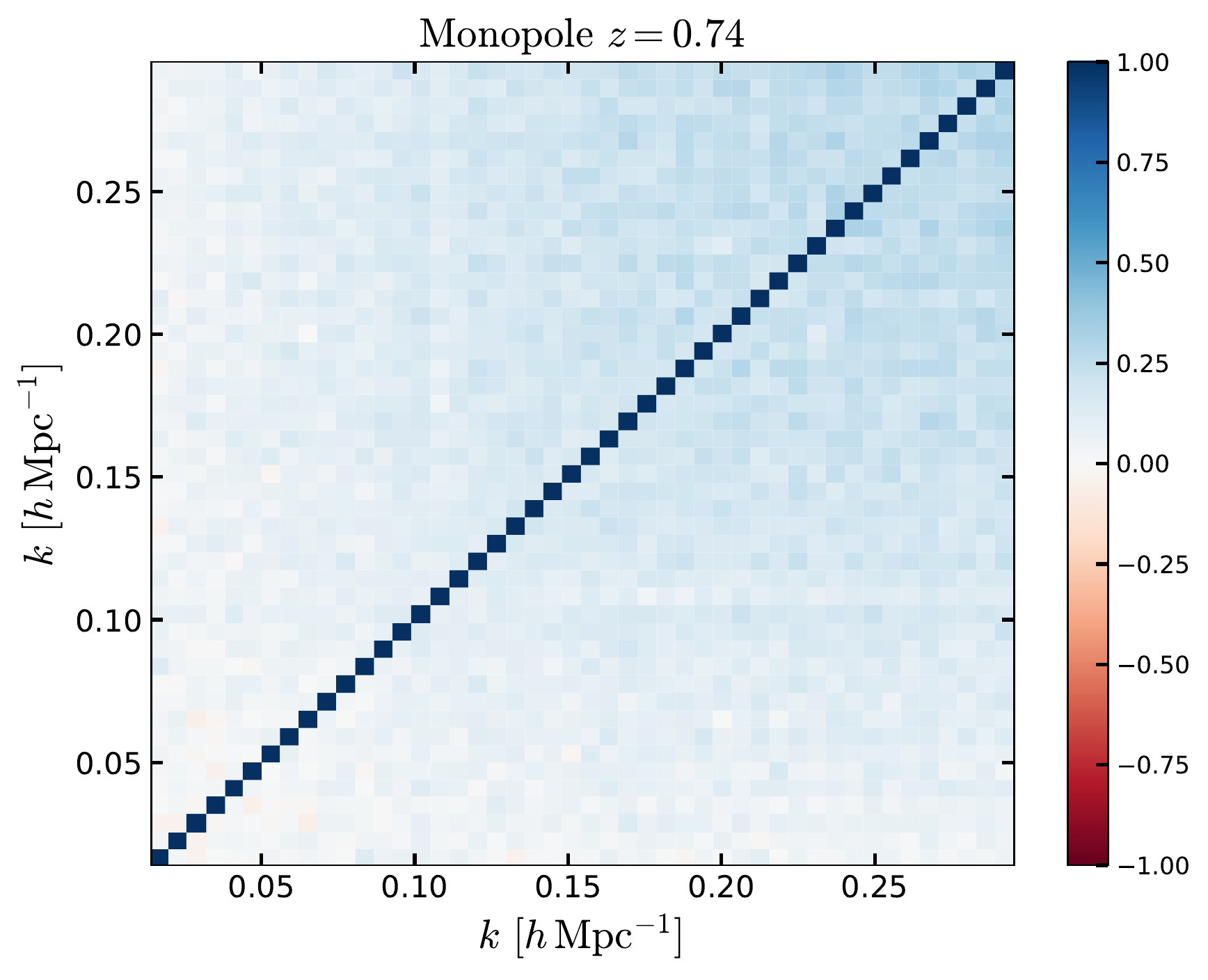}
\includegraphics[width=0.33\textwidth]{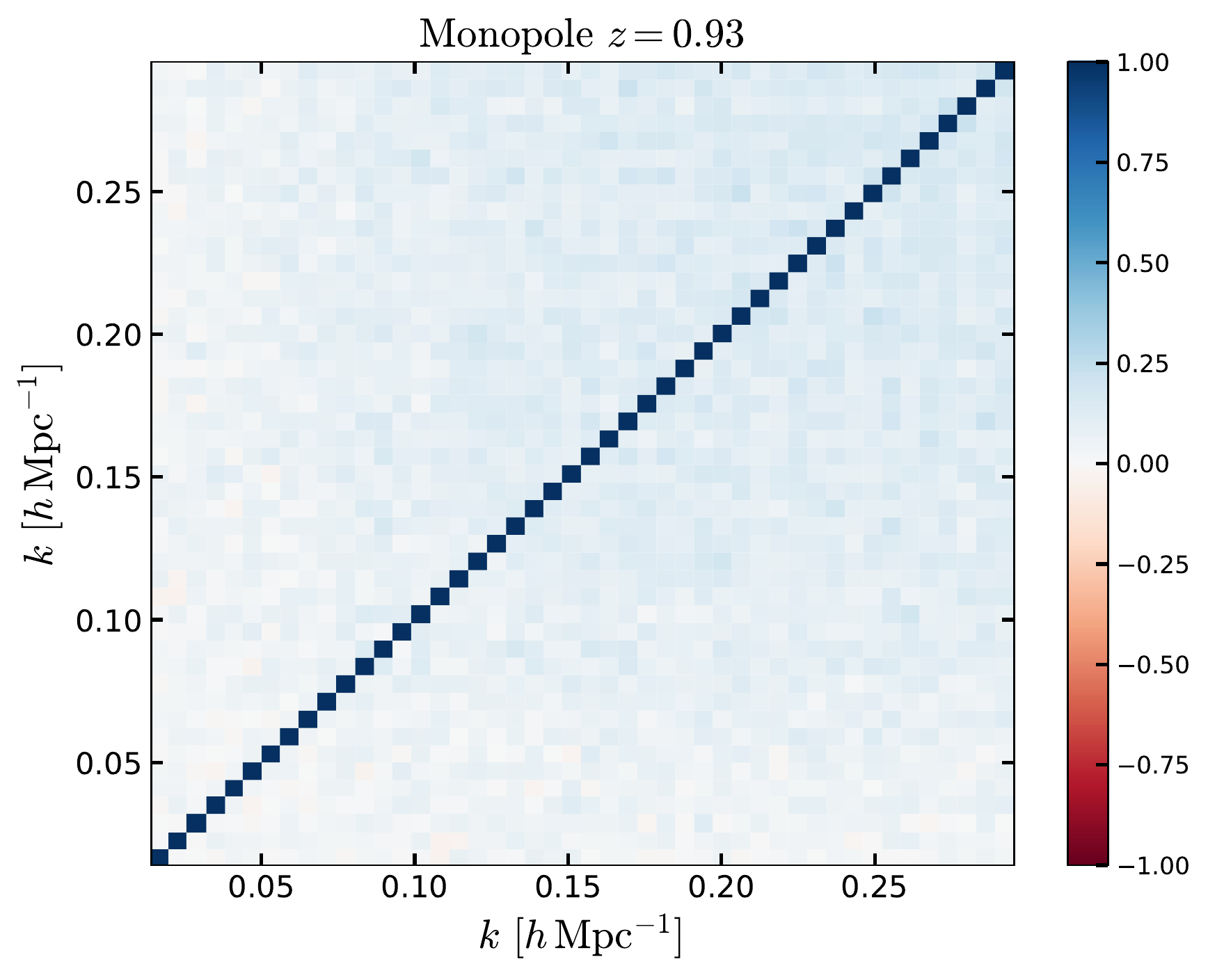}
\includegraphics[width=0.33\textwidth]{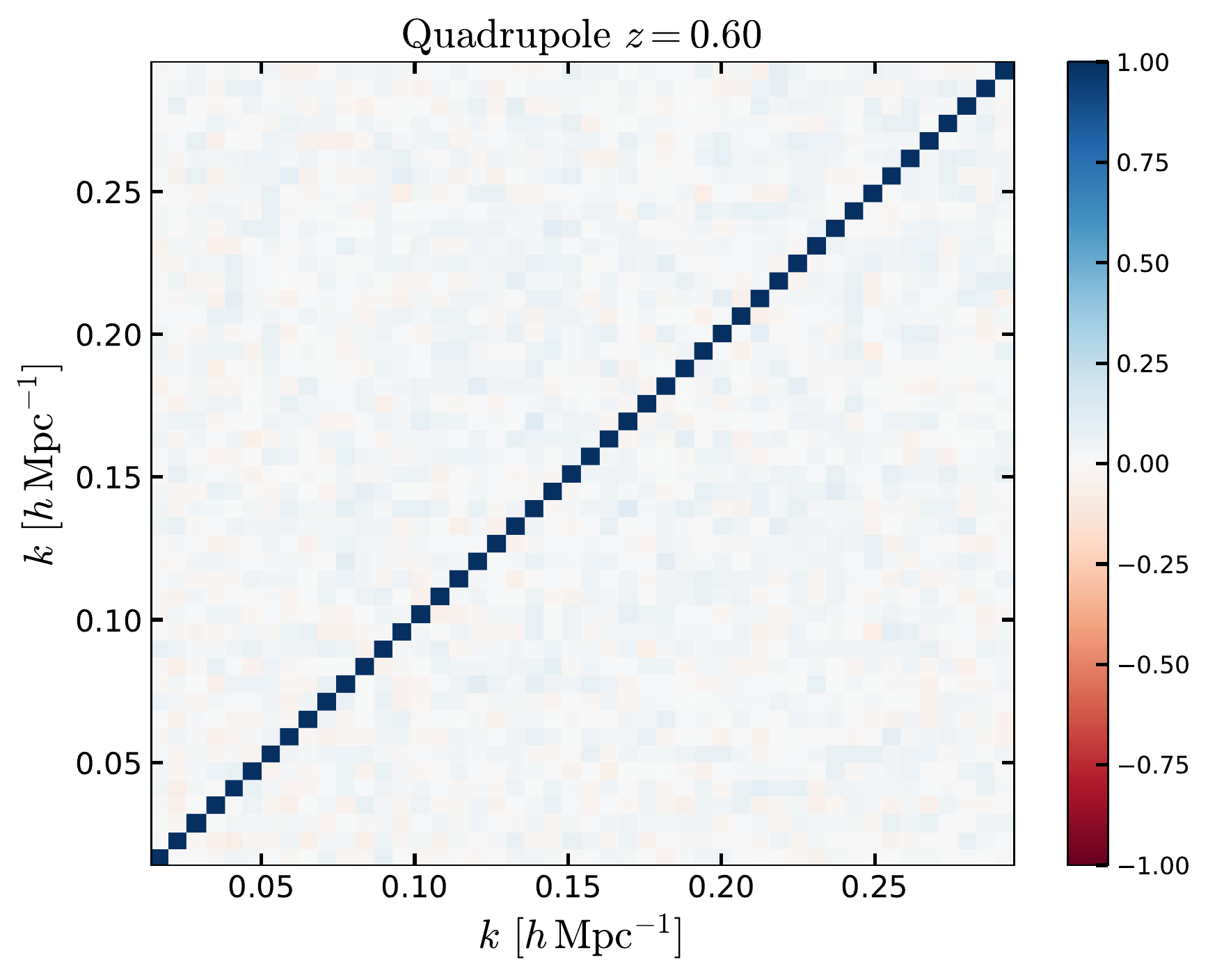}
\includegraphics[width=0.33\textwidth]{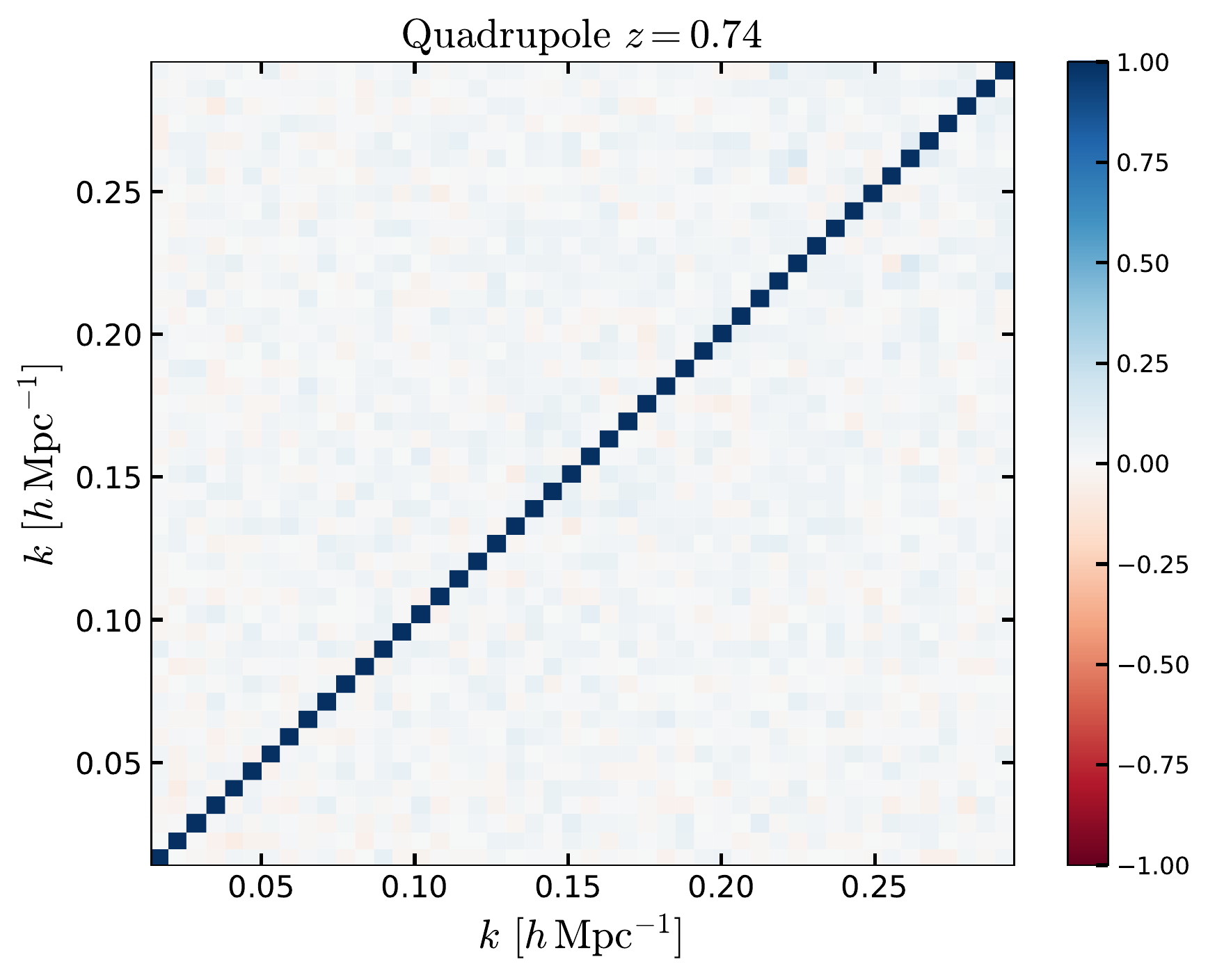}
\includegraphics[width=0.33\textwidth]{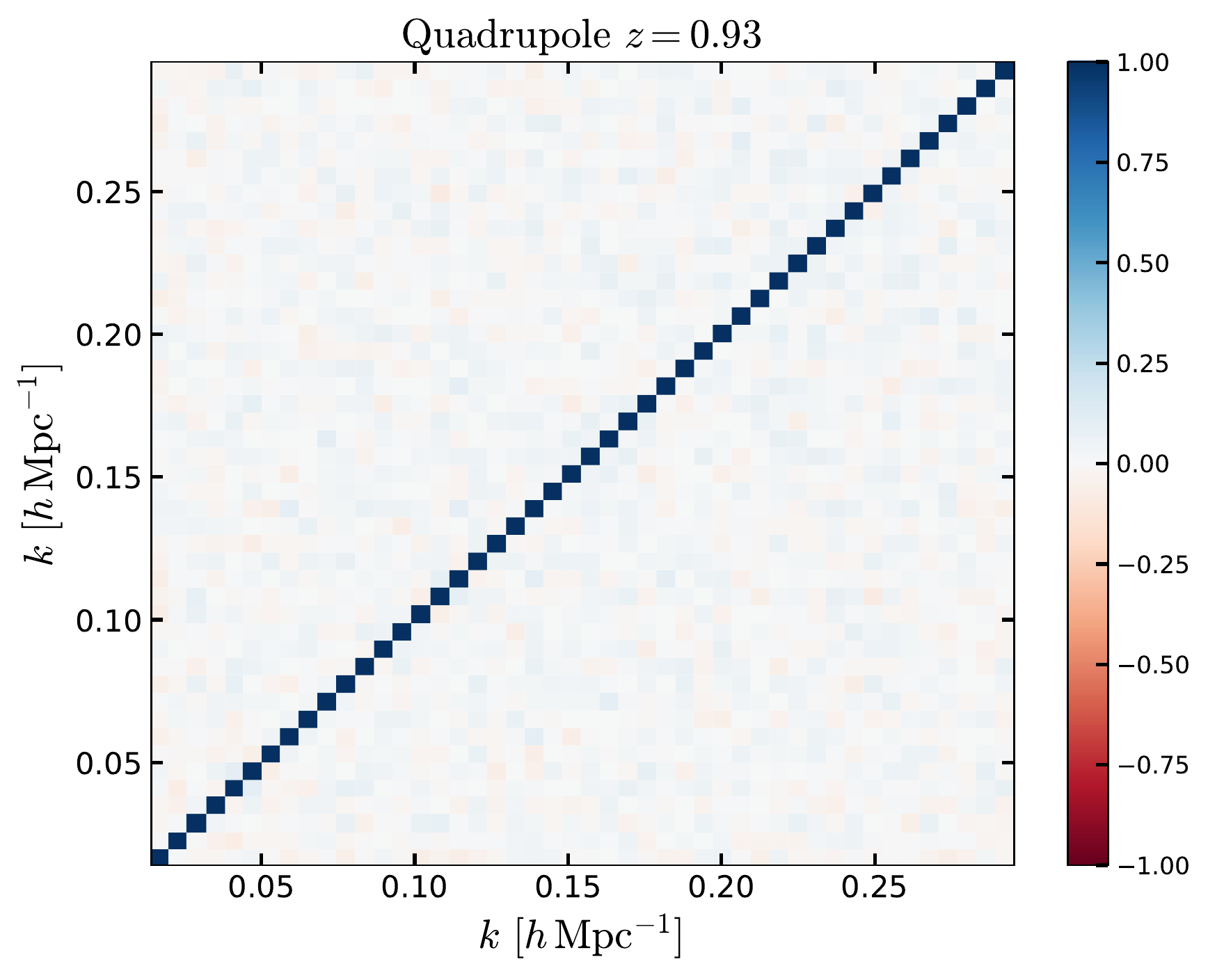}
\caption{Same as Figure~\ref{fig:xil_glam} but for the multipoles of the power spectrum.}
\label{fig:Pkl_glam}
\end{figure*}

%--------- Figure --------------
\begin{figure*}
 \centering
\includegraphics[width=0.33\textwidth]{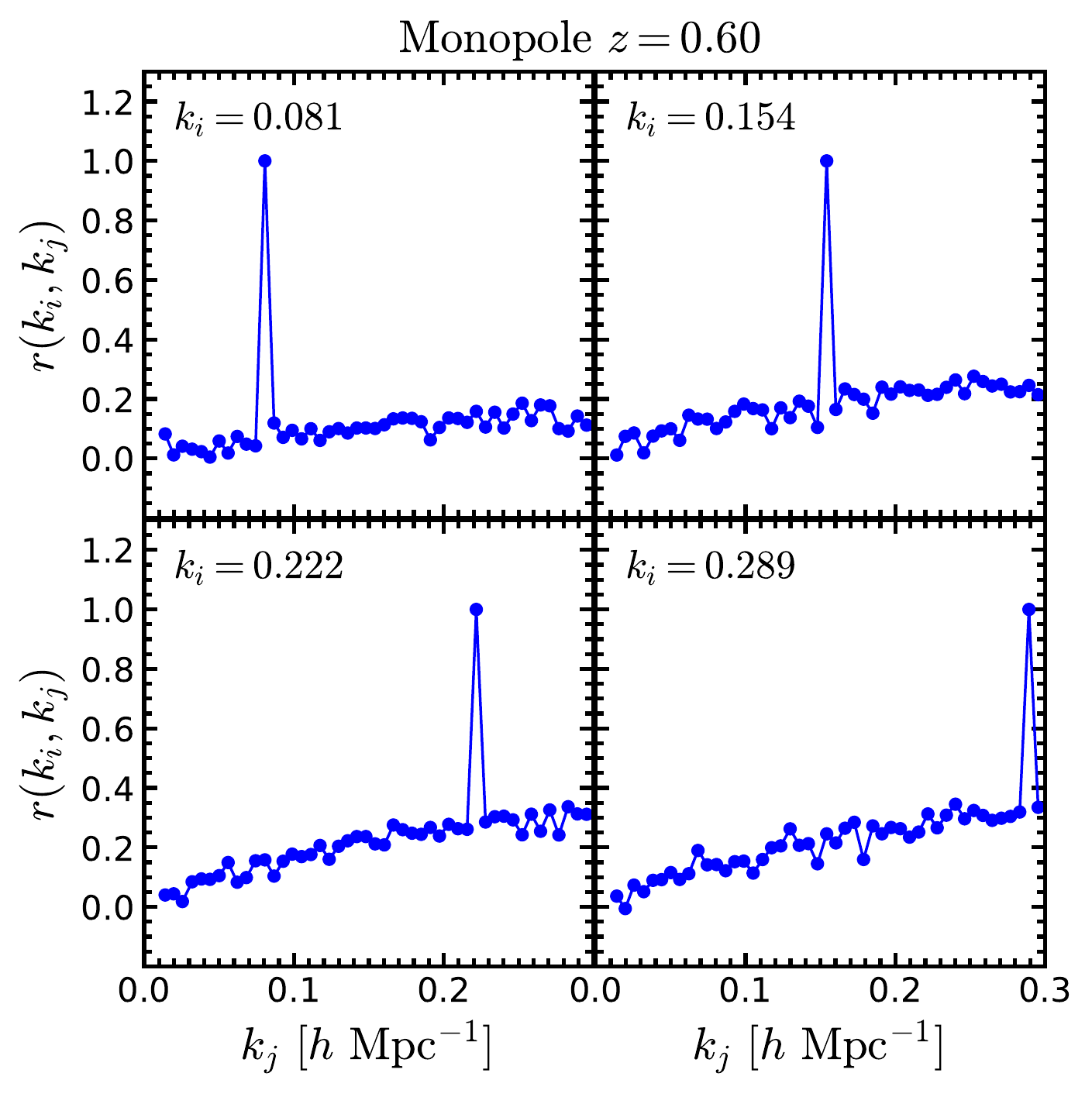}
\includegraphics[width=0.33\textwidth]{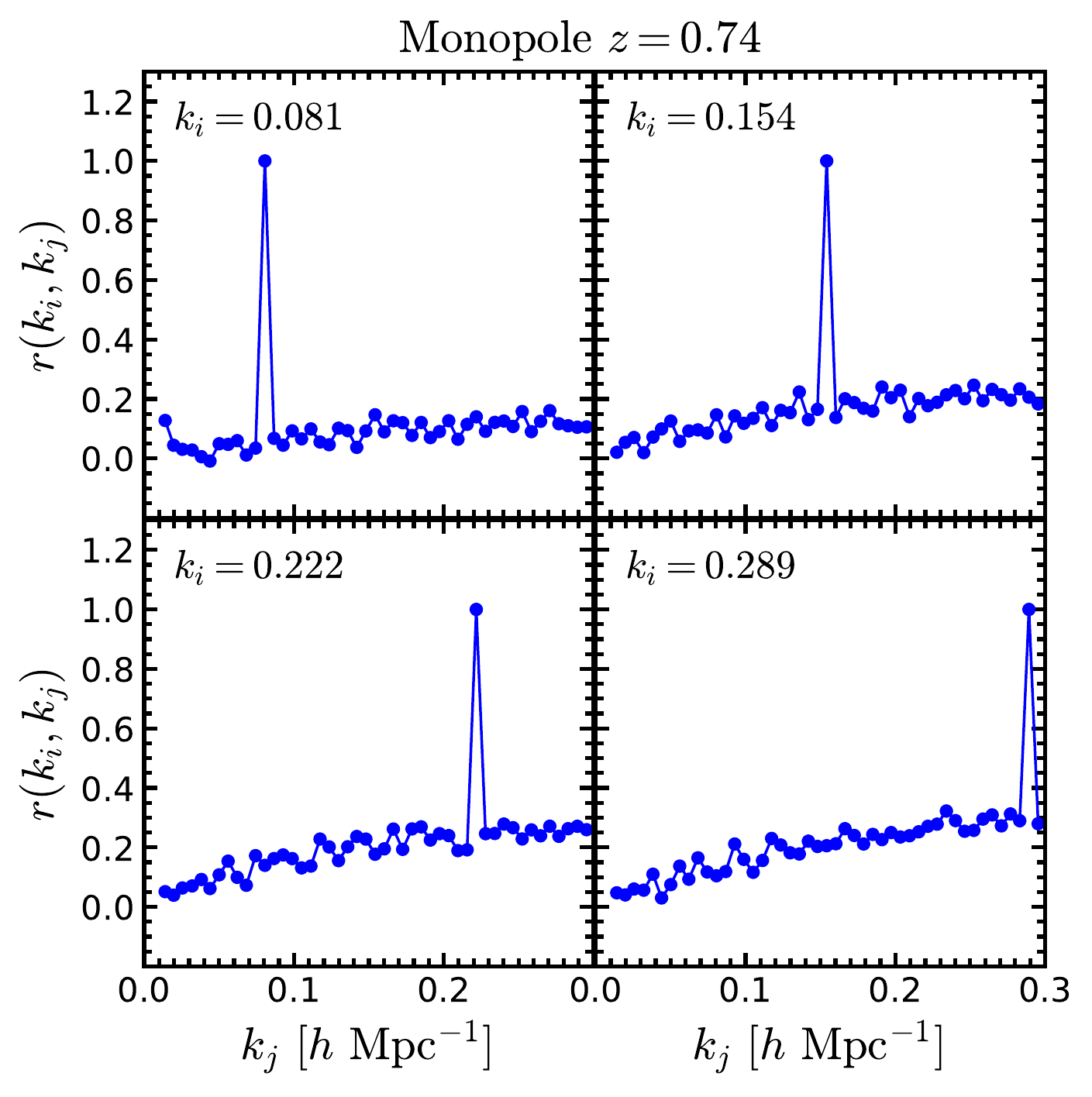}
\includegraphics[width=0.33\textwidth]{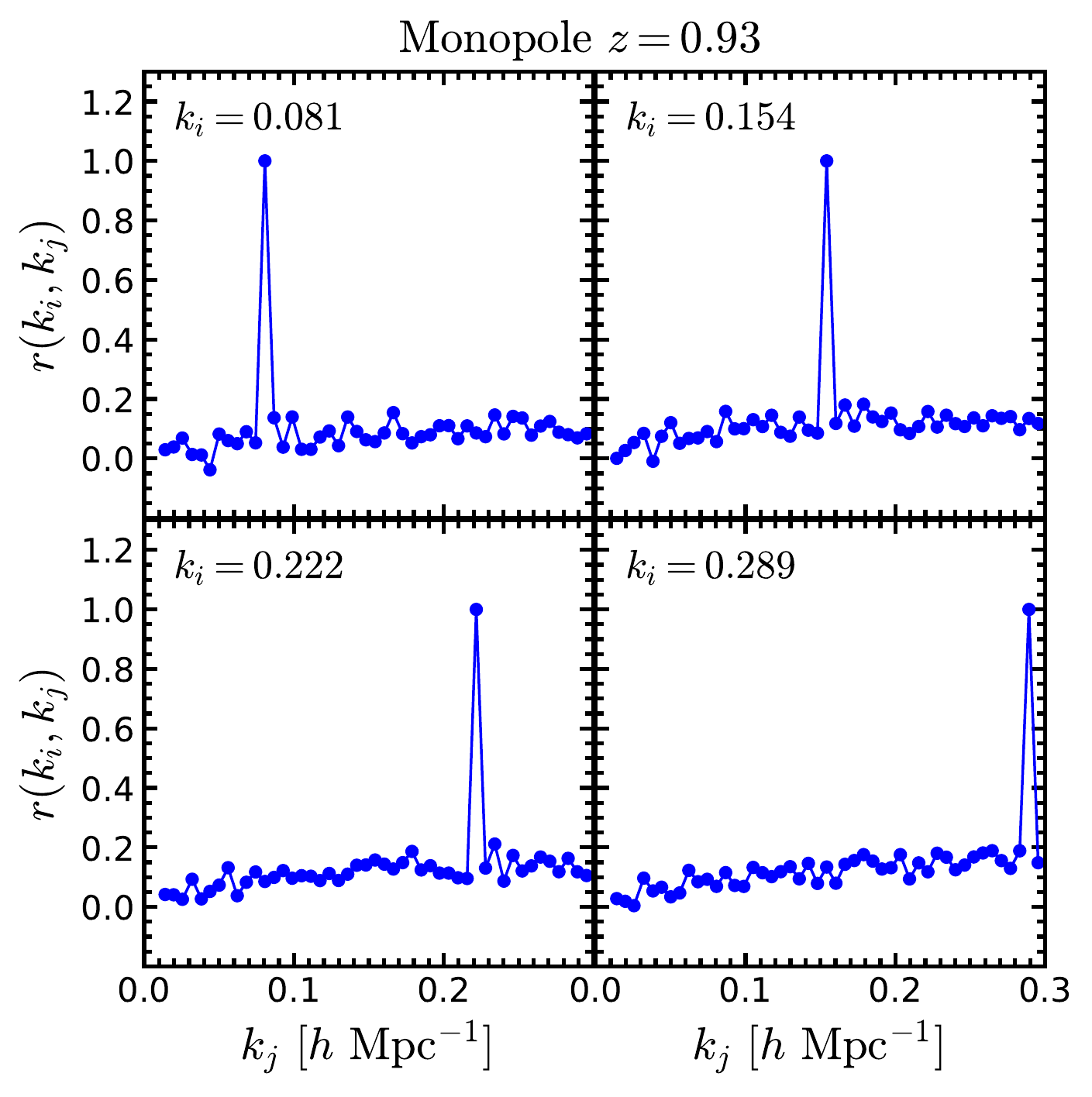}
\includegraphics[width=0.33\textwidth]{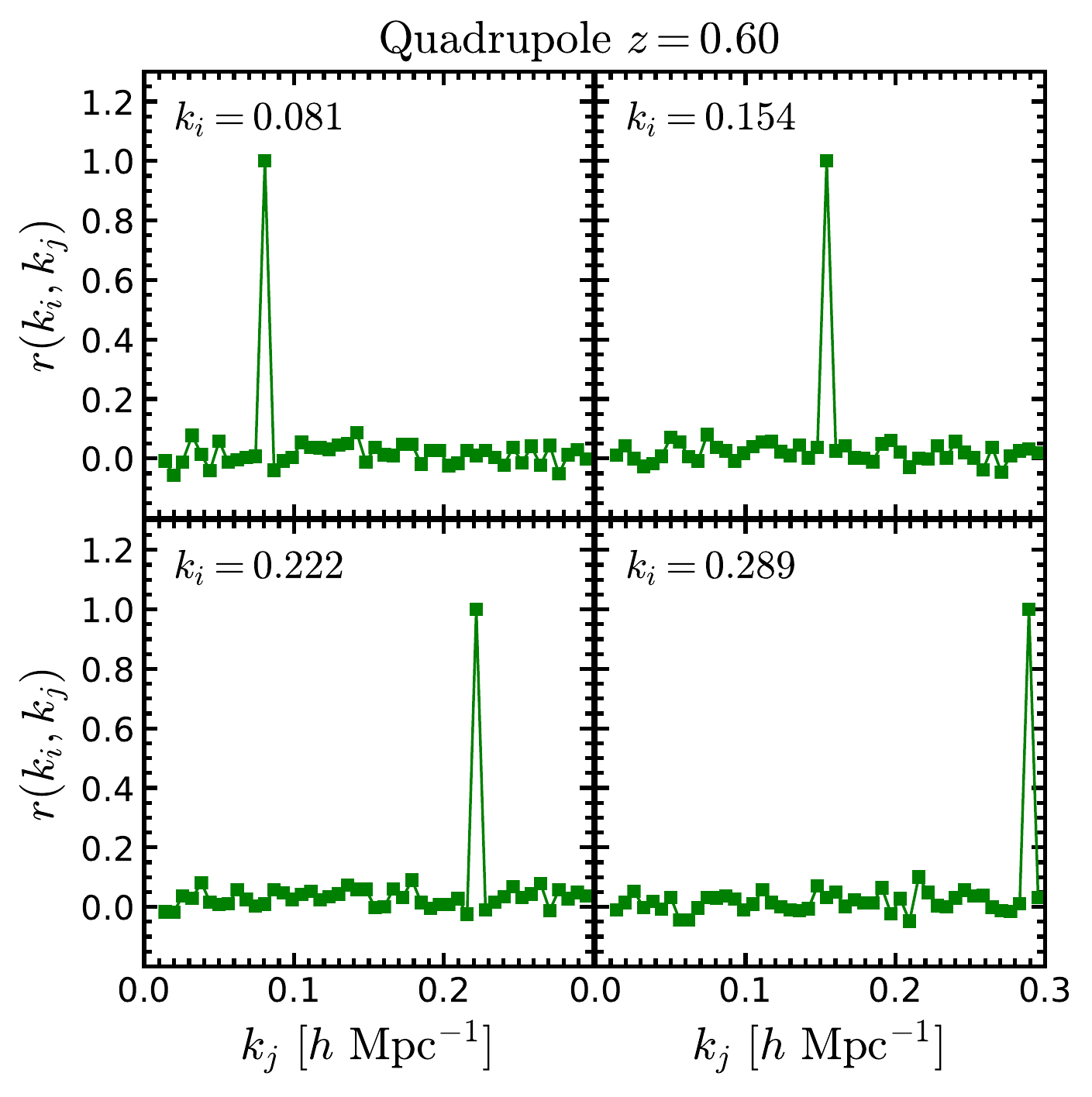}
\includegraphics[width=0.33\textwidth]{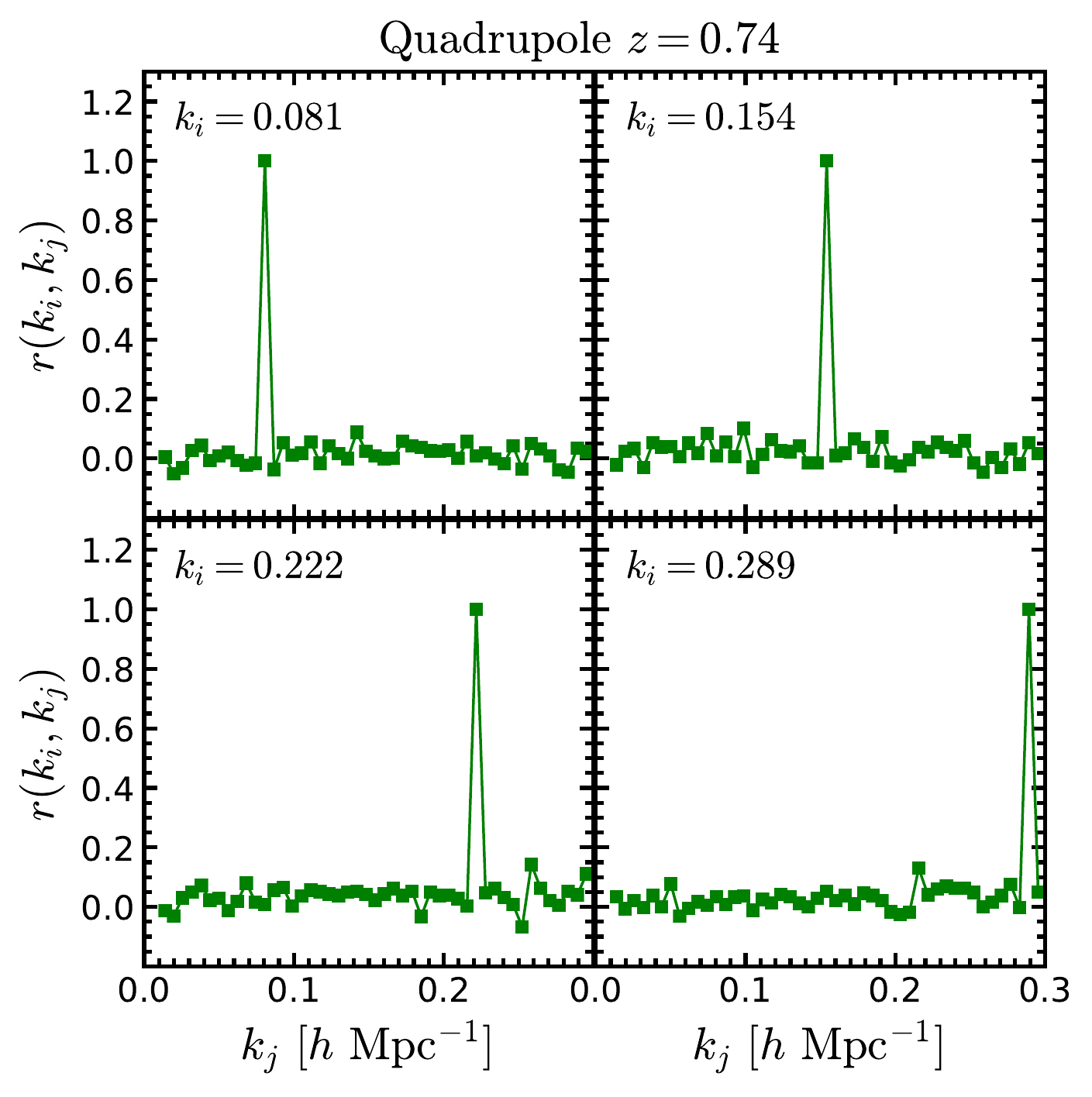}
\includegraphics[width=0.33\textwidth]{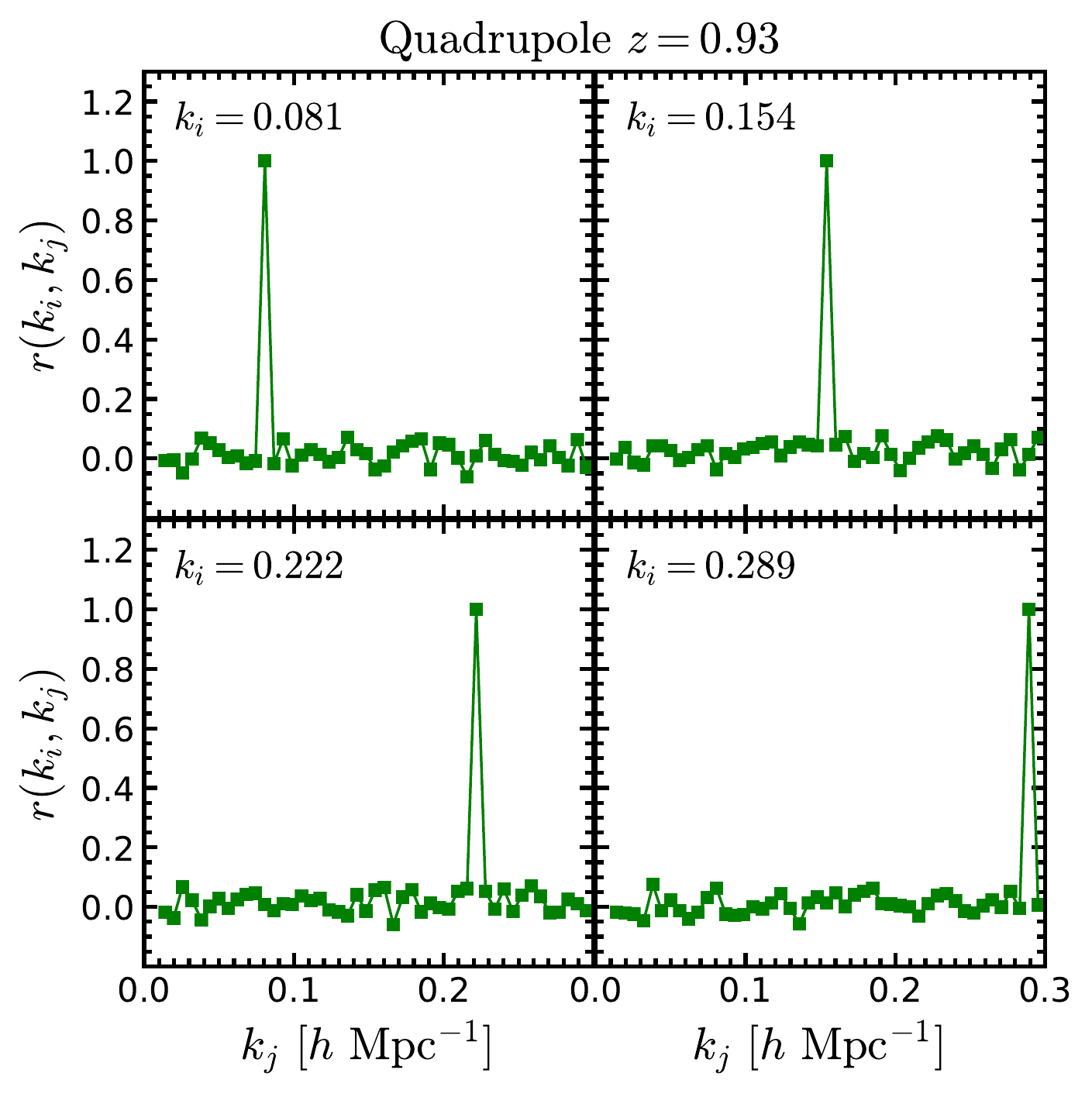}
\caption{Same as Figure~\ref{fig:xil_error} but for the multipoles of the power spectrum. In this case, we show the cuts through the correlation matrices at four different values of $k_i$ in units of $[\hMpc]$.}
\label{fig:Pkl_error}
\end{figure*}

%--------- Figure --------------
\begin{figure*}
 \centering
\includegraphics[width=0.33\textwidth]{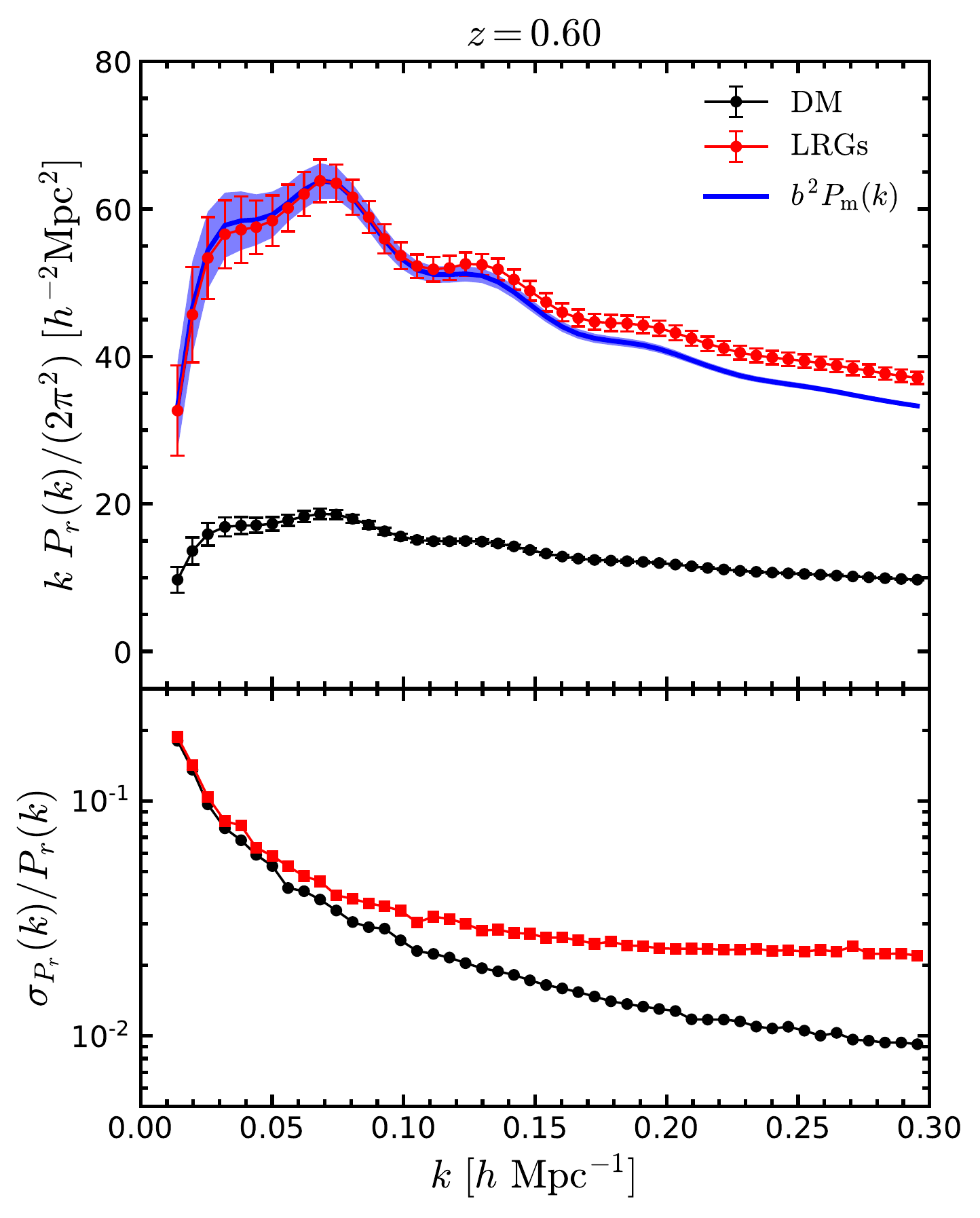}
\includegraphics[width=0.33\textwidth]{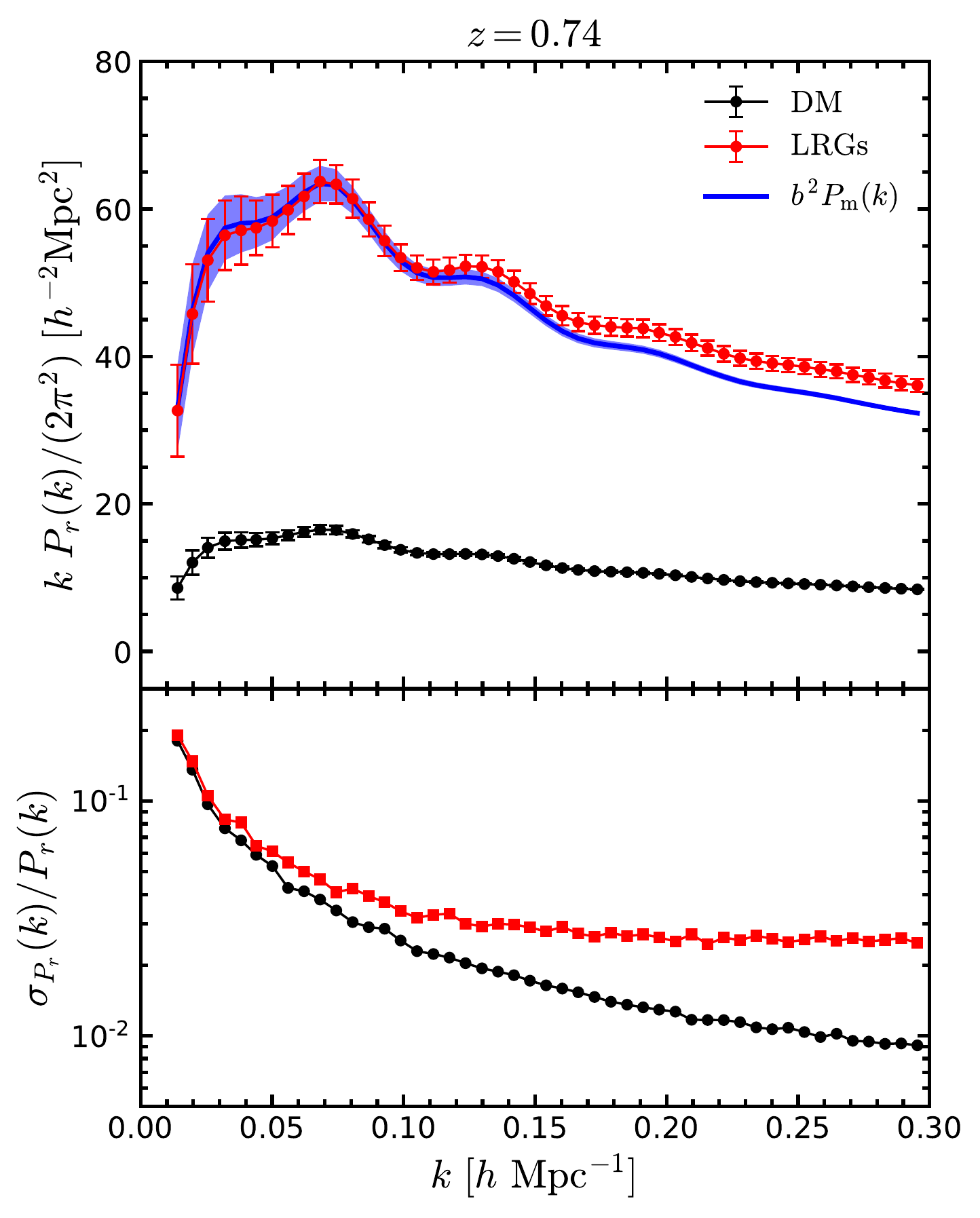}
\includegraphics[width=0.33\textwidth]{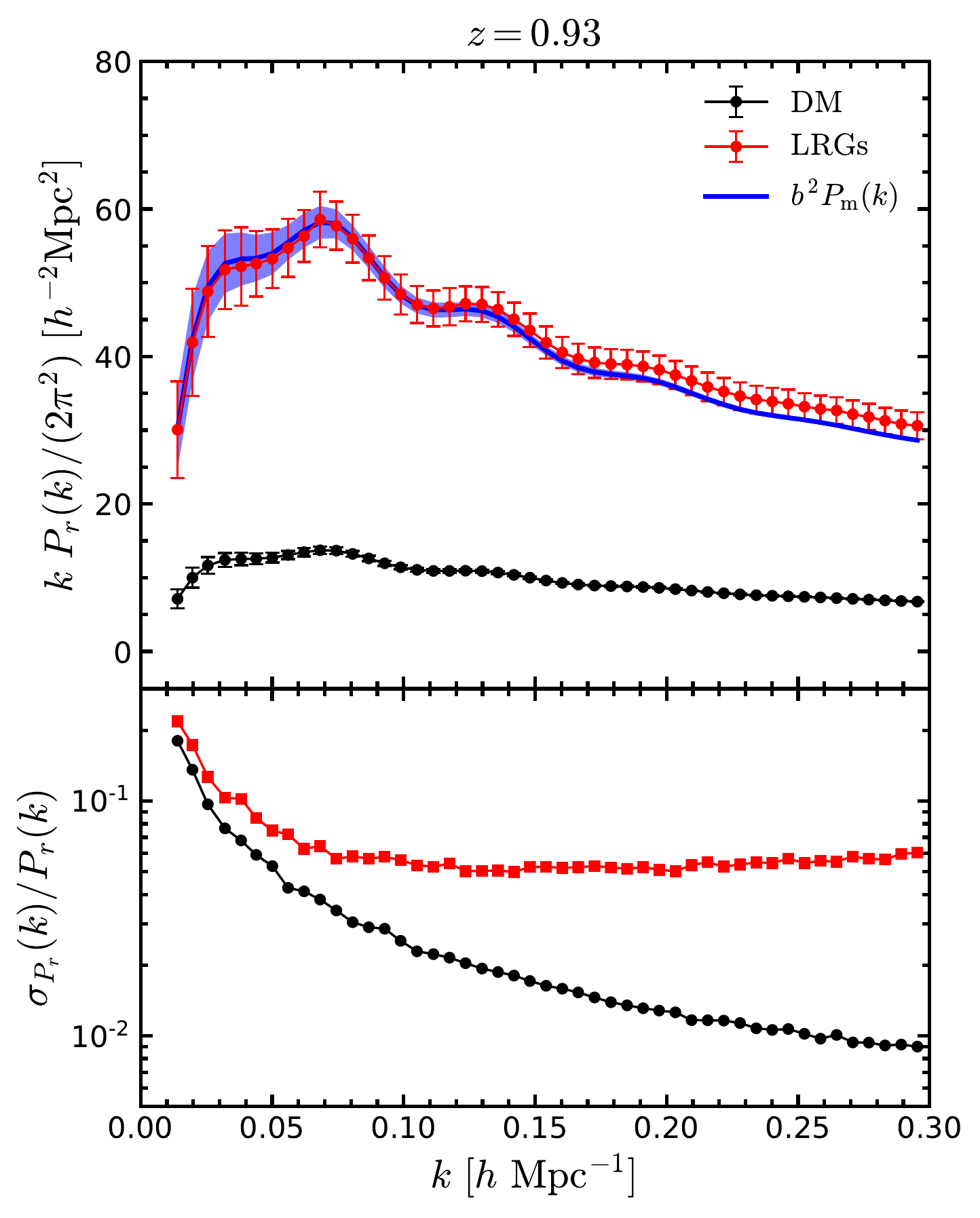}
\includegraphics[width=0.33\textwidth]{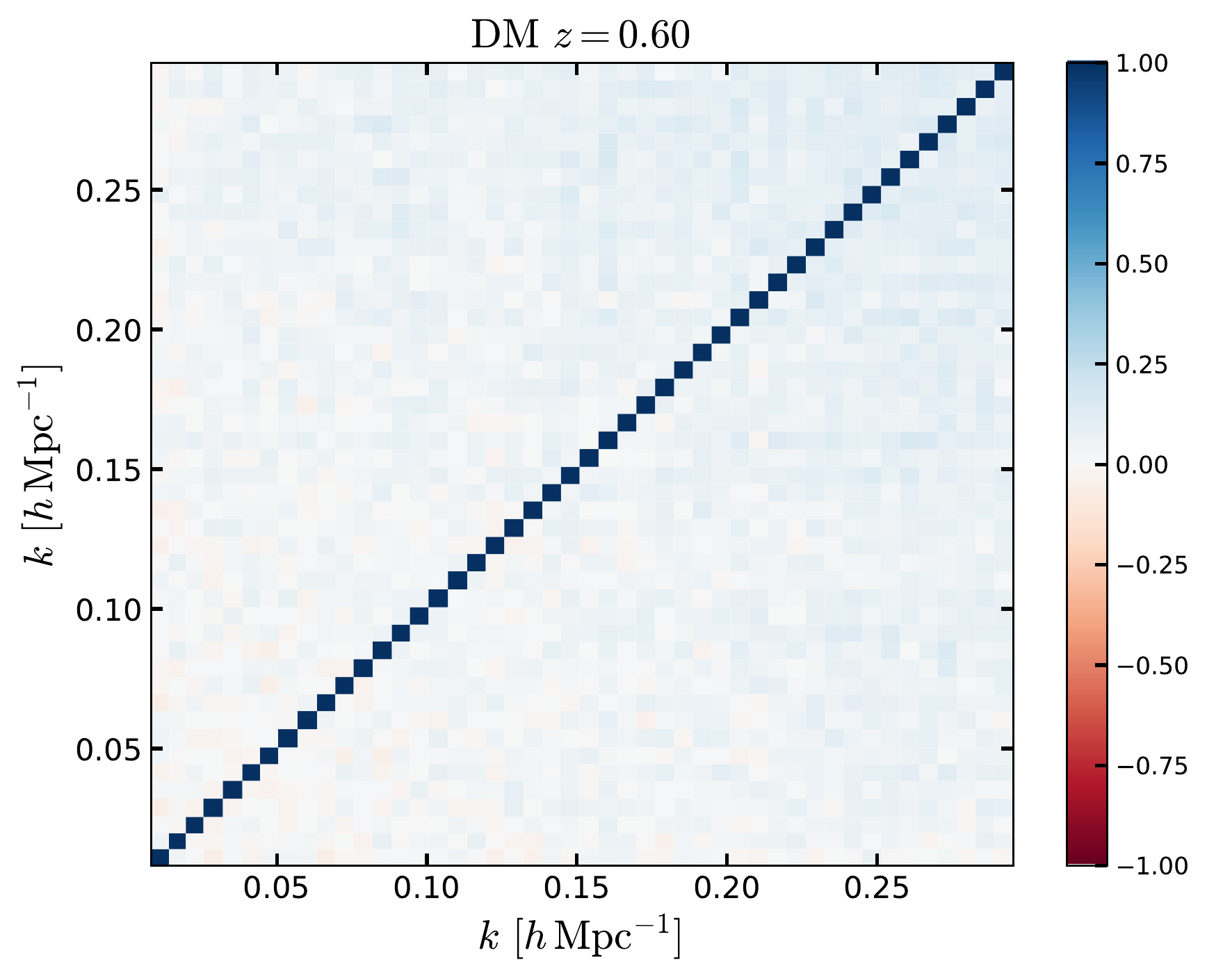}
\includegraphics[width=0.33\textwidth]{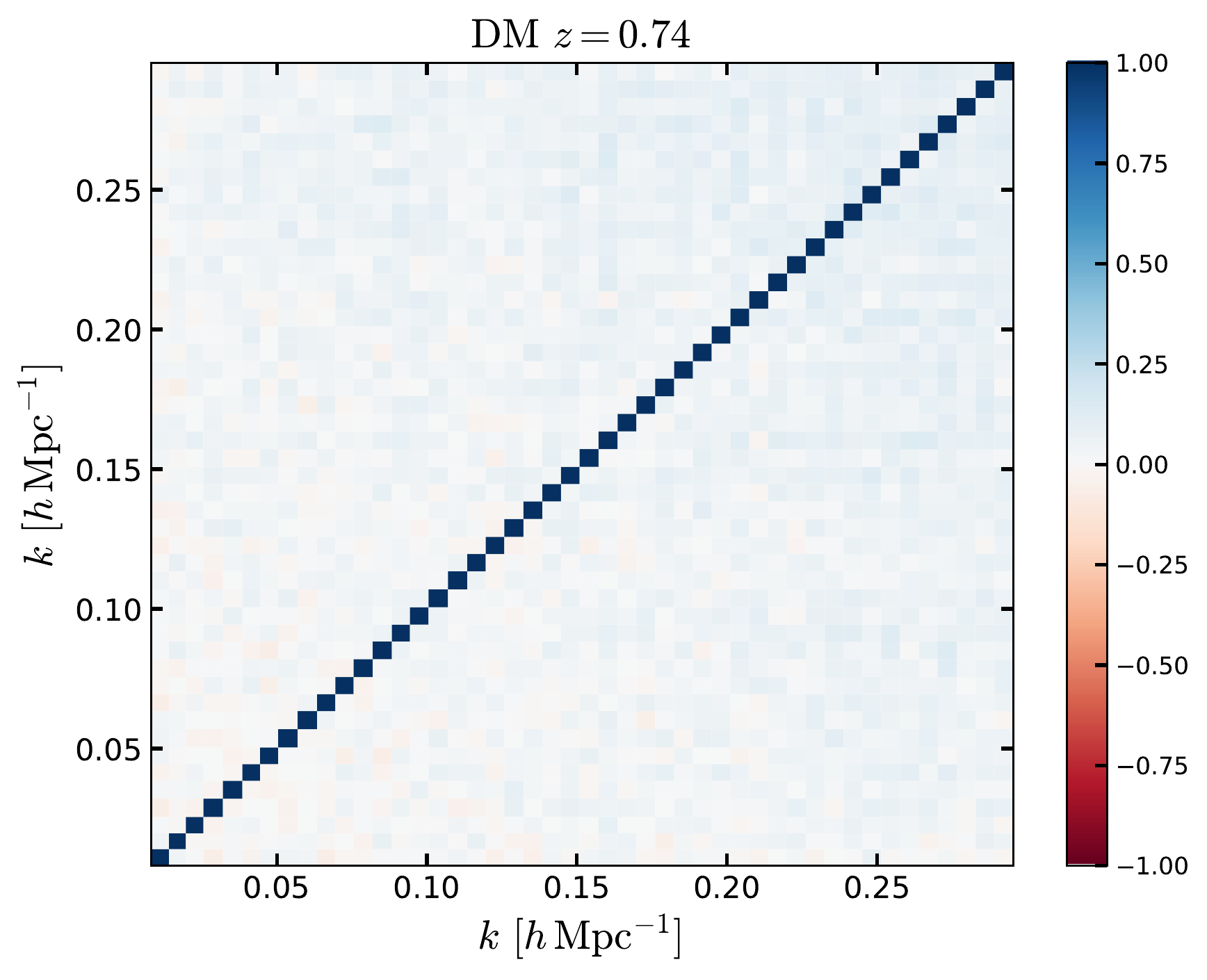}
\includegraphics[width=0.33\textwidth]{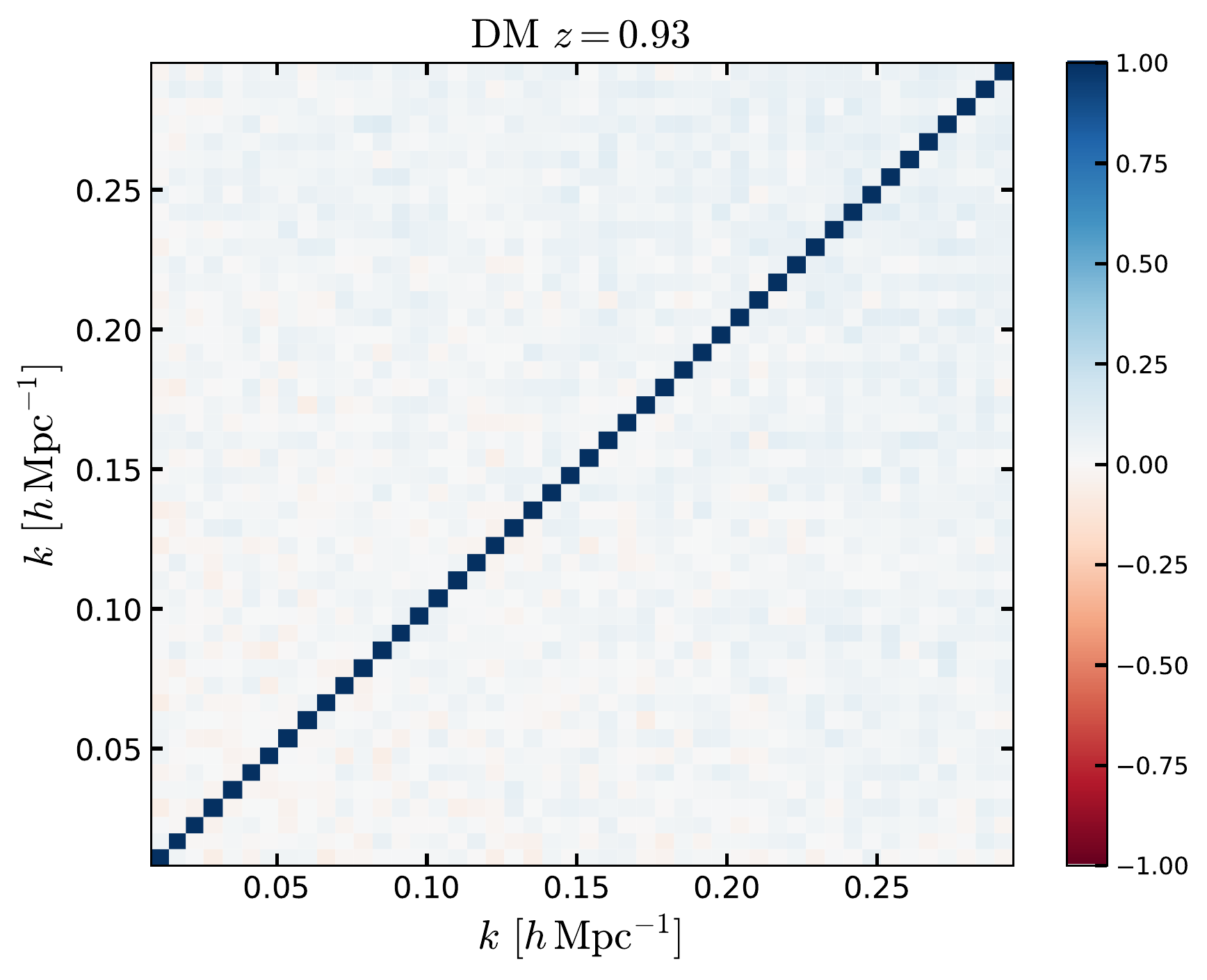}
\includegraphics[width=0.33\textwidth]{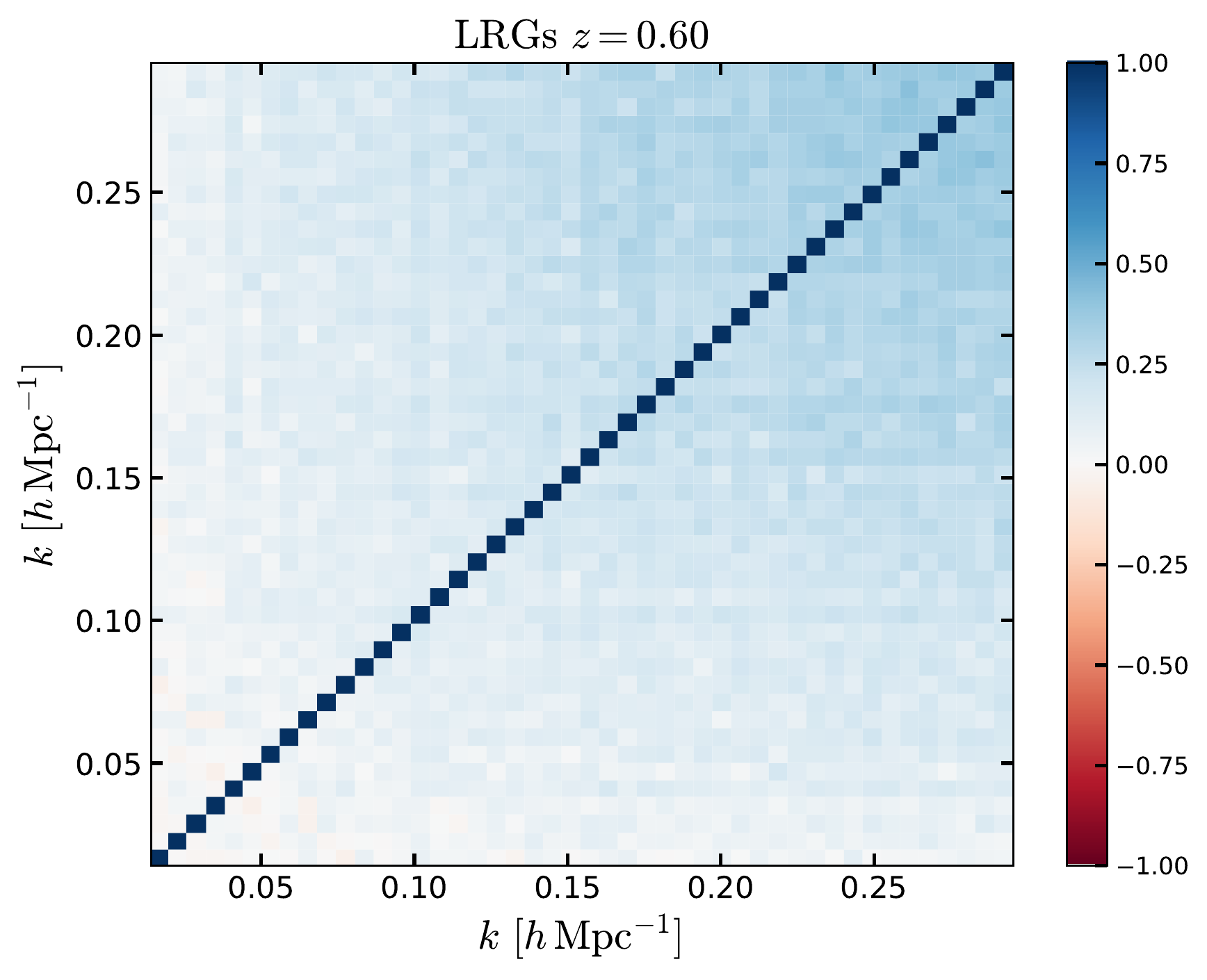}
\includegraphics[width=0.33\textwidth]{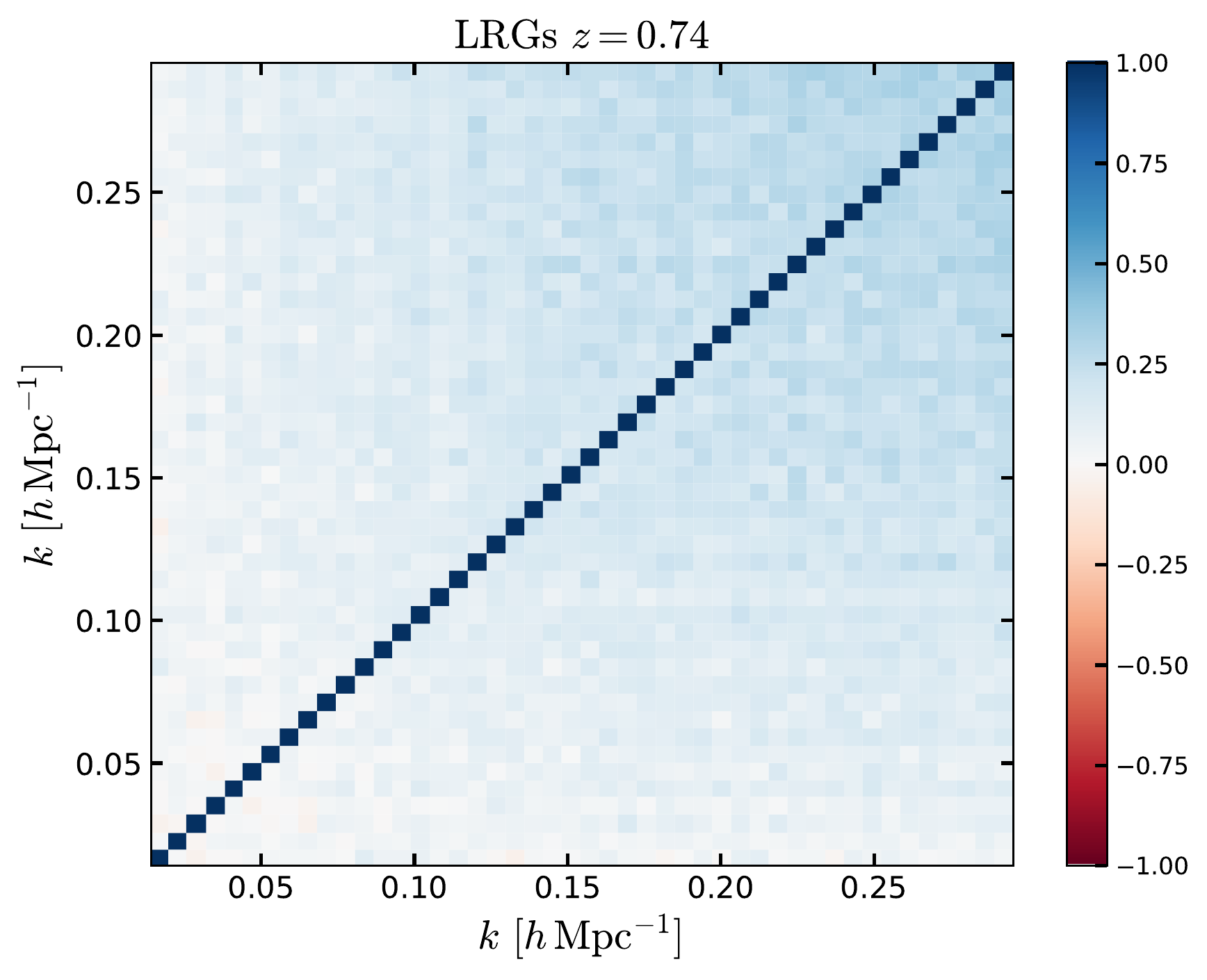}
\includegraphics[width=0.33\textwidth]{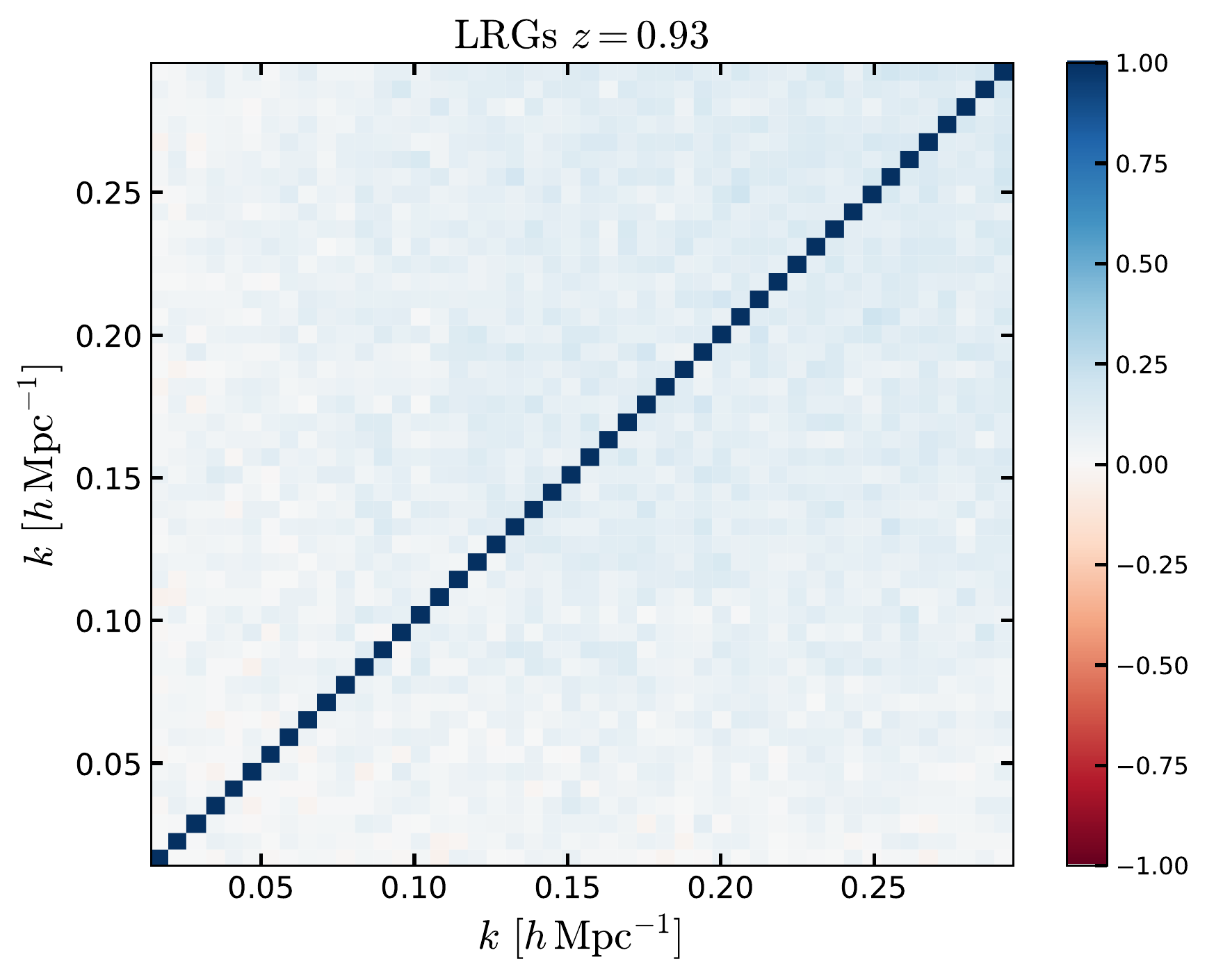}
\includegraphics[width=0.33\textwidth]{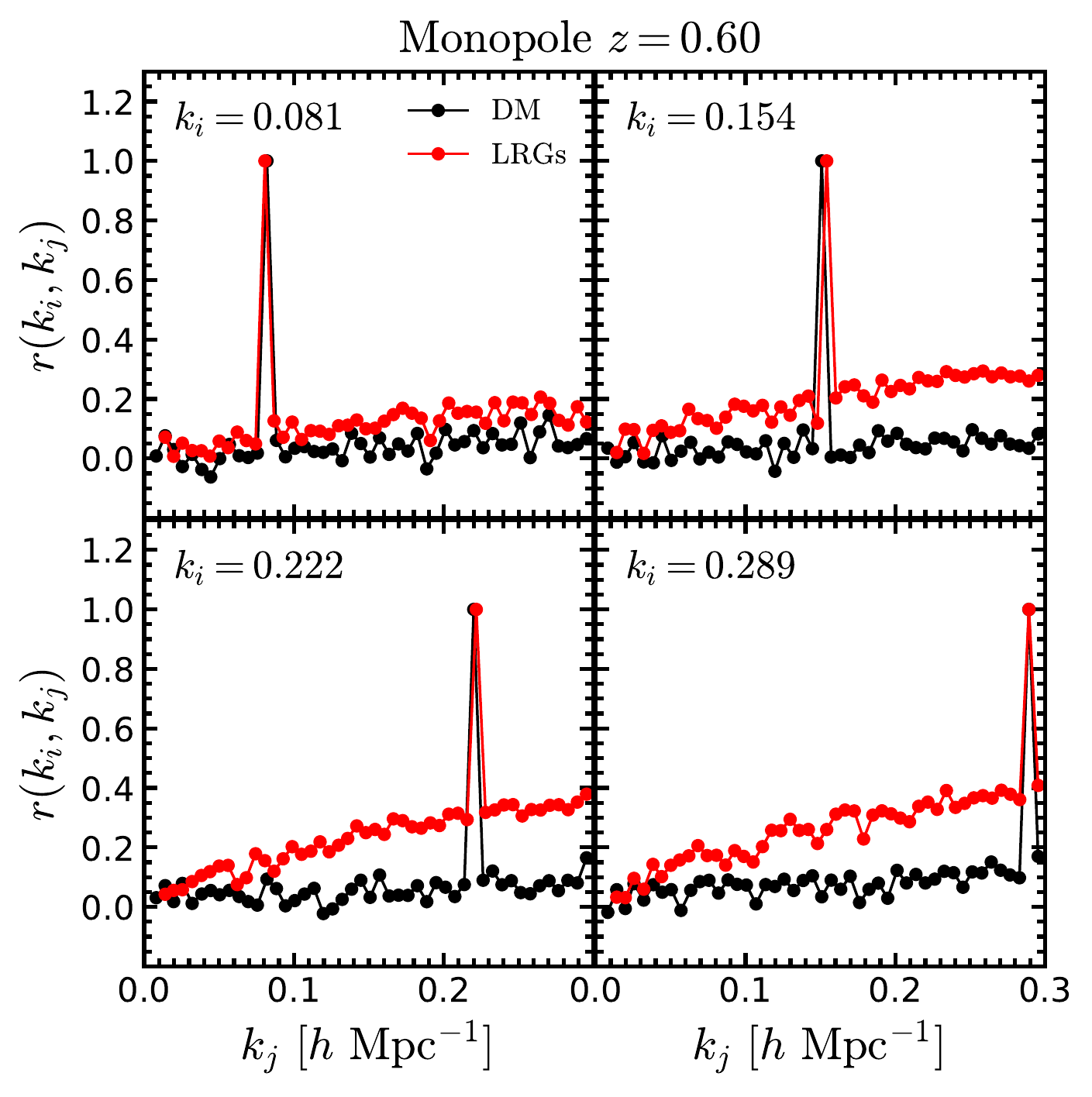}
\includegraphics[width=0.33\textwidth]{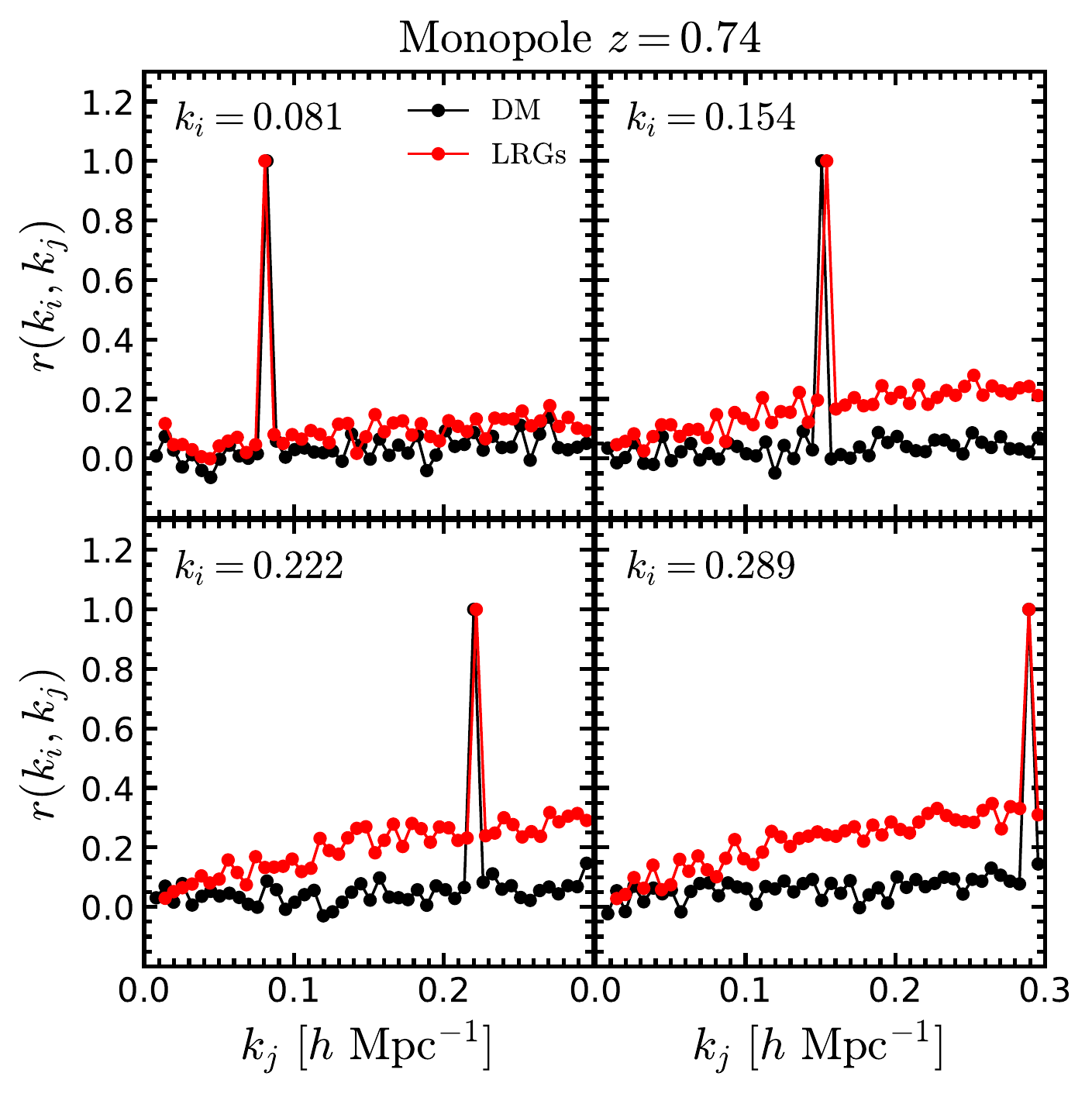}
\includegraphics[width=0.33\textwidth]{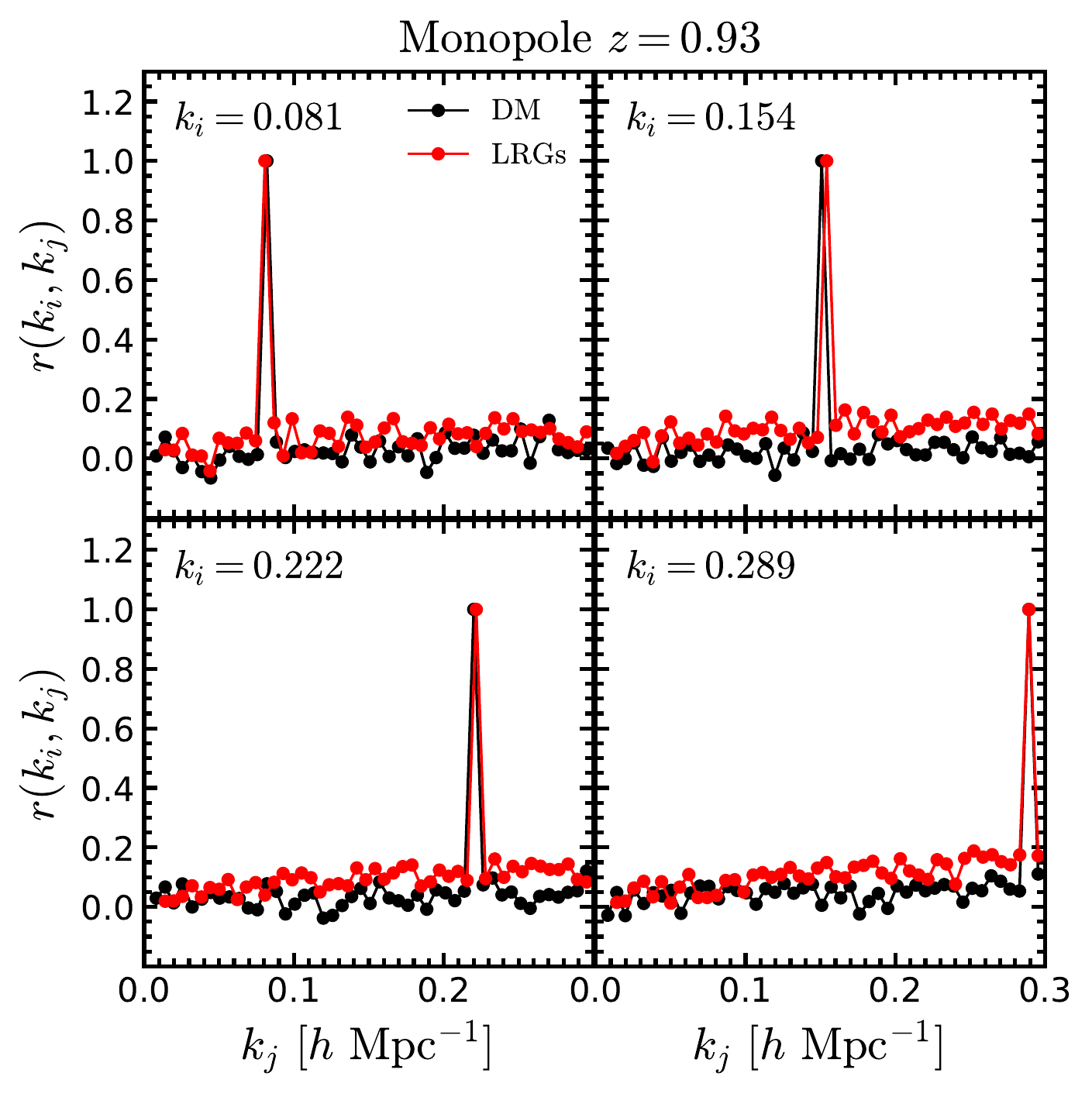}
\caption{Covariance analysis of the real-space dark matter and LRG power spectra at $z=0.6$, $0.74$ and $0.93$. {\it Upper panels:} Measured DM (black lines) and LRG (red lines) power spectrum from our \glam{} simulations together with the DM power spectrum multiplied by the bias squared (blue lines; Eq.~\eqref{eq:bias}), the {\it lower subpanels} show the error contribution. {\it Middle top} and {\it middle lower:} Correlation matrices of the real-space power spectrum for the DM density field and LRGs, respectively. {\it Bottom panels:} Slices through the correlation matrices at different values of $k_i$ in units of $[\hMpc]$.}
\label{fig:Pkm_error}
\end{figure*}

In general it is not possible to measure the three dimensional clustering of galaxies in real-space from observations. Some compromise involving projection is usually required to obtained a real-space statistic, such as the angular correlation function or the projected correlation function. The most direct three dimensional clustering measurements from surveys provide statistics in redshift-space, which are affected by peculiar velocities.
Moreover, future surveys like DESI aim to measure galaxy clustering on scales up to $\sim 200\Mpch$. Hence, taking advantage of our \glam{}-HOD machinery, here we present predictions for the large-scale galaxy clustering and covariance matrices of DESI-like LRGs for the correlation function and power spectrum. These quantities are fundamental for error estimates on the measurements of BAO and RSD \citep[see e.g.,][]{BOSS-DR12:2017sqa}.

In the following, we focus our attention on the large-scale clustering of DESI-like LRGs for pair separations in the range $0 < s/[\Mpch] < 150$ for the correlation function. For the power spectrum we show results in the wavenumber range $0.01 < k/[\hMpc] < 0.3$. 

%--------- Figure --------------
\begin{figure*}
 \centering
\includegraphics[width=0.47\textwidth]{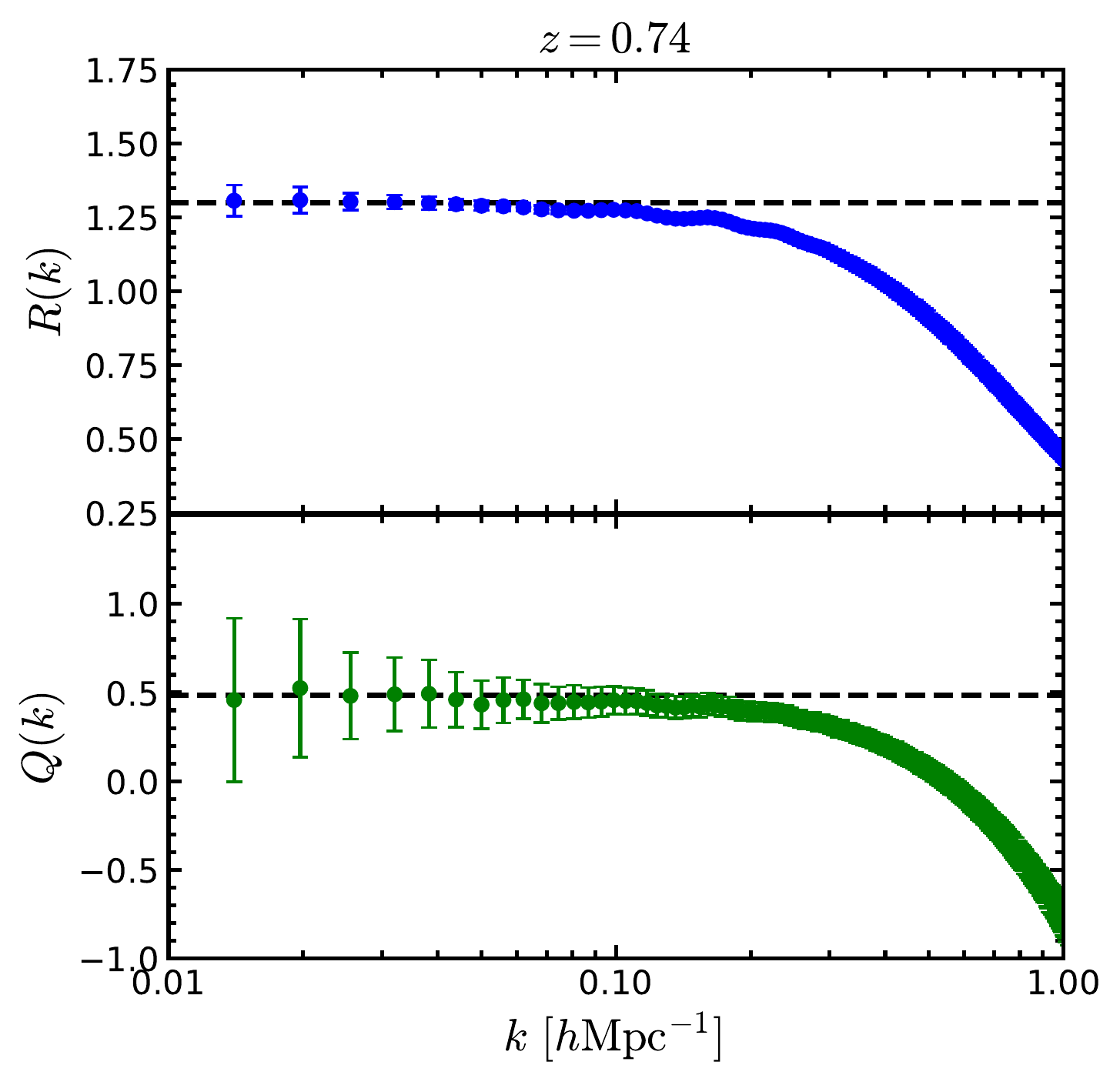}
\includegraphics[width=0.45\textwidth]{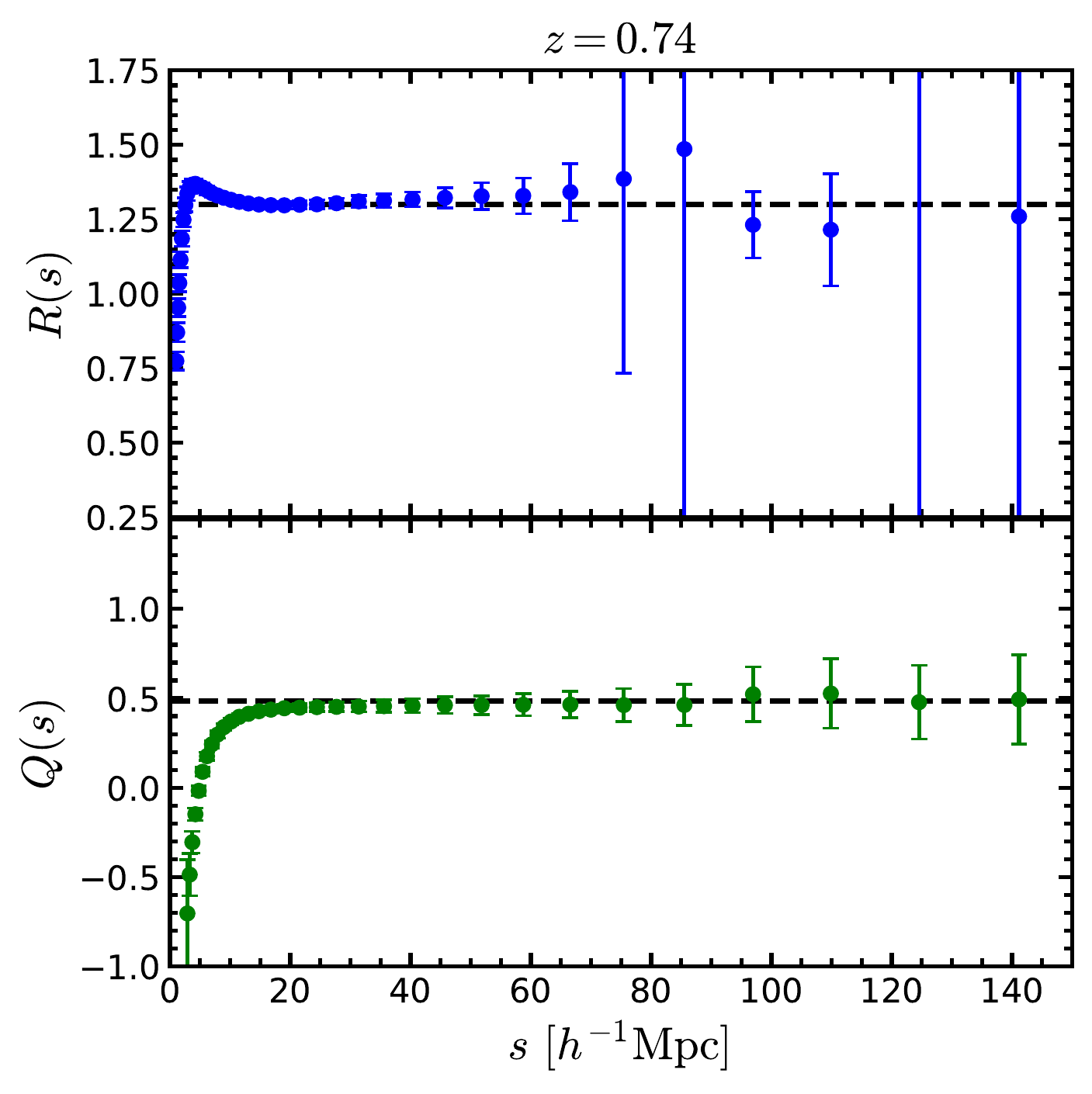}
\caption{Estimators $R$ ({\it upper subpanels}) (Eq.~\eqref{eq:Rs}) and $Q$ ({\it lower subpanels}), (Eq.~\eqref{eq:Qs}) as function of separation in Fourier ({\it left panels}) and configuration ({\it right panels}) space at $z=0.74$. Symbols with errorbars show the mean and standard deviation of the estimator measured from our 1000 \glam{} catalogues. The black dashed line in each panel represents the fiducial linear theory value.} 
\label{fig:rsd}
\end{figure*}

The upper panels of Figs.~\ref{fig:xil_glam} and \ref{fig:Pkl_glam} display the mean and standard deviation of the multipoles of the correlation function and the power spectrum calculated over 1000 \glam{} DESI-like LRGs realisations at $z=0.6$, $0.74$ and $0.93$. We also measure the covariance matrix, $\mathbf{C}$, of each estimator $\mathbf{E}$, as follows,
\begin{equation}\label{eq:cov}
    C_{ij} = \frac{1}{N_s - 1} \sum^{N_s}_{k=1}\left(E^k_i - \bar{E}_i\right)\left(E^k_j - \bar{E}_j\right)\,,
\end{equation}
where $N_s = 1000$ is the number of mocks, $\bar{E}_i = 1/N_s\sum_k E^k_i$ is the mean value of the estimator in the $i$-th separation bin, and $E^k_i$ is the corresponding measurement from the $k$-th mock. The standard deviation is estimated from the diagonal elements of the covariance matrix,
\begin{equation}\label{eq:sig}
    \sigma_i = \sqrt{C_{ii}}\,.
\end{equation}
We show the diagonal error contribution, $\sigma_{E_i}/E_i$, of the moments of the correlation function and power spectrum in the lower subpanels of the upper row of Figs.~\ref{fig:xil_glam} and \ref{fig:Pkl_glam}. We observe an increase in the size of the error contribution at large-scales, especially for the monopole and quadrupole in configuration space.

We display the correlation matrix,
\begin{equation}\label{eq:Rij}
    r_{ij} = \frac{C_{ij}}{\sqrt{C_{ii}C_{jj}}}\,,    
\end{equation}
in the middle (monopole) and bottom (quadrupole) panels of Figs.~\ref{fig:xil_glam} and \ref{fig:Pkl_glam} for the correlation function and power spectrum, respectively. The diagonal and non-diagonal components have different magnitudes and evolve differently with redshift. Figs.~\ref{fig:xil_error} and \ref{fig:Pkl_error} show cuts through the correlation matrices corresponding to our measurements in configuration and Fourier space, respectively. These diagrams help us to better display the level of correlation and the structure of the matrices. In the case of the moments of the correlation function (Fig.~\ref{fig:xil_error}), we show the cuts at four different separation bins, $s_i = (37.5,72.5,107.5,142.5) \Mpch$, while in Fourier space (Fig.~\ref{fig:Pkl_error}) we use $k_i = (0.081,0.154,0.222,0.289) \hMpc$. We see a strong correlation between the bins close to the diagonal elements in the monopole and quadrupole of the correlation function at $z=0.60$ and $z=0.74$; this correlation becomes weaker at $z=0.93$ (Fig.~\ref{fig:xil_error}). In the case of the multipoles of the power spectrum, the off-diagonal elements are much less correlated than the diagonal components, with values close to zero (Fig.~\ref{fig:Pkl_error}). This trend is strongest for the quadrupole of the power spectrum.

\citet{Klypin:2017jwl} carried out an extensive study of the covariance and correlation matrix associated with the dark-matter power spectrum of \glam{} simulations. Our results for the estimation of errors from the \glam{}-HOD catalogues extends the work of \citeauthor{Klypin:2017jwl} to galaxies and to the correlation function. In detail, Fig.~\ref{fig:Pkm_error} shows the covariance analysis of the real-space DM and LRG power spectra. We summarise our findings as follows. First, in the upper panels we display the measurements from our simulations, we observe that the size of the error is similar for both DM and LRGs at large-scales $(k<0.05\hMpc)$ but on smaller scales the amplitude of the error of the galaxy power spectrum becomes larger with increasing redshift. We also show the DM power spectrum and its errors scaled by the LRG bias squared (see Sec.~\ref{sec:comparison} for details) as a blue solid line (with a shaded region showing the $1\sigma$ error) in the upper panel of the first row of Fig.~\ref{fig:Pkm_error}. Second, the correlation matrices are shown in the middle panels (upper middle panels for DM and lower middle panels for LRGs), we find that the amplitude of the DM correlation matrices are consistent with those reported by \citet{Klypin:2017jwl}. On the other hand, the correlation amplitude of the LRG power spectrum is similar to its analogue in redshift space (see middle panels of Fig.~\ref{fig:Pkl_glam}). Lastly, the evolution of the non-diagonal terms of the correlation matrices are displayed in the bottom panels of Fig.~\ref{fig:Pkm_error}. We compare the level of correlation at four values of the separation bin, $k_i = (0.081,0.154,0.222,0.289) \hMpc$, finding a more complex behaviour from the LRGs correlation matrices with an increase amplitude at small scales, this behaviour is also consistent with our findings in redshift space (see Fig.~\ref{fig:Pkl_glam}). Moreover, the amplitude of the non-diagonal elements are similar for both DM and LRGs at $z=0.93$ (bottom right panel of Fig.~\ref{fig:Pkm_error}).

Finally, we can use the covariance matrix of each estimator to define a chi-squared to find the best-fitting cosmological parameters as follows,
\begin{equation}\label{eq:chi2}
    \chi^2 = \sum^{N_s}_{i,j=1} \left(E^{\rm th}_i - E^{\rm obs}_i\right) C^{-1}_{ij} \left(E^{\rm th}_j - E^{\rm obs}_j\right)\,,
\end{equation}
where $C^{-1}_{ij}$ is the inverse of the covariance matrix, Eq.~\eqref{eq:cov}, $E^{\rm th}$ is the theoretical expectation of the estimator that depends on the cosmological parameters and $E^{\rm obs}$ is the measured estimator from our \glam{}-HOD catalogues. This definition is used in Sec.~\ref{sec:rsd} and Sec.~\ref{sec:bao}. 

It is instructive to compare the errors we obtain in the \glam{} simulation boxes with the errors expected in the DESI measurements. DESI will measure the clustering of LRGs in a series of redshift shells over a solid angle of 14\,000 square degrees. We anticipate that DESI will sample a comoving volume of $V/[h^{-3}{\rm Gpc}^3] = 2.63$, $3.15$ and $4.10$ respectively on bins centred at redshifts of $z = 0.65$, $0.75$ and $0.95$ \citep{DESI:2016zmz}. Hence, to get a rough impression of how our error estimates (on the cosmological parameters) will scale to those expected for DESI, we can scale the \glam{} errors by the square root of the inverse volume ratio (e.g. \citealt{fkp}): $\sigma^\prime(z)=\sigma(z)\sqrt{V_{\rm GLAM}/V_{\rm DESI}(z)}$, where $V_{\rm GLAM} = 1\,h^{-3}{\rm Gpc}^3$. Note that we refrain from carrying out a more detailed comparison with the errors reported in \cite{DESI:2016zmz}, as these were obtained using a Fisher matrix method, which assumes Gaussian errors and no off-diagonal terms.
%---------------------------------------------------------------
\subsection{Linear redshift-space distortions}\label{sec:rsd}
%---------------------------------------------------------------
In large volume galaxy surveys we can extract information about the growth of structure through the linear growth rate, $f$, which is defined as the logarithmic derivative of the linear growth function of density perturbations, $D$, with respect to the scale factor, $a$, 
\begin{equation}\label{eq:f_lin}
f \equiv \frac{{\rm d}\ln D}{{\rm d}\ln a}\,.
\end{equation}

In linear perturbation theory, the relation between the redshift-space galaxy power spectrum, $P_{\rm s}$, and its real-space counerpart, $P_{\rm r}$, is given by \citep{Kaiser:1987}: 
\begin{equation}\label{eq:Pk_s_lin}
P_{\rm s}(k,\mu) = (1 + \beta \mu^2)^2 P_{\rm r}(k)\,.
\end{equation}
From Eq.~\eqref{eq:Pk_s_lin} we can see that the amplitude of the RSD is related to the distortion parameter $\beta$, defined as
 \begin{equation}\label{eq:beta}
 \beta(z) \equiv \frac{f(z)}{b(z)}\,,
 \end{equation}
where $f$ is the linear growth rate (Eq.\eqref{eq:f_lin}), and $b$ is the linear galaxy bias both of which vary with redshift, Eq.~\eqref{eq:bias}. 

The monopole and quadrupole moments of the power spectrum can be estimated from Eqs.~\eqref{eq:Pk_l} and \eqref{eq:Pk_s_lin},
\begin{eqnarray}
P_0(k) &=& \left( 1 + \frac{2\beta}{3} + \frac{\beta^2}{5} \right)P_{\rm r}(k)\,,\label{eq:Pk0_lin}\\
P_2(k) &=& \left( \frac{4\beta}{3} + \frac{4\beta^2}{7} \right)P_{\rm r}(k)\,,\label{eq:Pk2_lin}
\end{eqnarray}
where $P_{\rm r}(k)$ is galaxy power spectrum in real-space.

On the other hand, the redshift-space correlation function can be expressed as follows \citep{Hamilton:1992zz,Hamilton:1997zq}:
\begin{equation}
\xi(s,\mu) = [1 + \beta (\partial/\partial z)^2(\nabla^2)^{-1}]^2\xi(r)\,,
\end{equation}
In linear theory, the monopole and quadrupole of the correlation function can be estimated using \citep{Hamilton:1992zz}, i. e.,
\begin{eqnarray}
\xi_0(s) &=& \left( 1 + \frac{2\beta}{3} + \frac{\beta^2}{5} \right)\xi(r)\,,\label{eq:xi0_lin}\\
\xi_2(s) &=& \left( \frac{4\beta}{3} + \frac{4\beta^2}{7} \right)[\xi(r) - \bar{\xi}(r)]\,,\label{eq:xi2_lin}
\end{eqnarray}
where $\xi(r)$ is the galaxy correlation function in real-space and $\bar{\xi}$ is its volume integral out to pair separation $r$:
\begin{equation}
\bar{\xi}(r) = \frac{3}{r^3} \int^r_0 \xi(r')r'^2~{\rm d}r' \,.
\label{eq:volave}
\end{equation}

From Eqs.~\eqref{eq:Pk0_lin}$-$\eqref{eq:Pk2_lin} and Eqs.~\eqref{eq:xi0_lin}$-$\eqref{eq:xi2_lin} we can define two estimators to obtain the distortion parameter, $\beta$, or the linear growth rate, $f$, \citep{Cole:1993kh,Hawkins:2002sg},
\begin{equation}\label{eq:Rs}
R(k/s) = \frac{P_0(k)}{P_{\rm r}(k)} = \frac{\xi_0(s)}{\xi(r)} = 1 + \frac{2\beta}{3} + \frac{\beta^2}{5}\,,
\end{equation}
and
\begin{equation}\label{eq:Qs}
Q(k/s) = \frac{P_2(k)}{P_0(k)} =  \frac{\xi_2(s)}{\xi_0(s) - \bar{\xi}_0(s)} = \frac{(4/3)\beta + (4/7)\beta^2}{1 + (2/3)\beta + (1/5)\beta^2}\,,
\end{equation}
where $\mathcal{F}(k/s)$ indicates that the quantity $\mathcal{F}$ can be a function of $k$ or $s$ and
\begin{equation}
\bar{\xi}_0(s) = \frac{3}{s^3} \int^s_0 \xi_0(s')s'^{2}~{\rm d}s'\,,
\end{equation}
is the volume average of the monopole in redshift space, the analogue of Eq.~\ref{eq:volave}.

Fig.~\ref{fig:rsd} shows our measurements of the $R(k/s)$, and the $Q(k/s)$ estimators from our DESI-\glam{} LRG mock catalogues at the median redshift $z=0.74$. The black dashed line in each panel corresponds to the linear theory predictions. From the measurements in Fourier space (left panels of Fig.~\ref{fig:rsd}), we can see that both estimators become closer to the linear theory predictions at scales $k \lesssim 0.1 \hMpc$, this means that linear theory is only valid on sufficiently large-scales. On small scales, where the non-linear motions of galaxies dominate, we observe a downturn in the signal of each estimator. The trend is similar in configuration space (right panel of Fig.~\ref{fig:rsd}), where we observe that the linear theory limit is reached on scales $s>20\Mpch$. All panels in Fig.~\ref{fig:rsd} show the same range of values on the vertical axis, allow us to see that the errors are slightly different in Fourier and configuration space, especially in $R$. 

%--------- Figure --------------
\begin{figure*}
 \centering
\includegraphics[width=0.45\textwidth]{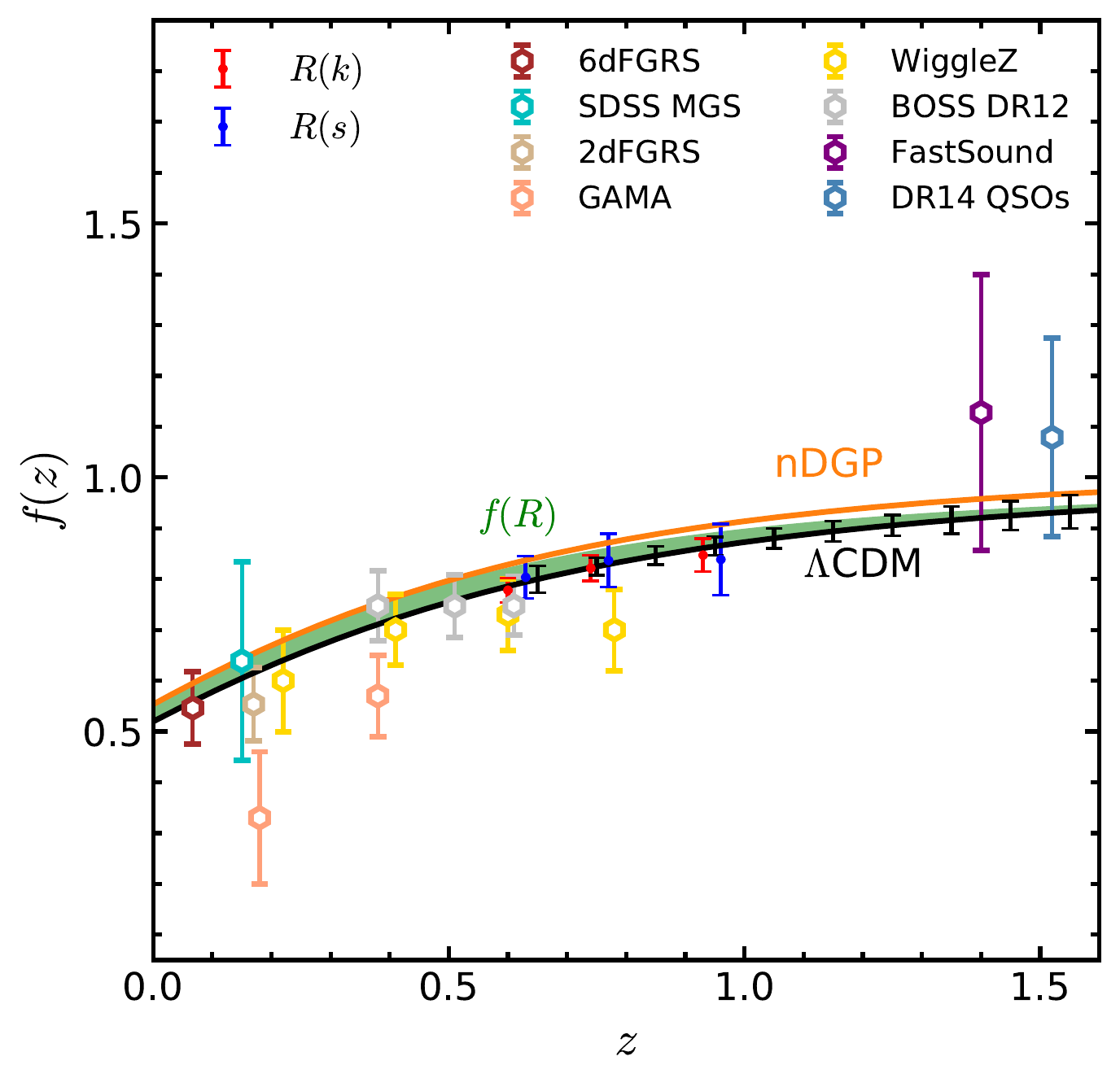}
\includegraphics[width=0.45\textwidth]{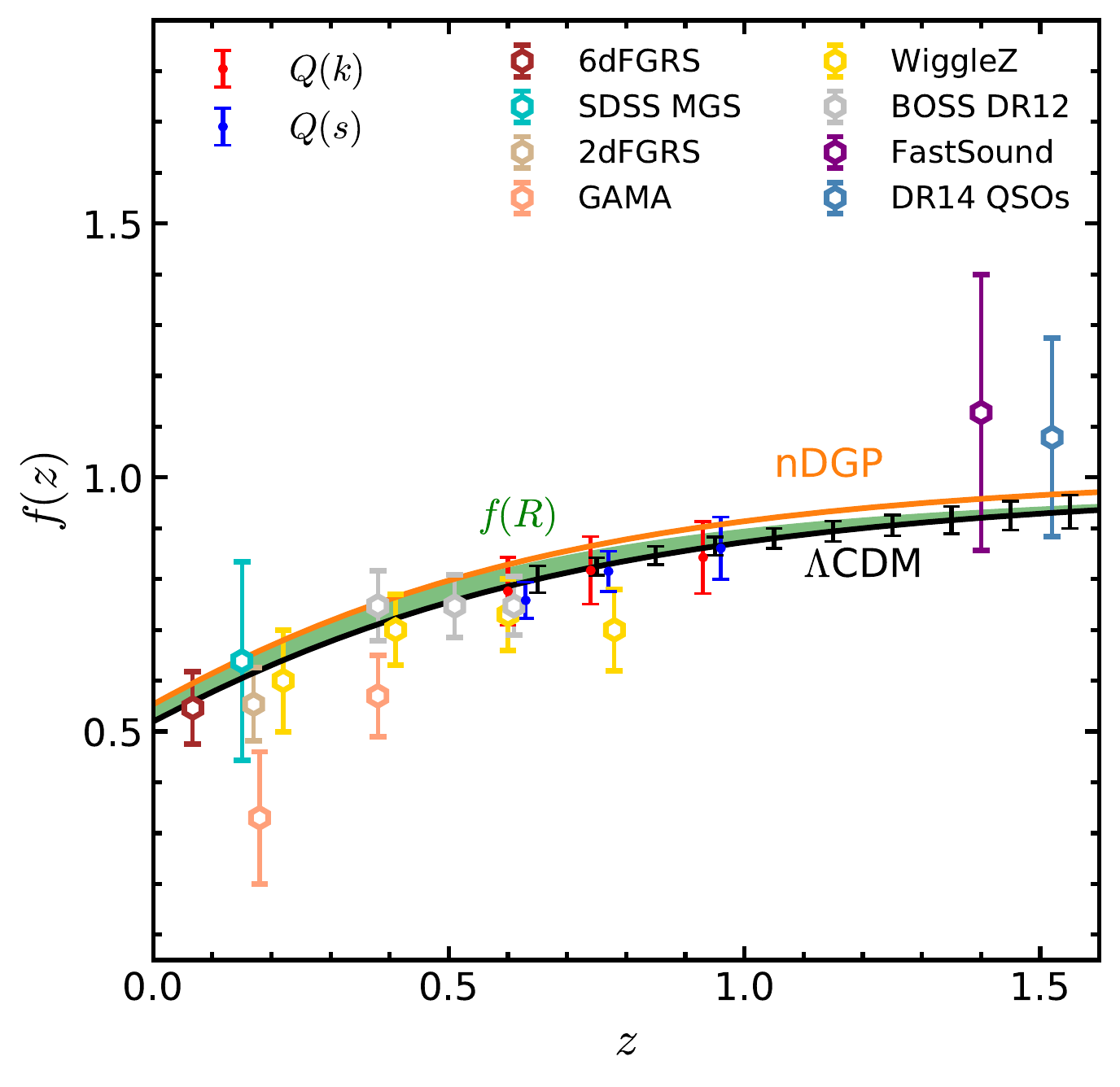}
\caption{Evolution of the linear growth rate, $f$, as a function of redshift. Our estimations from $R$ and $Q$ are shown in the left and right panel in configuration (red dots with errobars) and Fourier (blue dots with errorbars) space, respectively. The coloured symbols display measurements from different surveys at different redshifts as specified in the legend. Solid curves show the prediction for  $\Lambda$CDM (black), nDGP (orange) and $f(R)$-gravity (green shaded region that represent wavenumbers $0.01 \leq k/[\hMpc] \leq 0.1$) models. The black errorbars over the $\Lambda$CDM prediction represent the DESI 14K forecast for $k_{\rm max} = 0.1\hMpc$ \citep[see table 2.3 of][]{DESI:2016zmz}.}
\label{fig:f_g}
\end{figure*}
%--------- Table --------------
\begin{table*}
\caption{Results for the best-fitting values of the linear-growth rate, $f$, at redshifts $0.60$, $0.74$ and $0.93$ for our estimators $R(k/s)$, Eq.~\eqref{eq:Rs}, and $Q(k/s)$, Eq.~\eqref{eq:Qs}.}
\begin{tabular}{ccccccc}
\hline
\hline
Measurement & Redshift   & Fiducial & $R(k)$            & $Q(k)$            & $R(s)$            & $Q(s)$            \\ \hline
$f$         & $z = 0.60$ & $0.786$  & $0.778 \pm 0.024$ & $0.776 \pm 0.067$ & $0.803 \pm 0.041$ & $0.758 \pm 0.036$ \\
$f$         & $z=0.74$   & $0.823$  & $0.822 \pm 0.025$ & $0.817 \pm 0.066$ & $0.839 \pm 0.053$ & $0.815 \pm 0.039$ \\
$f$         & $z=0.93$   & $0.861$  & $0.847 \pm 0.033$ & $0.842 \pm 0.071$ & $0.838 \pm 0.070$ & $0.861 \pm 0.061$ \\ \hline \hline
\end{tabular}\label{tab:f_g}
\end{table*}

To extract the linear growth rate, $f$, from our measurements, we perform a likelihood analysis by minimising $\chi^2$ defined by Eq.~\eqref{eq:chi2} by fitting the measurements of $R(k/s)$, Eq.~\eqref{eq:Rs}, and $Q(k/s)$, Eq.~\eqref{eq:Qs} over the range of scales $k < 0.1 \hMpc$ in Fourier space and $s > 20 \Mpch$ in configuration space. We fix the galaxy bias and just allow the linear growth rate to vary. To do so, we employ the Monte Carlo Markov Chain (MCMC) technique implemented in the {\sc emcee} python package \citep{emcee:2013}. 

In Fig.~\ref{fig:f_g}, we compare the predictions for the linear growth rate $f(z)$ obtained from our DESI-\glam{} LRG mocks at $z=0.6$, $0.74$ and $0.93$ with the current observational measurements from large galaxy surveys, including 6dFGRS at $z = 0.067$ \citep{6dFGRS:2012px}, SDSS MGS at $z=0.15$ \citep{SDSS-MGS:2014opa}, 2dFGRS at $z=0.17$ \citep{2dFGRS:2004fs}, GAMA at $z=0.18$ and $0.38$ \citep{GAMA:2013nif}, WiggleZ at $z=0.22$, $0.41$, $0.6$ and $0.78$ \citep{WiggleZ:2011rj}, BOSS DR12 at $z=0.32$, $0.51$ and $z=0.61$ \citep{BOSS-DR12:2017sqa}, FastSound at $z=1.4$ \citep{Okumura:2015lvp} and the eBOSS DR14 QSO sample at $z=1.52$ \citep{Zarrouk:2018vwy}. The black errorbars over the $\Lambda$CDM predictions indicate the estimated error from the DESI forecast \citep[see table 2.3 of][]{DESI:2016zmz}. Note that in this work we do not include the light-cone and survey geometry effects on our mocks. These will be considered in a forthcoming paper.

Table~\ref{tab:f_g} summarises the best-fitting values of the linear growth rate, $f$, at $z=0.60$, $0.74$ and $0.93$ obtained from the estimators $R$ (Eq.~\eqref{eq:Rs}) and $Q$ (Eq.~\eqref{eq:Qs}) in configuration and Fourier space. We also show the values from the fidiciual cosmology. We find very good agreement between our estimations and the theoretical predictions. The largest errors come from the $R$ estimator in configuration space and $Q$ in Fourier space, this might be due to the size of the error contribution of our measurements (see Sec.~\ref{sec:cov} for details). The best case is $R$ in Fourier case, which estimates the linear growth rate with a precision better than $4$ per cent. As we mentioned above, we should expect that our errors differ up to a factor of two when comparing to the DESI forecast.

%--------- Figure --------------
\begin{figure*}
 \centering
\includegraphics[width=0.47\textwidth]{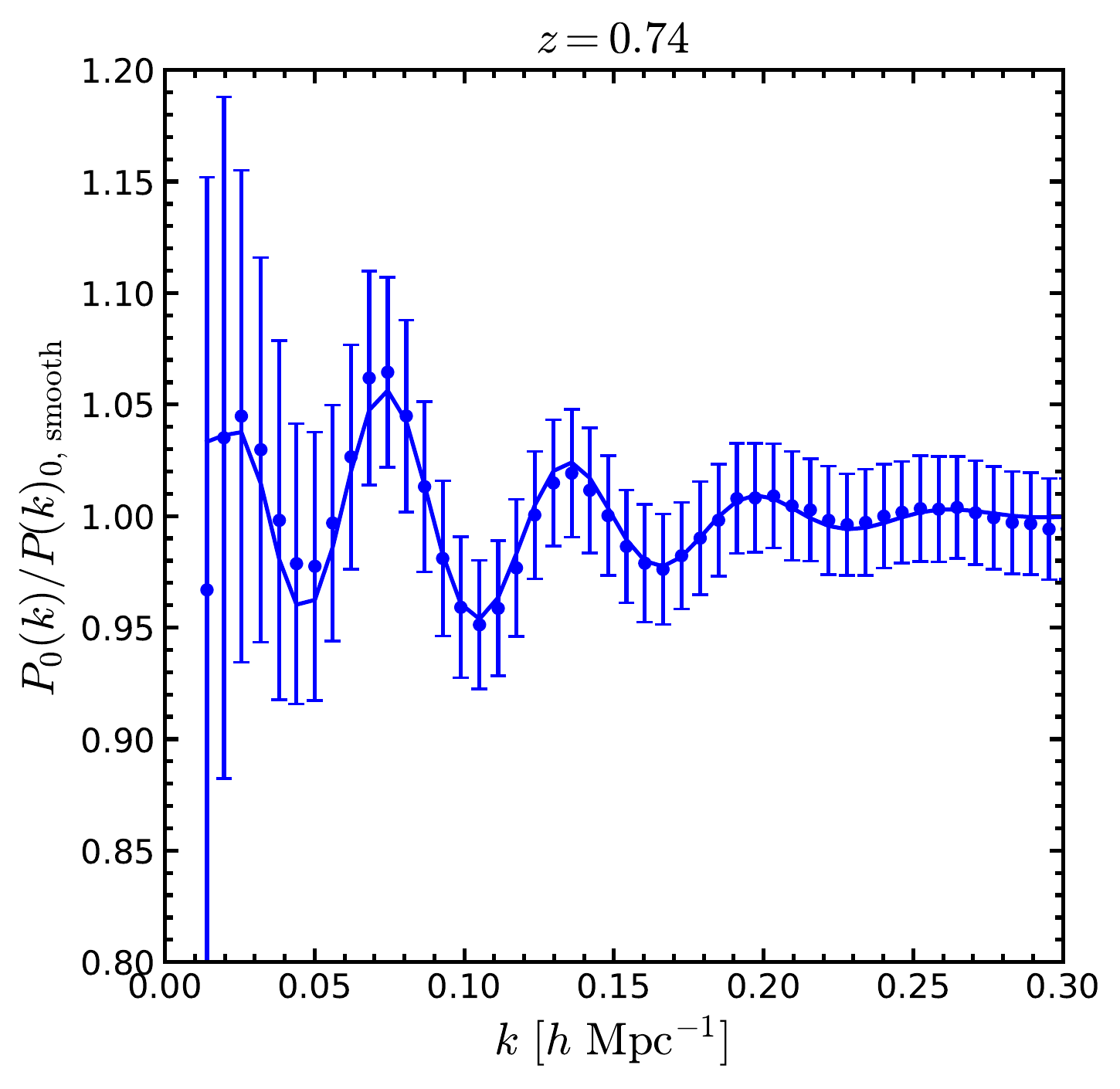}
\includegraphics[width=0.445\textwidth]{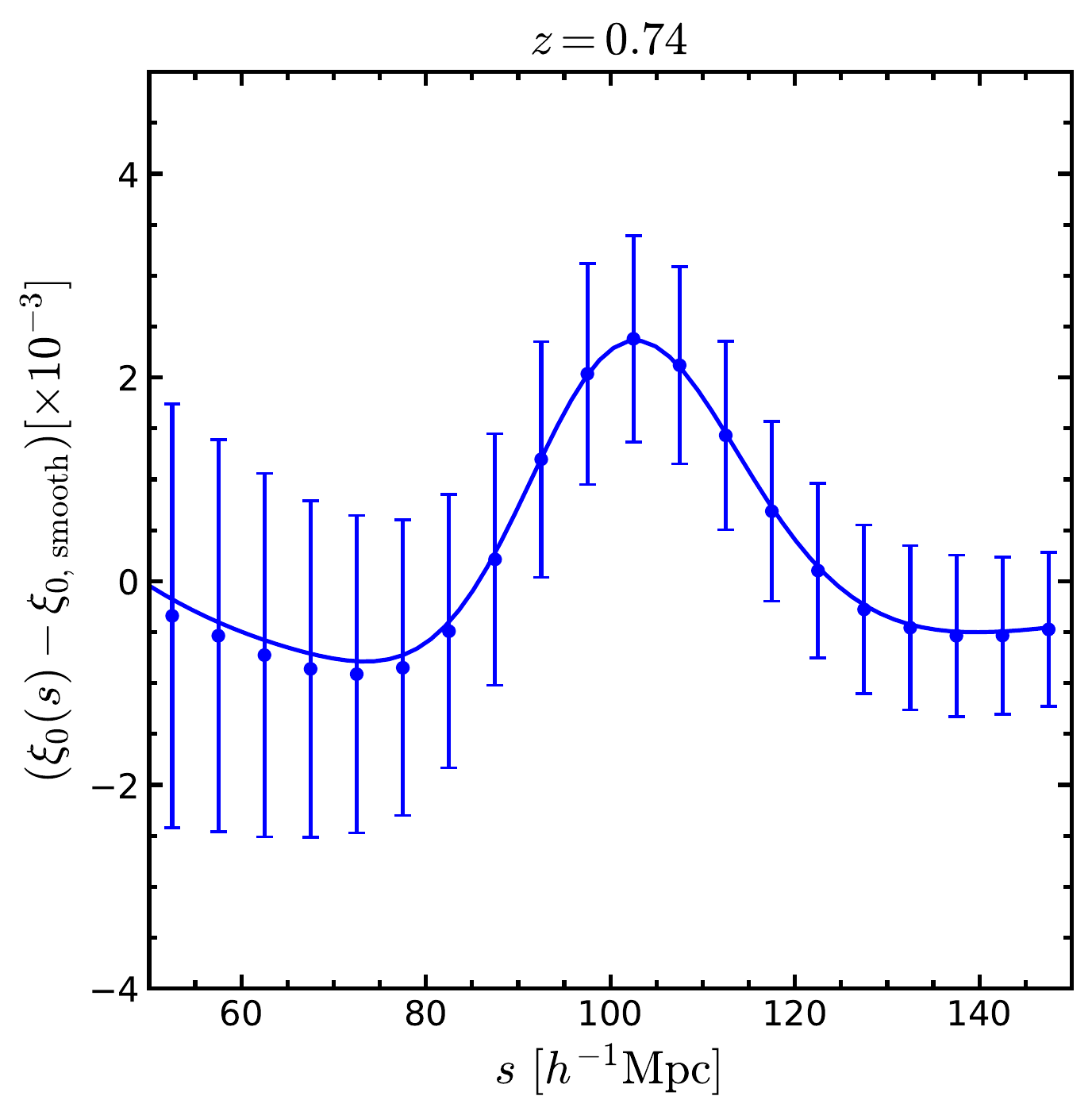}
\caption{BAO signals in the monopole of the power spectrum ({\it left panel}) and correlation function ({\it right panel}) at $z=0.74$. Blue dots with error bars come from the measurements from our \glam{}-LRG catalogues. The solid lines show the predictions from the best-fitting BAO models. In order to highlight the BAO features, we have divided the $P(k)$ measurements and the best-fitting  model by the no-wiggle power spectrum of the best-fitting model. In configuration space we have also subtracted the smooth component of the best-fitting model.}
\label{fig:bao}
\end{figure*}

It is expected that DESI will provide a means to distinguish between gravity models. For this reason, in Fig.~\ref{fig:f_g}, we also show the theoretical expectations from two representative modified gravity models: the $f(R)$ Hu-Sawicki model \citep{Hu:2007nk} and the normal branch of the DGP model \citep[nDGP;][]{Dvali:2000hr}. Previously, \cite{Hernandez-Aguayo:2018oxg} presented predictions for the linear and non-linear RSDs in configuration space for these models but for the {\sc boss}-{\sc cmass} sample at $z \leq 0.5$ \citep{Manera:2012sc}.

The linear growth for the matter fluctuations in these gravity models can be obtained by solving the equation of the linear growth factor, $D$,
\begin{equation}\label{eq:Dp}
D'' + \left(2 - \frac{3}{2}\Omega_{\rm m}(a) \right)D' - \frac{3}{2}\frac{G_{\rm eff}}{G}\Omega_{\rm m}(a) D = 0\,,
\end{equation}
where $^\prime$ denotes a derivative with respect to $\ln a$ and $G_{\rm eff}$ takes values of
\begin{equation}\label{eq:Geff}
\frac{G_{\rm eff}}{G} = 
\left \{
      \begin{array}{lc}
          1 + k^2/[3(k^2 + a^2m^2_{f_R}(a))] & f(R)\,,\\ 
          1 + 1/[3\beta_{\rm DGP}(a)] & {\rm nDGP} \,,
      \end{array}
   \right.
\end{equation}
where 
\begin{equation}\label{eq_ma}
m^2_{f_R}(a) = \frac{H_0^2(\Omega_{\rm m} + 4\Omega_\Lambda)}{2|f_{R0}|} \left(\frac{\Omega_{\rm m} a^{-3} + 4\Omega_\Lambda}{\Omega_{\rm m} + 4\Omega_\Lambda}\right)^{3}\,,
\end{equation}
and 
\begin{equation}\label{eq:beta_dgp}
\beta_{\rm DGP}(a) = 1 + \frac{\Omega_{\rm m}a^{-3} + 2\Omega_\Lambda}{2\sqrt{\Omega_{\rm rc}(\Omega_{\rm m}a^{-3} + \Omega_{\Lambda})}}\,,
\end{equation}
where $H_0$ is the present-day value of the Hubble parameter, $\Omega_{\rm m}$ and $\Omega_\Lambda$ are the current matter and dark energy density parameters, respectively. $f_{R0}$ and $\Omega_{\rm rc}$ are free parameters of each model that affect the deviation from the $\Lambda$CDM model. Note that $G^{f(R)}_{\rm eff}$ is a function of time and scale, which means that the linear growth of structure for $f(R)$ gravity is scale dependent, while for nDGP it is scale independent. In Fig.~\ref{fig:f_g} we show the theoretical values of the linear growth rate, Eq.~\ref{eq:f_lin}, of these models for the cases: $f_{R0} = -10^{-5}$ and the range of scales $0.01 \leq k/[\hMpc] \leq 0.1$ for $f(R)$-gravity and $\Omega_{\rm rc}=0.25$ for the nDGP model.

We see that the size of the errors from the DESI forecast is small enough to distinguish between the $\Lambda$CDM and the nDGP model. However, it is still unclear if we will be able to rule out $f(R)$ gravity models using RSDs.
%---------------------------------------------------------------
\subsection{Isotropic measurements of the baryon acoustic oscillations scale}\label{sec:bao}
%---------------------------------------------------------------
Another direct application of our \glam{}-HOD catalogues is the prediction of the BAO feature for DESI-like LRGs at different redshifts. This scale was not accessible in the \Pmill{} run due to its volume. We extract the BAO scale through the dilation parameter, $\alpha$, which is related to physical distances via \citep{Eisenstein:2005su}
\begin{equation}\label{eq:alpha}
    \alpha \equiv \frac{D_{\rm V}(z)r^{\rm fid}_{\rm d}}{D^{\rm fid}_{\rm V}(z)r_{\rm d}}\,,
\end{equation}
where
\begin{equation}\label{eq:DV}
    D_{\rm V}(z) = \left[cz(1+z)^2D^2_{\rm A}(z)H^{-1}(z)\right]\,,
\end{equation}
$D_{\rm A}(z)$ is the angular-diametre distance, $r_{\rm d}$ is the sound horizon at the baryon drag epoch $(z_{\rm d}\sim 1020)$ and the superscript `fid' indicates the value of the distances in our fiducial cosmology, i.e., the \Pmill{} cosmology (see Sec.~\ref{sec:Pmill}). In our fiducial cosmology, the values of $D_{\rm V}(z)$ and $r_{\rm d}$ are,
\begin{eqnarray}
&&D^{\rm fid}_{\rm V}(z=0.60) = 2141.07\,{\rm Mpc}\label{eq:DV_fid1}\\
&&D^{\rm fid}_{\rm V}(z=0.74) = 2502.62\,{\rm Mpc}\label{eq:DV_fid2}\\
&&D^{\rm fid}_{\rm V}(z=0.93) = 2926.11\,{\rm Mpc}\label{eq:DV_fid3}\\
&&r^{\rm fid}_{\rm d} = 148.13\,{\rm Mpc}\label{eq:rd_fid}\,.
\end{eqnarray}

The BAO scale can be extracted by fitting the monopole of the power spectrum (or correlation function) to a template that includes the dilation parameter. Therefore, the monopole of the power spectrum is modelled as the product of a smooth component and the BAO signal as (e.g.  \citealt{Anderson:2013zyy,Ross:2014qpa}),
\begin{equation}\label{eq:Pfit}
    P_{\rm 0,fit}(k) = P_{\rm sm}(k) {\rm O}_{\rm damp}(k/\alpha)\,,    
\end{equation}
where $P_{\rm sm}(k)$ is a smooth power spectrum, i.e., without any BAO feature, and ${\rm O}_{\rm damp}(k)$ represents the damped BAO signal (see below for the definitions of these quantities).

The smooth power spectrum component is modelled as \citep{Anderson:2013zyy,Ross:2014qpa,Hernandez-Aguayo:2019pua}
\begin{equation}\label{eq:Psm}
    P_{\rm sm}(k) = B^2_p P_{\rm nw}(k) + A_1k + A_2 + \frac{A_3}{k}\,,
\end{equation}
where $P_{\rm nw}(k)$ is a smooth ``de-wiggled'' template obtained using the fitting formula of \citet{Eisenstein:1997ik}, $B_p$ is a large-scale bias parameter, and $A_1$, $A_2$ and $A_3$ are further free parameters.

The oscillatory component of the power spectrum is given by,
\begin{equation}\label{eq:Odamp}
    {\rm O}_{\rm damp}(k) = 1 + \left(\frac{P_{\rm lin}(k)}{P_{\rm nw}(k)} - 1\right)e^{-\frac{1}{2}k^2\Sigma^2_{\rm nl} }\,,
\end{equation}
where $\Sigma_{\rm nl}$ is a damping parameter.

The monopole of the redshift-space correlation function is given by the model \citep{Anderson:2013zyy,Ross:2014qpa},
\begin{equation}\label{eq:xifit}
    \xi_{\rm 0,fit}(s) = B^2_s \xi_{\rm lin,\,damp}(\alpha s) + \frac{a_1}{s^2} + \frac{a_2}{s} + a_3\,,
\end{equation}
where $\xi_{\rm lin,\,damp}(s)$ is the Fourier transform of $P_{\rm nw}(k){\rm O}_{\rm damp}(k)$, $B_s$ is the equivalent of $B_p$ mentioned above, and $a_1$, $a_2$ and $a_3$ are polynomial free parameters.

To obtain the best-fitting $\alpha$ value, we use Bayesian statistics and maximise the likelihood, $\mathcal{L}\propto \exp(-\chi^2/2)$ (where $\chi^2$ is defined by Eq.~\eqref{eq:chi2}) by fitting the measurements of the monopole of the power spectrum on scales with $k < 0.3 \hMpc$ and on scales with $s > 40 \Mpch$ for the monopole of the correlation function. To find the best-fitting $\alpha$ value and its confidence levels we again use the MCMC technique via the package {\sc emcee}.

Fig.~\ref{fig:bao} displays the BAO feature in Fourier (left panel) and configuration space (right panel) at $z=0.74$ (similar trends were found at $z=0.60$ and $z=0.93$). The BAO feature was isolated by dividing the best-fitting model and measurements of the monopole of the power spectrum by the smooth component of the best-fitting model. In the case of the monopole of the correlation function, we subtract the smooth component of the best-fitting model to the best-fitting model and measurements. We can see a clear BAO signal in both cases.

%--------- Figure --------------
\begin{figure}
 \centering
\includegraphics[width=0.47\textwidth]{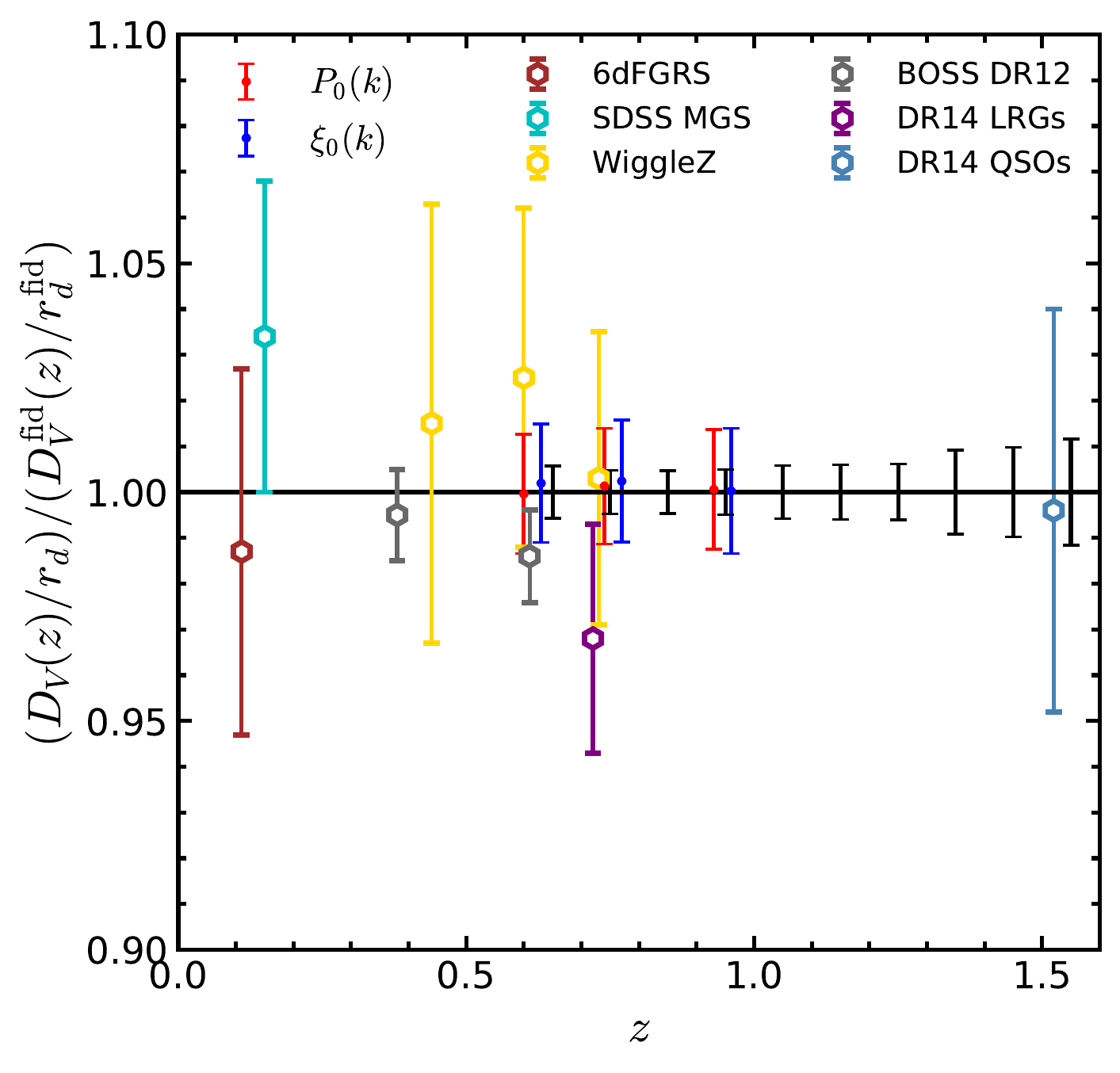}
\caption{Isotropic BAO measurements as a function of redshift. Our estimates from the monopole of the power spectrum are shown by the red dots with errorbars while the blue dots with errobars represent the best-fitting values from the monopole of the correlation function. We show the measurements from different galaxy surveys as labelled. The black errorbars show the DESI 14K forecast presented in \citet{DESI:2016zmz}.}
\label{fig:DV}
\end{figure}

Our estimates of the dilation parameter are shown in Fig.~\ref{fig:DV} together with isotropic BAO measurements from the 6dFGRS at $z=0.11$ \citep{Beutler:2011hx}, the SDSS MGC at $z=0.15$ \citep{Ross:2014qpa}, BOSS DR12 at $z = [0.38,0.61]$ \citep{BOSS-DR12:2017sqa}, WiggleZ at $z= [0.44,0.6,0.73]$ \citep{Blake:2011en}, eBOSS DR14 LRGs at $z=0.72$ \citep{Bautista:2017wwp} and eBOSS DR14 QSO sample at $z=1.52$ \citep{Ata:2017dya}. The black errorbars are from the DESI forecast \citep[see table 2.3 of][]{DESI:2016zmz}. At a first glance, our estimates of the errorbar bars (red and blue symbols) have almost the same amplitude as those predicted for DESI; the simple discussion above in terms of the comparison of the \glam{} simulation volume and the volume of the redshift shells to be probed by DESI suggest that the errors could differ by a factor of around two, although different assumptions are made in arriving at the two estimates.

Using the fiducial values of $D^{\rm fid}_{\rm V}$, Eqs.~\eqref{eq:DV_fid1}$-$\eqref{eq:DV_fid3}, we convert our best-fitting $\alpha$ values into distance measurements via Eq.~\eqref{eq:alpha},
\begin{equation}\label{eq:DV1}
    D_{\rm V}(z=0.60) = 
\left \{
      \begin{array}{lc}
          2140 \pm 28\,(r_{\rm d}/r^{\rm fid}_{\rm d})\,{\rm Mpc} & P_0(k)\,,\\ 
          2145 \pm 27\,(r_{\rm d}/r^{\rm fid}_{\rm d})\,{\rm Mpc} & \xi_0(s) \,,
      \end{array}
   \right.
\end{equation}
\begin{equation}\label{eq:DV2}
    D_{\rm V}(z=0.74) = 
\left \{
      \begin{array}{lc}
          2505 \pm 31\,(r_{\rm d}/r^{\rm fid}_{\rm d})\,{\rm Mpc} & P_0(k)\,,\\ 
          2508 \pm 33\,(r_{\rm d}/r^{\rm fid}_{\rm d})\,{\rm Mpc} & \xi_0(s) \,,
      \end{array}
   \right.
\end{equation}
\begin{equation}\label{eq:DV3}
    D_{\rm V}(z=0.93) = 
\left \{
      \begin{array}{lc}
          2927 \pm 38\,(r_{\rm d}/r^{\rm fid}_{\rm d})\,{\rm Mpc} & P_0(k)\,,\\ 
          2926 \pm 40\,(r_{\rm d}/r^{\rm fid}_{\rm d})\,{\rm Mpc} & \xi_0(s) \,.
      \end{array}
   \right.
\end{equation}

We find good agreement between our estimates and the fiducial values of $D_{\rm V}(z)$. The agreement is well within the $1\sigma$ level. In our case, the monopole of the correlation function gives slightly better constraints than the power spectrum. In general, we can  estimate the isotropic BAO distance to better than $1.3$ per cent in both spaces. 

%---------------------------------------------------------------
\section{Summary and conclusions}\label{sec:conc}
%---------------------------------------------------------------
We have  presented predictions for the properties and clustering of LRGs selected using the  colour-magnitude cuts in the $r, z, W1$ bands that will be applied in the DESI LRG survey \citep{DESI:2016zmz}. The predictions were made using the \Galform{} semi-analytic model of galaxy formation run on the \Pmill{} $N$-body simulation \citep{Baugh:2018kkh}  and a suite of low-resolution, larger volume simulations run with the Parallel-PM $N$-body code \glam{} \citep{Klypin:2017jwl}.

We made predictions for the abundance of DESI-like LRGs and explore how the target selection cuts affect which galaxies are selected and how these populate haloes and subhaloes. We find that a small but important fraction of the most massive galaxies (those with stellar mass $\log_{10}(M_*/\Msh) > 11.15)$ are {\it not} selected as LRGs (see  Fig.~\ref{fig:sMF_LF}). A similar trend is seen in the galaxy luminosity function, and is most pronounced at shorter wavelengths: essentially all bright galaxies in the $W1$-band luminosity function are LRGs, but only roughly half of the galaxies in the bright end of the $r$-band luminosity function are LRGs.
This shows that applying the full photometric selection is essential to reproduce LRGs in a galaxy formation model and that using a proxy, such as stellar mass, to select LRGs is at best an approximation.
We explored the galaxy-(sub)halo connection of LRGs through the halo occupation distribution and the subhalo mass function. We find that the shape of the HOD does not follow the canonical shape proposed by \cite{Zheng:2004id}; in particular, the occupation of central galaxies does not reach unity for the most massive haloes (see Fig.~\ref{fig:hod}), and drops with increasing mass. 

We compared the HOD and the subhalo mass functions of galaxies selected by their stellar mass with those measured for the LRGs (see Fig.~\ref{fig:sub_hod}). By doing this exercise, we reaffirm that the DESI-LRG cuts affect the selection of subhaloes that are populated by LRGs. Mass alone is not enough to determine if a subhalo hosts an LRG.
By comparing the clustering of these galaxy samples (Fig.~\ref{fig:xir}) we found a difference that ranges from $10\%$ at $z=0.6-0.74$ to up to $150\%$ at $z=0.93$. Hence, we conclude that using galaxy stellar mass as a proxy for selecting LRGs could change the expected clustering signal.

To prepare for the clustering measurements of DESI we ran 1000 \glam{} simulations. When comparing the halo statistics between the \glam{} simulation ensemble and the \Pmill{} high-resolution run, we found good agreement between the halo mass functions, but differences of $\sim 10\%$ in the halo clustering (see Fig.~\ref{fig:hmf}). This difference can be attributed to the different halo finder used in the \Pmill{} and \glam{} simulations. Despite the difference in halo clustering, the galaxy clustering statistics measured from the \glam{}-LRG catalogues are in good agreement with that in the \Pmill{}-\Galform{} LRG sample in both configuration and Fourier space. To populate the \glam{} halo catalogues with DESI-like LRGs we used the tabulated HODs obtained from \Galform{}.
We also found a good agreement between our clustering measurements in real-space with those reported by \citet{Kitanidis:2019rzi} and \citet{Rongpu:2020} (see upper panels of Figs.~\ref{fig:xi_glam} and \ref{fig:Pk_glam}).

We extended the analysis of covariance and correlation matrices of \glam{} simulations started by \citet{Klypin:2017jwl} to galaxies and correlation functions (see Figs.~\ref{fig:xil_glam}-\ref{fig:Pkm_error}). We found that the galaxy correlation matrix shows a different and more complex pattern than its dark-matter counterpart. 

We presented predictions for the large-scale clustering of DESI-like LRGs in configuration and Fourier space, by extracting the linear growth rate from the linear Kaiser RSD model and the BAO scale from the isotropic dilation parameter. In a follow-up project, we plan to extend this study to non-linear models of RSDs and an anisotropic analysis of the BAO scale, including the impact of the light-cone survey geometry and observational systematic.

Using our \glam{}-LRG catalogues we estimated the growth of structure from the ratio of the monopole in redshift space to the real-space power spectrum with a precision of $\sim 3-4\%$, and we can measure the BAO scale with a $1.3\%$ precision in both configuration and Fourier space. Nevertheless, if we want to compare the precision of our measurements with those expected from DESI \citep[table 2.3 of ][]{DESI:2016zmz}, our error estimations should take into account the contribution from the expected volume covered by DESI (see Sec.~\ref{sec:cov} for details).
However, the amplitude of the statistical errors estimated from our best-fitting search on the linear growth rate and BAO scale are consistent with the forecast presented by \citet{DESI:2016zmz}.

We conclude that the colour-magnitude cuts have a big impact on the properties and clustering of LRGs, showing that LRGs are different than stellar mass selected galaxies. But more importantly, the analysis presented in this paper provided accurate estimates on the galaxy clustering expected by DESI-LRGs thanks to our \glam{}-HOD pipeline. The \glam{}-LRG galaxy catalogues are made public at the \SU{} site\footnote{http://www.skiesanduniverses.org}. Moreover, our pipeline can be easily adapted to the specifications of other next generation surveys such as Euclid, the Vera Rubin Observatory (formerly the LSST), PFS and 4MOST.

%---------------------------------------------------------------
\section*{Acknowledgements}
%---------------------------------------------------------------
We wish to thank Rongpu Zhou for providing the data to generate Fig.~\ref{fig:nz}.
We thank Peder Norberg for useful discussions in the early stages of this work.
CH-A acknowledges support from the Mexican National Council of Science and Technology (CONACyT) through grant No. 286513/438352. 
FP and AK thank the support of the Spanish Ministry of Science funding grant PGC2018-101931-B-I00.
FP gratefully acknowledges the ICC at Durham for their warm hospitality and support during my summer visit of 2018, where this work was initiated.
This work used the DiRAC@Durham facility managed by the Institute for Computational Cosmology on behalf of the STFC DiRAC HPC Facility (\url{www.dirac.ac.uk}). The equipment was funded by BEIS capital funding via STFC capital grants ST/K00042X/1, ST/P002293/1, ST/R002371/1 and ST/S002502/1, Durham University and STFC operations grant ST/R000832/1. DiRAC is part of the National e-Infrastructure. We thank New Mexico State University (USA) and Instituto de Astrof\'isica de Andaluc\'ia CSIC (Spain) for hosting the \SU{} site (\url{www.skiesanduniverses.org}) for cosmological simulation products.

%%%%%%%%%%%%%%%%%%%%%%%%%%%%%%%%%%%%%%%%%%%%%%%%%%



\bibliographystyle{mnras}
\bibliography{ref} % if your bibtex file is called example.bib

%%%%%%%%%%%%%%%%%%%%%%%%%%%%%%%%%%%%%%%%%%%%%%%%%%

%%%%%%%%%%%%%%%%% APPENDICES %%%%%%%%%%%%%%%%%%%%%

%\appendix
%---------------------------------------------------------------
%\section{Appendix A}
%---------------------------------------------------------------

%%%%%%%%%%%%%%%%%%%%%%%%%%%%%%%%%%%%%%%%%%%%%%%%%%

% Don't change these lines
\bsp	% typesetting comment
\label{lastpage}
\end{document}